\newif\ifWSc\WScfalse
\newif\ifnoWSc\noWSctrue
\documentstyle[11pt,epsf,amssymb]{article}
\begin{document}
\newcommand{\calle}[1]{(\ref{#1})}
\newcommand{\beq}{\begin{equation}}
\newcommand{\eeq}{\end{equation}}
\newcommand{\beqn}{\begin{eqnarray}}
\newcommand{\eeqn}{\end{eqnarray}}
\newcommand{\bye}{\end{document}}
\newcounter{fig}[section]
\renewcommand{\thefig}{\thesection.\arabic{fig}}
\newcommand{\subtitle}[2]{\begin{center} 
                          \refstepcounter{fig}\label{#1}
                          {\footnotesize Figure \thefig : #2}
                          \end{center}
                          \vspace{5mm}
                          }
\newtheorem{definition}{Definition}
\newtheorem{theorem}{Theorem}
\def\BZ{ {\Bbb Z} }
\def\IC{ {\Bbb C} }
\def\IZ{ {\Bbb Z} }
\def\IP{ {\Bbb P} }
\def\IQ{ {\Bbb Q} }
\def\IR{ {\Bbb R} }
\def\AAA{{\cal A}} \def\cpR{{\cal R}}
\def\MM{{\cal M}}
\def\SS{{\cal S}}
\def\Pr#1#2#3{\Phi^{#1}_{#2,#3}}
\def\pr#1#2{\phi^{#1}_{#2}}
\def\PPr#1#2#3#4#5{ ({\textstyle #1{#2\atop #4}{#3\atop #5}})} 
\def\Pr#1#2#3#4{\Phi^{#1;#3}_{#2;#4}}
\def\chr#1#2{\hbox{$\chi^{#1}_{#2}$}}
\def\Tr#1{\hbox{{\bigrm Tr}\kern-1.05em \lower2.1ex \hbox{$\scriptstyle#1$}}\,} 
\def\s{super\-}
\def\zb{\overline z}
\def\qb{\overline q}
\def\lb{\overline l}
\def\sb{\overline s}
\def\mub{\overline \mu}
\def\pl{l'}
\def\pmu{\mu '}
\def\vl{\vec l}
\def\vq{\vec q}
\def\vlb{\vec{\lb}}
\def\vmu{\vec\mu}
\def\vmub{\vec{\mub}}
\def\vb{\vec\beta}
\def\vlp{\vec{\pl}}
\def\vmup{\vec{\pmu}}
\def\ex#1{\hbox{$\> e^{#1}\>$}}
\def\PP#1{\hbox{$\hbox{\IC \IP}^{#1}$}}
\def \eg{\hbox{e.g.~}}
\def\ee#1{{\rm e}^{^{\textstyle#1}}}
\def\d{{\rm d}}
\def\e{{\rm e}}
\def \LG{Landau-Ginzburg}
\def\ztim{\mathrel{\mathop \otimes\limits_{\BZ}}} \def\CP#1{{\IP}^{#1}}
\def \p{\partial}
\def\tilde{\widetilde}
\def\vbar{ \,|\, }
\def\Hom{\mathop{\rm Hom}}
\def\Spec{\mathop{\rm Spec}}
\def\X{X}
\def\W{Y}
\def\NN{\bf{\cal N}}
\def\DD{\bf{\bf \Delta}}
\def\VV{ {\cal M}_K }
\def\PS{ {\cal D}$^*$ }
\def\CY{Calabi-Yau}
\def\to{\rightarrow}
\def\tto{\longrightarrow}
\def\AA{{\AAA}}
\def\AAO{{\AAA_0}}
\def\AAO{{\Xi}}
\def\Psf{\Sigma^\prime}
\def\tablerule{\noalign{\hrule}}
\def\vol{\mathop{\rm vol}}
\def\spn{\mathop{\rm span}}
\font\bigrm=cmr10 scaled \magstephalf
\def\mm{minimal models}
\def\K{K\"ahler}
\def\C{$\cal C$}
\def\tC{$\cal C'$}
\font\bigrm=cmr10 scaled \magstephalf
\def\Pr#1#2#3{\Phi^{#1}_{#2,#3}} \def\pr#1#2{\phi^{#1}_{#2}}
\def\PPr#1#2#3#4#5{ ({\textstyle #1{#2\atop #4}{#3\atop #5}})}
\def\Pr#1#2#3#4{\Phi^{#1;#3}_{#2;#4}}
\def\chr#1#2{\hbox{$\chi^{#1}_{#2}$}}
\def\Tr#1{\hbox{{\bigrm Tr}\kern-1.05em \lower2.1ex \hbox{$\scriptstyle#1$}}\,}
\def\s{super}
\def\zb{\overline z}
\def\qb{\overline q}
\def\lb{\overline l}
\def\sb{\overline s}
\def\mub{\overline \mu}
\def\pl{l'}
\def\pmu{\mu '}
\def\vl{\vec l}
\def\vlb{\vec{\lb}}
\def\vmu{\vec\mu}
\def\vmub{\vec{\mub}}
\def\vb{\vec\beta}
\def\vlp{\vec{\pl}}
\def\vmup{\vec{\pmu}}
\def\ex#1{\hbox{$\> e^{#1}\>$}}
\def\cft{conformal field theory\ }
\def\cc{$(c,c)$} 
\def\ac{$(a,c)$}
\def\CFT{conformal field theory}
\def\CFT's{conformal field theories}
\def\ct{d}
\newcommand{\bfm}{\raisebox{0pt}{{\boldmath$m$\unboldmath}}}
\newcommand{\bfn}{\raisebox{0pt}{{\boldmath$n$\unboldmath}}}
\newcommand{\bfx}{\raisebox{0pt}{{\boldmath$x$\unboldmath}}}
\def\FC{${\cal F}(\cal C)$}
\newcommand{\newsection}[1]{
                            \setcounter{equation}{0}
                            \section{#1}
                            }
\renewcommand{\theequation}{\thesection.\arabic{equation}}

\ifnoWSc
\rightline{CU-TP-812}

\vspace{1cm}
\begin{center}
{\bf     
        
        \vspace{4mm}
        {\Huge STRING THEORY ON CALABI-YAU} 
   
        \vspace{1mm}
        {\Huge MANIFOLDS}
        }
\end{center}

\vspace{1truecm} 
\begin{center}
{\bf\Large Brian R. Greene}\footnote{On leave from:
F. R. Newman Laboratory of Nuclear Studies, Cornell University, Ithaca, 
NY  14853, USA}\\ 
\vspace{1truecm}
Departments of Physics and Mathematics\\
{Columbia University}\\
{New York,  NY 10027, USA}
\end{center}
\fi
\ifnoWSc
\vspace{1truecm}
{\footnotesize
These lectures are devoted to introducing some of the basic features
of quantum geometry that have been emerging from compactified string
theory over the last couple of years. The developments discussed include
new geometric features of string theory which occur even at the classical
level as well as those which require non-perturbative  effects.
These lecture notes are based on an evolving set of lectures presented
at a number of schools but most closely follow
a series of seven lectures
given at the TASI-96 summer school on {\it Strings, Fields and Duality}.
}
\fi

\tableofcontents
\newpage

\newsection{Introduction}
\label{sec:intro}
\subsection{The State of String Theory}

It has been about thirteen years since the modern era of string theory
began with the discovery of anomaly cancellation 
\cite{GS}. Even though the initial euphoria of having a
truly unified theory --- one that includes gravity --- led to premature
claims of solving all of the fundamental problems of theoretical 
particle physics, string theory continues to show ever increasing signs
of being the correct approach to understanding nature at its most
fundamental level. The last two years, in particular, have led
to stunning developments in our understanding. Problems that once
seemed almost insurmountable have now become fully within our
analytical grasp. The key ingredient in these developments
is the notion of {\it duality}: a given physical situation  may admit
more than one theoretical formulation and it can turn out that the respective
levels of difficulty in analysing these distinct formulations can be
wildly different.  Hard questions to answer from one perspective can
turn into far easier questions to answer in another.

Duality is not a new idea in string theory. Some time ago, for instance,
it was realized that if one considers string theory on a circle of
radius $R$, the resulting physics can equally well be described in
terms of string theory on a circle of radius $1/R$ (\cite{GPR}
and references therein).
Mirror symmetry
(\cite{MirrorbookI,MirrorbookII} and references therein),
as we describe in these lectures, is another
well known example of a duality. In this case, two topologically
distinct Calabi-Yau compactifications of string theory give rise
to identical physical models. The transformation relating these
two distinct geometrical formulations of the same physical model
is such that strong sigma model coupling questions in one 
can be mapped to weak sigma model coupling questions in the other.
By a judicious
choice of which geometrical model one uses, seemingly difficult physical
questions can be analysed with perturbative ease. 

During the last
couple of years, the  scope of duality in string theory has dramatically
increased. Whereas mirror symmetry can transform strong to weak
sigma model coupling, the new dualities can transform strong to weak
{\it string coupling}. For the first time, then, we can go
beyond perturbation theory and gain insight
into the nature of strongly coupled string theory
\cite{Wit2,HT}.
The remarkable thing is that all such strong coupling dynamics appears
to be controlled by one of  two structures: either the {\it weak}
coupling dynamics of a different string theory or by  a structure which
at low energies reduces to eleven dimensional supergravity. The latter
is to be thought of as the  low energy sector of an as yet incompletely
formulated nonperturbative theory dubbed  M-theory.

A central element in this impressive progress has been  played by
BPS saturated solitonic objects in string theory. These non-perturbative
degrees of freedom are oftentimes the dual variables which dominate
the low-energy structure arising from taking the strong coupling limit
of a familiar string theory. Although difficult to analyze in detail as
solitons, the supersymmetry algebra tells us a great deal about
their properties --- in some circumstances enough detail so as to
place duality conjectures on circumstantially compelling foundations.
The discovery of $D$-branes as a means for giving a microscopic 
description of these degrees of freedom
\cite{Pol-TASI} has subsequently provided a powerful
tool for their detailed investigation --- something that is at 
present being vigorously pursued.

Our intent in these lectures is to describe string compactification from
the basic level of perturbative string theory on through some of the most
recent developments involving the nonperturbative elements just mentioned.
The central theme running through our discussion is the way in which
a universe based on string theory is described by a geometrical
structure that differs from the classical geometry developed by mathematicians
during the last few hundred years. We shall refer to this structure as
quantum geometry.

\subsection{What is Quantum Geometry?}

Simply put, quantum geometry is the appropriate modification of standard 
classical geometry to make it suitable for describing the physics of string
theory. We are all familiar with the success that many ideas from classical
geometry have had in providing the language and technical framework for
understanding important structures in physics such as general relativity and
Yang-Mills theory. It is rather remarkable that the physical properties of these
fundamental theories can be directly described in the mathematical language of
differential geometry and topology. Heuristically, one can roughly understand
this  by noting that the basic building block of these mathematical
structures is
that of a
topological space --- which itself is a collection of {\it points}
grouped together in some particular manner. Pre-string theories of fundamental
physics are also based on a building block consisting of points --- namely,
 point
particles. That classical mathematics and pre-string physics have the same
elementary constituent is one rough way of understanding why they are so
harmonious. Thinking about things in this manner is particularly useful when we
come to string theory. As the fundamental constituent of perturbative string
theory is not a
point but rather a one-dimensional loop, it is natural to suspect that classical
geometry may not be the correct language for describing string physics. In fact,
this conclusion turns out to be correct. The power of geometry, however, is not
lost. Rather, string theory appears to be described by a {\it modified} form of
classical geometry, known as quantum 
geometry, with the modifications
disappearing as the typical size in a given system becomes large relative to the
string scale --- a length scale which is expected to be within a few 
orders of magnitude of the Planck scale, $10^{-33}cm$.

We should stress a point of terminology at the outset.
The term quantum geometry, in its most precise usage, refers to the geometrical
structure relevant for describing a fully quantum mechanical theory of strings.
In the first part of these lectures, though, our focus will be
 upon tree level string theory ---
that is, conformal field theory on the sphere --- which captures novel features
associated with the extended nature of the string, but does so at the classical
level. As such, the term quantum geometry in this context is a bit misleading
and a term such as ``stringy'' geometry would probably be more
appropriate. In the later lectures, we shall truly include quantum effects into
our discussion through some of the non-perturbative 
solitonic degrees of freedom, mentioned above.
Understanding the geometrical significance of
these quantum effects finally justifies using the term quantum geometry.
To understand this distinction a bit more completely, we note
that scattering amplitudes in perturbative string theory can be
organized in a manner analogous to the loop expansion in
ordinary quantum field theory 
\cite{Oog-TASI}. In field theory,
the loop expansion is controlled by $\hbar$, with an $L$-loop
amplitude coming with a prefactor of $\hbar^{L-1}$. In string theory,
the role of loops is  played by the genus  of the world sheet
of the string, and the role of $\hbar$ is played by the value of
the string coupling $g_s$. At any given genus in this expansion, we can analyze
the contribution to a string  scattering amplitude by means of
a {\it two}-dimensional auxiliary quantum field theory on
the genus-$g$ world sheet. This field theory is controlled by the inverse
string tension $\alpha'$ (or more precisely, the dimensionless
sigma model coupling $\alpha'/R^2$ with $R$ being
a typical radius of a compactified portion of space, as we shall discuss
below in detail). The limit of $\alpha' \rightarrow 0$ corresponds to an
infinitely tense string which thereby loses all internal structure and
reduces, in effect, to a structureless point particle.
Thus, in string theory there are really two expansions: the quantum
genus expansion and the sigma model expansion.

\begin{figure}[htbp]
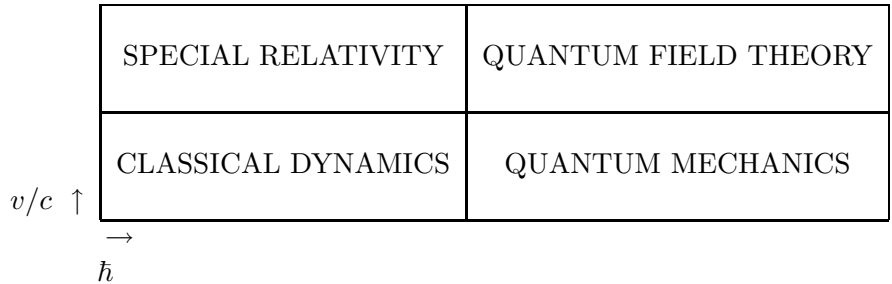

\begin{center}
\begin{tabular}{l|c|c|}\cline{2-3}
                       &    & \\ 
                       & SPECIAL RELATIVITY & QUANTUM FIELD THEORY \\ 
                       &    & \\ \cline{2-3} 
                       &    & \\ 
                       & CLASSICAL DYNAMICS & QUANTUM MECHANICS    \\ 
 ${v/c}~\uparrow$ &    & \\  \cline{2-3}
 \multicolumn{3}{l}{$~~~~~~~~~\rightarrow$} \\ 
 \multicolumn{3}{l}{$~~~~~~~~~\hbar$}       \\
\end{tabular}
\end{center}
\caption{The deformation from classical dynamics to quantum field theory.}
\label{box1}
\end{figure}

We can summarize these two effects through a  diagram
analagous to one relevant for understanding the relationship
between non-relativistic
classical mechanics and quantum field theory
\cite{Coleman}.
The latter is summarized by figure \ref{box1}.

\begin{figure}[htbp]
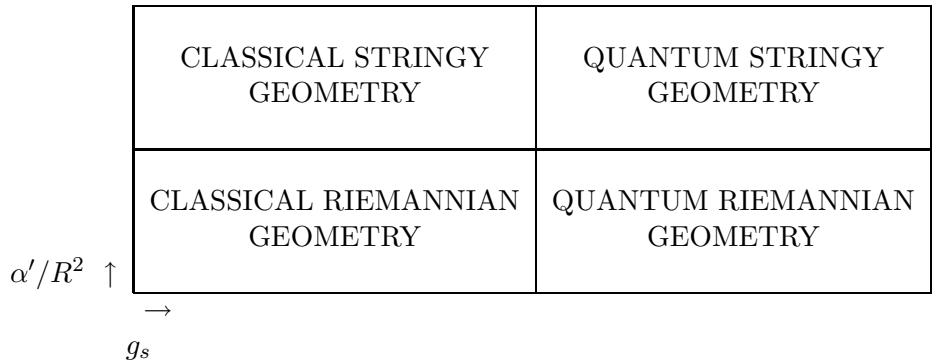

\begin{center}
\begin{tabular}{l|c|c|}\cline{2-3}
                       &    & \\
          & CLASSICAL STRINGY  & QUANTUM STRINGY  \\
          &  GEOMETRY          &  GEOMETRY \\
                       &    & \\ \cline{2-3}
                       &    & \\
          & CLASSICAL RIEMANNIAN  & QUANTUM RIEMANNIAN  \\
          &  GEOMETRY             &  GEOMETRY \\
 ${\alpha'/R^2}~\uparrow$ &    & \\  \cline{2-3}
 \multicolumn{3}{l}{$~~~~~~~~~~~~~\rightarrow$} \\
 \multicolumn{3}{l}{$~~~~~~~~~~~~g_s$}       \\
\end{tabular}
\end{center}
\caption{The deformation from classical Riemannian geometry to quantum stringy
geometry.}
\label{box2}
\end{figure}

We see that the horizontal axis corresponds to $\hbar$ (or more precisely,
$\hbar$ divided by the typical action of the system being studied),
while the vertical axis corresponds to $v/c$. In string theory
(which we take to incorporate relativity from the outset), as
just explained,
there are also two relevant expansion parameters, as shown in 
figure \ref{box2}.

Here we see that the horizontal axis corresponds to the value of
the string coupling constant, while the vertical axis corresponds
to the value of the sigma model coupling constant. In the extreme
$\alpha' = g_s = 0 $ limit, for instance, we recover relativistic particle dynamics.
For nonzero $g_s$ we recover point particle quantum field theory.
For $g_s = 0$ and nonzero $\alpha'$ we are studying classical string theory.
In general though, we need to understand the theory for arbitrary
values of these parameters.

An important implication of the duality results briefly discussed in
the last section is that any such decomposition into those physical
effects due to strong string coupling or strong sigma model coupling, etc.
is {\it not} invariant. Rather, strong/weak duality transformations
together with various geometric dualities can completely mix these
effects together. What might appear as a strong coupling phenomenon
from one perspective can then appear as a weak coupling phenomenon
in another. More generally, the values of string and sigma model
couplings can undergo complicated changes in the course of duality transformations.
The organization of figure \ref{box2} is thus highly dependent on which 
theoretical description of the underlying physics is used.

Having
made our terminology clear, we
will typically not be overly 
careful in our use of the term quantum vs. stringy geometry;
often we shall simply use the former allowing the context clarify 
the precise meaning. 

The discovery of the profound role played by extended solitonic
objects in string theory has in some sense questioned the nature of
the supposedly foundational position of the string itself. That is,
since string theory contains degrees of freedom with more (and less) than
one spatial dimension, and since  in certain   circumstances it is
these degrees of freedom which dominate the low-energy dynamics (as we
shall see later), maybe the term `string' theory is a historical misnomer.
Linguistic issues aside, from the point of view of quantum geometry it
is important to note that the geometrical structure which one sees
emerging from some particular situation can depend in part on
precisely which probe one uses to study it. If one uses a string
probe, one quantum geometrical structure will be accessed while if one uses,
for instance,
a $D$-brane of a particular dimension, another geometry may become manifest.
Thus, quantum geometry is an incredibly rich structure with diverse
properties of greater or lesser importance depending upon the detailed
physical situation being studied. In these lectures we will study
the quantum geometry which emerges from fundamental string probes and also
from nontrivial $D$-brane configurations. We shall not discuss the
quantum geometry that arises from using $D$-brane scattering dynamics;
for this the reader can consult the lectures of
Polchinski \cite{Pol-TASI} and
Shenker of this school.

As we shall see,
when the typical size in a string compactification does not meet the criterion
of being sufficiently large, the 
quantum geometry will shall find differs both
quantitatively and qualitatively from ordinary classical
geometry. In this sense, one can think of string theory as providing us with a
generalization of ordinary classical geometry which differs from it on short
distance scales and reduces to it on large distance scales. It is the purpose of
these lectures to discuss some of the foundations and properties of quantum
geometry.

\subsection{The Ingredients}

Recent developments in string theory have taken us much closer to understanding
the true nature of the fully quantized theory. Rather surprisingly, it appears
that perturbative tools --- judiciously used --- can take us a long way.
In these lectures we shall focus  on the tools necessary for
analysing string theory in perturbation theory as well as for
incorporating certain crucial nonperturbative elements. Our aim
in this section is to give a brief overview of
perturbative string theory in order to
have the language to describe a number of recent developments.

As discussed in Ooguri's lectures
\cite{Oog-TASI},
 it is most convenient to formulate first quantized string 
theory in terms of a 2-dimensional quantum field theory on the world sheet swept
out by the string. The delicate consistency of quantizing an extended object
places severe constraints on this 2-dimensional field theory. In particular, the
field theory must be a conformal field theory with central charge equal to
fifteen.

More precisely, we will
always discuss type IIA or IIB
 superstring theory, in which case the field theory must be 
superconformally invariant.
 (Without loss of clarity, we will often drop the
prefix {\it super}.) In fact,
 the study of these string theories
leads us naturally to focus on two-dimensional field theories with
two independent supersymmetries on the world sheet(on
the left and on the right) and hence we discuss $N = 2$ (or $(2,2)$ as
it is sometimes written)
superconformal field theories with central charge (both left and right) equal to
fifteen\footnote{Such theories can always be converted into more
phenomenologically
viable heterotic string theories but that will not be required for our
purposes.}. From a spacetime point of view, the corresponding effect
theory governing string modes has $N = 2$ supersymemtry as well.
The constrained structure of $N = 2$ theories
has allowed us to hone a number of powerful tools which greatly aid in their
understanding. We can thus push both our physical and our mathematical analysis
of these theories quite far.

Our study of perturbative string theories with space-time supersymmetry thus 
boils
 down to a study of 2-dimensional $N = 2$ superconformal field theories
with central charge fifteen. How do we build such field theories? We will study
this question in some detail in the ensuing sections; for now let us note the
typical setup. In studying these string models, we will generally assume that
the underlying $N = 2,~c = 15$ conformal theory can be decomposed as the
product of an $N = 2,~c = 6$ theory with an $N = 2$, $c = 9$ theory. The former
can then be realized most simply via a free theory of two complex chiral
superfields, (as we will discuss explicitly shortly) 
--- that is, a free theory of
four real bosons and their fermionic superpartners. We can interpret these free
bosons as the four Minkowski space-time coordinates of common experience. The $c
= 9$ theory is then an additional ``internal'' theory required by consistency of
string theory. Whereas we were directly led to a natural choice for the $c = 6$
theory, there is no guiding principle which leads us to a preferred choice for
the $c = 9$ theory from the known {\it huge} number of possibilities. The
simplest choice, again, is six free chiral superfields --- that is, six free
bosons and their fermionic superpartners. Together with the $c = 6$ theory, this
yields ten dimensional flat space-time 
--- the arena of the initial formulation of
superstring theory. In this case the ``internal theory'' is of the same form as
the usual ``external theory'' and hence, in reality, there is no natural way of
dividing the two. Thus, for many obvious reasons, this way of constructing a
consistent string model is of limited physical interest thereby supplying strong
motivation for seeking other methods. This problem --- constructing (and
classifying) $N = 2$ superconformal theories with central charge 9 to play the
role of the internal theory --- is one that has been vigorously pursued for a
number of years. As yet there is no complete classification but a wealth of
constructions have been found. 

The most
intuitive of these constructions are those in which six of the ten spatial 
dimensions in the flat space approach just discussed are ``compactified''. That
is, they are replaced by a small compact six dimensional space, say $M$ thus
yielding a space-time of the Kaluza-Klein type $M_4 \times M$ where $M_4$ is
Minkowski four-dimensional space. It is crucial to realize that most choices for $M$ will
not yield a consistent string theory because the associated two-dimensional
field theory --- which is now most appropriately described as a non-linear sigma
model with target space $M_4 \times M$ 
--- will not be conformally invariant.
Explicitly, the action for the internal part of this theory is  
%
%
\ifnoWSc 
\begin{eqnarray}
  S &=&
  {1 \over 4 \pi \alpha'} \int dz d{\overline z}\,
 \Big\lbrack g_{mn}(\partial X^m {\overline \partial}X^n +
    \partial X^n {\overline \partial}X^m)
      + B_{mn}(\partial
       X^m {\overline \partial}X^n -
    \partial X^n {\overline \partial}X^m)
    + \cdots \Big\rbrack~, ~~~~
\nonumber
\end{eqnarray}
\fi
where $g_{mn}$ is a metric on $M$, $B_{mn}$ is an antisymmetric tensor field,
and we have omitted additional fermionic terms required by supersymmetry whose
precise form will be given shortly. 

To meet the
criterion of conformal
invariance, $M$ must, to lowest order in sigma model perturbation 
theory\footnote{That is, to lowest order in $\alpha'/R^2$ where $R$ is a typical
radius of
the Calabi-Yau about which we shall be more precise later.}, admit a metric
$g_{\mu \nu}$ whose Ricci tensor $R_{\mu \nu}$ vanishes. In order to contribute
nine to the central charge, the dimension of $M$ must be six, and to ensure the
additional condition of $N = 2$ supersymmetry, $M$ must be a complex K\"ahler
manifold. These conditions together are referred to as the `Calabi-Yau'
conditions and manifolds $M$ meeting them are known as Calabi-Yau three-folds
(three here refers to three complex dimensions; one can more generally study
Calabi-Yau manifolds of arbitrary dimension known as Calabi-Yau $d$-folds).
We discuss some of the classical geometry of Calabi-Yau manifolds in the
next section. A
consistent string model, therefore, with four flat Minkowski
space-time
directions can be built using any Calabi-Yau three-fold as the internal target
space for a non-linear supersymmetric sigma model. If we take the typical radius
of such a Calabi-Yau manifold to be small (on the Planck scale, for instance)
then the ten-dimensional space-time $M_4 \times M$ will effectively look just
like $M_4$ (with the present level of sensitivity of our best probes) and hence
is consistent with observation \`a la Kaluza-Klein. We will have much to say
about these models shortly; for now we note that there are {\it many} Calabi-Yau
three-folds and
 each gives rise to different physics in $M_4$. Having no means to
choose which one is 
``right'', we lose  predictive power.

Calabi-Yau sigma models provide one
 means of building $N = 2$ superconformal
 models that can be taken as the internal part of a string theory. There are two
other types of constructions that will play a role in our subsequent discussion,
so we mention them here as well. A key feature of each of these constructions is
that at first glance neither of them has anything to do with the geometrical
Calabi-Yau approach just mentioned. Rather, each approach yields a quantum field
theory with the requisite properties but in neither does one introduce a curled
up manifold. The best way to think about this is that the central charge is a
measure of the number of degrees of freedom in a conformal theory. These degrees
of freedom can be associated with extra spatial dimensions, as in the Calabi-Yau
case, but as in the following two constructions they do not have to be. 

Landau-Ginzburg effective field theories have played a key role in a number of 
physical contexts. For our purpose we shall focus on Landau-Ginzburg theories
with $N = 2 $ supersymmetry. Concretely, such a theory is a quantum field theory
based on chiral superfields (as we shall discuss) that respects $N = 2$
supersymmetry and has a unique vacuum state. From our discussion above, to be of
use this theory must be conformally invariant. A simple but non-constructive way
of doing this is to allow an initial non-conformal Landau-Ginzburg theory to
flow towards the infrared via the renormalization group. Assuming the theory
flows to a non-trivial fixed point (an assumption with much supporting
circumstantial evidence) the endpoint of the flow is a conformally invariant $N
= 2$ theory. More explicitly, the action for an $N = 2 $ Landau-Ginzburg theory
can be written 
%
\begin{eqnarray}
  &&   \int d^2 z d^4 \theta\,  K(\Psi_1,\overline
      \Psi_1,...,\Psi_n,\overline \Psi_n)
     +\left( \int d^2 z d^2\theta\,
      W(\Psi_1,...,\Psi_n) + {\rm h.c.} \right)~,
\nonumber
\end{eqnarray}
%
where the kinetic terms are chosen so as
to yield conformal invariance\footnote{Usually the kinetic terms can only be
defined in this implicit form --- or in the slightly more detailed but no more
explicit manner of fixed points of the renormalization group flow, as we shall
discuss later.} and where the superpotential $W$, which is a holomorphic
function of the chiral superfields $\Psi_i$, is at least cubic so the $\Psi_i$
are massless. (Any quadratic terms in $W$
 represent massive fields that are
frozen out in the infrared limit.) 

By suitably adjusting the (polynomial)
superpotential $W$ governing these fields we can achieve central charge nine. 
More importantly, along renormalization group flows, the superpotential receives
nothing more than wavefunction renormalization. Its form, therefore, remains
fixed and can thus be used as a label for those theories which all belong to the
same universality class. The kinetic term $K$, on the contrary, does receive
corrections along renormalization group flows and thus achieving conformal
invariance amounts to choosing the kinetic term correctly. We do not know how to
do this explicitly, but thankfully much of what we shall do does not require
this ability.

The final approach to building suitable internal $N = 2$ theories that we shall 
consider is based upon the so called ``minimal models''. As we shall discuss in
more detail in the next section, a conformal theory is characterized by a
certain subset of its quantum field algebra known as primary fields. Most
conformal theories have infinitely many primary fields but certain special
examples --- known as minimal models --- have a finite number. Having a finite
number of primary fields greatly simplifies the analysis of a conformal theory
and leads to the ability to explicitly calculate essentially anything of
physical interest. For this reason the minimal models are often referred to as
being ``exactly soluble''. The precise definition of primary field depends, as
we shall see, on the particular chiral algebra which a theory respects. For
non-supersymmetric conformal theories, the chiral algebra is that of the
conformal symmetry only. In this case, it has been shown that only theories with
$c < 1$ can be minimal. One can take tensor products of such $c < 1$ theories to
yield new theories with central charge greater than one (since central charges
add when theories are combined in this manner). If our theory has a larger
chiral algebra, say the $N = 2$ superconformal algebra of interest for reasons
discussed, then there are analogous exactly soluble minimal models. In fact,
they can be indexed by the positive integers $P \in \BZ$ and have central
charges $c_P = 3P/(P+2)$. Again, even though these values of the central charge
are less than the desired value of nine, we can take tensor products to yield
this value. (In fact, as we shall discuss, it is not quite adequate to simply
take a tensor product. Rather we need to take an orbifold of a tensor product.)
In this way we can build internal $N = 2,~c = 9 $ theories that have the
virtue of being exactly soluble. It is worthwhile to emphasize that, as in the
Landau-Ginzburg case, these minimal model constructions do not have any obvious
geometrical interpretation; they appear to be purely algebraic in construction.

In the previous paragraphs we have outlined three fundamental and manifestly 
distinct ways
of constructing consistent string models. A remarkable fact, which will play a
 crucial role in our analysis, is that these three approaches are intimately
related. In fact, by varying certain parameters we can {\it smoothly }
interpolate between all three. As we shall see, a given conformal field theory
of interest typically lies in a multi-dimensional family of theories related to
each other by physically smooth deformations. The paramter space of such a
family is known as its moduli space. This moduli space is naturally divided into
various {\it phase } regions whose physical description is most directly given
in terms of one of the three methods described above (and combinations thereof)
as well as certain simple generalizations. Thus, for some range of parameters, a
conformal field theory might be most naturally described in terms of a 
non-linear
sigma model on a Calabi-Yau target space, for other ranges of parameters it
might most naturally be described in terms of a Landau-Ginzburg theory, while
for yet other values the most natural description might be some combined
version. We will see that physics changes smoothly as we vary the parameters to
move from region to region. Furthermore, in some phase diagrams there are
separate regions associated with Calabi-Yau sigma models on topologically
distinct spaces. Thus, since physics is smooth on passing from any region into
any other (in the same phase diagram), we establish that there are physically
smooth space-time topology changing processes in string theory.

The phenomenon of physically smooth
changes in spatial topology is one example 
of the way in which classical mathematics and string theory differ. In the
former, a change in topology is a discontinuous operation whereas in the latter
it is not. More generally,  many pivotal constructs of classical geometry
naturally emerge in the description of physical observables in string theory.
Typically such geometrical structures are found to exist in string theory  in a
``modified'' form, with the modifications tending to zero as the typical length
scale of the theory approaches infinity. In this sense, the classical
geometrical structures can be viewed as special cases of their string theoretic
counterparts. This idea encapsulates our earlier
discussion regarding what is meant by the phrase ``quantum
geometry''. Namely, one can seek to formulate a new geometrical discipline
 whose
basic ingredients are the observables of string theory. In appropriate limiting
cases, this discipline should reduce to more standard mathematical areas such as
algebraic geometry and topology, but more generally can exhibit numerous
qualitatively different properties. Physically smooth topology change is one
such striking qualitative difference. Mirror symmetry is aanother.
In the following discussion our aim shall be to cover some of the foundational
material needed for an understanding of quantum geometry of string theory.

Much of the discussion above has its technical roots in properties of the 
$N = 2$ superconformal algebra. Hence, in  section
\ref{TheN=2SCFA}
 we shall discuss this
algebra, its representation theory, and certain other key properties for later
developments. We shall also give some examples of theories which respect the $N
= 2$ algebra. In section \ref{FamiliesOfN=2Theories},
 we shall broaden our understanding of such theories
by studying examples which are smoothly connected to one another and hence form
a family of $N = 2$ superconformal theories. Namely, we shall discuss some
simple aspects of moduli spaces of conformal theories. In section 
\ref{InterrelationsBetweenSCFT}, we shall
further discuss some of the examples introduced in section 
\ref{TheN=2SCFA} and point out some
unexpected relationships between them. These results will be used in section
\ref{MirrorManifolds} 
to discuss mirror symmetry. In section \ref{Space-timeTopologyChange}
 we shall apply some properties of
mirror symmetry to establish that string theory admits physically smooth
operations resulting in the change of space-time topology. 
In section \ref{topologychange2},
 we shall go beyond the realm of perturbative string theory
by showing a means of augmenting the tools discussed above to capture
certain  non-perturbative effects that become important if space-time
degenerates in a particular manner. These effects involve
$D$-brane states wrapping around submanifolds of a Calabi-Yau compactification.
We shall see that these degrees of freedom mediate topology changing
transitions of a far more drastic sort than can be accessed with perturbative
methods.
Detailed understanding of the mathematics and physics of both
the perturbative and nonperturbative topology changing transitions
is greatly facilitated by the mathematics of toric geometry. In section 
\ref{toricintro},
therefore, we give an introduction to this subject that is closely alligned
with its physical applications. In section \ref{sec:web},
 we use this mathematical
formalism to extend the arena of drastic spacetime topology change
to a large class of Calabi-Yau manifolds, effectively linking them all
together through such transitions. In this way, quantum geometry
seems to indicate the existence of a universal vacuum moduli space for
type II string theory.

Even with the length of these lecture notes, they are
not completely self-contained.
Due to space and time limitations,
we assume some familiarity
with the essential features of conformal 
field theory. The reader uncomfortable with this material might want to consult,
for example, \cite{Gins} and 
\cite{Oog-TASI}. There are also a number of interesting and important
details which we quote without presenting a derivation, leaving the 
interested reader to find the details in the literature. Finally,
when discussing certain
important but well known background material in the following, we will content
ourselves with giving reference to various useful review articles rather than
giving detailed references to the original literature.

In an attempt to keep the mathemematical content of these lectures
fairly self-contained, we will begin our discussion in the next section
with some basic
elements of classical geometry. The aim is to lay the groundwork for
understanding the mathematical properties of Calabi-Yau manifolds.
 The reader familiar with real and complex
differential geometry can safely skip this section, and return to it
as a reference if needed.

\newsection{Some Classical Geometry}
\label{sec:Some Classical Geometry}

\subsection{Manifolds}

Our discussion will focus on compactified string theory, which as
we shall see, requires the compact portion of space-time to meet
certrain stringent constraints. Although there are more general solutions,
we shall study the case in which the extra ``curled-up'' dimensions
fill out an $n$-dimensional manifold that has the following properties:

\vspace{5mm}

\noindent
$\bullet$ it is compact,

\noindent
$\bullet$ it is complex, 

\noindent
$\bullet$ it is K\"ahler,

\noindent
$\bullet$ it has $SU(d)$ holonomy,

\vspace{5mm}
\noindent
where $d = n/2$. 
For much of these lectures, $n$ will be $6$ and hence $d = 3$.
Manifolds which meet these conditions are known as
Calabi-Yau manifolds, for reasons which will become clear shortly.
In this first lecture, we will discuss the meaning of these
properties and convey some of the essential geometrical and
topological features of Calabi-Yau manifolds. The reader
interested in a more exhaustive reference should consult
\cite{Hubsch}.

To start off gently,
we begin our explanation of these conditions
by going back to the the fundamental
concept of a {\it manifold}\footnote{An introductory course on manifolds
and the various structures on them, as well as physical situations
in which such structures are encountered, can be found in
the book \cite{EG}.}. 
We shall distinguish between
three kinds of manifolds, each having increasingly more refined
mathematical structure: topological manifolds, differentiable
manifolds and complex manifolds.

A {\it topological
manifold} consists of the minimal mathematical structure on a set of
points $X$ so that we can define a notion of continuity. Additionally,
with respect to this notion of  continuity, $X$ is required
to locally look like
a piece of $\IR^n$ for some fixed value of $n$.

More precisely, the
set of points $X$ must be endowed with a {\it topology} $\cal T$ which
consists of subsets $U_i$ of $X$ that are declared to be open. The axioms of a
topology require that the $U_i$ be closed under finite intersections,
arbitrary unions, and that the empty set and $X$ itself are members
of $\cal T$. These properties are modeled on the characteritics of
the familiar open sets in $\IR^n$, which can be
easily checked to precisely meet these conditions.
 $X$, together with $\cal T$, is
known as a {\it topological space}. The notion of continuity mentioned
above arises from declaring a function $\phi: X \rightarrow Y$
continuous if $\phi^{-1}(V_j)$ is an open set in $X$,
where $V_j$ is open in $Y$. For this to make
sense $Y$ itself must be a topological space so that we have 
a definition of open sets in the range and in  the domain. In the special
case in which the domain is $\IR^m$ and the range is $\IR^n$ with topologies
given by the standard notion of open sets, this definition of continuity
agrees with the usual `$\epsilon-\delta$' one from multi-variable calculus.

\begin{figure}[htbp]
\epsfysize=6cm
\centerline{\epsfbox{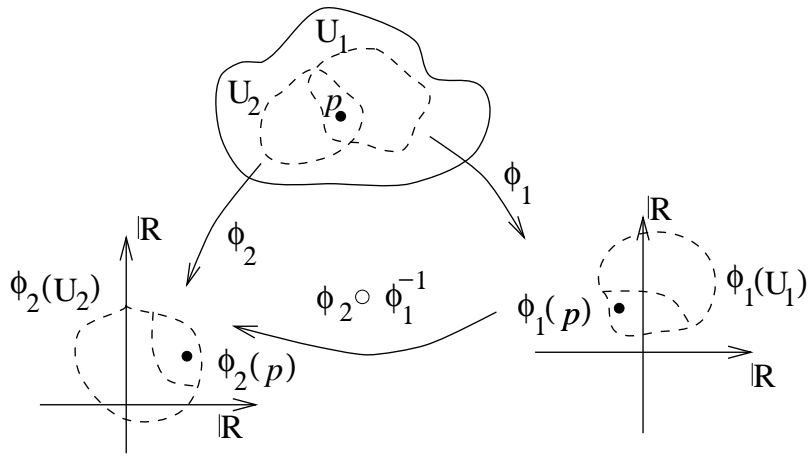}}
\caption{The charts of a manifold $X$.}
\label{fig:charts}
\end{figure}

The topological space $X$ is a topological manifold if it can be
covered with open sets $U_i \in \cal T$ such that for each $U_i$
one can find a continuous map $\phi_i: U_i \rightarrow \IR^n$ (for
a fixed non-negative integer $n$) with a continuous inverse
map $\phi^{-1}$. The pair $(U_i,\phi_i)$ is known
as a chart on $X$ since the $\phi_i$ give us a local coordinate
system for points lying in $U_i$. The coordinates
of a point in $U_i$ are given by its image under
$\phi_i$. Figure \ref{fig:charts} illustrates these charts
covering $X$ and the local coordinates they provide.


Using these local coordinates,
we can give a {\it coordinate representation} 
of an abstract function $f:X \rightarrow  \IR$ via considering the map
$f \circ \phi_i^{-1}: \phi_i(U_i) \rightarrow \IR$. It is easy to see
that $f$ is continuous according to the abstract definition in the
last paragraph if and only if its coordinate representation is
continuous in the usual sense of multi-variable calculus. Furthermore,
since each $\phi_i$ is continuous, the continuity properties of
$f$ are independent of which chart one uses for points $p$ that happen to
lie in overlap regions, $U_i \cap U_j$. The notion of the coordinate
representation of a function is clearly extendable to maps whose
range is an arbitrary topological manifold, by using the coordinate
charts on the domain and on the range.

With this background in hand, we can define the first property
in the definition of a Calabi-Yau manifold. $X$ is {\it compact}
if every collection of sets $V_j \in {\cal T}$ which covers $X$
(i.e. $X = \cup_j V_j$) has a {\it finite subcover}. If the index
$j$ only runs over finitely many sets, then this condition is
automatically met. If $j$ runs over infinitely many sets, this
condition requires that there exist a finite subcollection of
sets $\{W_k\} \subset \{V_j\}$ such that $X = \cup_k W_k$, $k$
now running over finitely many values. Compactness is clearly a property
of the choice of topology $\cal T$ on $X$. One of its virtues,
as we shall see, is that it implies certain mathematically and
physically desirable properties of harmonic analysis on $X$.

The next refinement of our ideas is to pass from topological manifolds
to {\it differentiable} manifolds.
Whereas a topological manifold is the structure necessary to
define a notion of continuity, a differentiable manifold has just
enough additional structure to define a notion of differentiation.
The reason why additional structure is required is easy to understand.
The differentiability of a function $f:X \rightarrow  \IR$ can
be analyzed by appealing to its 
coordinate representation
in patch $U_i$, $f \circ \phi_i^{-1}: \phi_i(U_i) \rightarrow \IR$ as
the latter is a map from $\IR^n$ to $\IR$. Such a coordinate
representation can be differentiated using standard multi-variable
calculus. An important consistency check, though, is that if $p$
lies in the overlap of two patches $p \in U_i \cap U_j$,
the differentiability of $f$ at $p$ does not depend upon which
coordinate representation is used. That is, the result should be
the same whether one works with $f \circ \phi_i^{-1}$ or
with $f \circ \phi_j^{-1}$. On their common domain of definition, we
note that 
$f \circ \phi_i^{-1} = f \circ \phi_j^{-1} \circ (\phi_j \circ
\phi_i^{-1})$. Now, nothing in the formalism of topological manifolds
places {\it any} differentiability properties on the so-called
{\it transition functions} $\phi_j \circ \phi_i^{-1}$ and hence,
without additional structure, nothing guarantees the desired patch
independence of differentiability. The requisite additional structure
follows directly from this discussion: a differentiable manifold
is a topological manifold with the additional restriction that
the transition functions are differentiable maps in the ordinary
sense of multi-variable calculus. One can refine this definition
in a number of ways (e.g. introducing $\IC^k$ differentiable manifolds
by only requiring $\IC^k$ differentiable transition functions), but
we shall not need to do so.

\begin{figure}[htbp]
\epsfysize=6cm
\centerline{\epsfbox{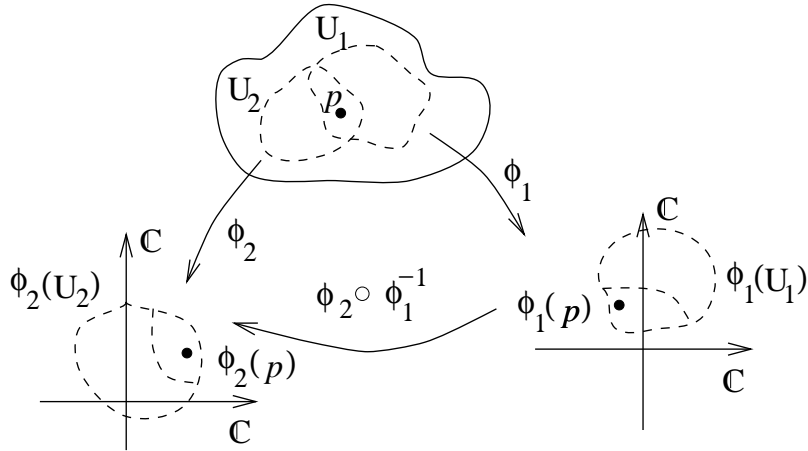}}
\caption{The charts for a complex manifold. Notice
that in this case the cooordinates are complex numbers.}
\label{fig:complex}
\end{figure}

The final refinement in our discussion takes us to the second
defining property of a Calabi-Yau manifold. Namely, we now discuss
the notion of a {\it complex manifold}.
Just as a differentiable manifold has enough structure to
define the notion of differentiable functions, a complex manifold
is one with enough structure to define the notion of holomorphic
functions $f: X \rightarrow \IC$. The additional structure required
over a differentiable manifold follows from exactly the same kind
of reasoning used above. Namely, if we demand that the transition
functions 
$\phi_j \circ \phi_i^{-1}$ 
satisfy the Cauchy-Riemann equations,
then the analytic properties of $f$ can be studied using its coordinate
representative 
$f \circ \phi_i^{-1}$ with assurance that the
conclusions drawn are patch independent. Introducing local
complex coordinates, the $\phi_i$ can be expressed as maps
from $U_i$ to an open set in $\IC^{n\over2}$, with $\phi_j \circ \phi_i^{-1}$
being a holomorphic map from $\IC^{n\over2}$ to $\IC^{n\over2}$. Clearly,
$n$ must be even for this to make sense. In local complex coordinates,
we recall that a function $h: \IC^{n\over2} \rightarrow \IC^{n\over2}$
is holomorphic if $h(z_1,\bar z_1,...,z_{n\over2},\bar z_{n\over2})$ is
actually independent of all the $\bar z_j$. In figure 
\ref{fig:complex}, 
we schematically
illustrate the form of a complex manifold $X$. In a given patch
on any even dimensional manifold, we can always introduce local
complex coordinates by, for instance,
forming the combinations $z_j = x_j + ix_{{n\over2} + j}$,
where the $x_j$ are local real coordinates. The real test is whether
the transition functions from one patch to another --- when expressed
in terms of the local complex coordinates --- are holomorphic maps.
If they are, we say that $X$ is a complex manifold of complex dimension
$d = n/2$. The local complex coordinates with holomorphic transition
functions provide $X$ with a {\it complex structure}.


Given a  differentiable manifold with real dimension $n$ being even,
it can be a difficult question to determine whether or not a
complex structure exists. For instance, it is still not known whether
$S^6$ --- the six-dimensional sphere --- admits a complex structure.
On the other hand, if some differentiable manifold $X$ does admit
a complex structure, nothing in our discussion implies that it
is unique. That is, there may be numerous inequivalent ways of defining
complex coordinates on $X$, as we shall discuss.

That takes care of the basic underlying ingredients in our discussion.
In a moment we will introduce additional structure on such manifolds
as ultimately required by our physical applications, but first we
give a few simple examples of complex manifolds as these may be
a bit less familiar.

\begin{figure}[htbp]
\epsfysize=4cm
\centerline{\epsfbox{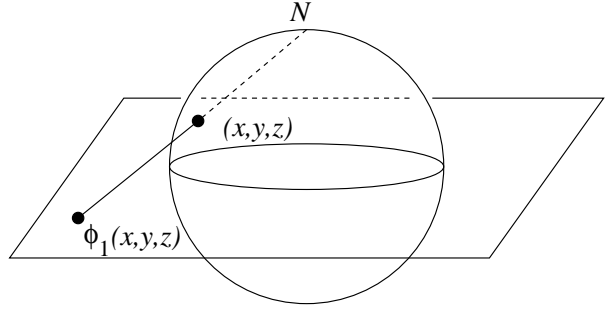}}
\caption{The stereographic projection of the 
          sphere from the north pole.}
\label{fig:stereographic} 
\end{figure}

For the first example, consider the case of the two-sphere
$S^2$. As a real differentiable manifold, it is most convenient
to introduce two coordinate patches by means of stereographic
projection from the north ${\cal N}$ and south
${\cal S}$ poles respectively. As is
easily discerned from figure \ref{fig:stereographic},
 if $U_1$ is the patch
associated with projection from the north pole, we have the local
coordinate map $\phi_1$ being
\beq
  \phi_1: S^2 \smallsetminus \{ {\cal N} \} \rightarrow \IR^2
  ~,
\eeq
with
\beq
  \phi_1(p) = (X,Y) = ({x \over 1-z}, {y \over 1-z})~.
\eeq
In this equation, $x,y,z$ are $\IR^3$ coordinates and the sphere
is the locus $x^2 + y^2 + z^2 = 1$.
Similarly, stereographic projection from the south pole yields
the second patch $U_2$ with 
\beq
\phi_2(p) = (U,V) = ({x \over 1+z}, {y \over 1+z})~.
\eeq
The transition functions are easily seen to be differentiable maps.

Now, define 
$$
  Z = X + iY~,
 ~~~ \bar Z = X - iY~,
 ~~~  W = U + iV~,
 ~~~ \bar W = U - iV~,
$$
as local complex coordinates in our two patches. A simple calculation
reveals that 
$$
   W = W(Z,\bar Z) = {1\over Z}
$$
 and hence our 
transition function (which maps
local coordinates in one patch to those of another) is holomorphic.
This establishes that $S^2$ is a complex manifold. 

\begin{figure}[htbp]
\epsfysize=4cm
\centerline{\epsfbox{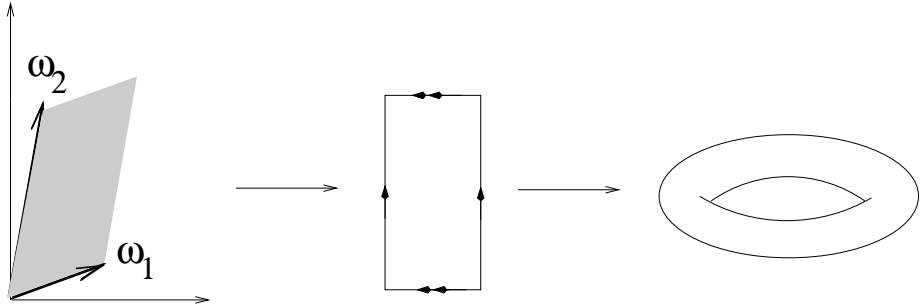}}
\caption{A torus is the quotient of $\IC$ by a 
         2-dimensional lattice $\Lambda$.}
\label{fig:torus} 
\end{figure}

As another example, consider a real two-torus $T^2$ defined by
$\IR^2/\Lambda$, where $\Lambda$ is a lattice $\Lambda = \{\omega_1 m +
\omega_2 n | m,n \in \BZ\}$. This is illustrated in figure 
\ref{fig:torus}.
Since $\IR^2$ is $\IC$, we can equally well think of $T^2$ as
$\IC/\Lambda$. In this way we directly see that the two-torus
is a complex manifold. It inherits complex coordinates
from the ambient space $\IC$ in which it is embedded.
The only distinction is that points labelled by $z$ are identified
with those labelled by $z + \lambda$, where $\lambda$ is any element
of $\Lambda$.

This second example is actually a prototype for how we will
construct Calabi-Yau manifolds. We will embed them in ambient
spaces which we know to be complex manifolds and in this way
be assured that we inherit a complex structure.

\subsection{Equivalences}

Given two manifolds $X$ and $Y$, it is important to have a definition
which allows us to decide whether they are ``different'' or ``the same''.
That is, $X$ and $Y$ might differ only in the particular way they
are presented even though fundamentally they are the same manifold.
Physically, then, they would be isomorphic and hence we would like
to have a framework for classifying the truly distinct possibilities.
The notions of {\it homeomorphism, diffeomorphism} and {\it biholomorphism}
provide the mathematics for doing so.

The essential idea is that whether or not $X$ and $Y$ are considered to
be ``the same'' manifold depends upon which of the structures, introduced
in the last section, we are considering. Specifically, if $X$ and $Y$ are
topological manifolds then we consider them to be the same if they
give rise to the same notion of {\it continuity}. This is embodied by
saying $X$ and $Y$ are {\it homeomorphic} if there exists a one-to-one
surjective map $\phi: X \rightarrow Y$ (and $\phi^{-1}: Y\rightarrow X$)
such that both $\phi$ and $\phi^{-1}$ are continuous (with respect to
the topologies on $X$ and $Y$). Such a map $\phi$ allows us to transport
the notion of continuity defined by $X$ to that defined by $Y$ and
$\phi^{-1}$ does the reverse. Since topological manifolds are characterized
by the definition of continuity they provide, $X$ and $Y$ are ``the same''
--- homeomorphic --- as topological manifolds. Intuitively, the map
$\phi$ allows us to continuously deform $X$ to $Y$.

If we now consider $X$ and $Y$ to be differentiable manifolds,
we want to consider them to be equivalent if they not only provide
the same notion of continuity, but if they also provide the same
notion of differentiability. This is ensured if the maps $\phi$
and $\phi^{-1}$ above are required, in addition, to be differentiable
maps. If so, they allow us to freely transport the notion of differentiability
defined on $X$ to that on $Y$ and vice versa. If such a $\phi$ exists,
$X$ and $Y$ are said to be {\it diffeomorphic}.

Finally, if $X$ and $Y$ are complex manifolds, we consider them to
be equivalent if there is a map $\phi: X \rightarrow Y$ which
in addition to being a diffeomorphism, is also a holomorphic map.
That is, when expressed in terms of the complex structures on $X$ and
$Y$ respectively, $\phi$ is holomorphic. It is not hard to show that
this necessarily implies that $\phi^{-1}$ is holomorphic as well and
hence $\phi$ is known as a biholomorphism.
Again, such a map allows us to identify the complex structures on
$X$ and $Y$ and hence they are isomorphic as complex manifolds.

These definitions do have content in the sense that there are pairs
of differentiable manifolds $X$ and $Y$ which are homeomorphic but
not diffeomorphic. And, as we shall see, there are complex manifolds
$X$ and $Y$ which are diffeomorphic but not biholomorphic. This means
that if one simply ignored the fact that $X$ and $Y$ admit local
complex coordinates (with holomorphic transition functions), and one
only worked in real coordinates, there would be no distinction between
$X$ and $Y$. The difference between them only arises from the way
in which complex coordinates have been laid down upon them.

\begin{figure}[htbp]
\epsfysize=4cm
\centerline{\epsfbox{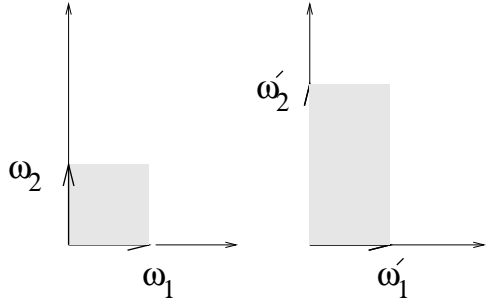}}
\caption{Two diffeomorphic but not biholomorphic tori.}
\label{fig:tori}
\end{figure}

Let us see a simple example of this latter phenomenon.
Consider the torus $T^2$ introduced above as an example of
a one-dimensional complex manifold (the superscriipt  denotes
the real dimension of the torus).  To be as concrete as possible,
lets consider two choices for the defining lattice $\Lambda$:
$(\omega_1,\omega_2) = ( (1,0), (0,1) )$ and 
$(\omega_1',\omega_2') = ( (1,0), (0,2) )$. These two tori are drawn
in figure \ref{fig:tori},
 where we call the first $X$ and
the second $Y$. 

\noindent
As differentiable manifolds, these two tori
are equivalent since the map $\phi$ provides an explicit diffeomorphism:
\beq
\label{torusdiff}
  \phi: X \rightarrow Y~, ~~~~~
  (y_1,y_2) = \phi(x_1,x_2) = (x_1, 2x_2)~, 
\eeq
where $(x_1,x_2)$ and $(y_1,y_2)$ are local coordinates on
$X$ and $Y$.
The map $\phi$ clearly meets all of the conditions of a diffeomorphism.
However, using local complex coordinates $w = x_1 + ix_2$ and
$z = y_1 +iy_2$, we see that
\beq
(z, \bar z) = \phi(w,\bar w) = {{ 3 \over 2 }} w - {{1 \over 2}} \bar w
\eeq
and the latter is {\it not} a holomorphic function of $w$.
Thus, $X$ and $Y$ are diffeomorhpic but not biholomorphic. They
are equivalent as differentiable manifolds but not as complex
manifolds. In fact, a simple extension of this reasoning shows that
for more general choices of $\Lambda$ and $\Lambda'$, the tori
have the same complex structure  if (
but not only if) the ratio $\omega_2 \over \omega_1$
equals $\omega_2' \over \omega_1'$. This ratio is usually called $\tau$.

\subsection{Tangent Spaces}

\begin{figure}[htbp]
\epsfysize=4cm
\centerline{\epsfbox{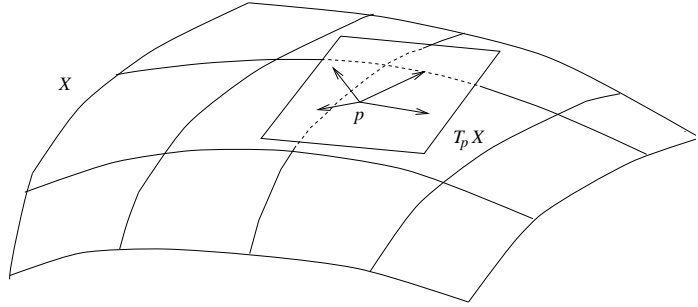}}
\caption{The tangent plane of $X$ at $p$.}
\label{fig:tangent}
\end{figure}

The tangent space to a manifold $X$ at a point $p$ is the closest
flat approximation to $X$ at that point. If the dimension of $X$ is
$n$, then the tangent space is an $\IR^n$ which just `grazes' $X$
at $p$, as shown in figure \ref{fig:tangent}.
By the familiar definition of tangency from multi-variable calculus,
the tangent space at $p$ embodies the ``slopes'' of $X$ at $p$ ---
that is, the first order variations along $X$ at $p$. For this reason,
a convenient basis for the tangent space of $X$ at $p$ consists
of the $n$ linearly independent partial derivative operators:
\beq
\label{realbasis}
  T_pX :
   \Big\{
   {\partial \over \partial x^1}\Big|_p,...,
   {\partial \over \partial x^n}\Big|_p
   \Big\}~.
\eeq
A vector $v \in T_pX$ can then be expressed as
$v = v^\alpha {\partial \over \partial x^\alpha}|_p$. At first sight,
it is a bit strange to have partial differential operators as our
basis vectors, but a moments thought reveals that this directly
captures what a tangent vector really is: a first order motion along
$X$ which can be expressed in terms of the translation operators
${\partial \over \partial x^i}|_p$.

Every vector space $V$ has a dual space $V^\star$ consisting of real valued
linear maps on $V$. Such is the case as well for $V = T_pX$ with the dual
space being denoted $T^\star_pX$. A convenient basis for the latter is
one which is dual to the basis in \calle{realbasis} and is usually denoted
by
\beq
\label{dualbasis}
T_p^\star X : \{dx^1|_p,...,dx^n|_p\} ~,
\eeq
where, by definition, $dx^i: T_pX \rightarrow \IR$ is a linear map
with $dx^i_p({\partial \over \partial x^j}|_p) =\delta^i_j$.
The $dx^i$ are called one-forms and we shall often drop the subscript
$p$, as the point of reference will be clear from context.

If $X$ is a complex manifold of complex dimension $d = n/2$, there
is a notion of the {\it complexified} tangent space of $X$,
$T_pX^{\IC}$. Concretely, $T_pX^{\IC}$ is the same as the real
tangent space $T_pX$ except that we allow complex coefficients
to be used in the vector space manipulations. This is often denoted
by writing $T_pX^{\IC} = T_pX \otimes \IC$. We can still take our basis
to be as in \calle{realbasis}\ with an arbitrary vector $v \in T_pX^{\IC}$
being expressed as $v = v^\alpha {\partial \over \partial x^\alpha}|_p$,
where the $v^{\alpha}$ can now be complex numbers. In fact, it is
convenient to rearrange the basis vectors in \calle{realbasis}\ to
more directly reflect the underlying complex structure. Specifically,
we take the following linear combinations of basis vectors in
\calle{realbasis}\ to be our new basis vectors:
%
%
%
\ifnoWSc
\beq
\label{complexbasis}
   T_pX^{\IC}: 
    \Big\{ 
   \Big( {\partial \over \partial x^1}  
   + i {\partial \over \partial x^{d+1}} \Big) \Big|_p,
   ...,
   \Big({\partial \over \partial x^d} 
   + i {\partial \over \partial x^{2d}} \Big)\Big|_p,
    \Big({\partial \over \partial x^1}  
   - i {\partial \over \partial x^{d+1}}\Big) \Big|_p,
   ...,
  \Big({\partial \over \partial x^d}  - 
  i {\partial \over \partial x^{2d}}\Big) \Big|_p
  \Big\}~.
\eeq
\fi
%
%
%
In terms of complex coordinates we can write this basis as
\beq
\label{basiccomplex}
    T_pX^{\IC}:  
    \Big\{
    {\partial \over \partial z^1}\Big|_p,...,
     {\partial \over \partial z^d}\Big|_p,
     {\partial \over \partial {\bar z^1}}\Big|_p,
      ...,{\partial \over \partial {\bar z^d}}\Big|_p\Big \}~.
\eeq
Notice that from the point of view of real vector spaces, 
${\partial \over \partial x^j}|_p$ and $i{\partial \over \partial x^j}|_p$
would be considered linearly independent and hence $T_pX^{\IC}$
has real dimension $4d$.

In exact analogy with the real case, we can define the dual to
$T_p
 X^{\IC}$, which we denote by $T_p^\star X^{\IC} = T_p^\star X \otimes \IC$, 
with basis
\beq
\label{dualbasiccomplex}
  T_p^\star
  X^{\IC}: \{dz^1|_p,...,dz^d|_p,d\bar z^1|_p,...,d\bar z^d|_p\}
  ~.
\eeq
For certain types of complex manifolds $X$ (Calabi-Yau manifolds
among these), it is worthwhile to refine the definition of
the complexified tangent and cotangent spaces. The refinement we have
in mind simply pulls apart the holomorphic and anti-holomorphic directions
in each of these two vector spaces. That is, we can write
\beq
\label{split}
   T_pX^{\IC} = T_pX^{(1,0)} \oplus T_pX^{(0,1)} 
   ~,
\eeq
where $T_pX^{(1,0)}$ is the vector space spanned by 
$\{{\partial \over \partial z^1}|_p,..., {\partial \over \partial z^d}|_p\}$
and $T_pX^{(0,1)}$ 
is the vector space spanned by 
$\{{\partial \over \partial \bar z^1}|_p,...,
 {\partial \over \partial \bar z^d}|_p\}$.
Similarly, we can write 
\beq
\label{split2}
 T_p^\star X^{\IC} = T_p^\star X^{(1,0)} \oplus T_p^\star X^{(0,1)} 
  ~,
\eeq
where $T_p^\star X^{(1,0)}$ is the vector space spanned by 
$\{dz^1|_p,...,dz^d|_p\}$ and 
$T_p^\star  X^{(0,1)}$ is the vector space spanned by 
$\{d{\bar z^1}|_p,...,d{\bar z^d}|_p\}$.
We call $T_pX^{(1,0)}$ the holomorphic tangent space; it has complex dimension
$d$ and we call $T_p^\star X^{1,0}$ the holomorphic cotangent space. It also
has complex dimension $d$. Their complements are known as the anti-holomorphic
tangent and cotangent spaces respectively. The utility of this decomposition
depends in part on whether it is respected by parallel translation on $X$;
this is a point we shall return to shortly.

\subsection{Differential Forms}

There is an important generalization of the one-forms we have
introduced above. In the context of real manifolds,
a one-form is a real valued linear map acting on $T_pX$.
One generalization of this idea is to consider a
$q$-tensor, $\alpha$,
 which is a real valued {\it multi-linear} map from
$T_pX \times T_pX \times ... \times T_pX$ (with $q$ factors):
\beq
\alpha(v_{(1)p},...,v_{(q)p}) \in \IR~.
\eeq
Multi-linearity here means linear on each factor independently.
For our purposes, it proves worthwhile to focus on a more constrained
generalization of a one-form called a $q$-form. This is a special
type of $q$-tensor which is {\it totally antisymmetric}. If $\omega$
is a $q$-form on $X$ at $p$, then
\beq
\omega(v_{(1)p},...,v_{(q)p}) \in \IR~,
\eeq
with
\beq
\alpha(v_{(1)p},v_{(2)p}...,v_{(q)p}) = - \alpha(v_{(2)p},v_{(1)p},...,v_{(q)p})
\eeq
and similarly for any other interchange of arguments.

We saw earlier that the $dx^j$ form a basis for the the one-forms on $X$
(in a patch with local coordinates given by $(x^1,...,x^n)$).
A basis for two-tensors can clearly be gotten from considering
all $dx^i \otimes dx^j$, where the notation means that
\beq
dx^i \otimes dx^j: T_pX \times T_pX \rightarrow \IR
\eeq
in a bilinear fashion according to
\beq
   dx^i \otimes dx^j({\partial \over \partial x^k},{\partial \over \partial x^l})
   = dx^i({\partial \over \partial x^k})
   \, dx^j({\partial \over \partial x^l}) =
   \delta^i_k\, \delta^j_l~.
\eeq
Now, to get a basis for two-forms, we can simply antisymmetrize the
basis for two-tensors by defining
\beq
dx^i \wedge dx^j = {1 \over 2}(dx^i \otimes dx^j - dx^j \otimes dx^i)~.
\eeq
By construction, $dx^i \wedge dx^j$ satisfies
\beq
dx^i \wedge dx^j(v_{(1)},v_{(2)}) = - dx^i \wedge dx^j(v_{(2)},v_{(1)})~.
\eeq
It is not hard to show that the $dx^i \wedge dx^j,~i<j$ are linearly
independent and hence form a basis for
two-forms. Any two-form  $\omega$, therefore, can be written
$\omega = \omega_{ij}dx^i \wedge dx^j$ for suitable coefficients
$\omega_{ij}$. The generalization to a $q$-form is immediate.
We construct a basis from all possible
\beq
  dx^{i_1} \wedge dx^{i_2} \wedge ... \wedge dx^{i_q}
  ~,
\eeq
where the latter is defined as
\beq
{{1 \over q!}}\sum_P {\rm sgn}P\,
 dx^{i_P(1)} \otimes dx^{i_P(2)} \otimes ... \otimes dx^{i_P(q)}~.
\eeq
Our notation is that $P$ is a permutation of $1,...,q$ and ${\rm sgn}P$
is $\pm 1$ depending whether the permutation is even or odd.
Then, any $q$-form $\omega$ can be written as
\beq
\label{realformm}
  \omega = \omega_{i_1...i_q}\,
   dx^{i_1} \wedge dx^{i_2} \wedge ... \wedge dx^{i_q}
   ~.
\eeq
The fact that $\omega$ is a totally antisymmetric map is sometimes
denoted by $\omega \in \wedge^q T^\star X$, where $\wedge^q$ denotes
the $q$-th antisymmetric tensor product.

All of these ideas extend directly to the realm of complex
manifolds, together with certain refinements due to the additional
structure of having local complex coordinates. If $X$ is a complex
manifold of complex dimension $d = n/2$, then we can define a
$q$-form as above, except that now we use $\wedge^q T^\star X^{\IC}$
instead of $\wedge^q T^\star X$. As the basis for $T^\star X^{\IC}$ is
given in \calle{dualbasiccomplex}\ we see that by suitable rearrangement
of indices we can write a $q$-form $\omega$ as
\beq
\label{complexform}
\omega = \omega_{i_1...i_q}\,
  dt^{i_1} \wedge dt^{i_2} \wedge ... \wedge dt^{i_q}~,
\eeq
where each $dt^j$ is an element of the basis in \calle{dualbasiccomplex}.
Any summand in \calle{complexform} can be labelled by the number $r$
of holomorphic one-forms it contains and by the number $s = q - r$
of anti-holomorphic one-forms it contains. By suitable rearrangement
of indices we can then write
\beq
\label{complexformhol}
\omega = \sum_r\,
 \omega_{i_1...i_r\bar\jmath_1...\bar\jmath_{q-r}}\,
  dz^{i_1} \wedge dz^{i_2} \wedge ... \wedge dz^{i_r}\wedge
  d\overline z^{\bar\jmath_1} \wedge d\overline z^{\bar\jmath_2}
  \wedge ... \wedge 
  d\overline z^{\bar\jmath_{q-r}}~.
\eeq
Each summand on the right hand side is said to belong to
$\Omega^{r,s}(X)$, the space of antisymmetric tensors with
$r$ holomorphic and $s$ anti-holomorphic indices. In this notation, then,
$\Omega^{r,s}(X) = \wedge^r T^{\star(1,0)}X \otimes \wedge^s T^{\star(0,1)}X$.

\subsection{Cohomology and Harmonic Analysis --- Part I}

There is a natural differentiation operation that takes a
$q$-form $\omega$ on a differentiable manifold $X$ to
a $(q+1)$-form on $X$. That is, there is a map
\beq
   d: \bigwedge^q T^\star X \rightarrow \bigwedge^{q+1} T^\star X~.
\eeq
Explicitly, in local coordinates this map $d$, known as {\it exterior
differentiation}, is given by
\beq
   d: \omega \rightarrow d\omega=
   {\partial \omega_{i_1...i_q} \over \partial x^{i_{q+1}}}\,
   dx^{i_{q+1}}\wedge dx^{i_1} \wedge dx^{i_2} \wedge ... \wedge dx^{i_q}
   ~.
\eeq
By construction, the right hand side is a $(q+1)$-form on $X$. We
will return to study the properties of forms and exterior differentiation.
First, though, we note that if $X$ is a complex manifold, there
is a refinement of exterior differentiation which will prove to be
of central concern.

Namely, let 
$\omega^{r,s} =\omega_{i_1...i_r\bar \jmath_1 ... \bar \jmath_s}
 dz^{i_1}\wedge ...  \wedge dz^{i_r}\wedge d\bar z^{\bar\jmath_1} \wedge ...  
\wedge d\bar z^{\bar\jmath_s}$
 be in $\Omega^{r,s}(X)$.  Then, since 
$\omega^{r,s}$ can certainly be thought of as a real $(r+s)$-form on $X$,
$d\omega^{r,s}$ is an $(r+s+1)$-form on $X$. This form may be decomposed
using the complex structure of $X$ into an element of 
$\Omega^{r+1,s}(X) \oplus \Omega^{r,s+1}(X)$. Explicitly,
\beqn
   d \omega^{r,s} \!\!\!\!&=&\!\!\!\! 
   {\partial \omega_{i_1...i_r \bar\jmath_1...\bar\jmath_s} \over 
   \partial z^{i_{r+1}}}
   dz^{i_{r+1}}\wedge dz^{i_1} \wedge dz^{i_2} \wedge ... \wedge dz^{i_r}
   \wedge d\bar z^{\bar\jmath_1}\wedge d\bar z^{\bar\jmath_2} \wedge
   ... \wedge d\bar z^{\bar\jmath_s}
   \nonumber \\
   \!\!\!\!&+&\!\!\!\! 
  {\partial\omega_{i_1...i_r \bar\jmath_1...\bar\jmath_s} \over 
   \partial\bar z^{i_{s+1}}}
  dz^{i_1} \wedge dz^{i_2} \wedge ... \wedge dz^{i_r}
  \wedge d{\bar z^{\bar\jmath_s+1}} \wedge d{\bar z^{\bar\jmath_1}} \wedge 
  d\bar z^{\bar\jmath_2} \wedge ... \wedge d\bar z^{\bar\jmath_s}~.
   \nonumber 
\eeqn
This equation is often summarized by writing
\beq
   d \omega^{r,s} = \partial \omega^{r,s} + \bar \partial \omega^{r,s}
   ~,
\eeq
where we are decomposing the real exterior differentiation operator
$d$ as $d = \partial + \bar \partial$, the latter two being
exterior differentiation in the holomorphic and anti-holomorphic
directions respectively.

There are many uses of exterior differentiation; we note one here.
The antisymmetry involved in exterior differentiation ensures
that $d(d\alpha) = 0$ for any form $\alpha$. That is, $d^2 = 0$. Now,
if it so happens that $\omega$ is a $q$-form for which $d \omega = 0$ ---
such an $\omega$ is called {\it closed} ---
there are then two possibilities. Either $\omega$ is {\it exact}
which means that it can be written
as $d \beta$ for a $(q-1)$-form $\beta$, in which case $d \omega = 0$
follows from the stated property of $d^2 = 0$, or $\omega$ cannot
be so expressed. Those $\omega$ which are closed but not exact provide
non-trivial solutions to the equation $d \omega = 0$ which motivates
the following defintion.

This $q$-th DeRham {\it cohomology} group $H^q_d(X)$ on a real
differentiable manifold $X$ is the quotient space
\beq
 H^q_d(X,\IR) = {{ \{\omega| d\omega = 0\} \over \{\alpha|\alpha= d \beta\}}}
 ~,
\eeq
where $\omega$ and $\alpha$ are $q$-forms.

Again, if $X$ is a complex manifold, there is a refinement of DeRham
cohomology into Dolbeault cohomology in which rather than using
the $d$ operator, we make use of the $\bar \partial$ operator. Since
$\bar \partial^2 = 0$, it makes sense to form the $(r,s)$-th Dolbeault
cohomology group on $X$ via
\beq
  H^{r,s}_{\bar \partial}(X,\IC) = 
 {{ \{\omega^{r,s}| \bar\partial\omega^{r,s} = 0\}
 \over \{\alpha^{r,s}|\alpha^{r,s}= \bar \partial \beta^{r,s-1}\}}}
 ~.
\eeq
This could also be formulated using $\partial$.

The cohomology groups of $X$ probe important fundamental information
about its geometrical structure, and will play a central role in
our physical analysis.

\subsection{Metrics: Hermitian and K\"ahler Manifolds}

A metric $g$ on a real differentiable manifold $X$ is a 
{\it symmetric} positive map
\beq
  g: T_pX \times T_pX \rightarrow \IR~.
\eeq
In local coordinates, $g$ can be written as $g = g_{ij}\, dx^i \otimes dx^j$
where the coeffecients satisfy $g_{ij} = g_{ji}$.
By measuring lengths of tangent vectors according to $g(v_p,v_p)$,
the metric can be used to measure distances on $X$.

If $X$ is a complex manifold, there is a natural extension of the
metric $g$ to a map
\beq
g: T_pX^{\IC} \times T_pX^{\IC} \rightarrow \IC
\eeq
defined in the following way. Let $r,s,u,v$ be four vectors in
$T_pX$. Using them, we can construct, for example, two vectors
$w_{(1)} = r + is$ and $w_{(2)} = u + iv$ which lie in $T_pX^{\IC}$. 
Then, we evaluate $g$ on $w_{(1)}$ and $w_{(2)}$ by linearity:
\beq
g(w_{(1)},w_{(2)}) = g(r + is,u + iv) = g(r,u) - g(s,v) + 
  i\, \lbrack g(r,v) + g(s,u) \rbrack ~.
\eeq
We can define components of this extension of the original metric
(which we have called by the same symbol) with respect to
complex coordinates in the usual way:
$g_{ij} = g({{\partial \over \partial z^i}},{{\partial \over \partial z^j}})$,
$g_{i \bar\jmath} = g({{\partial \over \partial z^i}},{{\partial \over 
\partial \bar z^{\bar\jmath}}})$ and so forth.
 The reality of our original metric
$g$ and its symmetry implies that in complex coordinates we have
$g_{ij}= g_{ji}$, $g_{i \bar \jmath} = 
g_{\bar \jmath i}$ and $\overline {g_{ij}} = g_{\bar\imath \bar\jmath}$, 
$\overline {g_{i \bar\jmath}} =
g_{\bar\imath  j}$.

So far our discussion is completely general. We started with a metric
tensor on $X$ viewed as a real differentiable manifold and noted
that it can be directly extended to act on the complexified tangent
space of $X$ when the latter is a complex manifold. Now we consider
two additional restrictions that prove to be quite useful.

The first is the notion of a {\it hermitian} metric on a complex
manifold $X$. In local coordinates, a metric $g$ is Hermitian if
$g_{ij} = g_{\bar i \bar j} = 0$. In this case, only the mixed type
components of $g$ are nonzero and hence it can be written as
\beq
  g = g_{i \bar\jmath}\, dz^i \otimes d{\bar z^{\bar\jmath}} 
  + g_{\bar\imath j}\, d{\bar z^{\bar\imath}}
  \otimes dz^j~.
\eeq
With a little bit of algebra one can work out the constraint this
implies for the original metric written
in real coordinates. 
(Abstractly, for those who are
a bit more familiar with these ideas, if ${\cal J}$ is a complex structure
acting on the real tangent space $T_pX$, i.e. ${\cal J}: T_pX \rightarrow T_pX$
with ${\cal J}^2 = -  I$, then the hermiticity condition on $g$
is $g({\cal J}v_{(1)},{\cal J}v_{(2)}) = g(v_{(1)},v_{(2)})$.)

The second is the notion of {\it k\"ahlerity},
which will define the third term in the
definition of a Calabi-Yau manifold. Given a hermitian metric
$g$ on $X$, we can build a form in $\Omega^{1,1}(X)$ --- that is,
a form of type $(1,1)$ in the following way:
\beq
J = ig_{i \bar\jmath}\,  dz^i \otimes d {\bar z^{\bar \jmath}} - 
ig_{\bar\jmath i}\, d {\bar z^{\bar\jmath}} \otimes dz^i~.
\eeq
By the symmetry of $g$, we can write this as
\beq
J = ig_{i \bar\jmath}\, dz^i \wedge d {\bar z^j}~.
\eeq
Now, if $J$ is closed, that is, if $dJ = 0$, then $J$ is
called a K\"ahler form and $X$ is called a K\"ahler manifold.
At first sight, this k\"ahlerity condition might not seem too
restrictive. However, it leads to remarkable simplifications in
the resulting differential geometry on $X$, as we indicate in the
next section.

\subsection{K\"ahler Differential Geometry}

In local coordinates, the fact that $d J = 0$ for a 
K\"ahler manifold implies
\beq
d J = (\partial + \bar \partial) ig_{i \bar\jmath}
 \, dz^i \wedge d \bar z^{\bar\jmath} = 0~.
\eeq
This implies that
\beq
\label{kahlermetric}
{{ \partial g_{i \bar\jmath} \over \partial z^l}} = 
{{ \partial g_{l \bar\jmath} \over \partial z^i}}
\eeq
and similary with $z$ and $\bar z$ interchanged.
From this we see that locally we can express $g_{i \bar\jmath}$ as
\beq
g_{i \bar\jmath} = {\partial^2 K\over\partial z^i\partial\bar z^{\bar\jmath}}~.
\eeq
That is, $J = i \partial \bar \partial K$, where $K$ is a locally
defined function in the patch whose local coordinates we are using,
which is known as the K\"ahler potential.

Given a metric on $X$, we can calculate the Levi-Civita connection
as in standard general relativity from the formula
\beq
\label{christ}
\Gamma^i_{jk} = {{1 \over 2}} g^{il}( {{\partial g_{lk} \over \partial x^j}}
+ {{\partial g_{lj} \over \partial x^k}} -
{{\partial g_{jk} \over \partial x^l}})~.
\eeq
Now, if $J$ on $X$ is a K\"ahler form, the conditions \calle{kahlermetric}
imply that there are numerous cancellations in \calle{christ}. In fact,
the only nonzero Christoffel symbols in {\it complex} coordinates
 are those of the form
$\Gamma^l_{jk}$ and $\Gamma^{\bar l}_{\bar\jmath\bar k}$, with all indices
holomorphic or anti-holomorphic.
Specifically, 
\beq
\label{Gamma}
\Gamma^l_{jk} = g^{l \bar s}\, {{\partial g_{k \bar s} 
\over \partial z^j}}
\eeq
 and 
\beq
\label{gammabar}
\Gamma^{\bar l}_{\bar\jmath \bar k} = g^{\bar l  s}\, {{\partial g_{\bar k  s} 
\over \partial {\bar z^{\bar\jmath}}}}~.
\eeq 

The curvature tensor also greatly simplifies. The only non-zero components
of the Riemann tensor, when written in complex coordinates, have the
form $R_{i \bar\jmath k \bar l}$ (up to index permutations consistent with
symmetries of the curvature tensor). And we have
\beq
R_{i \bar\jmath k \bar l} = g_{i \bar s} \,
{{\partial \Gamma^{\bar s}_{\bar\jmath \bar l} \over \partial z^k}}
 ~,
\eeq
as well as the Ricci tensor
\beq
\label{RicciKahler}
R_{\bar\imath j} = R^{\bar k}_{~\bar\imath \bar k j} = - 
{{\partial \Gamma^{\bar k}_{\bar\imath \bar k} \over \partial z^j}}~.
\eeq

\subsection{Holonomy}

Using the above results, we can now describe the final element in
the definition of a Calabi-Yau manifold. First, let $X$ be
a real differentiable manifold of real dimension $n$, and let
$v \in T_pX$. Assuming that $X$ is equipped with a metric $g$ and
the associated Levi-Civita connection $\Gamma$, we can imagine parallel
transporting $v$ along a curve $C$ in $X$ which begins and ends at $p$.
After the journey around the curve, the  vector $v$ will
generally {\it not} return to its original orientation in $T_pX$.
Rather, if $X$ is not flat, $v$ will return to $p$ pointing in another
direction, say $v'$. (Since we are using the Levi-Civita connection
for parallel transport, the length of $v$ will not change during this
process.) If $X$ is orientable, the vectors $v$ and $v'$ will be related
by an $SO(n)$ transformation $A_C$, where the subscript reminds
us of the curve we have moved around. That is
\beq
   v' = A_C v~.
\eeq
Now consider all possible closed curves in $X$ which pass through $p$,
and repeat the above procedure. This will yield a collection of
$SO(n)$ matrices $A_{C_1}, A_{C_2}, A_{C_3},...$, one for each curve.
Notice that if we traverse a curve $C$ which is the curve $C_i$ followed
by the curve $C_j$, the associated matrix will be $A_{C_j}A_{C_i}$ and
that if we traverse the curve $C_j$ in reverse, the associated matrix
will be $A_{C_j}^{-1}$. Thus, the collection of matrices generated
in this manner form a group --- namely, some subgroup of
$SO(n)$. Let us now take this one step further by following the same
procedure at all points $p$ on $X$. Similar reasoning to that just
used ensures that this  collection of matrices also
forms a group. This group describing how vectors change
upon parallel translation around loops on $X$ is called the
{\it holonomy} of $X$.

For a ``generic'' (orientable) differentiable manifold $X$, the holonomy
group will fill out all of $SO(n)$, but when $X$ meets certain
other requirements, the holonomy group can be a proper subgroup.
The simplest example of this is when $X$ is flat. In this case,
the orientation of parallel transported vectors does not change and
hence the holonomy group consists solely of the identity element.
In between these two extremes --- all of $SO(n)$ and the identity ---
a number of other things can happen. We will be interested in two of these.

First, if $X$ is a complex K\"ahler manifold, we have seen above
a number of simplifications which occur in the differential
geometry associated to $X$. In particular, the Levi-Civita connection,
as seen in \calle{Gamma} and \calle{gammabar}, only has nonzero components for
indices of the same type. As the connection controls parallel transport,
this implies that if $v$ is expressed in complex coordinates as
\beq
v = v^j {{\partial \over \partial z^j}} + 
v^{\bar\jmath} {\partial \over \partial \bar z^{\bar\jmath}}~,
\eeq
then the holomorphic components $v^j$ and anti-holomorphic components
$v^{\bar\jmath}$ do not mix together. In other words, the decomposition
at a point $p$ of $T_pX^{\IC} = T_p^{(1,0)}X \oplus  T_p^{(0,1)}X$ 
is unaffected by parallel translation away from $p$. The individual
factors in the direct sum do  not mix. 

In terms of the holonomy group, this implies that the holonomy
matrices can consistently be written in terms of their action on
the holomorphic or anti-holomorphic basis elements and hence
lie in a $U(d)$ subgroup of $SO(n)$ (where, as before, $d = n/2$).

The second special case are those complex K\"ahler manifolds whose
holonomy group is even further restricted to lie in $SU(d)$.
Although equally consistent as string compactifications, we will
typically not discuss $X$ whose holonomy is a proper
subgroup of $SU(d)$. Hence, we shall take Calabi-Yau to mean
holonomy which fills out $SU(d)$. In essence, as we shall discuss
shortly, having holonomy $SU(d)$ means that the $U(1)$ part
of the Levi-Civita connection $\Gamma$ vanishes. This can be
phrased as a topological restriction on $X$ which will greatly
aid in the construction of examples.

\subsection{Cohomology, Harmonic Analysis --- Part II}

We discussed previously the operation of exterior differentiation $d$ which
takes a differential $p$-form to a differential $(p+1)$-form. If the
manifold $X$ on which these forms are defined has a Riemannian metric,
then the operation $d$ has an adjoint $d^{\dagger}$, which maps
$p$-forms to $(p-1)$-forms, defined in the
following way:
\beq
\label{ddagger}
  d^{\dagger}: \omega \rightarrow d^\dagger\omega = 
  -{1\over (p-1)!}\,
   \omega^\mu_{~\mu_1\dots\mu_{p-1};\mu}\,
    dx^{\mu_1}\wedge\dots\wedge dx^{\mu_{p-1}}~,
\eeq
where $\omega^\mu_{~\nu\dots;\rho}$ denotes the covariant derivative
of $\omega^\mu_{~\nu\dots}~.$

To understand the meaning of this operation in greater detail,
it is necessary to introduce the notion of the Hodge star operation
$\star$, which is important in its own right.
This operator maps a $p$-form on a $n$-dimensional differentiable
manifold with metric $g$ to an $(n-p)$-form. Explicitly,
\beq
\label{Hodgestar} 
\omega \rightarrow \star\omega = 
 {1 \over (n - p)!p!}\,
  \epsilon_{i_1,...,i_n} \, \sqrt{|{\rm det} g|}
  \,g^{i_1j_1}...g^{i_pj_p}\, \omega_{j_1...j_p}
   \,dx^{i_{p+1}}  \wedge ... \wedge dx^{i_{n}}~.
\eeq
This map has the virtue of being bijective and coordinate independent.

Notice now that the composition $\star d\star$ maps a $p$-form first to
an $(n-p)$-form, then to a $(n-p+1)$-form and finally to a $(p-1)$-form.
In fact, 
\beq
\label{dh}
d^{\dagger} = (-1)^{np+n+1} \star d \star~.
\eeq

From a more abstract point of view, the Hodge star operator gives us
an inner product on $p$-forms via
\beq
\label{inner}
\langle \omega, \omega' \rangle = \int_X \omega \wedge \star\omega'~.
\eeq
We can then define the adjoint of $d$ from the requirment that
if $\beta$ is a $(p-1)$-form and $\omega$ is a $p$-form, then
\beq
\label{adjointabs}
\langle \omega, d \beta \rangle = \langle d^{\dagger} \omega, \beta \rangle 
 ~.
\eeq
When expressed in local coordinates, we obtain \calle{ddagger}.

There are numerous uses of $d$ and $d^{\dagger}$ in both mathematics
and physics. Here we focus on one --- the Hodge decomposition theorem ---
which states that any $p$-form on $X$ can be uniquely written as
\beq
\label{Hodgedecomp}
\omega = d \beta + d^{\dagger} \gamma +  \omega'~,
\eeq
where $\beta$ is a $(p-1)$-form, $\gamma$ is a $(p+1)$-form, and
$\omega'$ is a {\it harmonic} $p$-form. By definition,
a harmonic form is one that is annihilated by
$\Delta = d^{\dagger} d + d d^{\dagger}$, which
is the {\it Laplacian} acting on $p$-forms. It is easy to check by
writing $\Delta$ in local coordinates, that it is the curved space
generalization of the ordinary Laplacian on $\IR^n$.

In particular, if $\omega$ is closed then it is not hard to show
that $\gamma$ vanishes and hence we can write
\beq
\label{Hodgedecompclosed}
\omega = d \beta  + \omega'~.
\eeq
From our earlier discussion of cohomology, we now recognize
$\omega - d \beta$ as an element of $H^p(X,\IR)$ and hence we learn
that there is a {\it unique} harmonic $p$-form representative
in each cohomology class of $H^p(X,\IR)$.

As in the previous sections, if $X$ is a complex manifold and
$\omega^{r,s}$ is an $(r,s)$-form, the complex Hodge decomposition
allows us to write
\beq
\omega^{r,s} = \overline \partial \alpha^{r,s-1} + \overline \partial^{\dagger}
\beta^{r,s+1} + \omega'^{r,s}~,
\eeq
where $\omega'^{r,s}$ is harmonic with respect to the Laplacian
$\Delta_{\bar \partial} = \overline \partial^{\dagger} \overline \partial +
\overline \partial \overline \partial^{\dagger}$.

As in the real case, if $\omega^{r,s}$ is $\overline \partial$ closed,
then this decomposition gives us a unique $\Delta_{\bar \partial}$
harmonic representative for each class in $H^{r,s}_{\bar \partial}(X,\IC)$.

In the special case in which $X$ is a K\"ahler manifold, it is straightforward
to show that all of the Laplacians built from $d$, $\overline \partial$
and $\partial$, namely $\Delta$, $\Delta_{\overline \partial}$ and
$\Delta_{\partial}$ are related by
\beq
\Delta = 2\Delta_{\overline \partial} = 2\Delta_{\partial}~.
\eeq
In this case, then, the harmonic forms with respect to each operator
are the same.

We define $h^{r,s}_X$ to be the (complex) dimension of 
$H^{r,s}_{\bar \partial}(X,\IC)$ which is the same as the dimension
of the vector space of harmonic $(r,s)$-forms on $X$. The Hodge star
operator, with the obvious extension into the complex realm, ensures
\beq
h^{r,s}_X = h^{m-r,m-s}_X~.
\eeq
Using complex conjugation and K\"ahlerity, we also have
\beq
h^{r,s}_X = h^{s,r}_X~.
\eeq

K\"ahlerity also ensures the following relation between $d$ and
$\overline \partial$ cohomology:
\beq
H^p_d(X) = \bigoplus_{r+s=p}\, H_{\overline \partial}^{r,s}(X).
\eeq

\subsection{Examples of K\"ahler Manifolds}

The simplest example of a K\"ahler manifold is $\IC^m$. We can
write a K\"ahler form associated to the usual Euclidean metric
written in complex coordinates $ g = \sum_j dz^{j} \otimes d {\overline z}^j$
as $J = {{i \over 2}}\sum_jdz^{j} \wedge d {\overline z}^j$. Clearly $J$ is closed.

The next simplest examples of K\"ahler manifolds are Riemann surfaces.
These are orientable complex manifolds of complex dimension 1.
They are K\"ahler since {\it every} two-form is closed (as the real
dimension being two cannot support forms of higher degree).

Of most use in our subsequent discussions, are the examples of
ordinary and weighted complex projective spaces. Let us define these.

The ordinary complex projective $n$-space, ${\Bbb C}P^n$,
 is defined by introducing
$n+1$ {\it homogeneous} complex coordinates $z_1,...,z_{n+1}$ not all
of them simultaneously zero with
an equivalence relation stating that points labelled by
$(z_1,...,z_{n+1})$ are identified with points labelled by
$(\lambda z_1,...,\lambda z_{n+1})$ for any complex number $\lambda$.
In reality, the $z_i$ are therefore not local coordinates in the technical
sense. Rather, in the $j$-th patch, defined by $z^j \ne 0$, we can choose
$\lambda = 1/z^j$ and use local coordinates 
$$
 (\xi_{(j)}^1,\dots,\xi_{(j)}^n)=
({{z_1 \over z_j}},...,
{{z_{j-1} \over z_j}},{{z_{j+1}\over z_j}},\dots,{{z_{n+1} \over z_j}})~.
$$
We can see that ${\Bbb C}P^n$
is K\"ahler by defining $K_j$ in the $j$-th patch
to be $K_j = \sum|{{z^i\over z_j}}|^2$. Then $J = \partial \overline \partial
K_j$ is a {\it globally} defined closed 2-form class on $X$. To see this, since
we defined $J$ patch by patch,
one must only check that the differences in patch overlaps are exact. The
metric associated to $J$ is called the Fubini-Study metric.

For later use we point out that by suitable choice of $|\lambda|$, we can
always choose our representatives from the equivalence class of homogeneous
coordinates to satisfy
\beq
\label{symplectic1}
\sum_i|z_i|^2 = r
\eeq
for some arbitrarily chosen positive real number $r$. This ``fixes'' part of
the $\IC^*$ equivalence relation defining the projective space.
The rest --- associated with the phase of $\lambda$ --- is implemented
by identifying
\beq
\label{symplectic2}
(z_1,...,z_{n+1}) \sim (e^{i \theta} z_1,...,e^{i \theta} z_{n+1}).
\eeq

The weighted projective space $W{\Bbb C}P^n(\omega_1$, $\omega_2$,
$...$, $\omega_{n+1})$
is a simple generalization of ordinary projective space in 
which the homogeneous coordinate identification is
\beq
\label{wtedspace}
  (z_1,...,z_{n+1}) \sim
  (\lambda^{\omega_1} z_1,...,\lambda^{\omega_{n+1}} z_{n+1})~.
\eeq
It is again not hard to show that this space is K\"ahler. One
complication is that $W{\Bbb C}P^n(\omega_1,...,\omega_{n+1})$ is not
generally smooth because non-trivial fixed points under the
coordinate identification lead to singularities. Specifically,
if the weights $(\omega_1,...,\omega_{n+1})$ are not all relatively
prime, then there will be non-trivial values of $\lambda$ so
that $\lambda^{\omega_i} = \lambda^{\omega_j} = 1$ for some $i \ne j$.
By setting to zero all of the homogeneous coordinates whose weights
do not satisfy this equation, we will find a subspace of
the weighted projective space which is singular due to its being a
fixed point set of the coordinate identification.

In the rest of the lectures, we will be interested in compact
K\"ahler manifolds constructed as subspaces of larger
complex manifolds. We thus explain some notions and results
in this direction.

An analytic submanifold $N$ of the complex manifold $M$
is defined by a set of analytic equations
\beq
     N = \{ p\in M ~:~ f^\alpha(p)=0,~\alpha=1,\dots,m\le n\}
     ~,
\eeq
such that 
\beq
   {\rm rank}\lbrack{\partial f^\alpha\over\partial z^j}\Big|_p\rbrack
\eeq
where $z^j,~j=1,\dots,n$ are the coordinates on $M$,
is independent of the point $p\in M$ and equals  $m$.
When $f^\alpha$ are polynomials, we call $N$ an 
{\it algebraic variety} in $M$.

Then by letting
\beqn
   w^\alpha(p) &\equiv& f^\alpha(p)~,
   ~~~~~\alpha=1,\dots,m~, \\
   w^i(p)  &\equiv& z^i(p)~,
   ~~~~~i=m+1,\dots,n~,
\eeqn
we have a system of local coordinates such that $N$ is defined
by
$$
   w^1(p) = w^2(p) = \dots =w^m(p) =0~.
$$
Then $\zeta^i(p)=w^{i+m}(p),~i=1,\dots,n-m$ are the local coordinates of
$N$; therefore, the submanifold has (complex) dimension $n-m$.
(This construction of local coordinates will be explained in the example
below).

Now, let $M=\IC^n$. Under what conditions  is $N$  a compact
(analytic) submanifold? To answer this question, we recall the maximum
modulus theorem from complex analysis. According to this
theorem, an analytic function on a domain $D\subset\IC$ cannot
have an extremum on $D$ unless it is a constant. 
This theorem can be extended to the case of $\IC^n$.

If $N\subset\IC^n$ is a compact submanifold of $\IC^n$, the functions
$f^\alpha(\zeta)$ are analytic functions of $\zeta$ on $M$ and
since $N$ is compact each $f^i$ much be constant. We thus arrive
at the conclusion that any compact submanifold of $\IC^n$ has to
be a point!

With the result of the last pragraph at hand, we are led
to examine the construction of compact submanifolds
within  other complex manifolds. A natural next choice is
$\IC P^n$. In fact, $\IC P^n$ is compact and all
its complex submanifolds are compact. There is a famous
theorem due to Chow:

\begin{theorem}[Chow]\ \\
Any submanifold of $\IC P^n$ can be realized as the zero locus
of a finite number of homogeneous polynomial equations.
\end{theorem}

A well studied example of the above discussion, that will show up
in many places in the present lectures, is the set of
points in $\IC P^4$ given by the locus of zeros of the equation
\beq
\label{eq:fermat}
    \sum_{i=1}^5 \, (z_i)^5 = 0~.
\eeq
Let us denote it by $P$ and explain how
one places local coordinates upon it. Using the patches $U_j=\{ z\in\IC^5-{0},
z_j\ne 0\}$ and the
corresponding coordinates $\xi^i_{(j)},~j=1,2,3,4$ for $\IC P^4$
defined above, the equation \calle{eq:fermat} can be rewritten in the
form:
\beq
\label{eq:fermat1}
    1+\sum_{i=1}^4 \, (\xi_{(j)}^i)^5 = 0~, 
\eeq
on the patch $U_j$.

We now concentrate on the patch $U_1$ of $\IC P^4$.
Here, we make the following holomorphic change of coordinates
\beqn
     \eta^1 &=& 1+\sum_{i=1}^4 \, (\xi_{(1)}^i)^5~, \nonumber\\
     \eta^2 &=& \xi_{(1)}^2~, \nonumber\\
     \eta^3 &=& \xi_{(1)}^3~, \nonumber\\
     \eta^4 &=& \xi_{(1)}^4~. \nonumber
\eeqn
The new variables will be  good coordinates if the Jacobian
\beq
   {\partial(\eta)\over\partial(\xi)} = 5\,(\xi_{(1)}^1)^4 =
   5\, \left( {z_2\over z_1} \right)^4
\eeq
does not vanish. Therefore, as long as $z_2\ne 0$,
the part of $P$ found in $U_1$ is given by 
$\eta^1=0$. We have thus constructed a patch $V_1=P\cap U_1$
(actually is $P\cap U_1 \cap U_2$ since $z_2\ne 0$)
$$
   V_1 =\{ (\eta^1,\eta^2,\eta^3,\eta^4)\in \IC^4~,~\eta^1=0\}
$$
on $P$ and a local coordinate system
$$
       \zeta^1_{(1)}=\eta^2~,~~~
       \zeta^2_{(1)}=\eta^3~,~~~
       \zeta^3_{(1)}=\eta^4~.
$$

When $z_2=0$, the transformation $\xi\to\eta$ is not well defined;
in this case we introduce another transformation:
\beqn
     \eta^1 &=& \xi_{(1)}^2~, \nonumber\\
     \eta^2 &=& 1+\sum_{i=1}^4 \, (\xi_{(1)}^i)^5~, \nonumber\\
     \eta^3 &=& \xi_{(1)}^3~, \nonumber\\
     \eta^4 &=& \xi_{(1)}^4~. \nonumber
\eeqn
This transformation is well defined if $z_3\ne 0$. Continuing in this manner,
an atlas for $P$ can be constructed.

\subsection{Calabi-Yau Manifolds}

We have now defined all of the mathematical ingredients to understand
and study Calabi-Yau manifolds. Just to formalize things we write

\begin{definition}\ \\
A Calabi-Yau manifold is a compact, complex, K\"ahler manifold
which has $SU(d)$ holonomy.
\end{definition}

For the most part we shall study the case of $d = 3$. An equivalent statement
is that a Calabi-Yau manifold admits a Ricci-flat metric:
$$
   R_{i \bar\jmath} = -{{\partial \over \partial {\overline z^{\bar\jmath}}}}
 \Gamma_{ik}^k
  = 0 ~.
$$
It is not hard to show
that the vanishing of the $U(1)$ part of the connection,
effectively its trace, which ensures that the
holonomy lies in $SU(d)$ is tantamount to having
a Ricci-flat metric. We will generally take Calabi-Yau to mean
holonomy being precisely $SU(d)$ as mentioned in the definition.

An important theorem, both from the abstract and practical perspectives,
is that due to Yau who proved Calabi's conjecture that a complex
K\"ahler manifold of vanishing first Chern class admits a Ricci-flat
metric. We will not cover Chern classes in these notes in any detail;
for this the reader can consult, for instance,
\cite{GHa}. Briefly, though, the Chern classes of
$X$ probe basic topological properties of $X$. Specifically,
the $k$-th Chern class $c_k(X)$ is an element of $H^k_d(X)$ defined from
the expansion
\beq
\label{chern}
 c(X) = 1 + \sum_j c_j(X) = \det
 (1 + {\cal R}) = 1 + {\rm tr} {\cal R} + {\rm tr} ({\cal R} \wedge
 {\cal R} - 2 ({\rm tr}{\cal R})^2) + ...~,
\eeq
where ${\cal R}$ is the matrix valued curvature 2-form
$$
  {\cal R} = R^k_{~l i \bar \jmath}
   \,  d z^i \wedge d \bar z^{\bar\jmath}~.
$$

We should point out that the curvature tensor $R$ is really being thought
of as the curvature tensor of the tangent bundle $T_X$ of $X$, with
the matrix indices being those in the fiber direction. Thus, often
 $c(X)$ is also written as $c(T_X)$.

Rather remarkably, although constructed from the local curvature tensor,
the Chern classes only depend on far more crude topological
properties of $X$. We see directly that if $X$ has vanishing Ricci tensor,
then the first Chern class, being the trace of the curvature 
2-form, vanishes. Yau's theorem goes in the other direction and shows that
if the first Chern class vanishes (as a cohomology class in $H^2_d(X)$)
then $X$ admits a Ricci-flat metric.
More precisely,

\begin{theorem}[Yau]\ \\
If X is a complex K\"ahler manifold with
vanishing first Chern class and with K\"ahler form $J$, then there
exists a unique Ricci-flat metric on $X$ whose K\"ahler from $J'$ is in the
same cohomology class as $J$.
\end{theorem}

The utility of this theorem is that it is generally quite hard to
directly determine whether or not $X$ admits a Ricci-flat metric $g$.
In fact, no explicit Ricci-flat metrics are known on any Calabi-
Yau manifolds. On the other hand, it is a simple matter to compute
the first Chern class of $X$, and, in particular, to find
examples with vanishing first Chern class. Yau's theorem then ensures
the existence of a Ricci-flat metric.

In our subsequent discussions, we will need to know various things regarding
the cohomology of Calabi-Yau manifolds. There are a number of simplifications
which occur relative to the general K\"ahler manifold. In particular,
since the holonomy is $SU(d)$, it can be shown that
$h^{0,s} = h^{s,0}= 0$ for $1 < s < d$ and that $h^{0,d} = h^{d,0} =1$.
The latter is a holomorphic, nowhere vanishing differential form
of type $(d,0)$  on the Calabi-Yau, usually referred to as
$\Omega$.
 Using the fact
that the space $X$ is connected we also have $h^{0,0} =1$. We therefore
have the following form for the Hodge numbers (arranged in the
so-called Hodge diamond) for $d = 1,2,3$:

\ifnoWSc
\begin{center}
\begin{tabular}{|c|cccc cccc  cccc cccc|} 
\hline\hline
     & & & &  & & & & & 
     & & &  & & &  &   \\ 
{}&
     & & & $h^{0,0}$ & & & & & 
     & & & 1        & & &  &   \\ 
$d=1$&
     & & $h^{1,0}$ &  & $h^{0,1}$ & & & 
=  &
     & & 1 &  & 1 & & &  \\
{}&
     & & & $h^{1,1}$ & & & & &
     & & & 1        & & & &  \\ 
      & & &  & & & & & &
     & & &  & & &  &   \\ 
\hline
%
%
%
%
      & & &  & & & & & &
     & & &  & & &  &   \\ 
{}&
     & & & $h^{0,0}$ & & & & &
     & & & 1        & & &  &   \\
{}&
     & & $h^{1,0}$ &  & $h^{0,1}$ & & & &
     & & 0 &  & 0 & & &  \\
$d=2$&
     & $h^{2,0}$& & $h^{1,1}$ & & $h^{0,2}$ & & 
=  &
     & 1 & & 20        & & 1 &  &   \\
{}&
     & & $h^{2,1}$ &  & $h^{1,2}$ & & & &
     & & 0 &  & 0 & & &  \\
{}&
     & & & $h^{2,2}$ & & & & &
     & & & 1        & & &  &   \\
      & & &  & & & & & &
     & & &  & & &  &   \\ 
\hline
%
%
%
%
      & & &  & & & & & &
     & & &  & & &  &   \\ 
{}&
     & & & $h^{0,0}$ & & & & &
     & & & 1        & & &  &   \\
{}&
     & & $h^{1,0}$ &  & $h^{0,1}$ & & & &
     & & 0 &  & 0 & & &  \\
{}&
     & $h^{2,0}$& & $h^{1,1}$ & & $h^{0,2}$ & & &
     & 0 & &  $h^{1,1}$       & & 0 &  &   \\
$d=3$&
     $h^{3,0}$& & $h^{2,1}$ & & $h^{1,2}$ & &  $h^{0,3}$ & 
=  &
     1 & & $h^{2,1}$ & & $h^{2,1}$ & &  1 & \\
{}&
     & $h^{3,1}$& & $h^{2,2}$ & & $h^{1,3}$ & & &
     & 0 & &  $h^{1,1}$       & & 0 &  &   \\
{}&
     & & $h^{3,2}$ &  & $h^{2,3}$ & & & &
     & & 0 &  & 0 & & &  \\
{}&
     & & & $h^{3,3}$ & & & & &
     & & & 1        & & &  &   \\
     & & &  & & & & & &
     & & &  & & &  &   \\ 
\hline
\hline
\end{tabular}
\end{center}
\fi

Notice that in the two-dimensional case we have explicitly filled in
$h^{1,1} = 20$. We will derive this shortly. The key distinction,
relative to the three-dimensional case
 and higher is that there is a {\it unique}
two-dimensional Calabi-Yau manifold.  In the case of
complex dimension three, there are numerous possibilities
for the Hodge numbers.

To illustrate these ideas as concretely as possible, we now consider
a Calabi-Yau manifold  of complex dimension 3 which, in fact,
we have already encountered in \calle{eq:fermat}: the quintic
hypersurface in 
complex projective
four-space ${\Bbb C}P^4$ with homogeneous coordinates $(z_1,...,z_5)$
given by the locus $P(z_1,...,z_5) = 0$. In order to understand why
this is a Calabi-Yau manifold, lets first keep the degree of
$P$ unspecified.
We see that $P$ must be
homogeneous of some fixed degree $t$ in order that
$P(\lambda z_1,...,\lambda z_5)$ also vanishes, ensuring that $P$
is well defined on ${\Bbb C}P^4$. The locus $X$ given by
$P = 0$ is  K\"ahler, inheriting these properties
from ${\Bbb C}P^4$. As we now discuss, $t$ determines
the value of the first Chern class of the locus $\{ P = 0\} $ in
${\Bbb C}P^4$.

We do not have time nor space to fill in all the details of this
calculation, but we will give the essential ideas.
More details can be found, for instance, in 
\cite{Hubsch}.
 First,
we need to understand how to calculate the Chern classes
of ${\Bbb C}P^4$ itself. The basic ingredient is something called the
{\it splitting principle}. This is the statement that upon adding
a trivial line bundle ${\cal O}$ to the tangent bundle of ${\Bbb C}P^4$
(which has no bearing on Chern classes),
we obtain a bundle $T{\Bbb C}P^4 \oplus {\cal O}$ whose curvature
2-form, at least as far as calculating Chern classes goes,
can be diagonalized to the $5\times5$ matrix diag$(J,...,J)$,
where $J$ is the K\"ahler form on ${\Bbb C}P^4$. From \calle{chern}
 we learn that
\beq
\label{chernfour}
   c({\Bbb C}P^4) = (1 + J)^5~,
\eeq
where the righthand side is subject to $J^5 = 0$ as 
${\Bbb C}P^4$
is four-dimensional.
In particular, note that $c_1({\Bbb C}P^4) = 5J$.

Now, to calculate the Chern classes of $X=\{P = 0\} $ in ${\Bbb C}P^4$
we note that the tangent bundle $T{\Bbb C}P^4$ of ${\Bbb C}P^4$
when restricted to $\{ P = 0\}$ gives
\beq
  T{\Bbb C}P^4|_{P = 0} = TX \oplus NX
   ~,
\eeq
where $TX$ and $NX$ are the tangent bundles of $X$ and
the normal bundle of $X$ inside of ${\Bbb C}P^4$ respectively.
From the basic definitions, we have
\beq
  c(T{\Bbb C}P^4|_{P = 0}) = c(T_P) \wedge c(NX)~.
\eeq
Formally, we can solve for $c(T_P)$ and write
\beq
\label{adj}
   c(TX) = {{c(T{\Bbb C}P^4|_{P = 0}) \over c(NX)}}~.
\eeq
Now, the normal bundle $NX$ is a line bundle over $\{P = 0\}$ and it
can be shown that it has Chern form $(1 + tJ)$. The right-hand-side
of \calle{adj} is then interpreted as a formal power series
in $J$. So, we find
\beq
\label{quintic}
   c(TX) = {{(1 + J)^5 \over (1 + tJ)}} = 1 + (5-t)J + \cdots~.
\eeq
From this we see that $\{P = 0\}$ has vanishing first Chern class if it
is a homogeneous polynomial of degree $t = 5$.
Thus, a quintic hypersurface in ${\Bbb C}P^4$ is a Calabi-Yau manifold
with complex dimension 3.

We can briefly carry on this discussion in two ways. First,
the Euler characteristic of $X$ comes from $\int_X c_3(X)$.
By expanding \calle{quintic} to third order, we see that
$c_3 = -40 J^3$. As $J^3$ integrates to $5$ (a fact that
is most easily derived by counting the number of points
of intersection of three hyperplanes and the quintic --- the
homological dual of the integral), we see that the quintic
has Euler number $-200$. Second, we can equally well carry out
this discussion, say, for a hypersurface in ${\Bbb C}P^3$. Following
the same reasoning as above, we see that it will have vanishing
first Chern class if it is given by the vanishing locus of a
homogeneous degree $4$ polynomial. In this case, the second Chern
class is $6 J^2$ and hence we find Euler number $24$. This is
the unique two-dimensional
Calabi-Yau space mentioned above. It
is known as $K3$ and is the subject of Aspinwall's
lectures in this volume.

There are many generalizations of the construction presented here.
One can look at intersections of numerous constraints in higher dimensional
projective and weighted projective space, and products thereof.
Thousands of examples have been constructed in this manner.

\subsection{Moduli Spaces}
\label{sec:bacmod}

In the last section we constructed the simplest example of a Calabi-Yau
manifold with three complex dimensions. For most of what follows we
will stick to the three-dimensional case. In constructing the quintic
we specified, in fact, very little information about its detailed
structure. After all, we only determined that it is given by the vanishing
locus of a quintic polynomial, but we did not specify anything about
the detailed form of this polynomial. Furthermore, although we mentioned
that the quintic is K\"ahler by virtue of its being a submanifold of
${\Bbb C}P^4$, we did not actually specify the K\"ahler class chosen.
What these facts indicate is that the Calabi-Yau manifold we have
constructed is actually part of a continuous family of Calabi-Yau's,
each differing from the others by the particular choices made for
these data: the precise form of the defining equation and also the precise
form of the K\"ahler class. This is a general feature of Calabi-Yau manifolds:
they typically come in multidimensional families. In this section
we briefly discuss this point.

If $X$ is Calabi-Yau, then $X$ admits a metric $g$ such that
$R_{i \bar\jmath}(g) = 0$. Now, given such a $g$, can we continuously
perturb to a new metric $g + \delta g$ such that the Ricci tensor
still vanishes? This is a question studied at length  in, for example,
\cite{PC},
and the result is as follows.
There are two basic types of
perturbations $\delta g$ that we can consider:
 those with pure and those with mixed type
indices:
\beq
\label{perturb}
  \delta g = \delta g_{ij}\, dz^i dz^j +
  \delta g_{i \bar\jmath}\, dz^i d\bar z^{\bar\jmath} + 
{\rm \ \ c.c.}~.
\eeq
As $g$ is a Hermitian metric, 
the perturbations with mixed type indices preserve the original index
structure of $g$ while those of pure type do not. We will discuss the
meaning of this in a moment. Plugging these perturbations of the metric
into the curvature tensor and demanding preservation of Ricci-flatness
imposes severe restrictions on $\delta g$. In particular, it turns out that
$\delta g_{i\bar\jmath}\, dz^i \wedge d{\bar z^{\bar\jmath}}$
 must be harmonic and
hence is uniquely associated to an element of
$H^{1,1}_{\bar \partial}(X)$. Using the holomorphic three-form $\Omega$,
it can also be shown that $\Omega_{ijk}\, g^{k \bar k} \delta_{\bar k \bar l}
dz^i \wedge dz^j \wedge d{\bar z}^{\bar l}$ is an element of
$H^{2,1}_{\bar \partial}(X)$. Any two representatives in the same cohomology
class yield metric perturbations that can be undone by  
coordinate redefinitions.
Hence, the cohomology classes
 capture the non-trivial Ricci-flat metric deformations.

These two cohomology groups are therefore associated with the space
of deformations of an initial Ricci-flat metric on $X$ to a nearby
Ricci-flat metric. In fact, we can be a bit more precise. Deformations
to the metric with pure type indices yield a metric $g + \delta g$ which
is no longer Hermitian. However, by a suitable change of variables,
this new metric can be put back into Hermitian form --- with only
mixed type indices. This change of variables, however, is necessarily
not holomorphic as holomorphic coordinate changes cannot affect the index
structure of a tensor. What this means is that the new metric is
Hermitian with respect to a {\it different} complex structure on $X$ ---
a new set of complex coordinates which are {\it not} holomorphic functions
of the original coordinates. Those deformations of the metric of pure type
which are associated to elements of $H^{2,1}_{\bar \partial}(X)$ 
therefore correspond to {\it deformations of the complex structure} of $X$.

Deformations of mixed type are more easily interpreted: they simply
correspond to deformations of the K\"ahler class $J$ of $X$ to a new
element of $H^{1,1}_{\bar \partial}(X)$.

We can make contact with the discussion at the beginning of this subsection
by noting --- as discussed in detail in
\cite{PC}
 --- that
deformations of the complex structure of $X$ correspond to changes in
the defining polynomial(s) $P$, preserving the requisite degree of homogeneity
requirement(s). We refer the reader to
\cite{PC}
 for details but the
idea is clear: one can define three local complex coordinates on the Calabi-Yau
by taking the defining equation in a given patch and 
solving for some of the variables.
The form of the defining equation(s) clearly affects this choice and hence plays
a central role in determining the complex structure. Since changes 
to the form of the defining equation(s) (preserving its degree and
homogeneity properties) and elements of $H^{2,1}_{\bar \partial}(X)$
are associated with deformations of the complex structure of $X$,
there is a one-to-one map between them. Additionally
harmonic $(1,1)$-forms
are associated to deformations of the K\"ahler class of the Calabi-Yau.
The parameter space of those Calabi-Yau manifolds continuously connected
to some initial one $X$ thereby consists of the possible choices of complex
and K\"ahler structures on the underlying differentiable manifold.

For the quintic hypersurface, it turns out that $h^{2,1} = 101$ and
$h^{1,1} = 1$. As $(1,1)$-forms are naturally real, this gives us
a moduli space of real dimension $2\cdot 101 + 1 = 203$.
It is not hard to derive these numbers: there are $126$ distinct
monomials in five variables giving us $126$ adjustable coefficients
in the defining equation of the quintic. Twenty-five of these can be
set to zero by $GL(5,\IC)$ coordinate transformations, leaving us
with $101$ complex structure degrees of freedom.
Now, the famous {\it index theorem}  tells us that
\beq
\label{index}
\chi(X) = \sum_{r,s} \, (-1)^{r+s}h^{r,s}~.
\eeq
Using our earlier calculation that the Euler number of the
quintic is $-200$, and our present calculation that $h^{2,1} = 101$,
we can deduce $h^{1,1}= 1$. This is nothing but the statement that
the quintic hypersurface inherits its single K\"ahler degree of
freedom from the ambient ${\Bbb C}P^4$ in which it is embedded.

For the case of $K3$, the fact that the Euler number is $24$ and that
the general constraints on Hodge numbers  determine all
but $h^{1,1}$, \calle{index} then gives:
\beq
\chi(K3) = 24 = 1 + 1 + 1 + 1 + h^{1,1}
\eeq
thereby yielding $h^{1,1} = 20$ as given earlier.

There is one other aspect of Calabi-Yau moduli space which will come
in handy later on. First, it can be shown that both the complex structure
and K\"ahler structure moduli spaces are complex K\"ahler manifolds in
their own right. A very nice discussion of this can be found in
\cite{CdlO}. Explicitly, the K\"ahler potential
on the complex structure moduli space turns out to be
$-\ln(i\int_M \Omega \wedge \bar \Omega)$, while that for the
K\"ahler moduli space is $\int_M J \wedge J \wedge J$ (for a three-fold).
The latter, we shall see, suffers quantum corrections in string theory.
Even beyond being K\"ahler manifolds, these parameter spaces turn
out to be {\it special} K\"ahler manifolds which means that there is
a holomorphic function --- called the prepotential --- which
determines the K\"ahler potential (hence the prefix {\it pre}).
Explicitly, if we call  the prepotential ${\cal F}(z)$,
then the K\"ahler potential of a special K\"ahler manifold is determined
by
\beq
\label{prepotential}
  K = i\, \left(\bar w^j {\partial{\cal F} \over \partial w^j}
 -  w^j {\partial\overline{\cal F} \over \partial \bar w^j} \right)~.
\eeq
In our later discussions, we shall see that this highly restrictive
structure is quite powerful.

\subsection{Important Invariants}

We close our discussion of Calabi-Yau three-folds by noting two
important trilinear functions on their cohomology groups.
The first is known as the triple intersection form. It
takes three elements of $H^{1,1}_{\bar \partial}(X)$ and produces
a real number:
\beq
\label{triple}
I^{1,1}: H^{1,1}_{\bar \partial}(X) \times H^{1,1}_{\bar \partial}(X)
\times H^{1,1}_{\bar \partial}(X) \rightarrow \IR~,
\eeq
via
\beq
I^{1,1}(A,B,C) = \int_X A \wedge B \wedge C~.
\eeq
It is called the triple intersection form because it can be
equally well phrased in terms of homology in which case the integral
just counts the common points of intersection of the three four-cycles, dual
to these three two-forms.
This integral is a topological invariant of $X$. The homological
description makes this most evident as the number of points of intersection
will not change under smooth deformation.

The second invariant is as follows:
\beq
\label{triplecpx}
I^{2,1}: H^{2,1}_{\bar \partial}(X) \times H^{2,1}_{\bar \partial}(X)
\times H^{2,1}_{\bar \partial}(X) \rightarrow \IC~,
\eeq
with
\beq
I^{2,1}(A,B,C) = \int_x \Omega_{ljk} \tilde A^l \wedge \tilde B^j \wedge
\tilde C^k \wedge \Omega~,
\eeq
where 
\beq
\label{iso} 
  \tilde A^l\otimes
  {{\partial \over \partial z^l}} = \Omega^{lij}\, A_{ij \bar k}\,
   d\overline z^{\bar k} \otimes
  {{\partial \over \partial z^l}}~.
\eeq

This integral is a 
{\it pseudo}-topological invariant in that it does depend on the complex
structure of $X$.

A point worth mentioning isi that in this expression,
$\tilde A$, $\tilde B$, $\tilde C$ are actually elements of
$H^1_{\overline \partial}(X,T^{(1,0)})$, holomorphic tangent 
bundle-valued cohomology on $X$. That is, we have described differential
forms as being expressed as in
\calle{realformm} with the coefficients $\omega_{i_1...i_q}$ being
numbers. In fact, there is a powerful generalization in which the
coeffecients take on other kinds of values beyond numbers. For instance,
the coefficients at some point $p$ on $X$ can be elements of the vector space
 $V_p$ which is the fiber at $p$ of a vector bundle $V$
on $X$. Taking $V$ to be a trivial line
bundle, we recover the usual notion of differential forms. This is but the
simplest choice for $V$; other, nontrivial choices can be quite important.
What we see above is that by considering differential $(0,1)$-forms whose
coefficients lie in the holomorphic tangent bundle of $X$, we are
studying a structure that happens to be isomorphic to the space
of $(2,1)$-forms. Contraction with the holomorphic three-form $\Omega$
provides the explicit map, as we see in 
\calle{iso}. Along these lines, we can also mention that harmonic $(0,1)$-forms
taking values in $(T^{(1,0)})^\star$ (which we shall often write as $T^\star$
for short) --- $H^1_{\overline \partial}(X,T^\star)$ --- are manifestly
the same as $(1,1)$-forms. The $T^\star$ ``coefficients'' can equally well be
thought of as $(1,0)$-forms with numerical coefficients, yielding the
$(1,1)$-form interpretation. Following the same line of reasoning,
$(r,s)$ cohomology classes can be expressed as $(0,r)$-forms taking
values in $\wedge^r T^\star$, 
$H_{\overline \partial}^{0,s}(X,\wedge^r T^\star)$.
 More details on bundle-valued cohomology can be found in
\cite{GHa}.

We will see both of these invariants arising in our studies of the
physics associated with these manifolds in string theory.

In the next section we change gears completely and turn to quantum field
theory, in particular, supersymmetric two-dimensional quantum field theory which
respects the superconformal algebra. After a number of twists
and turns, our quantum field theory discussion will take us
back to the Calabi-Yau manifolds which have been the subject of
this section.

\newsection{The N = 2 Superconformal Algebra} 
\label{TheN=2SCFA}

In this section, we assume the reader has some working
familiarity with the basics of string theory and
with conformal field theory. If not, a good
reference to study first is \cite{Gins}
as well as \cite{Oog-TASI,Lyk-TASI}.
 Additionally,
good references for
the material in this
section are 
\cite{LVW},
the review articles of \cite{Warner,Gepnert,Schwartz} 
and  references therein.

\subsection{The Algebra}

It has long been known that the perturbative
 consistency of string theory demands that we 
describe its ground states in terms of two-dimensional conformal field theories
of particular central charge (depending on the local symmetry algebra of the
particular string being studied). For superstrings, this central charge must be
fifteen. A typical setting for studying superstring theory is to realize a
central charge of fifteen via $M_4 \times \{ {\rm an}~N=2,~c=9$
superconformal field theory$\}$,
 where by $M_4$ we mean Minkowski space --- or
more precisely, the 
$c = 6$ superconformal field theory of four free bosons and
their superpartners. Actually, this formulation would yield a theory whose
Hilbert space is larger than that of the string; rather, we should work in light
cone gauge in which a total central charge of twelve is realized by the above
construction with $M_4$ replaced by the $c = 3$ free conformal theory of two
free (transverse) bosons and their superpartners. As discussed, the restriction
to $N = 2$ theories is not fundamentally required but it does give rise to a
number of important properties such as perturbative stability, space-time
supersymmetry and enhanced analytic control.  For these reasons, we shall focus
exclusively on $N = 2$, $c = 9$ superconformal theories. 

To begin our quantitative study of these theories, let us first write down the 
superconformal algebra. As with more familiar algebras based on compact Lie
groups, the superconformal algebra can be expressed in terms of the
(anti)commutators of its generators. Unlike the case of a compact Lie algebra,
though, there are an infinite number of generators in this case. It is
especially convenient to write the algebra in terms of the operator products of
its generators. This contains the same information as the (anti)commutators of
the modes of the generators, but for completeness we will write both.

Let us start in the more familiar setting of the $(N=0)$ conformal algebra. 
For a given conformal theory, this algebra is generated by the 
energy-momentum
tensor $T(z)$ whose operator product with itself takes the following form
\begin{eqnarray*}
     T(z)T(w) = { c/2 \over (z-w)^4} +
      {2\, T(w) \over (z-w)^2} +
        {\partial_w T(w) \over z-w} + \cdots ~,
\end{eqnarray*} 
where $c$ is the central charge of the
theory. To write this in terms of commutators, we define the modes $L_n$ of
$T(z)$ according to 
\begin{equation}
       T(z) = \sum_n\, L_n\, z^{-n-2} ~.
\label{eq21}
\end{equation} 
Then, the operator
product expansion above implies 
\begin{eqnarray}
       [L_n,L_m] &=& (n-m)\, L_{n+m} +
        {c\over 12}\,  n(n^2-1)\,  \delta_{m+n,0}  ~.
\label{eq22}
\end{eqnarray}

To extend this algebra to the $N = 1$ superconformal algebra, we include an 
additional
 generator $G(z)$ which is the (worldsheet) superpartner of the 
energy-momentum
 tensor $T(z)$ and has conformal weight $3/2$. The operator product of $T$
with itself is unchanged and the algebra is thus determined by the operator
product expansion above  and the following two equations:
\begin{eqnarray}
     T(z)G(w) &=& {3/2\over (z-w)^2} \, G(w) 
             + {\partial_w G(w)\over z-w} + \cdots~,
\label{eq23}    \\
     G(z)G(w) &=& {2\, c/3 \over (z - w)^3} + {2\, T(w) \over z - w} + 
        \cdots~ .
\label{eq24}    
\end{eqnarray}

As discussed, our main interest is in the $N = 2$ superconformal algebra. 
We get this by including two weight $3/2$ supercurrents $G^1(z)$ and $G^2(z)$.
The operator product of $T(z)$ with itself is unchanged, the operator product of
$T(z)$ with each of $G^1(w)$ and $G^2(w)$ is as in (2.3). The new product
involves the two supercurrents and takes the form 
%
\ifnoWSc
\begin{eqnarray}
        G^1(z) G^2(w) &=& {{2\, c/3 \over (z - w)^3}}
                  + { 2\, T(w) \over z - w}
                    + i\,\left( {2\, J(w) \over (z - w)^2}
                  + {\partial_w J(w) \over z - w} \right) + \cdots~.
\label{eq25}
\end{eqnarray}
\fi
Notice that on the
right hand side yet another field $J$ has appeared. It has conformal weight one
and is a $U(1)$ current. Thus, the structure of the $N = 2$ algebra demands that
in addition to $T$, $G^1$
 and $G^2$ we must introduce $J$. The additional operator
products which determine the $N = 2$ algebra are as follows: 
\begin{eqnarray}
   T(z)J(w) &=&
         {{ J(w) \over (z - w)^2}} + {\partial_w J(w) \over z - w} + \cdots
         ~,
\label{eq26}  \\
   J(z)G^1(w) &=& {i\, G^2(w) \over z - w} + \cdots~,
\label{eq27} \\
   J(z)G^2(w) &=& {-i\, G^1(w) \over z - w} + \cdots~,
\label{eq28} \\
   J(z)J(w)   &=& {{ c/3 \over (z - w)^2}}+\cdots~. 
\label{eq29}
\end{eqnarray} 
The first of these just implies that $J$ is a primary field of
weight one, and the second and third just give the charges of the supercurrents
under this $U(1)$ charge. It proves worthwhile to diagonalize the action of $J$
on the supercurrents by introducing 
$$
   G^{\pm}(z) = {{1 \over \sqrt 2}}(G^1(z)\pm i G^2(z))~.
$$
Then \calle{eq27} and \calle{eq28} become 
\begin{equation}
      J(z)G^{\pm}(w) =
      \pm\, {G^{\pm}(w) \over z - w} + \cdots~. 
\label{eq210}
\end{equation} 

The complete list of the operator products, therefore, which determine the 
$N = 2$ algebra are as follows:
\begin{eqnarray*}
    T(z)T(w) &=& { c/2 \over (z-w)^4} + {2\, T(w) \over (z-w)^2} + 
         {\partial_w T(w) \over z-w} + \cdots ~,\\
    T(z)G^{\pm}(w) &=& {3/2 \over (z-w)^2}\, G^\pm(w) +
           {\partial_w G^{\pm}(w) \over z-w} + \cdots ~,\\
    T(z)J(w) &=& {J(w) \over (z-w)^2} + {\partial_w J(w) \over z-w} + \cdots~,\\
    G^{+}(z)G^{-}(w) &=& {2\,c/3 \over (z-w)^3}
                      + {2J(w) \over (z-w)^2} +
               {2T(w) + \partial_w J(w) \over z-w} + \cdots~,\\
    J(z)G^{\pm}(w) &=& \pm\, {G^{\pm}(w) \over z-w} + \cdots~,\\
    J(z)J(w) &=&  {c/3\over (z-w)^2} + \cdots~.
\end{eqnarray*}

We can again re-express this data in terms of modes by writing, in addition to 
\calle{eq21},
\begin{eqnarray}
     J(z) &=& \sum_n\, J_n\, z^{-n-1} ~,
\label{eq211} \\
     G^{\pm}(z) &=& \sum_n\, G^{\pm}_{n \pm a}\, z^{-(n \pm a) - 3/2}~.
\label{eq212}
\end{eqnarray} 
Notice that in the latter expression we have a parameter $a$ in the mode
expansion which lies in the range $0 \le a < 1$. This parameter controls the
boundary conditions on the fermionic currents. We see this directly by allowing
$ z \rightarrow e^{2 \pi i}\, z$ as then 
\begin{equation}
    G^{\pm}(e^{2 \pi i} z) = -
               e^{\mp 2 \pi i a}\, G^{\pm}(z)~.
\label{eq213}
\end{equation}  
For our purposes, we shall always choose either
integral or half-integral values for $a$ corresponding to anti-periodic or
periodic boundary conditions. The former are usually called Ramond boundary
conditions and the latter Neveu-Schwarz boundary conditions.

In terms of these modes, the $N = 2$ superconformal algebra takes the form: 
%
\ifnoWSc
\begin{eqnarray*}
     \left[ L_m, L_n\right] & = & (m - n) L_{m+n} +
             {c\over 12}\, m (m^2 - 1)
          \,\delta_{m+n,0}~,   \\
     \left[ J_m, J_n\right] & = &{c\over 3} m\, \delta_{m+n,0}~, \\
     \left[ L_n, J_m\right] & = &-m\, J_{m+n}~,  \\
     \left[ L_n, G^{\pm}_{m \pm a}\right] & = &
              \left( {n\over 2} - (m \pm a)\right) \, G^{\pm}_{m + n\pm a}~, \\
     \left[ J_n, G^{\pm}_{m \pm a}\right] & =&\pm\, G^{\pm}_{m + n \pm a}~,\\
    \left\{G^+_{n+a}, G^-_{m-a}\right\} &= &2\, L_{m+n} +
      (n-m+2a)\, J_{n+m}
      +{c\over 3} \left( (n+a)^2 - {1 \over 4} \right)
      \delta_{m+n,0}~.
\end{eqnarray*}
\fi


On first contact, the above algebra may appear
to be an impenetrable list of commutation and anti-commutation relations.
However, we shall shortly see that by carefully studying this algebra we can
learn much about theories which respect it.

\subsection{ Representation Theory of the N = 2 Superconformal Algebra} 

As with any algebra, understanding the physical properties of the $N = 2$ 
superconformal algebra requires an understanding of its representation theory.
Thankfully, the unitary irreducible representations of this algebra can be built
up in a systematic manner using the notion of highest weight states. That is,
just as in the more familiar case of a compact Lie algebra such as 
${\frak su}\,(2)$, we
can build up representations by dividing the generators of the algebra into
``creation'' and  ``annihilation'' operators (raising and lowering operators),
finding ``ground states'' which are killed by all of the annihilation operators
(these are the so-called highest weight states) and then building up the
representation by repeatedly applying all of the creation operators to this
state.

Let us first review how this is done in the ordinary $N = 0$ case as it is a 
simple matter to extend this to the $N = 2$ case. The commutation relations
\calle{eq22} imply that we can view the $L_n$ for $n$ positive as annihilation
operators. To see this, note that we can use the eigenvalues of $L_0$ to label
states in a representation. Consider, then, a state $|s\rangle$ such that
$L_0|s\rangle = s|s\rangle$. Notice that the state $|s'\rangle = L_m|s\rangle$
is such that $L_0|s'\rangle = (s-m)|s'\rangle$.  Thus, the $L_m$ with $m$
positive lower the $L_0$ eigenvalue of a state. Assuming that this eigenvalue is
bounded below (which is reasonable as $L_0$ is essentially the left moving part
of the Hamiltonian), we expect to reach a state $|\phi\rangle$ satisfying
$L_0|\phi\rangle = h_\phi|\phi\rangle$ for some eigenvalue $h_\phi$ and such
that $L_m|\phi\rangle = 0$ for all $m > 0$. Such a state $|\phi\rangle$ is known
as a highest weight state (although it might more naturally be called a lowest
weight state). We can then build a representation of the conformal algebra by
acting on $|\phi\rangle$ with all of the creation operators, i.e. by considering
all $\prod L_{-n_j}|\phi\rangle$. Let us note that from the operator product
viewpoint, we can think of the state $|\phi\rangle$ as being built from the
action of a field $\phi(z)$ according to $|\phi\rangle = \phi(0)|0\rangle$. The
constraint that $|\phi\rangle$ be a highest weight state is equivalent to
$\phi(z)$ satisfying 
\begin{equation}
       T(z) \phi(w) = {{h_{\phi} \over (z - w)^2}} +
       {{\partial_w \phi(w) \over z-w}} + \cdots~ .
\label{eq214}
\end{equation}

By a similar procedure, we can build up highest weight representations of the 
$N = 2$ superconformal algebra. The main difference is that instead of just
having the modes $L_n$ we also have the modes $J_m$ and $G^{\pm}_r$. Thus, we
also must divide these other kinds of modes into creation and annihilation
operators. By reasoning as above, we can again think of all modes with positive
indices as annihilation operators as they lower the value of the $L_0$
eigenvalue of a state. Since it is again reasonable to take these eigenvalues to
be bounded from below, we seek highest weight states $|\phi\rangle$ satisfying
$L_n|\phi\rangle =0$, $G^{\pm}_r|\phi\rangle = 0$ and $J_m|\phi\rangle = 0$ for
$n,r,m$ positive. If we are in the NS sector, then the only zero index modes are
$L_0$ and $J_0$ whose eigenvalues we can use to label states: $L_0|\phi\rangle =
h_{\phi}|\phi\rangle$ and $J_0|\phi\rangle = Q_\phi |\phi\rangle$. Given such a
state $|\phi\rangle$ we can build up a representation by acting with all
combinations of the creation operators, that is, with modes having negative mode
numbers. If we are in the R sector then we also have to contend with $G^{\pm}_0$
modes. If a state $|\phi\rangle$ in the Ramond sector of the
 theory is annihilated
by both $G^{\pm}_0$, then we say it is in the {\it Ramond ground state}. 
Let us again point out that a highest weight state $|\phi\rangle$ as discussed
here is created by a ``superconformal primary field'' $\phi$ according to
$|\phi\rangle = \phi(0) |0\rangle$ where $\phi$ satisfies 
\begin{eqnarray}
   T(z)\phi (w) &=& {{h_{\phi} \over (z - w)^2}} + {{\partial_w \phi(w) \over
                    z-w}} \cdots~,
\label{eq215} \\
   J(z)\phi(w)&=& {{Q_{\phi} \over z-w}}\, \phi(w) + \cdots~,
\label{eq216} \\
   G^{\pm}(z) \phi(w) &=& {{\tilde \phi^{\pm}(w) \over z-w}} +
        \cdots = 
             {{(G^{\pm}_{-1/2}\phi)(w) \over z-w}} + \cdots~. 
\label{eq217}
\end{eqnarray} 
In the latter
expression $\tilde \phi^{\pm}(w) = (G^{\pm}_{-1/2}\phi)(w)$ are the
``superpartners'' of $\phi(w)$. 

\subsection{ Chiral Primary Fields}

For a number of reasons which will become clear in the sequel, it proves 
worthwhile to further constrain the notion of a primary field in an $N = 2$
theory to a subset known as {\it chiral primary fields}. Simply put, a chiral
primary field is a primary field $\phi$ that creates a state $|\phi\rangle$
which is annihilated by the operator $G^+_{-1/2}$, that is
\begin{equation}
     G^+_{-1/2}|\phi\rangle = 0~.
\label{eq218}
\end{equation} 
In the operator product language,
this implies that 
\begin{equation}
    G^+(z) \phi(w) ={\rm reg} ~,
\label{eq219}
\end{equation} 
that is, there is no
singularity in the product. We can similarly define the notion of an 
{\it antichiral
primary field} by demanding the corresponding highest weight state to be
killed by
$G^-_{-1/2}$, translating into 
\begin{equation}
     G^-(z) \phi(w)= {\rm reg}~.
\label{eq220}
\end{equation} 
Furthermore,
let us note that in all of our discussion we have suppressed the
anti-holomorphic
 side of things. We can similarly define the notions of chiral
and antichiral primary by replacing $G^{\pm}$ with $\overline G^{\pm}$. More
precisely, then, we have primary fields which are chiral in the holomorphic sense
and chiral in the anti-holomorphic sense, denoted $(c,c)$; primary fields which
are antichiral in the holomorphic sense and chiral in the anti-holomorphic
 sense
denoted $(a,c)$ and their complex conjugates $(a,a)$ and $(c,a)$. 

To gain some initial understanding of why this notion of (anti)chiral primary 
fields is of interest, we now establish two important properties of these
fields. First, there are a finite number of them in any non-degenerate $N = 2$
conformal field theory and they close upon themselves in a non-singular operator
algebra.

Establishing these results will require nothing more than a repeated 
application of the $N = 2$ superconformal algebra. To begin, consider the
anticommutator 
\begin{equation}
    \{G^-_{1/2},G^+_{-1/2}\} = 2 L_0 - J_0 
\label{eq221}
\end{equation} 
sandwiched
between a chiral primary field $|\phi\rangle$: 
\begin{equation}
   \langle\phi |\{G^-_{1/2},G^+_{-1/2}\} |\phi\rangle = \langle\phi| 2 L_0 - J_0
    |\phi\rangle~.
\label{eq222}
\end{equation} 
Since $|\phi\rangle$ is assumed chiral primary, the left hand
side vanishes and hence we learn that 
$$
   h_{\phi} = {Q_{\phi}\over 2}
$$
 for a chiral
primary field. Furthermore, note that the left hand side of 
\calle{eq222} is always
non-negative since $(G^+_{-1/2})^{\dagger} = G^-_{1/2}$. Thus, we learn that for
any state $|\psi\rangle$,
$h_{\psi} \ge Q_{\psi}/2$ and equality occurs precisely
when $\psi$ is chiral primary\footnote{Actually we have not quite proven this 
as we
have not shown that satisfying $h_{\phi} = Q_{\phi}/2$ implies that $\phi$ is
chiral primary. It is true and the details can be found in \cite{LVW}.}.
With this
simple result in hand, let us now consider the operator product of two chiral
primary fields $\phi$ and $\chi$. By dimensional analysis we can write
\begin{equation}
     \phi(z) \chi(w) = \sum_i\, (z -w)^{h_{\psi_i}-h_{\phi}-h_{\chi}}\,
      \psi_i(w) ~,
\label{eq223}
\end{equation} 
where the $\psi_i$ are fields of weight $h_{\psi_i}$. Now, we just
learned that for any field $\psi$ we have $h_{\psi} \ge Q_{\psi}/2$ with
equality holding for chiral primary fields such as $\phi$ and $\chi$. Thus,
since $U(1)$ charges add upon operator product, we have $Q_{\psi} = Q_{\phi} +
Q_{\chi}$ and hence $h_{\psi_i} \ge h_{\phi} + h_{\chi} $. From this and 
\calle{eq223}
we see that there are no singular terms in the operator product of two chiral
primary fields. Furthermore, as $z \rightarrow w$ the only terms which survive
on the right hand side of 
\calle{eq223} are those for which $\psi$ is itself chiral
primary. Thus, the chiral primary fields yield a non-singular and closed ring
under the operation of operator product. 

We can go a bit further by exploiting the anticommutator 
\begin{equation}
       \{G^-_{3/2},G^+_{-3/2}\} = 2L_0 - 3J_0 + {2\over 3}\, c~.
\label{eq224}
\end{equation}
From this we
see, by sandwiching it between a chiral primary field,  that the conformal
weight of any chiral primary field is bounded above by $c/6$. Thus, for a
non-degenerate theory we see that we have a finite number of chiral primary
fields. This finiteness characteristic together with the simple ring structure
just discussed and, furthermore, the fact that much of the conformal field
theory is encoded in properties of the chiral primary fields accounts for their
great utility in studying $N = 2$ conformal theories.

As we mentioned before, in our discussion we have actually neglected three 
quarters of the story: we have focused on chiral primary fields in the
holomorphic sector and found a natural ring structure. In fact the analysis
clearly extends to antichiral fields and to fields in the anti-holomorphic
sector and hence we have four rings: the $(c,c), (a,c), (a,a)$ and $(c,a)$
rings. The latter two are complex conjugates of the first two. In terms of
charge eigenvalues, an antichiral primary field has $h_{\psi} = - Q_{\psi}/2$.

\subsection{ Spectral Flow and the U(1) Projection} 

We have seen that there is a free parameter, called ``$a$'' in the $N = 2$ 
\def\SCA{superconformal algebra} \SCA. In this section, we discuss the
mathematical and physical significance of this parameter. 

We have seen in equation (2.12) that the mode numbers of the supercurrents 
$G^{\pm}$ are written in the form $m \pm a$ with $m$ an integer and $a \in
[0,1)$. Mathematically, the value of $a$, as we have seen, specifies the
boundary conditions on the supercurrents. When $a$ is $0$ (or, equivalently, an
integer), $G^{\pm}$ are anti-periodic giving us the Ramond sector;
 when $a = {{1
\over 2}}$ (or, equivalently, half integral) $G^{\pm}$ are periodic giving
us the Neveu-Schwarz sector. 

At first glance, every choice of $a$ determines a different algebra with 
(slightly) different commutation relations. In fact, however, all of the
algebras parameterized by different values of $a$ are {\it isomorphic}. 

To see this, it proves convenient to introduce slightly different notation. 
Let us write 
$$
  G^+_{m+a} = G^+_{m + {1 \over 2} + \eta}~,
  ~~~~~
  G^-_{m-a} = G^+_{m - {1 \over 2} - \eta}~,
$$ 
i.e. we are setting 
\beq
  a \equiv  \eta + {{1 \over 2}}~.
\eeq
 It
also proves worthwhile
to note that the commutator $[L_n, G^{\pm}_{m\pm a}]$ can
be written 
$$
  [L_n, G^{\pm}_s] = ({{n \over 2}} - s) G^{\pm}_{n+s}
$$ 
for any
relevant choice of $s$ (integral or half-integral).

With these notational conventions, let us now define: 
\begin{eqnarray}
   L'_n &\equiv&  L_n + \eta J_n + {c \over 6} 
        \,\eta^2 \,  \delta_{n,0}~, \nonumber \\ 
   J'_n &\equiv& J_n + {c \over 3}\, \eta \, \delta_{n,0}~, 
   \label{eq225}   \\
       G^{\prime\pm}_r &\equiv&  G^{\pm}_{r \pm \eta} ~. \nonumber
\end{eqnarray}  
Notice that the primed objects are simply particular linear combinations of the
$N = 2$ generators for a particular choice of $\eta$. Therefore, the algebra
generated by the primed objects is isomorphic to that generated by the unprimed
objects, and the latter is the $N = 2$ algebra for a given value of $\eta$. We
claim that the algebra generated by the former, i.e. the primed generators, is
precisely that of the $N = 2$ algebra for $\eta = 0$. If this is true, we would
therefore have succeeded in showing that the algebras for any choice of $\eta$
are all isomorphic. 

Showing that the primed objects generate the $N = 2$ algebra for $\eta = 0$ can 
be accomplished by brute force calculation. We will do two steps and leave the
others for the reader.

Consider
\begin{eqnarray}
    [L'_n, L'_m] = [L_n + \eta J_n + {\eta^2 \over 6}c 
     \,\delta_{n,0}  ~ , ~
       L_m + \eta J_m + {\eta^2 \over 6}\,c\,\delta_{m,0}]~.
\label{eq226}
\end{eqnarray}
By using the $N = 2$ commutation relations, this equals 
\ifnoWSc
\begin{eqnarray}
   [L'_n, L'_m]=
  (n - m)\, L_{m+n} +  \eta\, (n-m)\, J_{n+m}
   +{c \over 12}\, n (n^2 - 1)\, \delta_{m + n,0}
    + {\eta^2 \over 3}\, c\, n\, \delta_{m+n,0}~.
\label{eq227}
\end{eqnarray}
\fi
Collecting terms,
we have 
%
\ifnoWSc
\begin{eqnarray}
  [L'_n, L'_m] = (n - m)(L_{n+m} + \eta J_{n+m} +
      {\eta^2 \over 6}\,c\,
     \delta_{m+n,0})
     + {c \over 12}\, n\, (n^2 - 1)\, \delta_{m+n,0}~.
\label{eq228}
\end{eqnarray}
\fi
This we see is
just 
\begin{equation}
   [L'_n, L'_m]=
      (n - m)\, L'_{m+n} + {c \over 12}\, n (n^2 - 1)\,
      \delta_{n+m,0}
\label{eq229}
\end{equation} 
and hence the $L_n'$ commute in the correct way to generate the
first relation of the $N = 2$ \SCA.

Of particular interest is what occurs when we consider 
$[L'_n, G^{\prime\pm}_r]$. 
Again, by explicit calculation we find
\begin{eqnarray}
    [L'_n, G^{\prime\pm}_r] = 
    ({n \over 2} - r)\, G^{\pm}_{r + n \pm\eta}
   =({n \over 2} - r)\, G^{\prime\pm}_{r + n}~.
\label{eq230}
\end{eqnarray}
Notice that the latter is the appropriate commutation relation for the $N = 2$ 
\SCA\ with $\eta = 0$. In fact, carrying on in this way one finds that all of
the (anti)commutation relations of the $N = 2$ \SCA\ with $\eta = 0$ are
satisfied by the primed generators. Thus, the unprimed generators which naively
yield a family of different algebras labeled by $\eta$, in fact yield algebras
isomorphic to each other and to the $N = 2$ \SCA\ with $\eta = 0$. Hence, we see
that although $\eta$ determines different boundary conditions on the
supercurrents, the resulting algebras are all isomorphic. 

Now, let us try to
take this one step further. Since the algebras for arbitrary choices of $\eta$ 
are isomorphic, and we have the explicit isomorphism in hand, we should be able
to extend this isomorphism to representations of the algebra as well. Namely,
given an (infinite) collection of states $|f\rangle$ that provide a
representation of the $N = 2$ \SCA\ with, say, $\eta = 0$, we should be able to
construct an (isomorphic) collection of states $|f_{\eta}\rangle$ that
constitute a representation of the $N = 2$ algebra for non-zero $\eta$. In the
language of linear algebra, let $U_{\eta}$ be the unitary  map, which on the
level of operators satisfies 
\begin{eqnarray*}
      L'_n &=& U_{\eta} L_n U_{\eta}^{-1}~, \\ 
      J'_n &=& U_{\eta} J_n U_{\eta}^{-1}~, 
\end{eqnarray*} 
and similarly for the other
generators. Then, at the level of states in the representation of the algebra,
\begin{equation}
    |f_{\eta}\rangle = U_{\eta}\, |f\rangle~.
\label{eq231}
\end{equation} 
Such a map is commonly
referred to as {\it spectral flow} by an amount $\eta$. 

Our goal, here, is to explicitly understand how this map $U_{\eta}$ acts on 
states. From 
\calle{eq225} above, we can immediately learn some simple properties of
this action. Namely, what is the new conformal weight $h_{\eta}$ and the new
$U(1)$ charge $q_{\eta}$ of the new state $|f_{\eta}\rangle$? This is easy to
work out. Let $h_{\eta}$ and $q_{\eta}$ be defined with respect to the $\eta =
0$ generators since it is with respect to the action of these generators on the
original state $|f\rangle$ that we wish to compare. Namely, 
\begin{eqnarray*}
    L_0|f_{\eta}\rangle & =& h_{\eta} |f_{\eta}\rangle~, \\
    J_0 |f_{\eta}\rangle & =&
     q_{\eta} |f_{\eta}\rangle~. 
\end{eqnarray*} 
Now, to determine $h_{\eta}$ and $q_{\eta}$ we note that
\begin{eqnarray*}
     L_0' |f_{\eta}\rangle & =& U_{\eta} L_0 U_{\eta}^{-1} U_{\eta}
         |f\rangle \nonumber\\
      &=& h |f_{\eta}\rangle~,
\end{eqnarray*}
and
\begin{eqnarray*}
    J_0' |f_{\eta}\rangle & =& U_{\eta} J_0 U_{\eta}^{-1} U_{\eta} 
    |f\rangle\\
          & = &q |f_{\eta}\rangle~,
\end{eqnarray*}
where $h$ and $q$ are the conformal weight and $U(1)$ charge of
$|f\rangle$.
 Now, using the explicit isomorphism of \calle{eq225}, we also have 
\begin{equation}
    L_0'|f_{\eta}\rangle = 
    (h_\eta + \eta\, q_\eta + {\eta^2 \over 6}\, c)
     \, |f_{\eta}\rangle~,
\label{eq232}
\end{equation} 
and 
\begin{equation}
     J_0' |f_{\eta}\rangle = (q_{\eta} + {\eta \over 3}
    \, c) |f_{\eta}\rangle~.
\label{eq233}
\end{equation} 
Solving for $h_{\eta}$ and $q_{\eta}$ we have
\begin{equation}
   \begin{array}{l}
      h_{\eta} = h - \eta\, q + {c \over 6}\, \eta^2  ~, \\
      q_{\eta} = q - {c \over 3} \eta ~.
   \end{array}
\label{eq234}
\end{equation} 

There is a subtle point to this discussion which we should emphasize. In 
general, when we have an isomorphism between two algebras, as above, it would
not even make sense to act with the original generators (the unprimed
generators) on the new representation (the $|f_{\eta}\rangle$). However, in this
particular case, changing the value of $\eta$ does not have any effect on the
$L_n$ or $J_n$ generators and hence it {\it does} make sense to have them acting
on both the original and the new representation. Connected with this, however,
let us also note that changing the value of $\eta$ does directly effect the modes
of the supercurrents. For them, therefore, it {\it only} makes sense to have the
non-zero $\eta$ modes of $G^{\pm}$ acting on the $|f_{\eta}\rangle$. A good way
to think about this is to note that, in general, a state $|s\rangle$ is said to
lie in the $\eta$-twisted sector if the corresponding operator $s(z)$ which
creates the state from the vacuum has an operator product expansion with
$G^{\pm}(w)$ involving terms of the form $(w - z)^{\rm integer \pm \eta}$.
Having an operator product expansion of this form implies that {\it we must}
impose $\eta$-twisted boundary conditions on the $G^{\pm}$, as in 
\calle{eq213}, in
order to have an overall single-valued operator product between $s$ and
$G^{\pm}$. Conversely, given a set of states on which the $\eta$-twisted modes
of $G^{\pm}$ act in a single value manner, we say that these states are in the
$\eta$-twisted sector. Thus, the states $|f_{\eta}\rangle$, that we have been
discussing, lie in the $\eta$-twisted sector.

Can we be more explicit about the relationship between the states $|f\rangle$ 
and $|f_{\eta}\rangle$? The answer to this is yes, and now let us see this. 

To do so, let us first note that the $U(1)$ current in the $N = 2$ \SCA\ can be 
bosonized and written in the form
\begin{equation}
     J(z) = i {\sqrt {{c \over 3}}}\, \partial_z \phi~,
\label{eq235}
\end{equation} 
where $\phi$ is a
free scalar boson. Then, any field $f$ which creates the state $|f\rangle$ with
$U(1)$ charge $q$  can be written as
\begin{equation}
     f = \hat f\,  e^{i q {\sqrt{{ 3 \over c}}} \phi} ~,
\label{eq236}
\end{equation} 
where $\hat f$ is
a neutral field. One immediately checks that the operator product of 
\calle{eq236}
with 
\calle{eq235} does imply that 
\calle{eq236} has charge $q$. To go the other way and
prove that {\it any} state of charge $q$ can be so written requires a bit more
work that we shall not cover here. 

Now, consider a field $f$ of charge $q$
which creates the state $|f\rangle$ in the $\eta = 0$ sector. The claim is that 
the field $f_{\eta}$ which creates the state $|f_{\eta}\rangle$ in the
$\eta$-twisted sector can be explicitly written as
\begin{equation}
     f_{\eta}(z) = \hat f(z)\, 
  e^{i\sqrt{3\over c}(q -{c\over 3} \eta)\phi}~.
\label{eq237}
\end{equation} 
Let's see if this checks with our previous analysis.
First, let us examine the conformal weight and $U(1)$ charge of such a state. By
construction, 
\begin{equation}
    q_{f_\eta} = q_f - {c \over 3}\, \eta~.
\label{eq238}
\end{equation} 
By direct
calculation we also see that 
\begin{equation}
    h_f = h_{\hat f} + { 1 \over 2} \, \left(q
       \sqrt{{ 3 \over c}}\right)^2
\label{eq239}
\end{equation} 
and 
\begin{equation}
     h_{f_\eta} = h_{\hat f} + { 1 \over 2}\,
       \left(q \sqrt{{ 3 \over c}} - \eta \sqrt{{ c \over 3}}\right)^2~.
\label{eq240}
\end{equation} 
In these expressions,
to avoid confusion, we have subscripted the eigenvalues in an obvious manner. We
see that these relations are precisely those found earlier 
\calle{eq234}. 

Furthermore, we claim that $f_{\eta}$ so defined, does create states in the 
$\eta$-twisted sector. To see this, let us write 
\begin{equation}
     G^{\pm}(z) = \hat
      G^{\pm}(z)\, e^{\pm i {\sqrt {{ 3 \over c}}} \phi(z) } 
\label{eq241}
\end{equation} 
and then use this form to
calculate the operator product $G^{\pm}(z) f_{\eta}(w)$.  From the operator
product of the bosonic exponentials, we see that we pick up  a factor involving
$(z - w)^{\pm \eta}$, with factors coming from the other terms being 
single-valued. Thus, this fits in precisely with our definition of being in the
$\eta$-twisted sector.

Hence, the field $f_{\eta}$ that we have constructed does indeed create a state 
$|f_{\eta}\rangle$ in the $\eta$-twisted sector with the correct conformal
weight and $U(1)$ charge to be identified with the state $U_{\eta} |f\rangle$.
From the form of \calle{eq237} we can read off, therefore, that
\begin{equation}
       U_{\eta} = e^{-i {\sqrt{{ c \over 3}}} \eta \phi}~.
\label{eq242}
\end{equation} 
Spectral
flow, therefore, is accomplished by shifting the bosonic exponential. 

There are a number of reasons why spectral flow plays an important role in 
$N = 2$ theories. Amongst these, probably the most important is that the GSO
projection and modular invariance, as is well known, requires that we include
both the Neveu-Schwarz sector ($\eta = 0$) and the Ramond sector ($\eta = 1/2$)
in the Hilbert space of our theory. We see now that given one of these, we can
determine the other by the operation of spectral flow by $1/2$ unit. Now, as the
NS sector gives rise to space-time bosons and the R sector gives rise to
space-time fermions, we see that spectral flow has a space-time interpretation as
the supersymmetry operator. We see, quite directly therefore, the relationship
between $N = 2$ supersymmetry and space-time supersymmetry as the two are linked
by the operation of spectral flow by $1/2$ unit. We note that the image of a
chiral primary state under spectral flow by $1/2$ unit yields a state that is
annihilated by $G^{\pm}_0$ --- that is,  a {\it Ramond ground state}. If we flow
by another half unit we get an antichiral primary field.

In fact, we can take this discussion one step further. Since we have naturally 
been led to build a space-time supersymmetry operator from $U_{\eta}$ for $\eta =
1/2$, we expect to only have a well defined theory if $U_{1/2}$ is (at worst)
{\it semi-local} with respect to all states in theory. That is, the
supersymmetry operator should create a square root branch cut on the world sheet
thereby exchanging space-time bosons and fermions.  Now, from 
\calle{eq236} we see
that an arbitrary field $f$ will have such an operator product expansion with
$U_{1/2}$ if its $U(1)$ charge $q$ is an odd integer. Thus, our discussion
appears to lead us to the conclusion that space-time supersymmetry will ensue if
we project our theory, (in the sense of conformal field theory quotients) onto
one with odd integral $U(1)$ charges. In fact this is true, as has been
established by more careful investigations \cite{SUSY1,SUSY2,SUSY3}.
  One important
point not to overlook is that this restriction on $U(1)$
charges is on the {\it whole} theory including the internal and the
four-dimensional part. The total central charge, in light cone gauge (so that we
do not need to discuss the effects of ghosts) is $12$. Hence, when we spectral
flow by $\eta = 1/2$ the charge of, say, a NS state is shifted by $c/6 = 2$.
Thus, if the original state has odd integral charge, so does its image in the R
sector. Furthermore, let us note that this projection onto odd integral $U(1)$
charge can be thought of as occurring in two steps. First, we can project onto
integral $U(1)$ charges. Then we can perform what amounts to a generalized GSO
projection onto odd integral charges. In much of the following, we will only
explicitly be concerned with the first of these projections and we will not
write down the second step of the GSO projection. To actually carry out this
projection amounts to orbifolding the given $N = 2$ theory by the operator $e^{2
\pi i J_0}$ (and by $e^{2 \pi i \bar J_0}$) \cite{Vafaorb}. 

In fact, we will be studying
theories in which the left and right moving charges differ by an integer and 
hence it suffices to quotient by either of these. As shown in \cite{Vafaorb}
together with \cite{SUSY1,SUSY2,SUSY3}
 if the central charge is a multiple of three, such a
projection (followed the GSO projection) does indeed yield the desired space-time
supersymmetric theory.

To summarize this section, we note that the $N = 2$ \SCA\ is really a family of 
algebras labeled by a parameter $\eta$ which determines the boundary conditions
on the supercurrents $G^{\pm}(z)$. For $\eta \ne 0$ we say that we are working
in the $\eta$-twisted $N = 2$ \SCA. All of these algebras are actually
isomorphic to one another, by an explicit isomorphism given above. Since they
are isomorphic, we can map a representation of the algebra for some chosen value
of $\eta$ (say $\eta = 0$) to another representation of the algebra for a
different value of $\eta$. The latter representation is said to lie in the
$\eta$-twisted sector. We have explicitly seen that the latter map is simply
multiplication (in the sense of operator product) by $e^{-i {\sqrt{{ c \over
3}}} \eta \phi}$ where $J(z) = i {\sqrt {{ c \over 3}} } \partial \phi$. The new
states are only local with respect to the $\eta$-twisted $N = 2$ \SCA\
generators, i.e. $L_n$, $J_n$ and $G^{\pm}_{r \pm \eta}$. Hence, these states do
indeed lie in the $\eta$-twisted sector. For the special case of $\eta = 1/2$ we
see that spectral flow takes us from NS states to R states and hence has the
earmark of a space-time supersymmetry operator. By suitably restricting the
$U(1)$ charges of states in the theory (in the precise manner described above
and henceforth called the $U(1)$ {\it projection}),
 this observation can be borne
out and hence one has a well defined procedure for building space-time
supersymmetric theories from such $N = 2$ superconformal models.

\subsection{Four Examples}
\label{FourExamples}

In this section we would like to make the previous abstract discussion more 
concrete by introducing four examples of theories which possess $N = 2$
superconformal symmetry.
 Each of these theories plays an important role in our
discussion and hence are interesting in their own right. 

\vspace{.2in} 
\noindent
{\it Example 1: Free Field Theory}

\vspace{.13in}
The simplest example we can write down is that of free field theory. 
So, consider a free two dimensional theory of a single complex boson $X = X^1 +
iX^2$ and a free complex fermion $\Psi = \Psi^1 + i\Psi^2$. As $\Psi$ is meant
to be the superpartner of $X$ and we are familiar that the equations of motion
for $X$ show that it splits into the sum of a left moving (holomorphic) and
right moving (anti-holomorphic) part, we can more precisely introduce a left
moving complex fermion $\psi(z)$ (with complex conjugate $\psi^*(z)$, and a
right moving fermion $\lambda(\overline z)$ (with complex conjugate
$\lambda^*(\overline z)$. The action for this theory can be written in the
familiar form (focusing just on the holomorphic part) 
\begin{equation}
      S = \int d^2z\, ( \partial X \overline \partial X^* +
      \psi^*\overline \partial \psi + \psi \overline \partial \psi^*
      +\lambda^*\partial \lambda + \lambda \partial \lambda^*)~.
\label{eq243}
\end{equation} 
Our claim is
that this theory has $N = 2$ superconformal symmetry. In fact, more precisely,
it has this symmetry in both the holomorphic and anti-holomorphic sectors 
and
hence has what is usually denoted $(2,2)$ worldsheet supersymmetry. Their are
many ways to see this, the most direct of which is to construct the generators
of the $N = 2$ algebra directly from the fields defining the theory. Thus, the
reader is urged to check that 
\begin{eqnarray}
      T(z) &=& -\partial X \partial X^* + {{1\over 2}} 
                \psi^* \partial \psi +
             {{1 \over 2}} \psi \partial \psi^*  ~,
\label{eq244}  \\
      G^+(z) &=& {{1 \over 2}} \psi^* \partial X   ~,
\label{eq245}  \\
      G^-(z) &=& {{1 \over 2}} \psi \partial X^*  ~,
\label{eq246}   \\
      J(z)   &=& {{1 \over 4}} \psi^*  \psi 
\label{eq247}
\end{eqnarray} 
do in fact have the correct $N= 2$
superconformal operator products given earlier.

This theory has central charge 3 (in both the holomorphic and the
anti-holomorphic 
sectors) coming from two bosonic degrees of freedom ($c=2$) and two fermionic
degrees of freedom ($c=1$). Let us explicitly work out the various chiral rings.
As discussed, quite generally, a field satisfying $h=\pm Q/2$ is a chiral or
antichiral primary. Notice that the fields $\psi,\psi^*, \lambda$ and
$\lambda^*$ all satisfy this relation (on both the holomorphic and
anti-holomorphic sides). 
Furthermore, appropriate products such as $\psi
\lambda$ which has $(h,\overline h) = (1/2,1/2)$ and $(Q,\overline Q) = (1,1)$
satisfies $h = Q/2$,$\overline h = \overline Q/2$ and hence is also a  (chiral,
 chiral)$=(c,c)$ ring field. Bearing in
mind that no  (chiral, chiral) field has conformal 
weights $(h,\overline
h)$ greater than $(c/6,\overline c/6)$ (where $c$ here is the central charge,
equal to three in this case), we see that we have exhausted fully the $(\rm
chiral,\rm chiral)$ ring. Hence the  
(chiral,chiral) in this example
consists of $\{1,\psi,\lambda, \psi \lambda\}$. Similarly the $({\rm
antichiral,chiral}) = (a,c)$ ring consists of $\{1,\psi^*,\lambda, \psi^*
\lambda\}$, and the other two rings are gotten from these by complex
conjugation.

Although a very simple theory, this example does play a key role in string 
theory.
Namely, we build string theory with four extended dimensions, as discussed, by 
the construction
$ M_4 \times \{ c = 9, N = 2$ conformal theory$\}$ where $M_4$ really 
refers to a $c = 6$, $N = 2$ free superconformal theory (free because of the
restriction to flat space-time).  In light cone gauge, the latter becomes a $c
=3$, $N = 2$ free theory --- that is, the theory just constructed. Hence, our
example corresponds to the extended part of space-time in string theory.

\vspace{.2in} 
\noindent
{\it Example 2: Non-linear Sigma Models} 

\vspace{.13in}

 Our second
example is that of an $N = 2$ superconformal non-linear sigma model. In reality,
this is nothing but a simple, yet rich, generalization of the previous free
field theory example. Namely, we include more bosonic fields, more fermionic
fields (the partners to the bosons) and we no longer require the theory to be
free. Rather, we imagine that the bosons are coordinates on a ``target space''
which might be a curved Riemannian manifold $M$
 with non-trivial metric. (In the
previous case, one can think of $X$ as a coordinate on the flat manifold 
${\Bbb C}^1$
with trivial Euclidean metric.) The fermions can then be viewed as sections of
the (pullback) of the tangent bundle of the target space. Concretely, we can
write the action for such a theory as  
%
\ifnoWSc
\begin{eqnarray}
  S ={ 1 \over 4 \pi \alpha'}
     \int_{\Sigma}\,d^2z\, \Bigg\lbrack
    {1 \over 2} g_{\mu \nu}(X) \partial_z X^{\mu}
     \partial_{\overline z}X^{\nu}
    +
     g_{\mu \nu}\,( \psi^{\mu}D_{\overline z}
    \psi^{\nu}
   + \lambda^{\mu} D_z \lambda^{\nu})
      +{1 \over 4}
      R_{\mu \nu \rho \sigma} \psi^{\mu} \psi^{\nu}
    \lambda^{\rho} \lambda^{\sigma} \Bigg\rbrack ~.~~
\label{eq248}
\end{eqnarray}
\fi
where $g_{\mu \nu}$ is the metric on the target manifold and
$R_{\mu \nu \rho \sigma}$ is its Riemann tensor. (We note that we have not
explicitly written the term involving the antisymmetric tensor field.) 
 We now set about to determine
under what conditions this theory has $(2,2)$ superconformal symmetry. Let us
begin with $(2,2)$ supersymmetry. In general, this theory does not possess this
symmetry, but as shown in \cite{Zumino} if the target manifold is {\it a complex
 K\"ahler} manifold then it does.
The easiest way to see this is to note that the action 
\calle{eq248} can be explicitly
written in a superspace formalism. As this is discussed in the lectures in
\cite{Distler}, we will not bother to do this here in detail but simply write 
the $N =2$ superspace version of {\calle{eq248}:} 
\begin{equation}
    S = { 1 \over 4 \pi^2 \alpha'}
      \int\, d^2 z d^4 \theta\, K(X^i,X^{\overline\jmath})   ~, 
\end{equation} 
where the $X^i$ are
chiral superfields whose lowest components are the bosonic coordinates above and
$$
 g_{i \overline\jmath} = {{ 2 i \over \pi}}{{\partial^2 K \over \partial X^i
 \partial X^{\overline\jmath}}}~.
$$

What about conformal invariance?
In general the action 
\calle{eq248} is not conformally invariant. A direct way to see 
this is to calculate the $\beta$ function for the metric $g$ viewing it as a
coupling ``constant'' in this two-dimensional theory. The well known result
(ignoring the dilaton and antisymmetric tensor fields) is that the $\beta$
function, to lowest order, is proportional to the Ricci tensor of the target
manifold. Thus, we can achieve conformal invariance by choosing our target
manifold to have a metric with vanishing Ricci tensor. This is a highly
restrictive constraint. Our conclusion is that $(2,2)$ superconformal symmetry
implies that the target manifold must be complex, K\"ahler and admit a metric of
vanishing Ricci tensor. These conditions should
ring a bell: they  are the defining properties of {\it
Calabi-Yau manifolds} as discussed in the 
first chapter. Thus, a non-linear sigma
model with a Calabi-Yau target
space gives us another method of building $(2,2)$ superconformal field theories.

How do we construct the $(c,c)$ and $(a,c)$
fields in these theories? The answer to 
this is quite beautiful and goes back to the work of Witten in his paper
\cite{Wittenmof}. As we will
now discuss, these fields are closely associated with
the cohomology groups on the
target Calabi-Yau space.

To understand this result, we will approach it in the manner employed in 
\cite{Wittenmof} taking into account that our target space is a complex K\"ahler
manifold.  As emphasized by Witten \cite{Wittenmof}, states which have non-zero
momentum in such a theory necessarily have non-zero energy. Thus, in our effort
to understand the zero-energy modes --- that is, Ramond ground states from which
we can get the $(c,c)$ and $(a,c)$ rings by spectral flow --- we should restrict
attention to zero momentum modes. The latter modes are those which have no
spatial dependence on the worldsheet. Hence, even before we quantize the theory,
we can simply drop the spatial dependence of the fields in the action, thereby
effectively reducing our model to supersymmetric quantum mechanics on a K\"ahler
manifold.

To analyze this theory,
first note that the Majorana-Weyl fermions in our action, restricted to their 
constant mode components
(assuming that the fermions are periodic, i.e. in the Ramond sector in the 
common string parlance) satisfy:
\begin{equation}
     \{\psi^i,\psi^j\} = \{\psi^{\overline\imath},\psi^{\overline\jmath}\} =
     0~;~~~~~
      \{\psi^i,\psi^{\overline\jmath}\} = g^{i \overline\jmath} ~,
\label{eq250}
\end{equation} 
and similarly for the
$\lambda$ fermions. We see, therefore, that the K\"ahler structure supplies us
with a natural polarization that allows us to think of, say, the
$\psi^{\overline j}$ as creation operators and the $\psi^j$ as destruction
operators. For ease of present notation, we will take the opposite convention
for the $\lambda$ fermions. Namely, we will take $\lambda^j$ to be creation
operators and $\lambda^{\overline\jmath}$
to be destruction operators. We will come
back to this point shortly.

Consider now
the world sheet supersymmetry operator $Q$. More precisely, since we have 
$(2,2)$ world sheet supersymmetry, there are two left moving and two right
moving supersymmetry operators $Q_{L,1}$, $Q_{L,2}$,
$Q_{R,1}$, $Q_{R,2}$. We will focus
on the left moving operators for most of our discussion. In terms of fields, we
have to lowest order, 
\beq
   Q_{L,1} = g_{\mu \overline \nu}\, \psi^{\mu} \partial_z
   X^{\overline \nu}~, 
   ~~~~~
   Q_{L,2} = g_{\overline\mu\nu} \, \psi^{\overline \mu} \partial_z X^{\nu}~.
\eeq
 In 
the $N = 2$ language we have developed, these two operators arise from taking
the contour integral of the worldsheet supercurrents $G^+(z)$ and $G^-(z)$ and
hence we write $Q_{L,1} = G_0^+$ and $Q_{R,1} = G_0^-$. Also, $\partial_z
X^{\overline \nu}$ is $\pi^{\overline \nu}$, the momentum conjugate to
$X^{\overline \nu}$, and hence can be thought of as the functional derivative 
${{\delta \over \delta X^{\overline \nu} }}$. When restricted to the zero mode
sector this becomes the ordinary (covariant) derivative, $\cal D_{\overline \nu}
$ with respect to $X^{\overline \nu} $. Thus, in this zero mode approximation we
have 
\begin{equation} 
     G_0^- = \psi^{\overline \nu} {\cal D}_{\overline \nu} 
     ~,~~~~~
     G_0^+ = \psi^{\nu} {\cal D}_{\nu} ~.
\label{eq251}
\end{equation} 

Now, carrying on with our interpretation of the Fermi zero modes in terms of 
creation and annihilation operators, let us choose a Fock vacuum $|0\rangle$ for
our zero mode sector of the Hilbert space of states such that 
\begin{equation}
      \psi^i|0\rangle = \lambda^{\overline\imath}|0\rangle = 0~.
\label{eq253}
\end{equation} 
Then, a general state can
be written 
\begin{equation}
    |\Phi\rangle = \sum_{r,s} b_{i_1...i_r\overline
    \jmath_1...\overline \jmath_s} 
   \lambda^{i_1} ... \lambda^{i_r} \psi^{\overline \jmath_1}
    \dots
     \psi^{\overline \jmath_s}  |0\rangle   ~,
\label{eq254}
\end{equation} 
where we also sum over all repeated indices.
Let us focus our attention on the building blocks of such states, namely those
with fixed values of the integers $r$ and $s$. Such a state has $U(1)_{\rm L}
\times U(1)_{\rm R}$ charges $(-r,s)$. 

Because of the anti-commutation properties of these Fermi operators, this state 
is completely antisymmetric under interchange of any two holomorphic, or any two
anti-holomorphic indices.
Therefore, we see that the space of such states is 
{\it isomorphic} to the space of $(r,s)$-forms 
on $M$. Now, as in previous sections,
let us demand that this state be annihilated by the two supercharges, i.e. by
$G_0^-$ and $G_0^+$. Such a state, as we have described, will lie in the Ramond
ground state and hence will be related to $(c,c)$ and $(a,c)$
 fields by appropriate
spectral flow.  From 
\calle{eq251} (and using the anti-commutation relations
of the Fermi fields) we see that the former, acting on such states, is
isomorphic to the operator $\overline \partial$ acting on the corresponding
differential form, and the latter, acting on such states, is isormorphic to the
operator $\overline \partial^{\dagger}$ acting on the corresponding differential
form. Thus, demanding these operators annihilate the state is mathematically
equivalent to finding {\it harmonic} $(r,s)$-forms on $M$. Therefore, we have
explicitly shown that the Ramond ground states in such theories are in one to
one correspondence with the elements of cohomology on $M$.

We have not yet completed our analysis, as 
the reader may have noted, because of the arbitrary choice made in 
\calle{eq253}. 
Namely, which operators are going to be interpreted in terms of creation vs.
destruction operators. When deciding, say, between
 $\psi^i$ and $\psi^{\overline\imath}$,
either choice is equivalent; it is just a matter of convention. However,
after making such a conventional choice, the distinction between $\lambda^i$ and
$\lambda^{\overline\imath}$ is now one of content. Thus, in addition to 
\calle{eq253}, we
should also consider 
\begin{equation}
   \psi^i|0\rangle = \lambda^i|0\rangle = 0~.
\label{eq255}
\end{equation}
Then, we consider states of the form 
\begin{equation} 
    |\Phi\rangle = \sum_{r,s} \,
    b^{i_1...i_r}_{\overline\jmath_1...\overline\jmath_s} \, \lambda_{i_1} 
  \dots  \lambda_{i_r} \,
    \psi^{\overline\jmath_1} \dots \psi^{\overline\jmath_s}  |0\rangle ~,
\label{eq256}
\end{equation} 
where $\lambda_i =
g_{i \overline\jmath} \lambda^{\overline\jmath}.$
 These states have $U(1)_L \times
U(1)_R$ charges $(r,s)$. The same analysis as before shows these states to be in
one to one correspondence with $(0,s)$ forms taking values in $\wedge^r T$
where $T$ is the holomorphic tangent bundle of $M$.
(We note that earlier we wrote this as $TM^{(1,0)}$; for ease of notation
we shall henceforth not explicitly write the manifold $M$.)
 Now, applying the conditions
that the supercharges annihilate such a state shows it to be harmonic, that is,
a member of the Dolbeault cohomology group $H_{\overline \partial}^{0,s}(M,
\wedge^r T)$. To make the notation symmetric, the $(r,s)$-forms we previously
found can be thought of as lying in the cohomology group $H_{\overline
\partial}^{0,s}(M, \wedge^r T^\star)$, where $T^\star$
is the holomorphic cotangent
bundle.

In our subsequent discussion of these models, we shall be focusing our 
attention on certain subsets of the $(c,c)$ and $(a,c)$
rings. More specifically, we
shall be looking at elements whose left and right charges have equal absolute
values. The reader should check that the 
$(c,c)$ states of this form arise, after
spectral flow, from $H_{\overline \partial}^{0,p}(M, \wedge^p T)$ and the 
$(a,c)$
states similarly arise from $H_{\overline \partial}^{0,p}(M, \wedge^p
T^\star)$, for $p
= 0,1,2,3$. 

Let us also note that any element in
$H_{\overline \partial}^{0,s}$
$(M, \wedge^r T)$ can be associated with a 
harmonic $(3-r,s)$-form on $M$ via 
contraction with the holomorphic $(3,0)$-form
$\Omega$ which all Calabi-Yau manifolds have by virtue of the triviality of the
canonical bundle.

An important and interesting question is to not only address the geometrical 
realization of the (anti)chiral primary fields, as we have done, but also to
understand the geometrical interpretation of the ring structure amongst these
fields. We will not discuss this now, but will return to it later. 

\vspace{.2in}
\noindent
{\it Example 3: Landau-Ginzburg Models} 

\vspace{.13in} 

Landau-Ginzburg effective
field theories have played a prominent role in many areas of physics. Using some
simple reasoning, we shall shortly see that they can be put to great use in the
present setting. First, let us see how one can use \LG\ theory to construct $N =
2$ superconformally invariant field theories.

Our basic ingredient is a field theoretic realization of the chiral and 
antichiral representations of the $N = 2$ superconformal algebra. Towards this
end, we introduce differential operator realizations of the $N = 2$
supersymmetry generators $G^{\pm}_{-1/2}$ and $\overline G^{\pm}_{-1/2}$. We do
this in a standard superspace formalism via  
\begin{equation}
        {\cal D }_{\pm} =
    {{\partial \over \partial \theta^{\pm}}} + \theta^{\mp} {{\partial \over
         \partial z}} 
\label{eq257}
\end{equation} 
and
similarly for its complex conjugate. In this representation, a chiral superfield
$\Phi=\Phi(z,\overline z,$ $\theta^{\pm}$, $\overline\theta^{\pm})$ is one that
satisfies 
\begin{equation} 
       {\cal D}_+ \Phi = \overline {\cal D}_+ \Phi = 0~.
\label{eq258}
\end{equation}

In terms of fields of this type, we can build an $N = 2$ supersymmetric 
quantum field theory by taking an action of the form 
%
\ifnoWSc
\begin{eqnarray}
    S &=& \int d^2 z d^4 \theta\,  K(\Phi_1,\overline \Phi_1,...,\Phi_n,
          \overline \Phi_n)
     +\Big( \int d^2 z d^2 \theta \, W(\Phi_1,...,\Phi_n) + {\rm h.c.}
     \Big)~.
\label{eq259}
\end{eqnarray}
\fi
In analogy with \LG\ effective scalar field theories, we can refer to a theory 
of this form as an $N = 2$ supersymmetric \LG\ theory. 

In order to generate a conformally invariant theory, we can follow 
renormalization group ideas 
 and allow the theory to
evolve by the renormalization group equations, hopefully encountering an
infrared fixed point. The fixed point theory does not further evolve with
changes in scale and hence is our desired conformally invariant theory. Of
course, one needs to prove that there is a fixed point under such a
renormalization group flow; this is a hard issue to address as it really is a
non-perturbative question. We will simply assume that there is a fixed point and
deduce the consequences. We will find that the delicate consistency and far
reaching nature of these ensuing consequences, lends strong support for the
veracity of our initial assumption. 

An important property of the renormalization group flow, which can be 
established at least at the level of perturbation theory, is that the
superpotential is not renormalized under such renormalization group 
flows\footnote{Actually, a more correct statement 
is that the only renormalization
suffered by the superpotential arises from wavefunction renormalization. If the
superpotential is quasi-homogeneous, as we have required, this renormalization
is absorbed by an overall rescaling that in effect leaves the superpotential
unchanged.}. On the other hand, the kinetic term in \calle{eq259} in general
undergoes substantial renormalization along the flow towards the conformally
invariant fixed point. Given this state of affairs, we see that we can {\it
label} various renormalization group trajectories by the superpotential of the
corresponding model. In this way, for instance, we can gain headway on the
classification of $N = 2$ conformally invariant theories as we have identified a
simple renormalization group invariant describing models of this type.

How do we generate the $(c,c)$ and $(a,c)$ fields in such a theory? As shown in 
\cite{LVW}, the $(c,c)$ ring is obtained from the Jacobian ring
$$
  {\IC[\Phi_1,...,\Phi_n]\over \partial_{\Phi_j} W(\Phi_1,...,\Phi_n) }~.
$$
Concretely,
this is all polynomials in the chiral fields modulo relations of the form
$\partial_{\Phi_j} W = 0$. For instance, 
if we have a theory with a single field
$\Phi$ appearing in $W$ to the $n$-th power, the $(c,c)$ ring has elements $\{1,
\Phi, \Phi^2, ..., \Phi^{n-2} \}$. In \cite{LVW} it is also shown that the
$(a,c)$
ring in such theories is trivial, consisting only of the identity element.
Later, we shall discuss orbifolds of these theories for which both the 
$(c,c)$ and
$(a,c)$ rings are non-trivial. 

\vspace{.2in} 
\noindent
{\it Example 4: Minimal Models} 

\vspace{.13in} 

In the context of
non-supersymmetric ($N=0$) conformal field theory, it is well known that the
necessary conditions for unitary highest weight representations of the Virasoro
algebra constrain the values of the central charge and the conformal weights of
the primary fields of the theory as follows: 
\begin{equation} 
       c \ge 1~,~~~ h \ge 0~,
\end{equation}
or 
\begin{eqnarray} 
         c &=& 1 - {{ 6 \over m(m+1) }}~, \\ 
        h_{p,q}(m) &=& {{ \lbrack (m + 1)p -
          mq\rbrack^2 - 1 \over 4m(m+1) }}~,
\end{eqnarray} 
where
\beq
      1 \le p \le m-1~,~~~ 1 \le q \le p~, ~~~m\ge 3 ~.
\eeq

In the latter case, these conditions are sufficient as well as necessary and 
the corresponding theories, labeled by the value of $m$, are called {\it minimal
models}. Notice that there are a finite number of primary fields; this is in
contrast to the fact that for $c > 1$ we necessarily have an infinite number of
primary fields. This fact, in conjunction with the conformal Ward identities,
allows for the complete and explicit solution of these theories. That is, we can
explicitly calculate correlation functions in these theories.

For the supersymmetric cases, the situation is quite similar. For either 
$N = 1$ or $N = 2$ we again have an infinite sequence of unitary theories,
labeled by a positive integer $m$, which have a finite number of primary fields
(with respect to the extended chiral algebra which includes the supercurrents).
These theories are also called minimal models. For $N = 1$, the minimal models
have 
\beq
  c = {3 \over 2}\, \left(1 - {8\over (m + 2)(m + 4)} \right)~,
\eeq
while for $N = 2$,
they have central charge 
\beq
   c = {3m \over m + 2}~.
\eeq
 We will return to a
discussion of the $N = 2$ minimal models later in these lectures.

\newsection{Families of N = 2 Theories}
\label{FamiliesOfN=2Theories}

\subsection{Marginal Operators}

In two dimensional field theory, an operator of conformal weight 
$(h, \bar{h})$ is said to be 
irrelevant if $h + \bar{h} > 2$,
relevant if $h + \bar{h} < 2$ and
marginal if $h + \bar{h} = 2$.
This terminology arises from studying what happens if we deform a given 
original theory by such operators, and then allow the renormalization group 
to act on the theory, driving it towards the infrared.
 If the operator has $h + \bar{h} > 2$, it will have no
 effect on the theory at its infrared fixed point --- the RG drives its
coefficient to zero. It is like adding a 
nonrenormalizable higher dimension term in four dimensions.  Such operators, by
dimensional analysis, are suppressed by some energy scale $E_0$ and hence
 at low
$E$ (the infrared limit) they are suppressed. If $h + \bar{h} < 2$, such an
operator, via the 
above reasoning can be dominant and have a significant effect on
the properties of the theory in the infrared limit --- in fact, they can make
the theory trivial in the infrared limit. 

Of most interest to our present study are operators with $h + \bar{h} = 2$, and 
in particular, we will study spinless operators with $h = \bar{h} = 1$. These
operators can deform a given conformal field theory to a ``nearby conformal
field theory''
 of the same central charge
and thereby generate a {\it family} of conformal field theories all continuously
related to one another. 

\begin{figure}[htbp]
\epsfxsize=3in
\centerline{\epsfbox{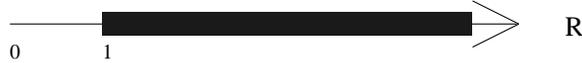}} 
\caption{The moduli space of conformal theory on a circle.}
\label{fig1}
\end{figure}

The simplest example of this is the case of $N=0,~c=1$ 
conformal field theory which can be 
realized, say, by a
free boson on a circle of radius $R_0$:
$$
  S_{R_0} = \int d^2z\, \partial X \bar{\partial}X
$$ with $ X \sim X + 2 
\pi R_0.$
\noindent Now, consider the (1,1) operator ${\cal{O}} = \partial X 
\bar{\partial}X$. We can deform $S_{R_0}$ by this operator: 
\begin{equation}
    S_{R_0} \rightarrow S_R= S_{R_0} + \epsilon \int d^2 z\,
    {\cal{O}} (z,\bar{z})~.
\end{equation} 
This is just
\begin{equation}
  S_R = (1+\epsilon)\, \int d^2 z\, \partial X \bar{\partial} X~.
\end{equation} 
Now, letting
$\tilde{X} = \sqrt{1+\epsilon} X $ this becomes 
\begin{equation}
        S_R  = \int d^2z\, \partial \tilde{X} \bar{\partial} \tilde{X}~,
\end{equation} 
with
$\tilde{X} \sim \tilde{X} + 2 \pi {R_0} (\sqrt{1 + \epsilon})$. 
So, the marginal operator ${\cal{O}} = \partial X \bar{\partial} X$ 
has the effect of changing the radius of the target space circle. The family of
conformal field theories  thereby generated consists of $c=1$
conformal field theories
of a free boson on a circle with arbitrary radius $R$. In fact, it 
can be shown that theories with radius $R$ and $1/R$ are isomorphic
 \cite{GPR} so need only
consider $R>1$.) 
The moduli space of such theories is  seen in figure \ref{fig1}.

 Actually, one should note that in an arbitrary conformal field 
theory not all
(1,1) operators can be used to deform the theory in the manner described. 
Rather, some (1,1) operators cease to be (1,1) operators after perturbation of
the original theory. Such operators can therefore not be used to move around the
moduli space as they spoil the conformal symmetry. The collection of operators
which continue to be (1,1) after perturbation by any other (in the collection)
are said to be {\it truly marginal} and it is these operators upon which we
focus our attention. 

By the superconformal
Ward Identities
\cite{Dixon,DG} it can be shown that
among these operators in the theories we study are:

\begin{enumerate}

\item
Let $\phi \in (c,c)$ ring with $h = \bar{h} = 1/2$, $Q = \bar{Q} = 1$. 
Then define $\hat{\phi}$ by 
\beq
  \hat{\phi} (w,\bar{w})\equiv\oint\, dz\, G^- (z) \phi (w,\bar{w}) ~.
\eeq
$\hat{\phi}$ has $ h= 1/2 + 1/2 = 1$, $\bar{h} = 1/2$; $Q=0$ and $\bar{Q} = 1.$
Then let 
\beq
  \Phi_{(1,1)} (w, \bar{w}) \equiv \oint d \bar{z}\, \bar{G}^- (\bar{z}) 
  \hat{\phi} (w,\bar{w})~.
\eeq
$\Phi_{(1,1)}$ has $h= \bar{h} = 1$, $Q= \bar{Q} = 0$ and
is a truly marginal operator. 

\item
Let $\phi \in (a,c)$ ring with $h = \bar{h} = 1/2$ and $Q=-\bar{Q} = 1$.
Following the above, let 
\beq
  \hat{\phi} (w,\bar{w})\equiv\oint\, dz\, {\overline G}^-(\bar{z}) 
  \phi(w,\bar{w})~,
\eeq
and
\beq
  \Phi_{(-1,1)} \equiv \oint\, dz\, G^+ (z) \hat{\phi}(w,\bar{w})
  = (G^+_{-1/2}{\overline G}^-_{-1/2} \phi)(w, \bar{w})
 ~.
\eeq
$\Phi_{(-1,1)}$ has $h= \bar{h} = 1$, $Q=\overline Q = 0$ and 
it is a truly marginal operator.

\end{enumerate}

We will focus on these truly marginal operators $\Phi_{(1,1)}$, 
$\Phi_{(-1,1)}$ 
and their (lower component) superpartners in the $(c,c)$ and $(a,c)$ rings.

\subsection{Moduli Spaces}
\label{sec:moduliI}

\begin{figure}[htbp]
\ifnoWSc\epsfxsize=6cm\fi 
\centerline{\epsfbox{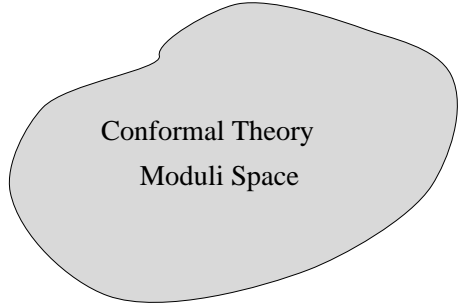}} 
\caption{Schematic drawing of the Moduli Space of an $N = 2$ theory.}
\label{fig2}
\end{figure}

In analogy with our drawing in figure 
\ref{fig1} for the one-dimensional moduli space 
for a free boson on a circle,
 with the radius of this circle being parameterized
by the moduli space variable $R$, we can draw the 
multi-dimensional moduli space
for a continuously connected family of $N=2$ superconformal field theories,
schematically, as in \ref{fig2}.

\begin{figure}[htbp]
\epsfxsize=7cm 
\centerline{\epsfbox{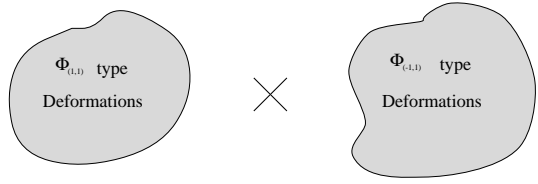}} 
\caption{The two sectors of an $N = 2$ moduli space.}
\label{fig3}
\end{figure}

Each point in this moduli space corresponds to one $N = 2$ superconformal 
theory. We can
deform any given theory into another by using the truly marginal operators. 
It can be shown that, at least locally, the conformal field
theory  Zamolodchikov metric
on this moduli space is block diagonal between the $\Phi_{(1, 1)}$-type and 
$\Phi_{(-1,1)}$-type marginal operators and hence we can think of moduli space
as being a metric product (at least locally) of the schematic form shown
in figure \ref{fig3}.

\noindent One of our main goals is to fully understand this picture in detail, 
especially when we are starting from an initial theory realized as an $N=2$
superconformal sigma model on some Calabi-Yau space.  So, to begin
imagine we build an $N=2$ superconformal field theory  as a non-linear sigma
model on a \CY\ target space $M$. Can we give a geometrical interpretation to the
two types of marginal operators (that we expect to exist if $c>3$ so
$h_{\rm{max}} = c/6 > 1/2$)?

The answer to this question comes immediately from the association between
$(c,c)$ and $(a,c)$ fields and harmonic differential forms. That is,
we already described the way in 
which
$(c,c)$ fields with $(h=1/2,\overline h = 1/2)$
 correspond to harmonic $(2,1)$-forms and how
$(a,c)$ fields with 
$(h=1/2,\overline h = 1/2)$ correspond to harmonic $(1,1)$-forms. 
This essentially establishes the geomtrically meaning of the abstract
conformal field theory operators, when the latter is constructed
as a nonlinear sigma model. To see this in a bit more detail,
let us look at the corresponding $\Phi_{(-1,1)}$ marginal operator. To lowest 
order, the $(a,c)$ field
can be written as
$b_{i \bar\jmath}\lambda^i \psi^{\bar\jmath}$
 with $b_{i \bar\jmath}$ being a harmonic 
$(1,1)$-form on $M$. Using the map between $(a,c)$ fields and marginal 
operators
given above, we map this to $ G^+_{-1/2} \bar{G}^-_{-1/2}$
$ (b_{i \bar\jmath}
\lambda^i \psi^{\bar\jmath})$. To lowest order again, this is 
$ b_{i \bar\jmath}
\partial X^i \bar{\partial} X^{\bar\jmath}$. 
This operator, as in the case of the circle, deforms the ``size'' of $M$, 
i.e., the K\"ahler form $ig_{i \bar\jmath} d X^i \wedge d X^{\bar\jmath}$
 on $M$. Hence, these conformal field theory marginal operators
correspond to geometrical deformations of the K\"ahler structure on the
Calabi-Yau manifold.
Similarly, the $\Phi_{(1,1)}$, which give pure-index type
metric perturbations, deform the ``shape'' 
of $M$ --- i.e.,
the complex structure of $M$. 
The notion of deforming the complex structure and K\"ahler structure of
a Calabi-Yau manifold, while still preserving the Calabi-Yau conditions,
thus has a direct conformal field theory manifestation in terms of
deformation by truly marginal operators.

For those readers less comfortable with the mathematics of such 
deformations,
the following figure illustrates their effect in the case of a
one-dimensional Calabi-Yau manifold, namely the torus. The original torus is
drawn with a solid line and its deformed versions are drawn with dotted lines. A
K\"ahler deformation
 leaves the shape (the angle between the cycles) fixed, but
changes the volume. A complex structure deformation leaves the volume fixed, but
changes the shape. 

\begin{figure}[htbp]
\epsfxsize=10cm
\centerline{\epsfbox{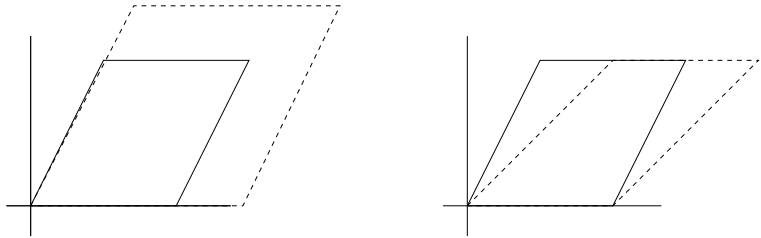}}
\caption{ K\"ahler and Complex structure deformations of a torus.}
\label{fig4}
\end{figure}

\begin{figure}[htbp]
\epsfxsize=7cm 
\centerline{\epsfbox{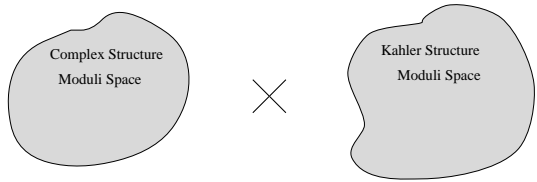}} 
\caption{Geometrical interpretation of the conformal field theory
         moduli space.}
\label{fig5}
\end{figure}

We have thus found that corresponding to the two types of operators that 
can deform our $N = 2$ superconformal theory to a nearby theory, there are two
geometrical operations that deform a Calabi-Yau manifold to a nearby manifold,
without spoiling the Calabi-Yau conditions. This gives the geometrical
interpretation of figure \ref{fig2} in the form given by figure 
\ref{fig5}. 

This is the geometrical interpretation of $N= 2$ superconformal field theory 
moduli space that
was subscribed to  for some time.

We will return to this picture repeatedly and see that the geometrical 
counterpart to the conformal field theory moduli space is {\it deficient} in a
number of ways. In other words, figure \ref{fig5}
 is quite incomplete as it presently
stands.  Eradicating these deficiencies will lead us to a number of 
remarkable
mathematical and physical features. 

As a presage to that discussion, let us note one fact: The distinction 
between the $\Phi_{(1,1)}$ and $\Phi_{(-1,1)}$ type marginal operators, at the
level of conformal field theory  is rather trivial, being just the
sign of a $U(1)$ charge. The distinction between their geometrical
counterparts --- harmonic (1,1)-forms and harmonic (2,1)-forms --- is
comparatively dramatic. These two types of forms are mathematically quite
different. This is a strange asymmetry to which we shall return.

\newsection{Interrelations Between Various N = 2 Superconformal Theories} 
\label{InterrelationsBetweenSCFT} 

At the end of section 
\ref{TheN=2SCFA}  we described three general ways of constructing 
$N = 2$ superconformal field theories: 
non-linear sigma models, Landau-Ginzburg
theories and minimal models.
Our goal now is to try to understand the relationship between these theories. 
We will proceed by finding pairwise connections. 

\subsection{Landau-Ginzburg Theories and Minimal Models}

Consider a Landau-Ginzburg theory with a single chiral superfield $\Psi$:

\begin{eqnarray*}
    S = \int d^2 z d^4 \theta\,  K (\Psi,\bar{\Psi}) +
        \left( \int d^2 z d^2 \theta\,  W 
      (\Psi) + c.c.\right)~.
\end{eqnarray*}
Take $W(\Psi) = \Psi^{P+2}$. In \cite{LVW} the authors calculate the central 
charge $c_P$ of this Landau-Ginzburg theory at its infrared fixed point to be 
$$
     c_P = 6 \left({1\over 2} - {1\over P+2}\right) = {3P\over P+2}~.
$$

Recall that the unitary $c<3,~N=2$ superconformal theories {\it are} the minimal 
models. Thus, we learn that a Landau-Ginzburg theory of this particular form is,
at its conformally invariant infrared fixed point (assuming such a point exists)
the $P$-th minimal model, $MM_P$. (In fact, to be a bit more precise, it is
the $P$-th minimal model with the so called A-modular invariant.) For more
information supporting this statement of equivalence and emphasis on subtleties
which arise, the reader should consult \cite{WittenLG}.  Thus, the
Landau-Ginzburg theories with
the simplest possible superpotential gives the Lagrangian realization of the
minimal models. 

\subsection{Minimal Models And Calabi-Yau Manifolds: A Conjectured
Correspondence} 

Conformally invariant non-linear sigma models with \CY\ target spaces have 
central charge $c = 3d$, where $d$ is the complex dimension of the \CY\ space.
As we have discussed, the $MM_P$ have central charges
 $c = {{3 P \over P + 2}} <
3$. Thus, for $ d > 1$ it would not seem that there could be any connection
between these types of conformal theories. However, given a collection of $r$
conformal theories with central charges $c_i,~i = 1,...,r$, one can build a new
conformal field theory --- {\it the tensor product theory} --- with central
charge $c = \sum_{i = 1}^r c_i$. The Hilbert space of this theory is the tensor
product of the Hilbert spaces of the constituent models and the energy-momentum
tensor takes the form 
\begin{equation}
  T = \sum_{i = 1}^r\, 1 \otimes \cdots \otimes 
  1 \otimes T_i \otimes 1\otimes \cdots \otimes 1~.
\end{equation} 
In fact, since
the operation of orbifolding by a finite discrete group does not change the
central charge of a conformal theory, a quotient of the above tensor product
will also have central charge $c = \sum_{i = 1}^r c_i$. Applying this to the
minimal models, we see that if we choose a collection of integers $P_i, i =
1,...,r$ such that 
$$
    \sum_{i=1}^r {{ 3 P_i \over P_i + 2}} = 3\, d~,
$$
then the
tensor product of these conformal theories and orbifolds thereof will have
central charge $3d$. In principle, then, there might be some relationship
between the $MM_P$ combined in this manner with \CY\ sigma models.

The first evidence that there is such a relationship was found by Gepner
\cite{GEP,GEP2}.
As we discussed,
it has long been known that a string
theory of the form we are studying, $M_4 \times\{ {\rm an}\ N = 2, ~c=9\ 
{\rm superconformal\ field\ theory\ }  \}$,
has space-time supersymmetry if and only if the superconformal theory has odd 
integral $U(1)_L$ and $U(1)_R$ charge eigenvalues. Following this lead, Gepner
considered a tensor product of minimal models with $\sum_{i=1}^r {{ 3 P_i \over
P_i + 2}} = 9$ orbifolded onto a spectrum with odd integral $U(1)$ charges. That
is, he considered $[MM_{P_1} \times \cdots \times MM_{P_r}] |_{\rm U(1)\ 
projected}$, which we denote by $(P_1,...,P_r)$. Gepner then compared the symmetries
and massless space-time spectra  of a particular case $(P_1,P_2,P_3,P_4,P_5)$ 
$=$
$(3, 3, 3, 3, 3)$  with that of the best studied
 \CY\ sigma model with $c = 9$ given
by the vanishing locus of $z_1^5 + z_2^5 + ... + z_5^5$ in 
$\IC P^4$. He found
these data to be essentially identical (up to some additional massless particles
in the minimal model formulation which were expected to generically become
massive under small perturbations). It was further shown in \cite{DG} that the
Yukawa couplings (whose form we will discuss) as computed in the minimal model
formulation and in the \CY\ formulation for this and a couple of other examples
also agreed. These results gave additional support to Gepner's conjecture that
the minimal model construction yields conformal theories interpretable as
non-linear sigma models on particular \CY\ manifolds. Again, this is quite
surprising as the minimal model formulation does not appear to have any
geometrical content.

This conjecture was initially put on firmer foundation in the works of 
\cite{GVW,Mart,rmany2}
 and more thoroughly \cite{WP}.  Each of these gives a procedure
for identifying which \CY\ manifold should correspond to a given minimal model
 construction; the paper of \cite{GVW} gives a heuristic path integral argument
establishing a direct link between the two types of constructions and \cite{WP}
provides a rigorous argument uncovering a rich phase structure (also found in
\cite{AGM}). 

We now briefly review these
arguments establishing a link between (orbifolds of) tensor product minimal 
model constructions and \CY\ sigma models. 

\subsection{Arguments Establishing Minimal-Model/Calabi-Yau Correspondence} 
\label{sec43}

Each of the papers \cite{GVW,Mart,rmany2,WP} makes use of the accepted
isomorphism  between
the minimal model theory at level $P$ and the \LG\ theory of a single chiral 
superfield\footnote{For ease of notation, we now call such a chiral
superfield
$X$ rather than $\Psi$ or $\Phi$.}
$X$  with superpotential $W = X^{P+2}$, described
above. In particular, since Hamiltonians (and hence Lagrangians) of tensor
product theories add, we immediately learn from our previous discussion that
$\otimes_{j=1}^r MM_j$ is isomorphic to the conformally invariant \LG\ model
with action  
\begin{eqnarray}
  S=\int d^2z d^4 \theta\, \sum_{j=1}^r K_j(X_j,\overline
     X_j) + 
    \left( \int d^2z d^2 \theta\, \sum_{j=1}^r X_j^{P_j+2} + h.c.\right)~.
\label{eq42}
\end{eqnarray}  
We reemphasize
that the explicit form of the kinetic terms consistent with conformal invariance
can not generally be written down, but should be thought of as being determined
by the fixed points of a renormalization group flow. 

We now specialize to the case of $r = d + 2$. A simple but crucial point
\cite{GVW}  to note is that the minimal model condition of 
$$
   \sum_{j=1}^{d + 2}{{
       3 P_j \over P_j + 2}} = 3 d
$$
 implies that $D$, the least common multiple of the
$P_j + 2$, satisfies
$$
  D = \sum_{j=1}^{d + 2}{{ D \over P_j + 2}} ~.
$$
This implies
that if we interpret the superpotential $W$ in 
\calle{eq42} to be an equation
$$
   \sum_{j=1}^{d+2}\, X_j^{P_j+2} = 0
$$
in
the weighted projective space
 $W\IC P^{d+1}(D \omega_1 \cdots D \omega_{d+2})$
(although nothing justifies this interpretation as yet), with
$$
     \omega_i \equiv {{1 \over P_i + 2}} ~,
$$
 then it is well defined (homogeneous of
degree $D$) and its  locus satisfies the condition that $D$ equals the
sum of the weights of the projective space. Using our earlier discussion
of classical geometry, we can see that this ensures
that the resulting space is
Calabi-Yau. To see this, we note that the weighted projective
space generalization of
\calle{chernfour} is
\beq
\label{chernfourwt}
   c(W{\Bbb C} P^{4}(w_1,w_2,w_3,w_4,w_5)) = \prod_{i=1}^5
  \, (1 + w_i J)~.
\eeq
Then, the generalization of
\calle{quintic} is
\beq
\label{CYwt}
   c(TW) = {{\prod (1 + w_i J) \over (1 + DJ)}} 
  = 1 + (\sum w_i - D)J + \cdots~.
\eeq
Vanishing of the first Chern
class thus requires the degree $D$ to be the sum
of the weights $w_i$ of the weighted projective space.
At the moment, though, nothing in our discussion justifies ascribing such an
interpretation to $W$. Following \cite{GVW} we can, however, give at least a
heuristic justification.

We consider \calle{eq42} and note that
since the K\"ahler potential is irrelevant we may choose it at will, and in 
particular we may choose it very small. To a first approximation, in fact, we
may ignore it. The path integral representing the partition function of the
theory now becomes
\def\d{{\cal D}}
\begin{eqnarray}
  \int\left[\d X_1\right] \ldots \left[\d X_r\right] \,
  e^{i\int\,d^2 \!z d^2 \!  \theta \, (W(X_1,\ldots, X_r) + {\rm c.c.})}~. 
\label{eq43}
\end{eqnarray}
In a patch of field space in which $X_1 \ne 0$, we can rewrite the path 
integral \calle{eq43} in terms of new variables 
\begin{equation}
  \xi_1^{\omega_1} = X_1~;
  ~~~~~ \xi_i = {X_i \over \xi_1^{\omega_i}}
    ~.
\label{eq44}
\end{equation}
Since
\begin{eqnarray}
   W(X_1, \ldots ,X_r) \equiv W'(\xi_1, \ldots ,\xi_r) =
       \xi_1 \,
        W'(1,\xi_2,\ldots , \xi_r)~,
\label{eq45}
\end{eqnarray}
\calle{eq43} becomes
\begin{eqnarray}
   \int\left[\d\xi_1\right] \ldots \left[\d\xi_r\right] \,
     J\, e^{i \int d^2 \! z 
  d^2 \! \theta \,\xi_1 (W'(1,\xi_2,\ldots, \xi_r) + {\rm c.c.})}~,
\label{eq46}
\end{eqnarray}
where $J$
is the Jacobian for the change of variables, and is given by 
\begin{equation} 
    J = \xi_1^\nu~,~~~~~
   \nu = 1 - \sum_{i = 1}^r \omega_i ~. 
\label{eq47}
\end{equation} 
If 
$\nu = 0$,
then $J$ is a constant, and the integration over $\xi_1$ in 
\calle{eq46} yields a 
delta function
$\delta(W'(1,\xi_2, \ldots, \xi_r))$,
constraining the remaining fields to lie on the 
manifold\footnote{We have 
ignored the fermionic components of the superfields here. Carrying them through
the calculation yields a delta function constraining them to lie tangent to the
manifold parameterized by the bosonic coordinates, as expected by
supersymmetry.} $W=0$.
 Proceeding in like fashion with the other fields, we can cover
field space with patches, in each of which we obtain a similar result. We note,
though, that the change of variables we have used to simplify the path integral
is not one-to-one. In fact, upon inspection we see that $\xi_i$ are invariant
under the transformation 
\begin{equation}
     X_i \to e^{2 \pi i \omega_i}\, X_i ~.
\label{eq48}
\end{equation} 
The
$X_i$ are thus naturally interpreted as homogeneous coordinates on 
$W\IC P^4($ $D\omega_1$,
 $\cdots$, $D\omega_{d+2})$.
 However, because of this invariance, the model we
have shown to be equivalent to propagation on the manifold $W = 0$ in this
projective space is not the theory 
\calle{eq42}
 but rather the quotient of this by the
transformation 
\calle{eq48}. Since the charge of $X_i$ is $\omega_i$, this is
precisely the quotient by $g_0 = e^{2 \pi i J_0}$ required to obtain a
consistent (space-time supersymmetric) string vacuum \cite{GEP,GEP2,Vafaorb},
 as we
discussed in section \ref{TheN=2SCFA}. 
Many properties of the resulting model may be extracted
from the superpotential alone, as discussed in \cite{LVW}. An important role in
this equivalence was played by the fact that the Jacobian for the transformation
to homogeneous coordinates was constant. For $r = 5$ this is simply the
condition
that the central charge $c = 9$. And, as we have seen,
 this is  precisely the condition
that the hypersurface $W = 0$ have vanishing first Chern class \cite{GVW}.

There are a few relevant remarks we should make. 

\begin{enumerate}
\item
 The connection between Landau-Ginzburg
 theory and \CY\ sigma models has been most easily 
achieved by first deforming away from the conformally invariant theories (by
manipulating the K\"ahler form), finding an isomorphism and then allowing the
renormalization group to appropriately adjust kinetic terms to reestablish
conformal invariance. Nothing in our argument assures that the resulting kinetic
term in the \CY\ sigma model will be sufficiently ``large'' to ensure that sigma
model perturbation theory will be valid. Hence, such a \CY\ sigma model would
only truly be defined via analytic continuation. This is precisely what happens
in the more rigorous approach to relating \LG\ theories to \CY\ sigma models
discussed in the next section.

\item
We have made a number of specializations in our discussions. First, we have 
focused attention on the case of $r$, the number of minimal models, being equal
to $d + 2$. Second, we have restricted attention to the A-invariants. Although
we shall not discuss it in detail here, both of those specializations can be
substantially relaxed.

\item
Although compelling, the argument of \cite{GVW} has some obvious deficiencies. 
Paramount amongst these is the manipulation of the kinetic energy terms in the
\LG\ action. As shown in the appendix of \cite{GVW} one can work with a more
conventional kinetic term, although the argument does become a bit cumbersome
and delicate.

\end{enumerate}

Finally, we note that subsequent to the
above arguments,  Witten  reexamined this correspondence and found a more 
robust and satisfying argument which also points out a number of important
subtleties. We now briefly
 review this approach. Our discussion will be limited
to the simplest possible case of a Calabi-Yau hypersurface with $h^{1,1} = 1$,
i.e. the quintic hypersurface. More general complete intersections are treated
in \cite{WP} and from the viewpoint of toric geometry in \cite{AG}.

Witten begins with neither the Calabi-Yau
sigma model nor the Landau-Ginsburg model.
Rather, he starts with an $N = 2$ supersymmetric gauge theory (called the 
``linear sigma model'') with gauge group $U(1)$ and action
\begin{equation}
 S = S_{\rm kinetic} + S_{W} + S_{\rm gauge} + S_{\rm FI-D\ term}~,
\label{eq49}
\end{equation}
 where the only terms whose precise form we need to explicitly write are
$S_W$ and $S_{\rm FI-D\ term}$. In particular,
\begin{equation} 
S_W = \int d^2z d^2 \theta\,  W(P,S_1,...,S_5) ~,
\label{eq410}
\end{equation} 
where $W$ is the
superpotential of the theory, $P,S_1,...,S_5$ are chiral superfields whose
$U(1)$ charges are $-5, 1,...,1$ respectively, and $W$ takes the $U(1)$
invariant form 
\begin{equation} 
    W = P\, G(S_1,...,S_5)~.
\label{eq411}
\end{equation} 
$G$ is a homogeneous transverse
quintic in the $S_i$. $S_{\rm FI-D\ term}$  is the supersymmetric
Fayet-Iliopoulos $D$-term. The important point for our present discussion is
that the bosonic potential of the theory takes the form 
%
\ifnoWSc
\begin{eqnarray}
   U =
  |G(s_i)|^2 + |p|^2 \sum_i\left|
  {\partial G \over \partial s_i} \right|^2 + {1 \over 2
   e^2}\, D^2 
     + 2|\sigma|^2\,\left( \sum_i\, |s_i|^2 + 25 |p|^2\right) ~,
\label{eq412}
\end{eqnarray} 
\fi
with
\begin{equation} 
   D = -e^2\, \left( \sum_i\, |s_i|^2 - 5 |p|^2 - r\right)~.
\label{eq413}
\end{equation} 
In this expression
lower case letters represent scalar components of the corresponding capital
letter superfields and $\sigma$ is a scalar field coming from the twisted chiral
multiplet whose presence is quite important but shall not play a central role in
our discussion.

 Our goal is to study the classical ground states of this theory
for various choices of the parameter $r$. 
It turns out that there are two qualitatively different answers depending upon 
the sign of $r$. Let us study the two possibilities in turn. 

First, let us take $r > 0$. Minimizing the $D$-term in $U$ then implies that not 
all $s_i$ can vanish. Assuming $G$ to be a transverse quintic polynomial then
implies that not all $|{{\partial G \over \partial s_i}}|$ vanish and hence
minimizing $U$ forces $p = 0$. Furthermore, with at least one $s_i$ non-zero we
learn that $\sigma = 0$ and finally, minimizing $U$ further implies $G = 0$ and
\begin{equation}
    \sum_i\, |s_i|^2 = r~.
\label{eq414}
\end{equation} 
We are not quite done in our identification of
the ground state because not all such configurations are distinct due to the
gauge symmetry of the model. Rather, we have the $U(1)$ identifications
\begin{equation}
   (s_1,...,s_5) \sim (e^{i \theta} s_1,...,e^{i \theta} s_5)~.
\label{eq415}
\end{equation}
 How can we
picture the meaning of the conditions we have found? At first sight, the fields
$s_1,...,s_5$ live in $\IC^5$. The combined constraints 
\calle{eq414} and \calle{eq415},
however, take us from $\IC^5$ to $\IC P^4$. This is nothing but the statements
from classical geometry embodied in \calle{symplectic1} and \calle{symplectic2}.
By way of review,
 note that the
equivalence relation of projective space $(z_1,...,z_5) \sim \lambda
(z_1,...,z_5)$ can be used to pick out one representative of each class. We can
do this by enforcing conditions on the coordinates which uniquely pick out a
value of $\lambda$ for a given choice of initial coordinates $(z_1,...,z_5)$.
Notice that 
\calle{eq414} and \calle{eq415}
 do precisely this and hence allow us to interpret
the $s_i$ as living in $\IC P^4$. \def\gt{>} Thus, the other condition of $G =
0$ yields the vanishing of a quintic polynomial in $\IC P^4$ --- the familiar
quintic Calabi-Yau hypersurface. Thus, for $r \gt 0$ the fields are constrained
to lie on this Calabi-Yau manifold and hence our original $U(1)$ gauge theory
reduces to this Calabi-Yau sigma model. We note, as discussed in \cite{WP} that
the original Lagrangian has other fields not present in the Calabi-Yau sigma
model whose masses are determined by the value of $r$. Thus, for $r \gt \gt 0$,
these fields play no role and hence in this limit we actually are recovering the
Calabi-Yau manifold not only as the ground state, but also as governing the
effective quantum field theory. In the infrared limit any non-zero mass field
drops out and hence in the conformal limit we as well regain the conformally
invariant non-linear sigma model. 

Let us also note that $r$ determines the ``size'' of the Calabi-Yau manifold 
from \calle{eq414}.
 In this sense, the variable $r$ can be thought of as determining
the K\"ahler modulus of the theory. We hasten to add, though, that as we let the
theory flow to the infrared and simultaneously integrate out the massive degrees
of freedom, the value of $r$ will change \`a la the Wilson renormalization
group. Hence, the actual value of the K\"ahler modulus at the infrared fixed
point will in general be determined by $r$ but will not be equal to it. The
parameter $r$ is often called the ``algebraic'' coordinate on the moduli space.
It is the natural variable for the linear sigma model. The value of the K\"ahler
modulus $\tilde r$ at the infrared fixed point is often called the
``sigma-model'' coordinate as it is the natural variable from the latter point
of view. For more discussion on these points see \cite{WP,AGMsd}.

Having discussed the case of $r \gt 0$, let us move on to the case of $r < 0$. 
Reasoning exactly as we did above, we find that all of the $s_i$ must vanish,
$p$ is constrained to be ${\sqrt {{-r \over 5}}}$, and there is an unbroken
$\IZ_5$ symmetry group (because $p$ has charge 5). Hence, unlike the case $r \gt
0$, the vacuum state is not an extended space, but rather is unique:
geometrically it is a point. Furthermore, from the form of the potential, the
$s_i$ are massless fluctuations about this vacuum state. The {\it configuration}
space, therefore is $\IC^5/\IZ_5$. Now, from our earlier discussion, this is an
{\it orbifold} of a Landau-Ginzburg theory since the latter can be described as
a theory with a unique vacuum state with some number of massless fields. The
$\IZ_5$ identifications coming from the unbroken gauge group gives rise to the
stated orbifolding. We note that this $\IZ_5$ action is in fact nothing but the
action of $e^{2 \pi i J_0}$ and hence constitutes the required $U(1)$ projection
that we have discussed earlier. It plays exactly the same role as the required
identifications, from the non-linear change of variable, in the path integral
argument given previously.

Thus, by varying the parameter $r$ in \calle{eq49}, it has been shown \cite{WP} 
that we interpolate between a linear sigma model on a \CY\ space and a \LG\
orbifold. By allowing the renormalization group to act, we thereby interpolate
between the conformally invariant limits of these two types of theories. We
might re-emphasize here the important point that the reason we should take $|r|$
to be large in each regime is to suppress the massive excitations which would
cause the theory obtained to differ from a non-linear sigma model or a \LG\ model.
After we flow to an infrared fixed point by the renormalization group, though,
any initial non-zero mass, no matter how small, becomes effectively infinite.
Thus, so long as we ultimately flow to an infrared fixed point, the size of
$|r|$ can be arbitrarily small.

As we mentioned, from 
\calle{eq413}, we see that the
actual value of $r$, for $r$ positive, sets the overall size of the ambient
projective space --- i.e. $r$ determines its K\"ahler form; by restriction to
the \CY\ hypersurface $r$ determines its K\"ahler form as well. For $r$
negative, its actual value sets the expectation of twist fields in the \LG\
theory \cite{WP}. Hence, in the language of section 3, $r$ may be thought of as a
\K\ moduli space parameter and the moduli space (for this simple 
discussion) consisting of $\IR$ divided into two regions $r > 0$ and $r < 0$.
Physically, the former region has a point (the ``deep interior point'') with $r$
being infinite corresponding to an infinite volume \CY\ space. The latter region
of $r < 0$ contains its own deep interior point of $r$ being negative infinity
which we have identified as the \LG\ orbifold point. (It is at this point, for
example, that the theory has an enhanced quantum symmetry \cite{Vafaorb,WP}.)
We call the first region of the moduli space the \CY\ sigma model region and the
second region the \LG\ orbifold region.

\begin{figure}[htbp]
\vspace{3mm}
\epsfxsize=7cm 
\centerline{\epsfbox{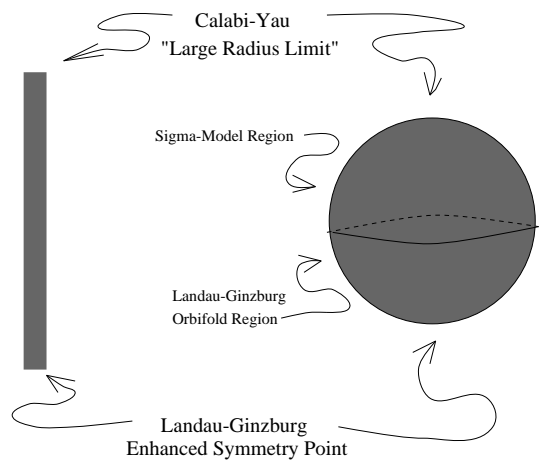}} 
\caption{The K\"ahler moduli space for the example discussed.}
\label{fig6}
\end{figure}

The left hand side of figure 
\ref{fig6} shows the $r$ moduli space. It is just the real 
line.
Now, as is well known, string theory instructs us to complexify the variable 
$r$ by combining it with an antisymmetric tensor field $r \rightarrow t = b +
ir$. A special feature of $b$ is that shifts of its value by an integer, in the
domain of large $r$,  do not affect the theory, and hence the natural complex
variable to use is $w = e^{2 \pi i (b + ir)}$. In this way we can map the sigma
model region of the moduli space to the upper hemisphere of a sphere. One can
give similar arguments for the Landau-Ginzburg region
\cite{AGM,WP,ASP} and in this way obtain a natural compactification of the
complexified moduli space, as shown in the right hand side of figure 
\ref{fig6}. We see
that this moduli space consists of two regions or ``phases'' and it can be shown
that there is no obstruction to smoothly varying the complex parameter $t$ to
move from one phase to the other.

Points in the first region correspond to \CY\ sigma models with K\"ahler class 
determined by the precise location of the point; points in the second region
correspond to 
\LG\ orbifolds with value of the twist field being determined by
the location of the point. As we try to move from the first region to the second
(or vice versa) conformal perturbation theory about the deep interior point in
region one (or region two, going in
 the other direction) breaks down. In the sigma
model region this is merely the statement that if the Calabi-Yau gets too small,
the expansion parameter $(\alpha')^2/r$ gets big and perturbation theory will be
invalid. However, using the results of \cite{AGM} and \cite{WP} we know that the
conformal theories corresponding to almost all points in the moduli space are
perfectly well defined, even if a perturbative understanding of them breaks
down, and hence we can smoothly continue our journey along such a path in the
moduli space. As a matter of convention, if a theory can be described via
conformal perturbation theory around one of our deep interior points, then we
categorize it as belonging to the same type of theory as this deep interior
point. This justifies the names we have given to the two regions above:  points
within the region in the $r > 0$ sector of figure 
\ref{fig5} are called \CY\ sigma
models while those in a similar region in the other sector are called \LG\
orbifold theories. An important point is that if 
we allow for analytic continuation, then we can
make sense of a perturbative expansion about the deep interior point of the $r >
0$ sector for essentially any point in the moduli space, even with $r < 0$.
Thus, in this sense, we can even think of the deep interior point in the \LG\
orbifold sector as being (the analytic continuation of a) \CY\ sigma model with
a particular (identifiable) K\"ahler class. In terms of the parameter $r$, we
see that this special choice seems to require a negative K\"ahler class, or more
precisely, an analytic continuation to a negative K\"ahler class. We should
note, though, that in \cite{AGMsd} it was shown that the physics of the
situation
implies that physical radii $\tilde r$ (and their analytic continuations) which
arise from integrating out massive modes in the linear sigma model, are
non-trivial functions of the $r$ parameter which appear to always be {\it
non-negative}. Thus, in this sense, one can interpret the result of \cite{WP} to
say that a \LG\ orbifold conformal model (of the type considered here) is
equivalent to the analytic continuation of a conformally invariant non-linear
sigma model on a \CY\ space to a particular (and identifiable
\cite{Vafaorb,WP,AGMsd}) ``small and positive'' value of the K\"ahler class.

\begin{figure}[htbp]
\vspace{3mm}
\epsfxsize=10cm
\centerline{\epsfbox{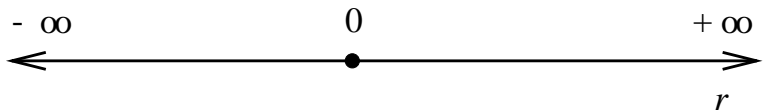}}
\caption{A slice of the linear sigma model moduli space.}
\label{xxa}
\end{figure}

\begin{figure}[htbp]
\vspace{3mm}
\epsfxsize=7cm
\centerline{\epsfbox{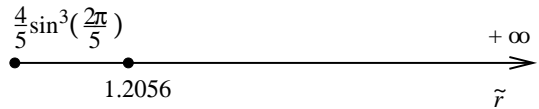}}
\caption{A slice of the non-linear sigma model moduli space.}
\label{xxb}
\end{figure}

Pictorially, the situation we are describing is illustrated by figures \ref{xxa} and
\ref{xxb} where
we show the one-parameter moduli space of the quintic hypersurface in
algebraic and then in non-linear sigma model variables. Details
can be found in \cite{AGMsd,CDGP}.

We see, therefore,
that this relatively rigorous argument lays out quite clearly the relationship 
between \CY\  conformal field theories and conformally invariant \LG\ orbifold
theories. We first pass to non-conformal members in the same universality class
as these theories,
which we see to be different ``phases'' of the {\it same} overarching theory
\calle{eq49} 
related by different values for the parameter $r$ (more precisely,
$t$). As we pass to the conformally invariant limit, the parameter
$r$ turns into the K\"ahler modulus, $\tilde r$. When $\tilde r$ is
sufficiently large, nonlinear sigma model perturbation theory converges,
and we have a direct geometrical interpretation of a string moving through
an ambient spacetime. If $\tilde r$ is not in this range of convergence,
the resulting model, at first sight, loses a geometrical interpretation.
However, the results of \cite{AGMsd} indicate that even these
regions have a geometric interpretation, so long as we appropriately
analytically continue from the large radius geometric region. In this way,
for instance, we see above that the Landau-Ginsburg point in the quintic
moduli space can be interpreted as a nonlinear sigma model with real
K\"ahler class  ${4\over5}\,\sin^3({2\pi\over5})   >0$
(see \cite{CDGP} for details of the derivation of this value).
 This is beyond the range
of non-linear sigma model perturbation theory, but can be fully
analysed via analytic continuation.

In our discussion of \cite{WP} we have limited our attention to theories with
one  \K\ modulus, and hence one $r$ parameter. For \CY\ manifolds with an
$h^{1,1}$ dimensional \K\ moduli space, there will be $h^{1,1}$ $r$ parameters,
$r_1,...,r_{h^{1,1}}$. The moduli space will again naturally divide itself into
phase regions, however the structure will typically be far richer than the
two phase regions found in the one-dimensional setting. We can again describe
theories in each region in terms of the most natural interpretation of their
corresponding
deep interior point. There will typically be a \LG\ orbifold
region, numerous smooth \CY\ regions (with the various \CY\ spaces being
birationally
 equivalent but possibly topologically distinct), \CY\ orbifold
regions and regions consisting of hybrids of these. Again, the work
of \cite{AGMsd} indicates that {\it all} of these regions --- in
the conformally invariant limit --- have the property that they can be
thought of as a nonlinear sigma model on a Calabi-Yau target space
with a sensible K\"ahler modulus. By sensible we mean that it is non-negative
with respect to at least one  Calabi-Yau topology.
We will discuss some of this in later sections;
for further details the reader
should see \cite{WP,AGM,AGMsd}.

\newsection{Mirror Manifolds}
\label{MirrorManifolds}

In our discussion to this point, we have seen that there are certain abstract 
properties of $(2,2)$ superconformal theories that can be realized by a variety
of field theoretic constructions. For instance, there is a
nice correspondence between geometrical constructs in the non-linear sigma model
formulation and their abstract conformal field theoretic counterparts. Furthermore,
 in general there are two types of marginal operators in
an abstract $c = 9,\ (2,2)$ superconformal theory and that these correspond to
the two geometrical ways of deforming a Calabi-Yau manifold without spoiling the
Calabi-Yau conditions. 

We note, however, that
there is an uncomfortable asymmetry between the abstract conformal field theory 
description and the geometrical realization. Namely, the two kinds of conformal
field theory marginal operators differ only in a rather trivial way: the
conventional sign of a $U(1)$ charge. On the other hand, their geometrical
counterparts differ far more significantly: the cohomology groups $H^1(M,T)$ and
$H^1(M, T^\star)$
(that is, $(d-1,1)$-forms, after contracting with $\Omega$ 
 and $(1,1)$-forms) are vastly different mathematical objects. It is surprising that
such a pronounced geometrical distinction finds such a trivial conformal field
theory manifestation. This led the authors of \cite{Dixon} and \cite{LVW} to
speculate on a possible resolution: if for each \CY\ manifold $M$ there was a
second \CY\ $\tilde M$ {\it corresponding to the same conformal field theory}
but with the association of $H^1(\tilde M,T)$ and $H^1(\tilde M, T^\star)$
 to
conformal field theory marginal operators {\it reversed} relative to that of
$M$, the asymmetry would be resolved. Each conformal field theory marginal
operator would then be interpretable geometrically either as a \K\ or as a
complex structure deformation, provided one chooses the \CY\ manifold
realization judiciously.

Although an interesting idea, at the time of these speculations there was no 
evidence for the existence of such pairs of \CY\ manifolds corresponding to the
same conformal field theory. Subsequently, two simultaneous independent
developments changed this. In \cite{CLS}\ the authors generated, via
computer, numerous \CY\ manifolds embedded in weighted projective 
four-dimensional space.
The data so generated consisted almost completely of pairs of manifolds $(M,
\tilde M)$ satisfying
\begin{equation}
    \begin{array}{c}
    \dim H^1(M,T)    = \dim H^1(\tilde M, T^\star)~,\\
    \dim H^1(M, T^\star) = \dim H^1(\tilde M,T)~.
    \end{array}
\label{eq51}
\end{equation} 
This, of
course, is a necessary condition for $M$ and $\tilde M$ to be identified with
the same conformal field theory as above, however it is far from sufficient. Two
quantum field theories can have sectors with the same {\it number} of fields,
but yet be otherwise completely unrelated. In \cite{GP}, on the other hand,
an explicit
construction of pairs of \CY\ manifolds $M$ and $\tilde M$ satisfying 
\calle{eq51} {\it
and corresponding to the same conformal field theory} was given. Using the fact
that $H^1(M,T) \cong H^{1,1}(M)$ and $H^1(M, T^\star) \cong H^{d-1,1}(M)$,
 one can
phrase \calle{eq51} as 
\begin{equation}
    \begin{array}{c}
      h^{1,1}(M) = h^{d-1,1}(\tilde M)~,\\
      h^{d-1,1}(M)= h^{1,1}(\tilde M) ~,
    \end{array}
\label{eq52}
\end{equation} 
where $h^{i,j}(M) = \dim H^{i,j}(M)$.
In particular, this implies that the {\it Hodge diamond} for $\tilde M$ is a
{\it mirror reflection} through a diagonal axis of the Hodge diamond for $M$.
For this reason, we chose \cite{GP}
 the term {\it mirror manifolds} for such pairs
$(M,\tilde M)$. The construction of \cite{GP}, therefore, established
conclusively that mirror manifolds exist. As of this writing, this is the only
established construction of mirror manifolds
 and hence shall be the focus of the
present section. There have been a number of other conjectured constructions of mirror
manifolds from both
physicists and mathematicians
 (including the work of \cite{CLS} alluded to above) and these are
discussed in the review paper of Berglund and Katz \cite{BK}. 
Recently, Strominger, Yau and Zaslow \cite{SYZ} have proposed
that mirror symmetry can be thought of as T-duality on toroidal
fibers of particular Calabi-Yau spaces.

\subsection{Strategy of the construction}
\label{sec51}

Before discussing the details of the mirror manifold construction of \cite{GP},
we  now briefly describe the general strategy of our approach.
\def\C{${\cal C}$}
Let \C\ be a conformal field theory associated with a \CY\ manifold $M$. \C\
 may be thought of as the non-linear sigma model with $M$ as a target space or,
somewhat more generally, \C\ may be thought of as the equivalence class of
conformal theories (regardless of the details of their particular construction)
which are isomorphic to this non-linear sigma model. There are many operations
one can perform upon \C\ to generate a new conformal theory \def\tC{${\cal C'}$}
\tC\ (or a new equivalence class \tC\ of conformal theories). For instance,
earlier we discussed the operations of deformation by truly marginal operators
which yield a new conformal theory from some chosen initial conformal model. In
general, not all operations which take one conformal theory to another have a
geometrical interpretation. In such a case, the resulting theory \tC\ may no
longer be associated with a \CY\ manifold. In other words, in the equivalence
class \tC, there need not be a conformal theory constructed from a non-linear
sigma model. As our interest here is in the geometrical interpretation of
conformal field theories, we will henceforth restrict our attention to
operations $\Gamma$ taking \C\ $\rightarrow$ \tC\ which have a functorial
geometric realization. In plain language, we focus on $\Gamma$ such that if the
non-linear sigma model on $M$ is in the class
\C, then the non-linear sigma model
on $\Gamma(M) = \tilde M$ is in the class $\Gamma($\C$) =$ \tC.

Amongst the operations $\Gamma$ which have this property is the operation of 
taking the quotient of \C\ by a discrete symmetry group $G$, so long as $G$ has
a geometrical interpretation as a holomorphic isometry preserving the
holomorphic $(d,0)$ form on the associated \CY\ space. In common parlance this
is referred to as {\it orbifolding} \cite{DHVW,DHVW2}
by $G$. Equivalently, this
operation amounts to gauging the discrete group
$G$ as we shall discuss later.  From the geometrical viewpoint, orbifolding by
the group $G$ has the  effect of yielding a new space $\Gamma(M) = M/G$ which is
characterized by the identification of all points $x,y$ in $M$ which are related
by 
\begin{equation}
 x = g(y) ~ {\rm with}~ g \in G~.
\end{equation} 
If there are points $x$ in $M$
such that $x = g(x)$, then $x$ is called a fixed point and $M/G$ (usually)
acquires a singularity at $x$. 

Now, let us imagine that we can find an operation
$\Gamma$ which has a geometric 
realization {\it and} such that: 

\begin{enumerate}

\item
$\Gamma({\cal C})\ {\rm is\ isomorphic\ to}\ {\cal C}.$ This would imply that the 
non-linear sigma models on $M$ and $\Gamma(M) $ are {\it
isomorphic as conformal field theories}. Such distinct spaces $M$ and 
$\Gamma(M)$ which nonetheless give rise to the same conformal field theory are
known as {\it classically string equivalent} \cite{GPNE}.

\item
The
explicit map realizing the isomorphism between \C\ and
 $\Gamma({\cal C})$ is changing
the sign of the right moving $U(1)_R$ charge of each operator in \C.

\item
The map between operators in the conformal field theory and geometrical
constructs in the associated \CY\ space (as discussed in section 
\ref{TheN=2SCFA}) is
independent of $\Gamma$. 

\end{enumerate}

\noindent
We claim that if such an operation $\Gamma$
meeting conditions (1)-(3) can be found, then $M$ and $\Gamma(M) = \tilde M$
would constitute a mirror pair of \CY\ spaces.

To see why, let us consider \C\ and $M$ as described with the property, say, 
that marginal operators with $U(1)_L \times U(1)_R$ charges $(-1,1)$ are
associated with the cohomology group 
$H^{1,1}(M)$ and marginal operators with
charges $(1,1)$ are associated with the cohomology group 
$H^{d-1,1}(M)$. Now,
let us apply $\Gamma$. By property $(1)$, $\Gamma$ has a geometrical
interpretation and hence we can construct a new Calabi-Yau space $\Gamma(M) =
\tilde M$  corresponding to the conformal theory $\Gamma({\cal C})$. By property
$(3)$, the cohomology groups $H^{1,1}(\tilde M)$ and $H^{d-1,1}(\tilde M)$
correspond to the marginal
 operators in $\Gamma({\cal C})$ with $U(1)_L \times U(1)_R$
charges $(-1,1)$ and $(1,1)$ respectively. By property $(2)$, we see therefore
that the cohomology groups $H^{1,1}(\tilde M)$ and $H^{d-1 ,1}(\tilde M)$ are
associated with marginal operators in \C\ with $U(1)_L \times U(1)_R$ charges
$(1,1)$ and $(-1, 1)$ respectively. Hence, $M$ and $\tilde M$ are string
equivalent (they correspond to isomorphic conformal field theories) and they
satisfy \calle{eq52}.
 This means that they constitute a mirror pair. 
In
the following sections we shall show that a suitable operation $\Gamma$ can in
fact be constructed. Before doing so, however, we should emphasize one subtle
point. As we have discussed, the $N = 2$ superconformal field theories we study
are often part  of continuous families of theories.
We let \FC\ denote the family of theories which are all related to \C\ via
deformation by truly marginal operators. Now, note that if we can find an
operation $\Gamma$ meeting conditions $(1)-(3)$
 for any theory ${\cal C}' \subset$
\FC, then such an operation exists for all other theories in \FC. The reason for
this is as follows. Let ${\cal C}'$ and ${\cal C}$ both belong to \FC\ and be
related via 
\begin{equation}
        {\cal C}' = {\cal U}({\cal {C}} ) ~,
\label{eq54}
\end{equation}
where ${\cal U }$ denotes the appropriate marginal operator
deformation. Now, if $\Gamma$ is the operation meeting conditions (1)$-$(3) for
${\cal C}' $, then the operation 
\def\circle{\circ} 
\begin{equation}
        { \tilde {\cal
           U} }^{-1} \circle \Gamma \circle {\cal U} 
\label{eq55}
\end{equation}  
meets conditions $(1)-(3)$ when
acting on \C. By definition, 
${\tilde {\cal U}}^{-1}$ is the
inverse deformation of ${\cal U}$ composed with explicitly changing the sign of
all $U(1)_R$ eigenvalues. Thus, so long as we can find an operation $\Gamma$
meeting $(1)-(3)$ for one theory in \FC, we are assured of such an operation for
every theory in the family. 
We note, further, that oftentimes
only some subset of all of the theories in \FC\ will have a natural geometric
interpretation in terms of a \CY\ sigma model. The meaning of conditions (1)
and $(3)$ for these theories is that when a candidate operation $\Gamma$ is
transported via \calle{eq55} to theories in \FC\ with a non-linear sigma model
interpretation (assuming such points exist in the family), then conditions (1)
and $(3)$ are met there.

\subsection{Minimal Models and their Automorphisms} 

As discussed in the last section, an important ingredient in the construction 
of mirror manifolds is an understanding of the minimal model conformal field
theories. We now turn to a more detailed discussion of these models.

The superconformal primary fields of the $P$-th minimal model are labeled by 
six integers, $l,m,s,\bar l, \bar m, \bar s$ and are typically 
written\footnote{We note, as discussed
below, that
these fields are not, strictly speaking, superconformal primary fields for all
values of the $s$ and $\overline s$ indices
 (which themselves are defined modulo 4).}
$\Phi_{s,\bar s}^{l,m,\bar l, \bar m}(z, \bar z)$.
  The meaning of these indices is made clear by recalling that the $N = 2 $
$P$-th minimal model is isomorphic to the coset of an $SU(2)$ WZW model at
level $P$ by a $U(1)$ subgroup together with a free boson. That is,
\begin{equation}
     MM_P = {{ SU(2)_P \over U(1) }} \times {\rm free~boson} ~.
\label{eq56}
\end{equation} 
In
other words, we remove a free boson at one radius 
by dividing out the $U(1)$ and we put a free boson back at a different radius.
Now, primary fields of an $SU(2)$ WZW model are labeled, in part, by their usual
$SU(2)$ angular momentum quantum numbers $l, m$ with $ |m| \le l$; this is the
meaning of the indices $l, m$ (and $\bar l, \bar m$ ) in $\Phi_{s, \bar
s}^{l,m,\bar l, \bar m}(z, \bar z)$, and these indices satisfy the same
inequality. In fact, to avoid dealing with half-integral values of spin, $l$ and
$m$ are defined to be twice their $SU(2)$ counterparts and hence $m$ can range
from $-l$ to $l$ in steps of two. With this convention, the value of $l$ can
range up to $P$. The index $s$ arises as a convenient bookkeeping device.
Namely, it proves convenient to split the Verma module built upon a given
superconformal primary field into those states which differ from the primary
field by the action of an even versus an odd number of supercurrents $G^{\pm}$.
In the NS sector, we take the value of $s$ to be zero or two; the former
referring to states which differ from the highest weight state by the action of
an even number of supercurrents (therefore including the bona fide original
primary field) and the latter referring to states obtained by the action of an
odd number of supercurrents. In the R sector, we define $s$ to be one or three
with these values playing an analogous role (we can equivalently replace three
by $-1$
 since this index is only defined modulo four). The index $s$ is referred
to as the `fermion' number.

More concretely, if we temporarily ignore the $s$ index and all of the 
anti-holomorphic dependence, superconformal primary fields in the NS sector can
be labeled $\Phi_{l,m}$ with conformal weights 
$$
  h = {{ l(l+2) \over 4 (P+2) }} -
 {{ m^2 \over 4 (P+2)}}
$$
 and $U(1)$ charge 
$$
  Q = {{ m \over P + 2 }}
$$
(and
similarly for the suppressed anti-holomorphic sector). The {\it chiral} primary
fields, i.e. those for which $G^+_{-1/2}$ also annihilates the corresponding
state, have $m = l$ and the antichiral primary fields ($G^-_{-1/2}$ annihilates
the corresponding state) have $m = -l$. In the Ramond  sector, our primary
fields can be written as $\Psi^{\pm}_{l,m}$ where $\pm$ refers to whether the
state is annihilated by $G_0^+$ or by $G_0^-$.
Then, the conformal weights of such states are 
$$ h = {{ l(l+2) \over 4
   (P+2) }} - {{ (m \pm 1)^2 \over 4 (P+2)}} + {{1 \over 8}}
$$
 and their $U(1)$
charges are 
$$
  Q = {{m \pm 1 \over P + 2}} \pm {{ 1 \over 2}}~.
$$
Ramond states annihilated by
both operators $G_0^\pm$ form the Ramond ground states and have 
$$
  h = {P\over 8 (P +2)}~.
$$
 The Ramond ground
states are $\Psi^+_{l,l}$ and $\Psi^-_{l,-l}$.
\def\sb{\overline s}
\def\mb{\overline m}
 Now let us reintroduce the $s$ index. We define
$\Phi^{l,m}_s$ to be: 
\beqn
    \Phi_{l,m}~,\quad &{\rm for}&~ s = 0~,\nonumber \\
   \Psi_{l,m}^+~,\quad &{\rm for}&~ s = 1~,\nonumber \\
    \Psi_{l,m}^-~,\quad &{\rm for}&~ s = -1~, \nonumber \\
     G^{\pm}\Phi_{l,m}~,\quad &{\rm for}&~ s = 2~,\nonumber
\eeqn
 and the index $s$
is defined modulo four, so this is a complete list. Now, the last definition
might seem ambiguous, but in fact we will only use this notation to describe the
Verma module (modulo fermion number as discussed) ${\cal H}^{l,m}_s$ to which
these states belong and both $G^{\pm}\Phi_{l,m}$ lie in the same Verma module
(and have the same fermion number mod 2).

Now, given such a field $\Phi_{s}^{l,m}(z, \bar z)$, 
we consider the character $\chi^{l,m}_s$ defined by 
\begin{eqnarray}
    \chr{l}{m,s}(\tau,z,u)\equiv\ex{-2\pi iu}\,
    \Tr{{\cal H}^{l}_{m,s}} \!\!\!\!\!\!\!
     ~\left\lbrack
    \ex{2\pi izJ_0}\ex{2\pi i\tau(L_0-c/24)}\right\rbrack~.
\label{eq57}
\end{eqnarray} 
We re-emphasize that
the trace is taken over a projection ${\cal H}^{l}_{m,s}$ to 
definite fermion number (mod 2) of a highest weight representation of the
(right-moving) $N=2$ algebra with highest weight vector
$\Pr{l}{q,s}{\lb}{\qb,\sb} (0)$, with fermion number 1 assigned to the 
superpartners $G^{\pm}$ of the energy-momentum tensor, 
as discussed in the last
paragraph. 

Our goal is to put together such chiral
characters in a modular invariant way to construct a consistent partition 
function.
To do so, we must first understand how the characters transform under modular 
transformations. This was worked out in \cite{GEP,GEP2} and the results are:
\ifnoWSc
\begin{eqnarray}
    \chr{l}{q,s}(\tau +1) &=&
    e^{\pi i {l(l+2)\over 2(P+2)} } \,
    e^{\pi i\left( -{q^2\over 2(P+2)}+{s^2\over 4} \right) }\,
      \chr{l}{q,s}(\tau)~, 
\label{eq58a}  \\
   \chr{l}{q,s}(-1/\tau) &=& {1\over \sqrt2 (P+2)}\,
   \sum_{l'+q'+s'=0\,{\rm mod}\, 2}\sin \left\lbrack \pi {(l+1)(l'+1)
   \over P+2} \right\rbrack \,
    e^{\pi i\left( {qq'\over P+2}- {ss'\over 2} \right) }\,
    \chr{l'}{q',s'}(\tau)~ .~~
\label{eq58b}  
\end{eqnarray}
\fi

An important observation is that the modular transformation properties 
factor into three pieces, each of which only acts on precisely one of the three 
indices labeling the characters. In fact, this is the real motivation for
introducing the auxiliary $s$ index in the first place. We recognize that the
index $l$ transforms under the representation of the modular group carried by
level $P$ affine $SU(2)$ characters, while the indices $m$ and $s$ transform
under the representations carried by level $-(P+2)$ and level $2$ theta
functions respectively. We can, therefore, construct modular invariant
combinations of these characters by combining known modular invariants for these
three types of objects. Up to discrete quotients, the general modular invariant
combination takes the form: 
\begin{equation}
    Z = {1 \over 2}\!\!\!\!\!\!\! \sum_{{\scriptstyle
   l,\lb,m,s}\atop {\scriptstyle l+m+s=0\, {\rm mod}\, 2}}\!\!\!\!\!\!\!
   \!\!\! \! A_{l,\lb} ~ \chr{l}{m,s}\,
    \chr{\lb \, *}{m,s}~,
\label{eq59}
\end{equation} 
where $A_{l,\lb}$ is any one of the $ADE$ classified 
affine modular invariants at level $P+2$. (The factor of $1\over 2$ reflects the
identifications on the fields discussed in, for example, \cite{GEP,GEP2}.)

We would now like to note the following important facts for the minimal model 
conformal theories.

\begin{enumerate}

\item
 $MM_P$ is invariant under a $\BZ_2 \times \BZ_{P+2} \times
 {\overline {\BZ}}_{2}\times {\overline {\BZ}}_{P+2}$ 
discrete symmetry group.

\item
The orbifold ${{MM_P \over \BZ_{P+2} }}$ of $MM_P$
 with respect to the diagonal group $\BZ_2 \times \BZ_{P+2}$
 is a new conformal theory ${\tilde { MM_P}}$ which 
is {\it isomorphic} to $MM_P$. The map from ${\tilde { MM_P}}$ to $MM_P$ is:
change the sign of the $U(1)_R$ eigenvalue associated with each field.

\end{enumerate}

These facts were first noted in
\cite{GQ} and, in reality, go back to \cite{FZ}.
 In the latter paper, the authors
realized the $MM_P$ as a combination of level $P$ parafermions and a free boson.
The level $P$ parafermions have a $\BZ_P$ discrete symmetry the action of which
via orbifolding is to produce an isomorphic theory. When combined with the free
boson, this $\BZ_P$ is promoted to $\BZ_{P+2}$; the action of orbifolding by the
latter is again to produce an isomorphic theory. For $P = 2$, the parafermion
system is an ordinary fermion and the isomorphism of this theory with a 
${\Bbb Z}_2$
orbifold of itself is nothing other than familiar order-disorder duality in the
Ising model. Hence one can think of property (2) above as a generalization of
this well known duality.

Explicitly, the discrete symmetry action can be written as 
\begin{eqnarray}
   g_q\cdot\Pr{l}{q,s}{\lb}{\qb,\sb}&=& 
   e^{2\pi i {q\over  P+2}} \, \Pr{l}{q,s}{\lb}{\qb,\sb}~,
\label{eq510a}   \\
   g_s\cdot\Pr{l}{q,s}{\lb}{\qb,\sb}&=&
   e^{2\pi i{s \over  2}}\, 
   \Pr{l}{q,s}{\lb}{\qb,\sb}\> ~.
\label{eq510b}  
\end{eqnarray} 

The second of these is a $\IZ_2$ symmetry which such theories also respect. 
In essence it is charge conjugation.

As property (2) is of central importance, let us look at it
 more closely. We can 
establish property $(2)$ in two ways. The first is via direct calculation and
the second is via manipulation of a path integral representation of these
theories. We will go through the first approach. 

\subsection{Direct Calculation}
\label{sec53}

Let us define $Z$ by 
\calle{eq59}. More generally, we define $Z[x,y]$ to be a twisted
 counterpart to the theory described by 
\calle{eq59}, where the boundary conditions of
all fields are twisted by the operator $e^{2 \pi i (J_0 + \overline J_0)x}$ in
the time direction and by $e^{2 \pi i (J_0 + \overline J_0)y}$ in the space
direction. For fields with these boundary conditions, the partition function
$Z[x,y]$ can be shown to be given by\footnote{For ease of presentation,
the action
we use here ignores the $s$ index and hence is not quite orbifolding by the
$U(1)$ charge; we will correct this below.} 
\begin{eqnarray}
     Z[x,y] = \!\!\!\!\!\!\!
     \sum_{{\scriptstyle l,\lb,m,s}\atop {\scriptstyle l+m+s=0\,{\rm mod}\,2}}
     \!\!\!\!\!\!\!\!\!\!
       e^{-2 \pi ix {m +\overline m\over 2 (P+2)}}
      \, A_{l,\lb} \, \chr{l}{m,s}\, \chr{\lb \, *}{m-2y,s}
      ~.
\label{eq511}
\end{eqnarray}
How does one get this result? The simplest way is to directly write
down $Z[x,0]$ --- this is easy since it simply involves inserting a factor
depending on the charge of each highest weight state into 
\calle{eq59}. Then, by
successive modular transformations, using 
\calle{eq58a} and \calle{eq58b}, we can derive \calle{eq511}. In
what follows, we shall drop explicit reference to the sums over 
$\scriptstyle l,s$ as they are not involved in any non-trivial way. 

We now create a new theory from \calle{eq59}
 by taking the {\it quotient}, in the 
sense of conformal theory, of 
\calle{eq59} by the left-right symmetric $\BZ_{P+2}$
symmetry group. By standard results in conformal field theory, 
\begin{eqnarray}
     Z_{\rm new} =
     {1\over P + 2 }\, \sum_{x,y = 0,...,P+1}\,
      Z[x,y]~.
\label{eq512} 
\end{eqnarray} 
Using our results from above,
\begin{eqnarray}
    Z_{\rm new} = {1\over P + 2}\,
    \sum_{x,y,m} e^{-2 \pi i x {(m + 
    \tilde m)\over 2(P+2)}}
  \, \chr{l}{m,s}\, \chr{\lb \, *}{\tilde m,s}~,
\label{eq513}
\end{eqnarray} 
where $\tilde m=m-2y$.
Now, performing the sum on $m$ we learn $2m -2y = 0\,{\rm mod}~(2 (P+2) )$
 to get a 
non-zero contribution. Hence, $\tilde m = -m\, {\rm mod}~ (2 (P+2) )$.
 By the symmetries
above we can thus write  
\begin{eqnarray}
    Z_{\rm new} &=& {1\over P + 2}\,
    \sum_{y,m}\, \chr{l}{m,s}\, \chr{\lb \, *}{-m,s} 
    \nonumber \\
    &=&\sum_m \chr{l}{m,s}\, \chr{\lb
       \, *}{-m,s}~. 
\label{eq514}
\end{eqnarray}

Now, to actually divide by $e^{2 \pi i (J_0 + \overline J_0)y}$, as noted, we 
need to include the $s$ dependence of the charge as well. This amounts to
dividing by the $\IZ_2$ action given earlier; the calculation is essentially
identical to that given and yields the final result\footnote{This extra $\BZ_2$
becomes part of the generalized GSO projection when such minimal models are used
to form string theories and hence is usually accounted for in that manner; see
\cite{GPN}.} 
\begin{equation}
          Z_{\rm new} = \sum_m \chr{l}{m,s}\,
             \chr{\lb \, *}{-m,-s}~. 
\label{eq515}
\end{equation}

Now, let us examine this new theory. Notice that it differs from the original 
theory \calle{eq59}
 only in that the $U(1)$ charges in the anti-holomorphic sector
all have opposite sign to those in 
\calle{eq59}. But the overall sign of this charge
is simply a matter of {\it convention}. That is, when we built the invariant
\calle{eq59}, we could equally well have written 
\begin{eqnarray}
    Z = {1 \over 2}\!\!\!\!\!
       \sum_{{\scriptstyle l,\lb,m,s}\atop {\scriptstyle
     l+m+s=0\,{\rm mod}\,2}} \!\!\!\!\!\!\!\!\!
     A_{l,\lb} \,\, \chr{l}{m,s}\, \chr{\lb \, *}{-m,-s}
\label{eq517}
\end{eqnarray}
and gotten an isomorphic theory (with the explicit isomorphism mapping a field 
in the theory based on 
\calle{eq59} to the field in \calle{eq517} with opposite sign of
right-moving $U(1)$ charge.

Thus we conclude that $Z_{\rm new}$ is {\it not} a new theory, after all. 
It is, in fact, isomorphic to the original minimal model theory $MM_P$
represented by the partition function $Z$ that we began with. Notice that this
is true regardless of which of the $ADE$ invariants we use in
constructing $Z$.

Having established that orbifolding $MM_P$ by its $\BZ_{P+2}$ discrete symmetry 
group yields an
isomorphic theory with the isomorphism simply being a change in the sign of all 
$U(1)_R$ eigenvalues, we see that we are part way along the path of realizing
the strategy for constructing mirror manifolds described previously. We now
extend this result to products of minimal models and hence to the Calabi-Yau
spaces that we can associate to them.

\subsection{Constructing Mirror Manifolds}
\label{sec:conmir}

In this subsection we will combine some of the results of the preceding two 
sections to realize the strategy, outlined previously, for the construction of
mirror manifolds.

We have seen that the $N = 2$ minimal models admit an operation, orbifolding, 
which produces an isomorphic conformal theory related to the original by a
change in the sign of all $U(1)_R$ eigenvalues. We have also seen that there is
an intimate connection between minimal models and \CY\ sigma models. The point
of this section is to show that through the minimal-model/Calabi-Yau connection,
we can transport the orbifolding operation on individual minimal models to an
operation $\Gamma$ meeting the three conditions discussed in section 
\ref{sec51} and
thereby yield a construction of mirror manifolds. We should emphasize at the
outset of our discussion that we will be making use of the comment made at the
end of section 
\ref{sec51} regarding the fact that we need only demonstrate the
existence of a suitable operation $\Gamma$ at one point in the moduli space of
theories $\cal F(\cal C)$ related to $\cal C$ by truly marginal deformation to
ensure that such an operation exists at all points. Furthermore, we can even
work at a point in the moduli space whose natural interpretation is not in terms
of a \CY\ sigma model (such as a \LG\ orbifold region). The meaning of 
conditions
(1)-(3) of section 
\ref{sec51} in this setting, as discussed, is that when
transported \`a la 
\calle{eq55} to a \CY\ sigma model region in the moduli space,
the resulting operation satisfies conditions (1)-(3). 

We will focus our detailed remarks on moduli spaces which contain a smooth 
\CY\ region corresponding to a
complex $d$-dimensional \CY\ hypersurface of Fermat type in weighted projective 
space. Such moduli spaces, as we have seen, also contain a region corresponding
to one of the $c = 3d$ minimal model constructions discussed in section 
\ref{sec43}
represented most conveniently as a \LG\ orbifold theory.  As we shall mention at
the end, the analysis directly extends to any \CY\ theory whose moduli space
contains a minimal model region.

Following our remarks above, our explicit analysis will take place at the deep 
interior point of
the \LG\ orbifold region in the \K\ moduli space and at the Fermat point in the 
complex structure moduli space. In other words, our calculations will take place
at the conformal field theory corresponding to the minimal model construction. 
We will see that we can construct a suitable operation $\Gamma$ at this point,
and by transporting it to the smooth \CY\ region in the moduli space construct a
mirror to the original manifold. (We should note that we could also interpret
the minimal model point as the analytic continuation of a \CY\ sigma model, as
discussed above and then our explicit analysis constructs the analytic
continuation of a sigma model on the mirror \CY\ space.) 

Consider, then, the tensor product of $s$ minimal model theories orbifolded 
onto integral $U(1)$ charges, $(P_1,...,P_s)$, with
$$
  \sum_{j=1}^s {{ 3 P_j \over
   P_j + 2}} = 3 d ~.
$$
We recall our notation: 
\ifnoWSc
\begin{eqnarray}
      (P_1,...,P_s) =
            [MM_{P_1}
             \otimes...\otimes MM_{P_s}]|_{ U(1)\ \rm projected}~ .
\label{eq518}
\end{eqnarray} 
\fi
By the result of section
\ref{sec53}, we can write 
\ifnoWSc
\begin{eqnarray}
     (P_1,...,P_s) \equiv 
           \left\lbrack {MM_{P_1} \over
    \BZ_{P_1 + 2}} \otimes...\otimes {MM_{P_s}\over \BZ_{P_s + 2}}
  \right\rbrack\Bigg|_{ U(1)\ \rm
     projected}~,
\label{eq519}
\end{eqnarray} 
\fi
with the explicit isomorphism being the changing of the sign of
all $U(1)_R$ eigenvalues. Now, we claim: 
\ifnoWSc
\begin{eqnarray} 
    \left\lbrack {MM_{P_1} \over \BZ_{P_1+2}}
     \otimes...\otimes
                 {MM_{P_s}\over \BZ_{P_s + 2}}
    \right\rbrack\Bigg|_{ U(1)~\rm projected} =
   {{[MM_{P_1} \otimes...\otimes MM_{P_s}]
  |_{ U(1)\ \rm projected}} \over G}
\label{eq520}
\end{eqnarray}
\fi
where $G$ is the maximal subgroup of 
$\BZ_{P_1 + 2} \times ... \times \BZ_{P_s +2}$
by which one can orbifold and preserve the integrality of the $U(1)$ charges
of the theory. In particular, the action of $G$ is
\begin{equation}
      (\Phi_1,...,\Phi_s) \rightarrow (e^{2 \pi i {n_1\over q_1}}\,
      \Phi_1,...,e^{2 \pi i {n_s\over q_s}}\, \Phi_s)~,
\label{eq521}
\end{equation} 
for arbitrary integers $(n_1,...,n_s)$
such that $\sum_{j=1}^s n_j/q_j$ is an integer.

Establishing this claim requires a calculation that can be 
found in \cite{GP,GPN}.
For our discussion here we simply note that it is a familiar
fact in conformal field theory that successive quotients of a theory can undo
each other if the subsequent quotients are quantum versions of
the previous one. All we have here is
an example of this phenomenon. The $U(1)$ projection has the effect of undoing
those quotients of the theory that do not respect 
\calle{eq521}.

Thus, we have shown that
\begin{equation}
 (P_1,...,P_s) \equiv {(P_1,...,P_s)\over G}~,
\label{eq522}
\end{equation} 
with the isomorphism between 
the two theories being a reversal in the sign of all $U(1)_R$ eigenvalues of the
fields in the left hand side relative to the right hand side.

Since this operation of orbifolding is independent of the \K\ modulus of the 
theory, it is trivial to transport it to a smooth \CY\ region. The action on the
\CY\ manifold $M$, 
\begin{equation}
   z_1^{P_1 +2} + \dots + z_{P_s}^{P_s + 2} = 0~,
\label{eq523}
\end{equation} 
in the
weighted projective space
 $W {\Bbb C} P^{s-1}({D \over P_1 + 2},...,{D \over P_s +2})$
 with arbitrary \K\ form is given by
\begin{equation}
     (z_1,...,z_s) \rightarrow (e^{2 \pi i {n_1\over q_1}}\, z_1,...,
     e^{2 \pi i {n_s\over q_s}}\, z_s)~,
\label{eq524}
\end{equation} 
for arbitrary integers $(n_1,...,n_s)$ such that
$\sum_{j=1}^s n_j/q_j$ is an integer. This condition, which defines $G$, is
interpretable in the \CY\ region as the condition of preserving the holomorphic
$d$-form $\Omega$ on $M$. 

Now, this operation of orbifolding by $G$ meets conditions $(1)$ and $(2)$ of 
section \ref{sec51}:
 since it is true at the minimal model point it is true everywhere in the 
moduli space, as discussed. This operation also meets condition (3) as can most
quickly be seen in the following way: consider a conformal field $\Lambda$
associated with a geometrical harmonic form $a_{\Lambda}$, such that $\Lambda$
and hence $a_{\Lambda}$ are invariant under the action of $G$ (there always will
be at least one such field: the restriction of the K\"ahler class of the ambient
projective space to $M$). Then, in the theory based on $M/G$, $\Lambda$ and
$a_{\Lambda}$ again correspond. This implies that if marginal operators of
charges $(1,1)$ and $(1,-1)$ are associated to elements of $H^1(M,T^\star)$ and
$H^1(M,T)$ respectively (or vice versa), then the same association holds in the
theory based on $M/G$. Namely,
 marginal operators of charges $(-1,1)$ and $(1,1)$
are associated to elements of $H^1(M/G,T^\star)$
 and $H^1(M/G,T)$ respectively (or
vice versa). This is so because the association of conformal fields to
geometrical cohomology can only take two possible forms (as explicitly noted).
One established association of a conformal field and a geometrical harmonic form
distinguishes between these two possibilities. Since we have shown that in both
$M$ and $M/G$ we have at least one identical association, we are done. Hence,
our operation meets condition $(3)$ as well.

Thus, we have shown that the \CY\ hypersurface $M$ given by \calle{eq523} 
(for arbitrary choice of \K\ form) has mirror given by $M/G$. Let us note that
following our discussion of the phase structure of the moduli space of these
theories, it is more appropriate to say the following. Let $M$ be a \CY\
manifold. It belongs to a moduli space of conformal theories. Consider $M/G$. It
is a \CY\ space (it is not smooth) which also belongs to a family of conformal
theories. We have shown that for each point in the first moduli space there is a
corresponding point in the second moduli space giving rise to an isomorphic
theory. Thus, mirror symmetry is more precisely a statement of pairs of {\it
families} of conformal theories.

\subsection{Examples}

In this section we give a few examples which illustrate the construction of the 
last section. The following two tables show mirror pairs of theories constructed
via the orbifolding operation above. The column `symmetries' denotes the group
action by which we quotient. For instance, in the first table, $[0,0,0,1,4]$
indicates that we take the quotient by the $\IZ_5$ action 
\begin{equation}
(z_1,z_2,z_3,z_4,z_5) \rightarrow (z_1,z_2,z_3, \alpha  z_4, \alpha^4 z_5)
  ~,
\label{eq525}
\end{equation}
where $\alpha$ is a fifth root of unity. Mirror pairs reside in symmetric
positions in the tables with respect to the horizontal axis through the center. 

\begin{table}
\caption{Orbifolds of the theory 
          $(P_1,P_2,P_3,P_4,P_5)=(3,3,3,3,3)$.} 

$$\vbox{\offinterlineskip
\hrule
\halign{&\vrule#&\strut\quad\hfil#\hfil\quad& \vrule#&\strut\quad\hfil#\hfil
\quad&
\vrule#&\strut\quad\hfil#\hfil\quad&
\vrule#&\strut\quad\hfil#\hfil\quad&
\vrule#&\strut\quad\hfil#\hfil\quad&\vrule#\cr height2pt&\omit&&\omit&&\omit
&&\omit&&\omit&\cr &Theory&&Symmetries&&$h^{2,1}$&&$h^{1,1}$&&$\chi$&\cr
height2pt&\omit&&\omit&&\omit&&\omit&&\omit&\cr \noalign{\hrule}
height3pt&\omit&&\omit&&\omit&&\omit&&\omit&\cr &&&&&101&&1&& $-200$ &\cr
height3pt&\omit&&\omit&&\omit&&\omit&&\omit&\cr
height0pt&\omit&\multispan8\hrulefill\cr
height3pt&\omit&&\omit&&\omit&&\omit&&\omit&\cr &&&[0,0,0,1,4]&&49&&5&& $-88$&
\cr
height3pt&\omit&&\omit&&\omit&&\omit&&\omit&\cr
height0pt&\omit&\multispan8\hrulefill\cr
height3pt&\omit&&\omit&&\omit&&\omit&&\omit&\cr &&&[0,1,2,3,4]&&21&&1&& $-40$ &
\cr
height3pt&\omit&&\omit&&\omit&&\omit&&\omit&\cr
height0pt&\omit&\multispan8\hrulefill\cr
height3pt&\omit&&\omit&&\omit&&\omit&&\omit&\cr &&&[0,1,1,4,4]&&&&&&&\cr &
$(3,3,3,3,3)$&&[0,1,2,3,4]&&21&&17&& $-8$ &\cr
height3pt&\omit&&\omit&&\omit&&\omit&&\omit&\cr
height0pt&\omit&\multispan8\hrulefill\cr
height3pt&\omit&&\omit&&\omit&&\omit&&\omit&\cr &or&&[0,1,1,4,4]&&17&&21&&8&\cr
height3pt&\omit&&\omit&&\omit&&\omit&&\omit&\cr
height0pt&\omit&\multispan8\hrulefill\cr
height3pt&\omit&&\omit&&\omit&&\omit&&\omit&\cr &&&[0,1,3,1,0]&&&&&&&\cr
&$z_1^5+\cdots+z_5^5=0$&&[0,1,1,0,3]&&1&&21&&40&\cr
height3pt&\omit&&\omit&&\omit&&\omit&&\omit&\cr
height0pt&\omit&\multispan8\hrulefill\cr
height3pt&\omit&&\omit&&\omit&&\omit&&\omit&\cr &in
${\Bbb C}P^4$ &&[0,1,4,0,0]&&&&&&&\cr &&&[0,3,0,1,1]&&5&&49&&88&\cr
height3pt&\omit&&\omit&&\omit&&\omit&&\omit&\cr
height0pt&\omit&\multispan8\hrulefill\cr &&&[0,1,2,3,4]&&&&&&&\cr
&&&[0,1,1,4,4]&&1&&101&&200&\cr &&&[0,0,0,1,4]&&&&&&&\cr
height2pt&\omit&&\omit&&\omit&&\omit&&\omit&\cr} \hrule}$$
\end{table}

\begin{table}
\caption{Orbifolds of theory
       $(P_1,P_2,P_3,P_4)=(3,8,8,8)$.} 

$$\vbox{\offinterlineskip
\hrule
\halign{&\vrule#&\strut\quad\hfil#\hfil\quad& \vrule#&\strut\quad\hfil#\hfil
\quad&
\vrule#&\strut\quad\hfil#\hfil\quad&
\vrule#&\strut\quad\hfil#\hfil\quad&
\vrule#&\strut\quad\hfil#\hfil\quad&\vrule#\cr height2pt&\omit&&\omit&&\omit
&&\omit&&\omit&\cr &Theory&&Symmetries&&$h^{2,1}$&&$h^{1,1}$&&$\chi$&\cr
height2pt&\omit&&\omit&&\omit&&\omit&&\omit&\cr \noalign{\hrule}
height3pt&\omit&&\omit&&\omit&&\omit&&\omit&\cr &&&&&145&&1&& $-288$ &\cr
height3pt&\omit&&\omit&&\omit&&\omit&&\omit&\cr
height0pt&\omit&\multispan8\hrulefill\cr
height3pt&\omit&&\omit&&\omit&&\omit&&\omit&\cr &&&[0,0,5,5]&&99&&3&& $-192$ 
 &\cr
height3pt&\omit&&\omit&&\omit&&\omit&&\omit&\cr
height0pt&\omit&\multispan8\hrulefill\cr
height3pt&\omit&&\omit&&\omit&&\omit&&\omit&\cr &&&[0,2,2,6]&&47&&11&&
 $-72$ &\cr
height3pt&\omit&&\omit&&\omit&&\omit&&\omit&\cr
height0pt&\omit&\multispan8\hrulefill\cr
height3pt&\omit&&\omit&&\omit&&\omit&&\omit&\cr &&&[0,0,1,9]&&39&&15&&
 $-48$ &\cr
height3pt&\omit&&\omit&&\omit&&\omit&&\omit&\cr
height0pt&\omit&\multispan8\hrulefill\cr
height3pt&\omit&&\omit&&\omit&&\omit&&\omit&\cr &
$(3,8,8,8)$&&[0,0,2,8]&&37&&13&&
 $-48$ &\cr
height3pt&\omit&&\omit&&\omit&&\omit&&\omit&\cr
height0pt&\omit&\multispan8\hrulefill\cr
height3pt&\omit&&\omit&&\omit&&\omit&&\omit&\cr &or&&[0,1,2,7]&&29&&17&&
 $-24$ &\cr
height3pt&\omit&&\omit&&\omit&&\omit&&\omit&\cr
height0pt&\omit&\multispan8\hrulefill\cr
height3pt&\omit&&\omit&&\omit&&\omit&&\omit&\cr
&$z_1^5+z_2^{10}+\cdots+z_4^{10}+z_5^2=0$&&[0,5,4,1]&&17&&29&&24&\cr
height3pt&\omit&&\omit&&\omit&&\omit&&\omit&\cr
height0pt&\omit&\multispan8\hrulefill\cr
height3pt&\omit&&\omit&&\omit&&\omit&&\omit&\cr &in
$W{\Bbb C}P^4(2,1,1,1,5)$   &&[0,8,1,1]&&15&&39&&48&\cr
height3pt&\omit&&\omit&&\omit&&\omit&&\omit&\cr
height0pt&\omit&\multispan8\hrulefill\cr
height3pt&\omit&&\omit&&\omit&&\omit&&\omit&\cr &&&[0,5,5,0]&&&&&&&\cr
&&&[0,8,1,1]&&13&&37&&48&\cr height3pt&\omit&&\omit&&\omit&&\omit&&\omit&\cr
height0pt&\omit&\multispan8\hrulefill\cr
height3pt&\omit&&\omit&&\omit&&\omit&&\omit&\cr &&&[0,5,5,0]&&&&&&&\cr
&&&[0,1,9,0]&&11&&47&&72&\cr height3pt&\omit&&\omit&&\omit&&\omit&&\omit&\cr
height0pt&\omit&\multispan8\hrulefill\cr
height3pt&\omit&&\omit&&\omit&&\omit&&\omit&\cr &&&[0,0,1,9]&&&&&&&\cr
&&&[0,8,0,2]&&3&&99&&192&\cr height3pt&\omit&&\omit&&\omit&&\omit&&\omit&\cr
height0pt&\omit&\multispan8\hrulefill\cr
height3pt&\omit&&\omit&&\omit&&\omit&&\omit&\cr &&&[0,0,1,9]&&&&&&&\cr
&&&[0,1,1,8]&&1&&145&&288&\cr height3pt&\omit&&\omit&&\omit&&\omit&&\omit&\cr}
\hrule}$$
\end{table}

\subsection{Implications}

Having reviewed the initial speculations and subsequent work which established 
the existence of mirror symmetry, we would now like to turn to a discussion of
the implications of this phenomenon, 
as well as some important work applying mirror
symmetry to interesting and explicit examples.

The general lesson learned from mirror manifolds is that there is a duality
in Calabi-Yau moduli space since we have two distinct geometrical
descriptions of a single physical situation. Moreover, perturbation theory
in each model is governed by the respective K\"ahler parameters as the
coupling expansion is in terms of $\alpha'/R^2$ (where, more precisely,
$R^2$ denotes the area of rational curves as measured by the relevant
K\"ahler form.) We have seen, though, that K\"ahler parameters
 of one Calabi-Yau
determine complex structure parameters of the other, and vice versa. Thus, if
 the K\"ahler parameters of one Calabi-Yau are
``small'' thereby making perturbation theory  suspect, its complex structure can
be adjusted so that
the K\"ahler structure of its mirror will be nice and
``big'', thereby ensuring the efficacy of perturbation theory. In other words,
mirror symmetry gives rise to  a strong/weak sigma-model coupling duality.
We will exploit this in an important physical context in the next chapter.
For now, we would like to briefly indicate one other specific
implication of mirror symmetry.

Let $M$ and $\tilde M$ be mirror Calabi-Yau manifolds each corresponding to the 
conformal field theory ${\cal C}$.
Consider a (non-vanishing) three-point function of
conformal field theory operators corresponding to $(2,1)$-forms on $M$.
We will not explicitly do the calculation here, but 
as shown in \cite{StromWitten}
(and rederived in \cite{DG} in a slightly different way)
this  correlation function is given by the simple
integral on $M$ \calle{triplecpx} 
\begin{equation}
   \int_{M} \Omega^{abc}\, \tilde b^{(i)}_a
   \wedge \tilde b^{(j)}_b \wedge \tilde b^{(k)}_c \wedge \Omega~,
\label{eq526}
\end{equation} 
where, as explained earlier, the $\tilde
b^{(i)}_a$ are $(2,1)$-forms (expressed as elements of $H^1(M,T)$ with their
subscripts being tangent space indices).
 Due to the non-renormalization theorem proved in \cite{DG}, we know that
this expression 
\calle{eq526} is the exact conformal field theory result. By mirror
symmetry, these same conformal field theory operators correspond to particular
and identifiable $(1,1)$-forms on the mirror
 $\tilde M$, which we can label $b^{(i)}$.
Mathematically, due to the absence of a non-renormalization theorem, the
expression for such a coupling in terms
of geometric quantities on $\tilde M$ is
comparatively complicated: 
\ifnoWSc
\begin{eqnarray}
   \int_{{\tilde M}} b^{(1)} \wedge b^{(2)} \wedge b^{(3)} + \sum_{m,\{u \}}
    \ex{-\int_{{\Bbb C}P^1 }u_m^*(J)}
    \left(\int_{{\Bbb C}P^1 } u^*(b^{(1)}) 
          \int_{{\Bbb C}P^1 } u^*(b^{(2)})
          \int_{{\Bbb C}P^1 } u^*(b^{(3)}) \right)~,
\label{eq527}
\end{eqnarray}
\fi
where
(as derived in
\cite{Strom,DSWW,DSWW2,CDGP,CDGP2,AM2})
the $b_{(i)}$ are elements of $H^1(\tilde M,T^\star)$, $\{u \}$
 is the set of holomorphic
 maps to rational curves on $\tilde M$, 
$u: {\Bbb C}P^1 \rightarrow \Gamma $ (with $\Gamma$
such a holomorphic curve), $\pi_m$ is an $m$-fold cover 
${\Bbb C}P^1 \rightarrow {\Bbb C}P^1$
and $u_m = u \circ \pi_m$. $J$ refers to the K\"ahler form on $\tilde M$. 
The infinite series of corrections in this expression arise from
string world sheet configurations that wrap around rational curves
(essentially two-spheres) embedded in $\tilde M$. Just like instantons
in the more familiar setting of gauge theories, these string configurations
are topologically nontrivial and contribute nonperturbative corrections to 
correlation functions. These corrections are an explicit examples of
how `stringy' geometry differs from ordinary classical geometry.
The first term in this correlation function is a familiar
mathematical construct --- the intersection form of  $\tilde M$.
The infinite series of instanton corrections move us
away from classical geometry; they are stringy in origin as they arise
because the extended nature of the string allows it to encircle
homologically nontrivial cycles as it moves.

Now, since both 
\calle{eq526} and \calle{eq527} correspond to the {\it same} conformal 
field theory correlation function, they must be equal; hence we have \cite{GP}
\ifnoWSc
\begin{eqnarray}
\label{eq528}
   \int_{{M}} \Omega^{abc}
    \tilde b^{(i)}_a \wedge \tilde b^{(j)}_b \wedge \tilde b^{(k)}_c \wedge 
     \Omega \!\!\!\!&=&\!\!\!\! 
    \int_{{\tilde M}} b^{(i)} \wedge b^{(j)} \wedge b^{(k)} 
    \\
  &+& \!\!\!\!
    \sum_{m,\{ u \}}\, \ex{\int_{ \IC P^1 }u_m^*(J)} 
   \left(\int_{\IC P^1}u^*(b^{(i)})\right) 
   \left(\int_{\IC P^1}u^*(b^{(j)})\right)
   \left(\int_{\IC P^1}u^*(b^{(k)})\right)
    ~. 
   \nonumber 
\end{eqnarray}
\fi

Notice the crucial role played by the underlying conformal field theory in 
deriving this equation. If we simply had two manifolds whose Hodge numbers were
interchanged we could not, of course, make any such statement. This result
\cite{GP} is rather surprising and clearly very powerful. We have related
expressions on {\it a priori} unrelated manifolds which probe rather intimately
the structure of each. Furthermore, the left hand side of 
\calle{eq528} is directly
calculable while the right hand side requires, among other things, knowledge of
the rational curves of every degree on the space.

The equality \calle{eq528} is another striking example of quantum geometry. 
First, as mentioned, the coupling 
\calle{eq527} has as its leading term a familiar expression from
classical geometry. 
However, there are corrections arising from the extended
nature of the string as this allows for homotopically non-trivial field
configurations sensitive to the rational curves on the manifold. Second, the
fact that the resulting expression is equal to 
\calle{eq526} on a different space is a
hallmark feature of quantum geometry: the distinctions of classical geometry
`smear' and can be erased in the quantum setting. We emphasize that {\it every}
correlation function in the underlying conformal theory now has two geometrical
interpretations. The equality \calle{eq528} is but {\it one} such example. In
principle, we could write down scores of others. 

Since written, 
\calle{eq528} has been verified in several illuminating examples 
\cite{M,Fon,KT,LT,BS,COFKM,HKTY}
pioneered by the papers \cite{CDGP,CDGP2}.
These authors showed that one could use
\calle{eq528} to determine 
the number of rational curves of arbitrary degree on $\tilde M$, a
question of mathematical interest that was previously unsolved. This brought to
the forefront the tremendous calculation power of mirror symmetry.
 In the next
section we shall see an impressive physical consequence of string theory that
mirror symmetry allows us to establish.

\newsection{Space-Time Topology Change --- The Mild Case}
\label{Space-timeTopologyChange}

This section is based in part on \cite{AGMB}. 

\subsection{Basic Ideas}

The essential lesson of general relativity is that the geometrical structure of 
space-time is governed by dynamical variables. That is, the metric changes in
time according to the Einstein equations. In the usual formulations of general
relativity, the space-time metric
 is defined on a space of fixed topological type
--- the ``size'' and ``shape'' of the space can smoothly change, but the
underlying topology does not. A natural question to ask is whether this
formulation is too restrictive; might the topology of space itself be a
dynamical variable and hence possibly change in time? This issue has long been
speculated upon. Heuristically, one suspects that topology might be able to
change by means of the violent curvature fluctuations which would be expected in
any quantum theory of gravity. Just as the fluctuations of the magnetic field in
a box of size $L$ are on the order of $(\hbar c)^{1/2}/L^2$,  those of the
curvature of the gravitational field are on the order of $({\hbar G \over
c^3})^{1/2}/L^3$.  Thus, on extremely small scales, say $L \sim L_{\rm Planck}$,
huge curvature fluctuations are unsuppressed. One can imagine that such
curvature fluctuations could ``tear'' the fabric of space resulting in a change
of topology. The expected discontinuities in physical observables accompanying
the discontinuous operation of a change in topology would be hidden, one hopes,
behind the smoothing effects of quantum uncertainty. Of course, without a true
theory of quantum gravity, one cannot make quantitative sense of such
hypothesized processes.

With the advent of string theory, we are led to ask whether any new quantitative
 light is shed on the issue of topology change. A number of works have addressed
aspects of this question, and in this and the next section, we shall review the results
in three of them \cite{WP,AGM,rGMS}. In the first two of these papers, the first
definitive evidence that
there are {\it physically smooth}  processes in
string theory which result in a change in the topology of space-time was given.
These processes occur at string tree level and hence are essentially classical
phenomena. Their unusual character therefore arises from the extended nature
of the string.
The type of topology change in these works is relatively mild in that
the Hodge numbers of the Calabi-Yau stay fixed while only more subtle
invariants (such as the cubic intersection form) change.
In \cite{rGMS}, on the other hand, quantum effects are taken into account ---
in fact, non-perturbative effects having to do with solitonic degrees of
freedom. In this more robust setting, it is shown that there are physical
processes that result in rather drastic topology changing transitions ---
transitions which change even the Hodge numbers of the Calabi-Yau compactification.
Here we discuss the case of mild topology change and later return
to the more drastic case.

\subsection{Mild Topology Change}

In this subsection we focus on the results of \cite{WP,AGM}. We will
see the first evidence of physically smooth topology changing processes
in string theory.
Furthermore, as phenomena in string theory, these processes are not at all
exotic. Rather, they correspond to the most basic kind of operation arising in
conformal field theory: deformation by a truly marginal operator. From a
space-time point of view, this corresponds to a slow variation in the 
vacuum expectation value of a
scalar field which has an exactly flat potential\footnote{To avoid
confusion, we
remark that the present discussion focuses on static vacuum solutions to string
theory. One expects that configurations involving the generic slow variation of
such scalar fields are solutions as well.}.
 It is crucial to emphasize, as remarked above, that these
physically smooth topology changing processes occur even at the level of 
classical  string theory. It is not, as had been suspected from point particle
intuition, that quantum effects give rise to topology change, but, 
rather, it is
the extended structure of the string which bears responsibility for this
effect. 

\begin{figure}[htbp]
\epsfxsize=2.5cm
\centerline{\epsfbox{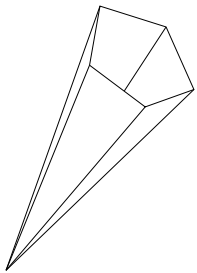}}
\caption{A K\"ahler cone.}
\label{fig7a}
\end{figure}

We can immediately summarize here the essential content of \cite{WP} and
\cite{AGM}.
 From the viewpoint of classical general relativity or the classical non-linear
sigma model, we know that there are constraints on the metric tensor
which
appears in the action. Namely, since the metric is used to measure lengths,
areas, volumes, etc., it must satisfy a set of positivity conditions. For
instance, if we have a non-linear sigma model on a K\"ahler target space $M$ with
metric  $g_{\mu \overline \nu}$, we can write the K\"ahler form of the metric as
$J = ig_{\mu \overline \nu} dX^{\mu} \wedge dX^{\overline \nu}$.
 The latter must satisfy 
\begin{equation}
 \int_{M_r } J^r > 0~,
\label{eq61}
\end{equation} 
where
$M_r$ is an $r$-dimensional (complex) submanifold of $M$ and $J^r$ represents
the $r$-fold wedge product of $J$ with itself. The set of real closed $2$-forms
which satisfy \calle{eq61} is a subset of $H^2(X,\IR)$ known as the {\it
K\"ahler}\/ cone and is schematically depicted in figure 
\ref{fig7a}.
This figure should
be thought of as a more precise version of the drawing of the K\"ahler moduli
space in figure
\ref{fig5}, in which we did not pay attention to details such as
positivity. Such K\"ahler forms manifestly span a cone because if $J$ satisfies
\calle{eq61}, then so does
 $sJ$ for any positive real $s$.

\begin{figure}[htbp]
\epsfysize=3cm
\centerline{\epsfbox{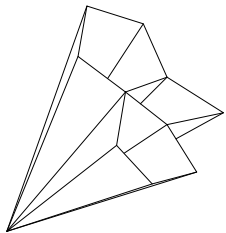}}
\caption{Adjoining K\"ahler cones.}
\label{fig:manycones}
\end{figure}

\noindent The burden of \cite{WP} and
\cite{AGM} is that, in string theory, 
\calle{eq61} can be relaxed and still result in
perfectly well behaved physics. As discussed, the K\"ahler form of a target
Calabi-Yau space is one of the moduli fields of the associated conformal field
theory. Investigation of the conformal field theory moduli space reveals that
the corresponding geometrical description {\it necessarily} involves
configurations in which the (supposed) K\"ahler form lies outside of the
K\"ahler cone of the particular \CY\ being studied. In fact, {\it any and all}
choices of an element of $H^2(X,\IR)$ give rise to well defined conformal field
theories. In \cite{WP,AGM} it was shown that some of these configurations
can be interpreted as non-linear sigma models on \CY\ manifolds of topological
type distinct from the original. With respect to this \CY\ of new topology, the
\K\ modulus satisfies 
\calle{eq61} and hence may be thought of as residing in a
new K\"ahler cone which shares a common wall with the original (figure
\ref{fig:manycones}). Furthermore,
there is no physical obstruction to continuously deforming the underlying
conformal field theory so that its geometrical description passes from one \K\
cone to another and hence results in a change in topology of the target space 
 --- i.e. of space itself.

Philosophically, this result is providing us with another important way in
 which classical geometry fails to capture qualitative properties of string
physics. A change in topology is, by its basic mathematical definition, a
process which is not smooth. However, {\it quantum geometry}, the proper
geometry for describing string physics, allows for topology change in a
perfectly smooth manner. 

To understand this result, we need to gain a deeper understanding of moduli 
spaces than has been presented to this point. Hence, in the next subsection 
we will give a discussion of
moduli spaces of both conformal theories and \CY\ manifolds in order to fill in 
a bit more detail required for the discussion of topology change. We will see
that this discussion raises an interesting puzzle whose resolution, as we will
discuss  directly, leads to the necessity of physically smooth topology changing
processes. We shall then go on to verify the abstract discussion of the
preceding sections in an explicit example which provides a highly sensitive
confirming test of the picture we present.

\subsection{Moduli Spaces}

Quite generally, as discussed previously, the conformal field theories we study 
here
come in continuously connected families related via deformations by truly 
marginal operators. 
When an $N$=2 conformal theory arises from a non-linear sigma
model with a Calabi-Yau target space, the marginal operators have geometrical
counterparts. The two types of 
marginal operators correspond to the two types of
deformations of the Calabi-Yau space which preserve the Calabi-Yau condition (of
Ricci flatness). As our analysis will involve a close study of these moduli
spaces, let us now describe each in a bit more detail. 

\subsubsection{K\"ahler Moduli Space}
\label{sec:complexJ}

\begin{figure}[htbp]
\epsfxsize=2.5cm
\centerline{\epsfbox{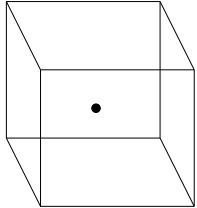}}
\caption{Domain of $w_l$'s.}
\label{fig7b}
\end{figure}

Given a K\"ahler metric $g_{\mu \overline \nu}$,
 we can construct the K\"ahler 
form
$J = ig_{\mu \overline \nu} dX^{\mu} \wedge dX^{\overline \nu}$. As discussed 
earlier, the set of allowed $J$'s forms a cone known as the K\"ahler cone of
$M$. One additional important fact is that string theory instructs us to work
not just with $J$ but also with $B = B_{\mu \overline \nu}$ the antisymmetric
tensor field. The latter, which is a closed two-form, combines with $J$ in the
form $B + iJ$ to yield  the highest component of a complex chiral multiplet we
shall call $K$. $K$ can therefore be thought of as a {\it complexified}\/
K\"ahler form. The precise way in which $B$ enters the conformal field theory is
such that if $B$ is replaced by $B + Q$,
 with $Q \in H^2(M,\BZ)$, then the
resulting physical model does not change. Thus, a convenient way to parameterize
the space of allowed and physically distinct $K$'s is to introduce 
\begin{equation}
    w_l = e^{2 \pi i (B_l + iJ_l)}~,
\label{eq62}
\end{equation} 
where we have expressed 
\begin{equation} 
    B + iJ = \sum_l
      \, (B_l + iJ_l)\, e^l~,
\label{eq63}
\end{equation} 
with the $e^l$ forming an integral basis for
$H^2(M,\BZ)$. The
$w_l$ have the invariance of the antisymmetric tensor field under integral
shifts built in; the constraint that $J$ lie in the K\"ahler cone bounds the
norm of the $w_l$. Thus, the K\"ahler cone and space of allowed and distinct
$w_l$ are schematically shown in figures 
\ref{fig7a} and \ref{fig7b}. Notice that any choice of
complexified K\"ahler form in the interior of figure \ref{fig7b} is physically
admissible. Choices of $K$ which correspond to points on the walls in figure 
\ref{fig7b}
(or \ref{fig7a}) correspond to metrics on $M$ which fail to meet 
\calle{eq61} and hence
are degenerate in some manner.

\subsubsection{Complex Structure Moduli Space}

\begin{figure}[htbp]
\epsfxsize=7cm
\centerline{\epsfbox{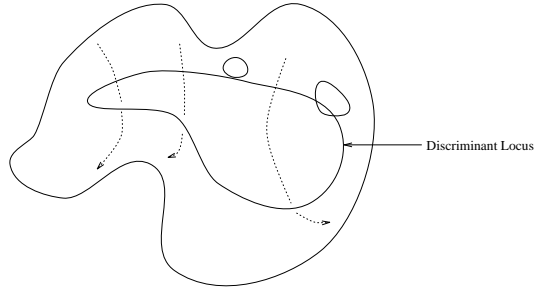}} 
\caption{The moduli space of complex structures.}
\label{fig8}
\end{figure}

All of the Calabi-Yau spaces we shall concern ourselves with here are given by 
the vanishing locus of homogeneous polynomial constraints in some projective
space (or possibly a weighted projective space and products thereof). For ease
of discussion, and in preparation for an explicit example we will examine
shortly, let us assume we are dealing with a Calabi-Yau manifold given by the
vanishing locus of a homogeneous polynomial $P$ of degree $d$ in weighted
projective four-dimensional space $W {\Bbb C} P^4(k_1,k_2,k_3,k_4,k_5)$.
 The Calabi-Yau condition
translates into the requirement that $d = \sum_i k_i$. Let us call the
homogeneous weighted projective space coordinates $(z_1,\ldots,z_5)$ and write
down the most general form for $P$: 
\begin{equation} 
      P = \sum_{i_1,i_2,i_3,i_4,i_5}\, a_{i_1 i_2 i_3 i_4 i_5}
    \,  z_1^{i_1} \, z_2^{i_2} \, z_3^{i_3}\, z_4^{i_4}\, z_5^{i_5} ~,
\label{eq64}
\end{equation} 
where $\sum_j k_j\, i_j = d$. Different choices for
the constants $a_{i_1 i_2 \ldots i_5}$ correspond to different 
choices for the
complex structure of the underlying Calabi-Yau manifold. There are two important
points worthy of emphasis in this regard. First, not all choices of the $a_{i_1
i_2 \ldots i_5}$ give rise to distinct complex structures. For instance,
distinct choices of the $a_{i_1 i_2 \ldots i_5}$ which can be
 related by a
rescaling of the $z_j$ of the form $z_j \rightarrow \lambda_j z_j$ with
$\lambda_j \in {\Bbb C}^*$
manifestly correspond to the same complex structure (as
they differ only by a trivial coordinate transformation). The most general
situation would require that we consider $a_{i_1 i_2 \ldots i_5}$'s related by
general linear transformations on the $z_j$'s. Second, not all choices of
$a_{i_1 i_2 \ldots i_5}$ give rise to smooth Calabi-Yau manifolds.
Specifically,
if the $a_{i_1 i_2 \ldots i_5}$ are such that $P$ and $\partial P \over \partial
z_j$ have a common zero (for all $j$), then the space given by the vanishing
locus of $P$ is not smooth. One can understand this by noting that the derivatives
of $P$ fill at the tangent space directions; if they simultaneously
 vanish on $P$ than the
tangent space has collapsed in some manner.
 The set of all choices of the coefficients $a_{i_1
i_2 \ldots i_5}$ which correspond to such singular spaces comprise the {\it
discriminant locus}\/ of the family of Calabi-Yau spaces associated with $P$.
The precise equation of the discriminant locus is generally quite complicated;
however, the only fact we need is that it forms a complex codimension one
subspace of the complex structure moduli space. From the viewpoint of
conformal field theory, the non-linear sigma model associated to points on the
discriminant locus appears to be ill defined. For example, the chiral ring
becomes infinite dimensional. Later, we shall
take up the  interesting and important question of
whether there might be some way of making sense of such
theories. For the present purposes, though, all we need to know is that at worst
the space of badly behaved physical models is complex codimension one in the
complex structure moduli space. We illustrate the form of the complex structure
moduli space in figure \ref{fig8}.

\subsection{Implications of Mirror Manifolds: Revisited} 
\label{sec:implications}

Locally the moduli space of Calabi-Yau deformations is a product space of the 
complex and K\"ahler deformations (in fact, up to subtleties which will not be
relevant here, we can think of the moduli space as a global product). Thus, we
expect, as in figure \ref{fig5}, 
\begin{eqnarray} 
     {\cal M}_{\rm CFT} &\equiv& 
      {\cal M}_{complex~structure}
                \times 
      {\cal M}_{K\ddot ahler~ structure} ~,
\label{eq65}
\end{eqnarray} 
with ${\cal M}_{(\ldots)}$ denoting
the moduli space of $(\ldots)$. Pictorially, we can paraphrase this by saying
that the conformal field theory moduli space is expected to be the product of
figure \ref{fig7b} and figure \ref{fig8},
which is a more accurate version of figure \ref{fig5}.

This, in fact, is the picture which had emerged from much work over a number of 
years and was generally accepted. The advent of mirror symmetry, however,
raised a
 serious puzzle related to this description.
Let $M$ and $\tilde M$ be a mirror pair of Calabi-Yau spaces. As we discussed
before, such a pair corresponds to isomorphic conformal theories with the
explicit isomorphism being a change in sign of, say, the right moving $U(1)$
charge. From our description of the moduli space, it then follows that the
moduli space of K\"ahler structures on $M$ should be isomorphic to the moduli
space of complex structures on $\tilde M$ and vice versa. That is, both $M$ and
$\tilde M$ correspond to the same family of conformal theories and hence yield
the same moduli space on the left hand side of 
\calle{eq65}. Therefore, the right hand
side of 
\calle{eq65} must also be the same for both $M$ and $\tilde M$. The explicit
isomorphism of mirror symmetry shows this to be true with the two factors on the
right hand side of 
\calle{eq65} being interchanged for $M$ relative to $\tilde M$.

The isomorphism of the K\"ahler moduli space of one Calabi-Yau and the complex 
structure of its mirror is a statement which appears to be in direct conflict
with the form of figure \ref{fig7b} and that of figure 
\ref{fig8}. Namely, the former is a bounded
domain while the latter is a quasi-projective variety. More concretely, the
subspace of theories which appear possibly to be badly behaved are the boundary
points in figure \ref{fig7b}
 (where the metric on the associated Calabi-Yau fails to meet
\calle{eq61}) 
and the points on the discriminant locus in figure \ref{fig8}. The former are
real codimension 1 while the latter are real codimension 2. Therefore, how can
these two spaces be isomorphic as implied by mirror symmetry?

\subsection{Flop Transitions}
\label{sec:flop}

As the puzzle raised in the last section was phrased in terms of those points 
in the moduli space which have the potential to correspond to badly behaved
theories, it proves worthwhile to study the nature of such points in more
detail. We will first do this from the point of view of the K\"ahler moduli
space of $M$. 

\begin{figure}[htbp]
\ifnoWSc\epsfxsize=5cm\fi 
\centerline{\epsfbox{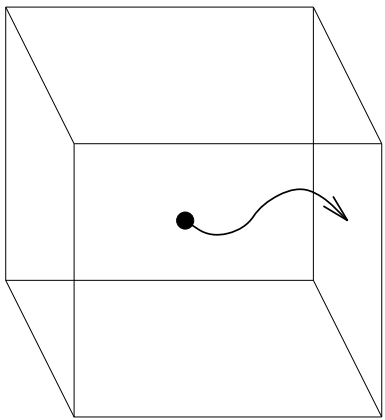}} 
\caption{A path to the wall.}
\label{fig9}
\end{figure}

Consider a path in the K\"ahler moduli space which begins deep in the interior 
and moves towards and finally reaches a boundary wall as illustrated in figure
\ref{fig9}.
 More specifically, we follow a path in which the area of a 
${\Bbb C}P^1$ (a rational
curve) on $M$ is continuously shrunk down to zero, attaining the latter value on
the wall itself. The question we ask ourselves is: does this choice for the
K\"ahler form on $M$ yield an ill defined conformal theory and furthermore, what
would happen if we try to extend our path beyond the wall where it appears that
the area of the rational curve would become negative? (We note the
linguistically awkward phrase ``area of a curve'' arises since we are dealing
with complex curves which therefore are real dimension two.)

\begin{figure}[htbp]
\ifnoWSc\epsfxsize=6cm\fi 
\centerline{\epsfbox{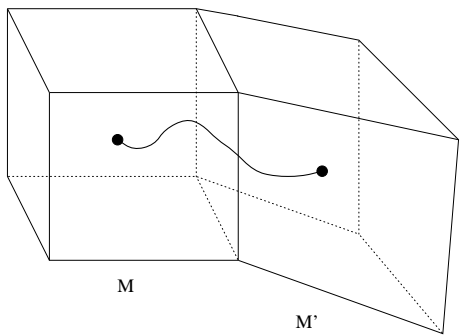}} 
\caption{A topology changing path.}
\label{fig10}
\end{figure}

As a prelude to answering this physical question, we note that precisely this 
operation is well known and thoroughly studied from the viewpoint of
mathematics. Namely, in algebraic geometry there is an operation called a {\it
flop}\/ in which the area of a rational curve is shrunk down to zero ({\it blown
down}) and then expanded back to positive volume ({\it blown up}) in a
``transverse'' direction. Typically (although not always),
 this operation results
in a change of the topology of the space in which the curve is embedded. Thus,
when we say that the blown up curve has positive volume we mean positive with
respect to the K\"ahler metric on the new ambient space. That is, the flop
operation involves first following a path like that in figure 
\ref{fig9} which blows the
curve down, and then continuing through the wall (as in figure 
\ref{fig10}) by blowing
the curve up to positive volume on a new Calabi-Yau space. The latter space,
$M'$ also has a K\"ahler cone whose complexification in the exponentiated $w_l$
coordinates is another bounded domain. Thus, the operation of the flop
corresponds to a path in moduli space beginning in the K\"ahler cone of $M$,
passing through one of its walls and landing in the {\it adjoining}\/ K\"ahler
cone of $M'$. Although $M$ and\footnote{To avoid confusion, we note
that the
mirror to $M$ is called $\tilde M$,
 not $M'$.} $M'$ can be topologically distinct, their
Hodge numbers are the same; they differ in more subtle topological invariants
such as the intersection form governing the classical homology ring.
Mathematically, they are said to be topologically distinct but in the same
birational equivalence class. 

\begin{figure}[htbp]
\epsfysize=5cm
\centerline{\epsfbox{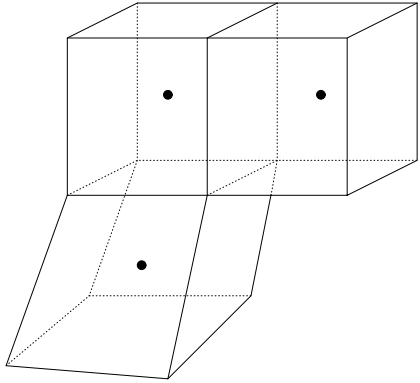}}
\caption{Adjoining K\"ahler moduli spaces.}
\label{fig:manyboxes}
\end{figure}

The mathematical formulation of what it means to pass to a wall in the 
K\"ahler moduli space has led us to a more detailed framework for studying the
corresponding description in conformal field theory. We see that from the
mathematical point of view, distinct K\"ahler moduli spaces naturally adjoin
along common walls (see figure \ref{fig:manyboxes}).
 We can rephrase our initial motivating question of two
paragraphs ago as: does the operation of flopping a rational curve (and thereby
changing the topology of the Calabi-Yau under study) have a physical
manifestation? That is, does a path such as that in figure 
\ref{fig10} correspond to a
family of well behaved conformal theories?

This is a hard question to answer directly because our main tool for analyzing 
non-linear sigma models is perturbation theory. The expansion parameters of such
perturbative studies are of the form $\alpha'/R^2$, where $R$ refers to the
set of K\"ahler moduli on the target manifold or, more
precisely, $\alpha'/\int_{\Sigma} u^*(K)$.
Now, when we approach or reach a
wall in the K\"ahler moduli space, at least one such moduli field $R$ is going
to zero, namely the one which sets the size of the blown down rational curve.
Hence, we are in the realm
of a strongly coupled sigma model;
 perturbation theory breaks down and we are hard pressed to
answer directly whether the associated conformal theory makes non-perturbative
sense. 

This situation --- one in which we require a non-perturbative understanding of 
observables on $M$ --- is tailor-made for an analysis based upon mirror
symmetry. Perturbation theory breaks down on $M$ because of the degenerate
(or nearly degenerate) choice of its K\"ahler structure. Note that all of
our discussion could be carried through for any convenient (smooth)
choice of its complex structure. Via mirror symmetry, this implies that the
relevant analysis for answering the question raised two paragraphs ago
should be carried out on $ \tilde M$ for
a particular form of the {\it complex structure}\/ (namely, that which is
mirror to the degenerate K\"ahler structure on $M$) but for any convenient
choice of the K\"ahler structure. The latter, though, determines the
applicability of sigma model perturbation theory on $\tilde M$. 
Thus, we can choose this K\"ahler structure to be arbitrarily ``large'' 
(that is, distant from any walls in the K\"ahler cone) and hence arrange
things so that we can completely trust perturbative reasoning. In other words,
by using mirror symmetry we have rephrased the difficult and necessarily
non-perturbative question of whether conformal field theory continues to make
sense for degenerating K\"ahler structures in terms of a purely perturbative
question on the mirror manifold. 

This latter perturbative question is one which is easy to answer and, in fact, 
we have already done so in our discussion of the complex structure moduli 
space.
For large values of the K\"ahler structure (again, this simply means that we 
are
far from the walls of the K\"ahler cone), the only choices of the complex
structure which yield (possibly) badly behaved conformal theories are those
which lie on the discriminant locus. As noted earlier, the discriminant locus is
complex codimension one in the moduli space (real codimension two). Thus, the
complex structure moduli space is, in particular, path connected. Any two points
can be joined by a path which only passes through well behaved theories; in
fact, the generic path in the complex structure moduli space has the latter
property. This is the answer to our question. By mirror symmetry, this
conclusion must hold for a generic path in K\"ahler moduli space and hence it
would seem that a topology changing path such as that of figure 
\ref{fig10} (by a
suitable small jiggle of the path, at worst)
 is a physically well behaved process. Even
though the metric degenerates, the physics of string theory continues to make
sense. We are already familiar from the foundational work on orbifolds
\cite{DHVW,DHVW2} that degenerate metrics can
lead to sensible string physics. Now we see that physically sensible
degenerations of other types (associated to flops) can alter the topology of the
universe. In fact, the operation being described --- deformation by a truly
marginal operator --- is amongst the most basic and common physical processes in
conformal field theory. 

\begin{figure}[htbp]
\epsfysize=10cm
\centerline{\epsfbox{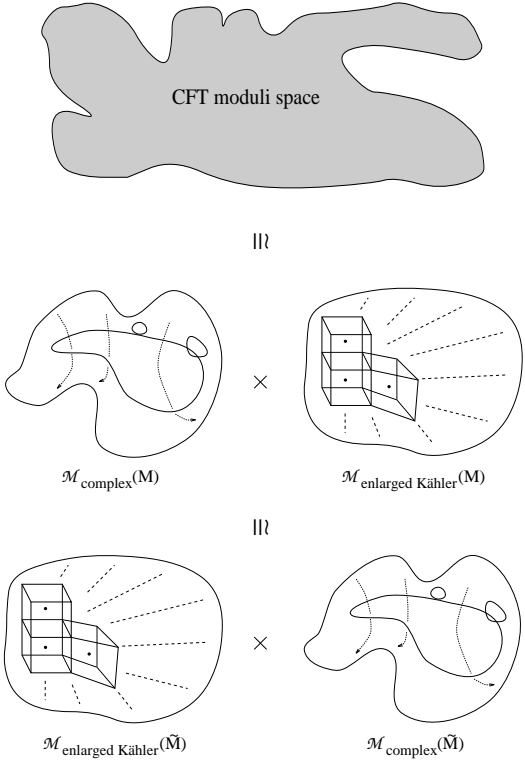}} 
\caption{The conformal field theory moduli space.}
\label{fig11}
\end{figure}

To summarize the picture of moduli
 space which has emerged from this discussion,
 we refer to figure \ref{fig11}. This is the picture which replaces the old and
incomplete version of figure \ref{fig5}.
  The conformal field theory moduli space
is geometrically interpretable in terms of the product of a complex structure
 moduli space and an enlarged K\"ahler moduli space 
${\cal M}_{\rm enlarged~ K\ddot ahler}$.
The latter contains numerous complexified K\"ahler cones
of birationally
equivalent yet topologically distinct Calabi-Yau manifolds
adjoined along common walls\footnote{The union of such regions constitutes
what we
call the ``partially enlarged'' K\"ahler moduli space. The enlarged K\"ahler
moduli space includes additional regions as we shall mention shortly.}.  There
are two such geometric interpretations, via mirror symmetry, with the roles of
complex structure and K\"ahler structure being interchanged. This is also
indicated in figure \ref{fig11}.

We should stress that from an abstract point of view this is a very satisfying
picture. As we will discuss more fully
in section \ref{sec:toricapl},  the
augmentation of the K\"ahler moduli space in the manner presented (and, more
precisely, as we will generalize shortly), gives it a mathematical structure
which is {\it identical}\/ to that of the complex structure moduli space of its
mirror. In the important case of Calabi-Yau manifolds
 which are {\it toric} hypersurfaces,
both of these moduli spaces are realized as identical compact {\it toric
varieties}. (The mathematics of toric geometry will be discussed in 
section \ref{toricintro}.)
 Hence, the picture presented resolves the previous troubling
asymmetry between the structure of 
these two spaces which are predicted to be
isomorphic by mirror symmetry. 

Although compelling, we have not proven that the picture we are presenting is 
correct. We have found a natural mathematical structure in algebraic geometry
which, {\it if}
 realized by the physics of conformal field theory, resolves some
thorny issues in mirror symmetry. We have not established, as yet, whether
conformal field theory makes use of this natural mathematical structure. If
conformal field theory does avail itself of this structure, though, there is a
very precise and concrete conclusion we can draw: every point in the (partially)
enlarged K\"ahler moduli space of $M$ must correspond under mirror symmetry to
some point in the complex structure moduli space of $\tilde M$. This implies, of
course, that any and all observables calculated in the theories associated to
these corresponding points must be identically equal. Let us
 concentrate on the
three-point functions we introduced earlier in 
\calle{eq528}. As we discussed, if we
choose a point in the K\"ahler moduli space for which the instanton corrections
are suppressed, the correlation function approaches the topological intersection
form on the Calabi-Yau manifold. For ease of calculation, we shall study the
correlation functions of 
\calle{eq527} in this limit. Unlike the analysis of
 \cite{ALR},  there is not a single unique ``large radius'' point of
the sort we are looking for.
 Rather, {\it every} cell in the (partially) enlarged
moduli space supplies us with one such point. Since these cells are the
complexified K\"ahler cones of topologically distinct spaces, the intersection
forms associated with these large radius points are different. If the moduli
space picture we are presenting in figure 
\ref{fig11} is correct, then there must be
points in the complex structure moduli space of the mirror whose correlation
functions exactly reproduce each and every one of these intersection forms. This
is a sharp statement whose veracity would provide a strong
verification of the picture presented in figure \ref{fig11}.
 In the next section we carry
out this verification in a particular example.

\subsection{An Example}
\label{sec67}

In this section,
 we briefly carry out the abstract program discussed in the last 
few sections in a specific example. We will see that the delicate predictions
just discussed can be explicitly verified. The mathematics of
toric geometry is required to carry out the details of this calculation.
As toric geometry can be a bit technical, we shall not discuss it
until section \ref{toricintro}. Therefore, in this section we
shall outline the calculation; some more details will then be given in
section \ref{sec:anexample}.

We focus on the Calabi-Yau manifold $M$ given by the vanishing locus of a 
degree $18$ homogeneous polynomial in the weighted projective space
$W{\Bbb C}P^4(6,6$, $3,2,1)$
 and its mirror $\tilde M$. For the former we can take the
polynomial constraint to be 
\begin{equation}
 z_0^3+z_1^3+z_2^6+z_3^9+z_4^{18}
+a_0 z_0z_1z_2z_3z_4 = 0 ~,
\end{equation}
where the $z_i$ are the homogeneous weighted space coordinates and $a_0$ is a 
large and positive constant (whose value, in fact, is inconsequential to the
calculations which follow). The mirror to this family of Calabi-Yau spaces is
constructed via the  method discussed, by taking an orbifold of $M$ by the
maximal scaling symmetry group $\IZ_3 \times \IZ_3 \times \IZ_3$. 

A study of the K\"ahler structure of $M$ reveals that there are five cells in 
its (partially) enlarged K\"ahler moduli space, each corresponding to a sigma
model on a smooth topologically distinct Calabi-Yau manifold. In each of these
cells there is a large radius point for which instanton corrections are
suppressed and hence the correlation functions of 
\calle{eq527} are just the
intersection numbers of the respective Calabi-Yau manifolds. These can be
calculated
for each of the five birationally equivalent, yet topologically distinct,
Calabi-Yau
spaces and we record the results in table 3. To avoid having
to deal with issues associated with normalizing fields in the subsequent
discussion, in table \ref{table:3}.
we have chosen to list the results in terms of ratios
of correlation functions for which such normalizations are irrelevant. (The
$D_i$ and $H$ are divisors on $M$, corresponding to elements in 
$H^1(M,T^\star)$
by Poincar\'e duality.)

\begin{table}
\caption{Ratios of intersection numbers.}
\label{table:3}
\begin{center}
\def\ilspace{\omit&height2pt&&&&&&&&&&&&\cr}
\def\tablerule{\noalign{\hrule}}
 $$\vbox{\tabskip=0pt 
\offinterlineskip
\halign{\strut#&
\vrule#&\hfil\quad$#$\quad\hfil&\vrule#&\hfil\quad$#$\quad\hfil& 
\vrule#&\hfil\quad$#$\quad\hfil&\vrule#&\hfil\quad$#$\quad\hfil&
\vrule#&\hfil\quad$#$\quad\hfil&\vrule#&\hfil\quad$#$\quad\hfil& \vrule#\cr
\tablerule\ilspace &&\hbox{Resolution}&&\Delta_1&&\Delta_2&&\Delta_3
&&\Delta_4&&\Delta_5&\cr\ilspace\tablerule\ilspace
&&{(D_1^3)(D_4^3)\over(D_1^2D_4)(D_1D_4^2)} && -7 && 0/0 && 0/0 && \infty && 9
&\cr\ilspace &&{(D_2^2D_4)(D_3^2D_4)\over(D_2D_3D_4)(D_2D_3D_4)} && 2 && 4 && 0
&& 0/0 && 0/0 &\cr\ilspace &&{(D_2D_3D_4)(HD_2^2)\over(D_2^2D_4)(HD_2D_3)} && 1
&& 1 && 1 && 0 && 0/0 &\cr\ilspace
&&{(D_2D_3D_4)(HD_1^2)\over(D_1^2D_4)(HD_2D_3)} && 2 && 1 && \infty && 0/0 && 0
&\cr
\ilspace \tablerule
}}$$ 
\end{center}
\end{table}

Following the discussion of the last section, the goal now is to find five 
limit points in the complex structure moduli space of $\tilde M$ such that
appropriate ratios of correlation functions yield the same results as in 
table \ref{table:3}.
To do so, we note that the most general complex structure on $\tilde M$ 
can
be written 
\ifnoWSc
\begin{eqnarray}
    W=z_0^3+z_1^3+z_2^6+z_3^9+z_4^{18}
   +a_0\,
    z_0z_1z_2z_3z_4 + a_1 z_2^3z_4^9 + a_2 z_3^6z_4^6 +
    a_3\, z_3^3z_4^{12} + a_4\,
    z_2^3z_3^3z_4^3 =0~.~~~
\label{eq67}
\end{eqnarray} 
\fi
We will describe these limit points by parameterizing
the complex structure as $a_i = s^{r_i}$ for real parameters $s$ and $r_i$ and
we send $s$ to infinity, the mirror operation of going to
large volume. The limit points are therefore distinguished by the
{\it rates}\/ at which the $a_i$ approach infinity. The task, therefore, is to
find appropriate values for the $r_i$, if they exist, such that we obtain
mirrors to the five large radius Calabi-Yau spaces of the last paragraph. The
technique we use to do this is to describe both the complex structure moduli
space of $\tilde M$ and the enlarged K\"ahler moduli space of $M$ in terms of
toric geometry. This description, as we shall see, makes it manifest that
these two moduli spaces are isomorphic.  For the present
purpose we note that a direct outcome of this analysis --- to
be covered in section \ref{sec:anexample} --- is a prediction
for five
choices of the vector $(r_0,\ldots,r_4)$ which should yield the desired mirrors.
As we have discussed, a sensitive test of these predictions is to calculate the
mirror of the ratios of correlation functions in table 3 (using 
\calle{eq526} and the
method of \cite{PC}) for each of these
complex structure limits and see if we get the same answers. This has been done 
and  the results are shown in tables \ref{table:4a}
 and \ref{table:4b}.
 Note that in the limit $s$ goes to infinity
we get precisely the same results. (The $\varphi_i$ are elements of $H^1(\tilde
M,T)$.) 

\begin{table}
\caption{Asymptotic ratios of 3-point functions.}
\label{table:4a}
\begin{center}
\begin{tabular}{|c|c|c|c|}\hline
 Resolution & $\Delta_1$ & $\Delta_2$ & $\Delta_3$ \\ \hline
 \strut
%
 Direction &
 $({11\over9},1,{4\over3},{5\over3},{5\over3})$ &
 $({7\over6},{1\over2},1,1,{5\over2})$ &
 $({3\over2},{1\over2},2,3,{3\over2})$ \\ \hline
%
%
 ${\langle \varphi_1^3\rangle\langle \varphi_4^3\rangle\over\langle
 \varphi_1^2\varphi_4\rangle\langle \varphi_1\varphi_4^2\rangle}$ &
 $-7-181s^{-1}+\dots$ &  & \\ \hline
%
%
 ${\langle \varphi_2^2\varphi_4\rangle\langle
 \varphi_3^2\varphi_4 \rangle\over\langle  
 \varphi_2\varphi_3\varphi_4\rangle\langle \varphi_2\varphi_3\varphi_4\rangle}$&
 $2-5s^{-1}+\dots$ & $4-22s^{-1}+\dots$ & $0+2s^{-1}+\dots$ \\ \hline
%
%
${\langle \varphi_2\varphi_3\varphi_4\rangle\langle \varphi_0\varphi_2^2
\rangle\over\langle \varphi_2^2\varphi_4\rangle\langle \varphi_0\varphi_2
\varphi_3\rangle}$ &
$1+{1\over2}s^{-1}+\dots$ & $1+{3\over2}s^{-2}+\dots$ &
$1+4s^{-1}+\dots$ \\ \hline
%
%
${\langle
\varphi_2\varphi_3\varphi_4\rangle\langle \varphi_0\varphi_1^2
\rangle\over\langle \varphi_1^2\varphi_4\rangle\langle \varphi_0\varphi_2
\varphi_3\rangle}$ &
 $2+27s^{-1}+\dots$ & $1-{1\over2}s^{-1}+\dots$ &
 $-2s-33+\dots$ \\ \hline 
\end{tabular}
\end{center}
\end{table}
%
%
%
%
%
%
%
%
\begin{table}
\caption{Asymptotic ratios of 3-point functions.}
\label{table:4b}
\begin{center}
\begin{tabular}{|c|c|c|}\hline
 Resolution & $\Delta_4$ & $\Delta_5$  \\ \hline
 \strut
%
 Direction &
 $({13\over9},2,{5\over3},{4\over3},{4\over3})$ &
 $({11\over6},{7\over2},1,1,{3\over2})$ \\ \hline
%
%
 ${\langle \varphi_1^3\rangle\langle \varphi_4^3\rangle\over\langle
 \varphi_1^2\varphi_4\rangle\langle \varphi_1\varphi_4^2\rangle}$ &
 $-{2\over5}s^2-{129\over 250}s+\dots$ &
   $9+289s^{-1}+\dots$  \\ \hline

%
%
 ${\langle \varphi_2^2\varphi_4\rangle\langle
 \varphi_3^2\varphi_4 \rangle\over\langle
 \varphi_2\varphi_3\varphi_4\rangle\langle \varphi_2\varphi_3\varphi_4\rangle}$&
 & \\ \hline
%
%
${\langle \varphi_2\varphi_3\varphi_4\rangle\langle \varphi_0\varphi_2^2
\rangle\over\langle \varphi_2^2\varphi_4\rangle\langle \varphi_0\varphi_2
\varphi_3\rangle}$ &
$0-2s^{-1}+\dots$ &   \\ \hline
%
%
${\langle
\varphi_2\varphi_3\varphi_4\rangle\langle \varphi_0\varphi_1^2
\rangle\over\langle \varphi_1^2\varphi_4\rangle\langle \varphi_0\varphi_2
\varphi_3\rangle}$ &
 & $0-2s^{-1}+\dots$ \\ \hline
\end{tabular}
\end{center}
\end{table}

This, in conjunction with the abstract and general isomorphism we find between
the complex structure moduli space of a Calabi-Yau and the {\it enlarged}\/
K\"ahler moduli space of its mirror, provides us with strong evidence that
 our understanding
 of Calabi-Yau conformal field theory moduli space is correct. In particular, as
our earlier discussion has emphasized, this implies that the basic operation of
deformation by a truly marginal operator (from a space-time point of view, this
corresponds to a slow variation in the vacuum expectation value of a scalar
field with an exactly flat potential) can result in a change in the topology of
the Calabi-Yau target space. This discontinuous mathematical change, however, is
perfectly smooth from
 the point of view of physics. In fact, using mirror
symmetry, such an evolution can be reinterpreted as a smooth, topology
preserving, change in the ``shape'' (complex structure) of the mirror space.

There are two important points we need to mention. First, for ease of 
discussion, we have focused on the case in which the only deformations are those
associated with the K\"ahler structure of $M$ and, correspondingly, only the
complex structure of $\tilde M$. This may have given the incorrect impression
that the topology changing transitions under study can always be reinterpreted
in a topology preserving manner in the mirror description. The generic
situation, however, is one in which the complex structure {\it and}\/ the
K\"ahler structure of $M$ and $\tilde M$ both change. Again, from a space-time
point of view, this simply corresponds to a slow variation of the expectation
values of a set of scalar fields with flat potentials. Under such circumstances,
topology change can occur in both the original and the mirror description. Our
reasoning will ensure that such changes are physically smooth. Clearly there is
{\it no} interpretation 
 --- on the original or on the mirror manifold --- which can avoid the
topology changing character of the processes. 

Second, we have used the terms ``enlarged'' and ``partially enlarged'' 
in our discussion of the K\"ahler moduli space. We now briefly 
indicate the 
distinction. The central result of this discussion
 is that the proper geometric interpretation of conformal field theory
moduli space requires
 that we augment the previously held notion of a single
complexified K\"ahler cone associated with a single topological type of
Calabi-Yau space. In the previous sections, we have focused on {\it part}\/ of
the requisite augmentation: we need to include the complexified K\"ahler cones
of Calabi-Yau spaces related to the original by flops of rational curves (of
course, it is arbitrary as to which Calabi-Yau we call the original). These
K\"ahler cones adjoin each other along common walls. The space so created is the
{\it partially}\/ enlarged 
K\"ahler moduli space. It turns out, though, that
conformal field theory moduli space requires that even more regions be added.
We saw  evidence of this in section \ref{sec43}.
Equivalently, the partially enlarged K\"ahler moduli space is only a subregion
of the moduli space which is mirror to the complex structure moduli space of the
mirror Calabi-Yau manifold. The extra regions which need to be added arise
directly from the toric geometric description and were first identified in the
two dimensional supersymmetric gauge
 theory approach of \cite{WP}, as discussed in section \ref{sec43}. These
regions correspond to the moduli spaces of conformal theories on orbifolds of
the original smooth Calabi-Yau, Landau-Ginzburg orbifolds, 
gauged
Landau-Ginzburg theories and hybrids of the above. The union of all of these
regions (which also join along common walls) constitutes the {\it enlarged}\/
K\"ahler moduli space.
In section \ref{sec43}, the examples we
studied only had a single Calabi-Yau
region and a single Landau-Ginzburg region. More complicated examples with
larger $h^{1,1}$ have richer phase structures.
  For instance, in the example studied in section \ref{sec67}, we
found that there were five regions in the partially enlarged K\"ahler moduli
space. The enlarged K\"ahler moduli space, as it turns out, has $100$ regions.
One of these is a Landau-Ginzburg orbifold region, $27$ of these are sigma
models on Calabi-Yau orbifolds, and $67$ of these are hybrid theories
consisting of Landau-Ginzburg models fibered over various compact spaces. It
is worthwhile emphasizing that in contrast with previously held notions,
orbifold theories are not simply boundary 
points in the moduli space of
smooth Calabi-Yau sigma models but, rather, they have their own regions in the
enlarged K\"ahler moduli space and hence are more on an equal footing with the
smooth examples.

\section{Space-Time Topology Change --- The Drastic Case}
\label{topologychange2}

In this section we discuss the results of \cite{rGMS} in which physically
smooth transitions between Calabi-Yau manifolds with {\it different} Hodge
numbers was established. Contrary to the above analysis,
 non-peturbative effects
play a crucial role.

\subsection{Basic Ideas}

A particularly useful way of summarizing the discussion
of the last section is as follows: classical reasoning
suggests that our physical models will be badly behaved if the complex
structure is chosen to lie on the discriminant locus or if the K\"ahler class
is chosen to lie on a wall of the classical K\"ahler moduli space. The fully quantum
corrected conformal field theory corresponding to such points 
(yielding genus zero string theory), though, proves to be generically 
{\it non-singular}
 on walls in the K\"ahler moduli space.
The pronounced distinction between the classical and stringy conclusions arises
because
such points are strongly
coupled theories (as the coupling parameter $\alpha'/R^2$ gets big as
we shrink down $R$ --- the radius of an $S^2$). Analyzing such strongly coupled
theories directly is hard; however, by mirror symmetry we know they are equivalent
to weakly coupled field theories on the mirror Calabi-Yau space where we
can directly show them to be well behaved. 

So much for the generic point on a wall in the K\"ahler parameter space:
classically they look singular but in fact they are well defined. What
about choosing the complex structure to lie on the discriminant locus (which
by mirror symmetry corresponds to a non-generic point on a wall in the K\"ahler 
parameter space of the mirror)? Might it be that these theories are well behaved too?
At first sight the answer seems to be ``no". By taking the K\"ahler class
to be deep inside a smooth phase 
(i.e. a smooth large radius Calabi-Yau background),
we trust perturbation theory and can directly compute conformal field theory
correlation functions. Some of them  diverge as we approach the
discriminant locus. This  establishes that the conformal field theory
is badly behaved. It is, however, important to distinguish between conformal
field theory and string theory. Conformal field theory is best thought of as
the effective description of string degrees of freedom which are light in the
$g_s \rightarrow 0$ limit, with $g_s$ being
the string coupling constant. This includes all of the familiar perturbative
string states, but effectively integrates out 
  non-perturbative states whose most direct description is
in terms of solitons in the low energy effective string 
action. As discussed in \cite{Pol-TASI}
a powerful microscopic way of describing these states is in
 in terms of Dirichlet-branes. For the most part of our discussion,
the details of such a description will not be essential.

 We are thus faced with the moduli space for an
effective string description that contains points where
 physics appears to be singular.
A close analog of this situation plays a central role in the celebrated work
of Seiberg and Witten \cite{rSW},
where it is argued that the apparent singularity is
due to the appearance of new massless non-perturbative
degrees of freedom at those singular moduli space points.
A natural guess in the present setting, then, is that the apparent singularity
encountered on the discriminant locus is due to previously
massive non-perturbative string states
becoming massless. This solution was proposed by Strominger and we review its
success in the next section.

\subsection{Strominger's Resolution of the Conifold Singularity}

To quantitatively understand the proposed resolution of conifold singularities,
we must introduce coordinates on the complex structure moduli space.
 A convenient
way to do this --- described in some detail 
in \cite{CdlO} ---
is to introduce a symplectic homology basis of $H^3(M,\IZ)$.
This means that we  find three-cycle representatives in $H^3(M,\IZ)$.
 denoted by
$\{A_I,B^J\}$, where $I,J = 0,...,h^{2,1}(M)$ such that
$$
  A_I \cap B^J = \delta_I^J~,
 ~~~ A_I \cap A_J = B^I \cap B^J = 0~.
$$
As we wander around the complex structure moduli space, the holomorphic
three-form will vary, and we can use this variation as a means of
establishing local coordinates. To do so, 
we let 
\beq
   z^J \equiv \int_{B^J} \Omega~,
\eeq
 and 
\beq
   G_I = \int_{A^I} \Omega~,
\eeq
where
$\Omega$, as usual,
 is the holomorphically varying three-form on the family of Calabi-Yau
spaces
being studied. It is well known that the $z^J$ provide a good set of local
projective coordinates on the moduli space of complex structures and that
the $G_I$ can be expressed as functions of the $z^J$.

In terms of these coordinates,
 a conifold point in the moduli space can roughly be
thought of as a point where some $z^J$'s
 vanishes (we will be more precise on this
in the next section). 
The corresponding $B^J$ is called a vanishing cycle as the period of $\Omega$
over it goes to zero. Intuitively, this homology cycle is collapsing as we
approach the point $z^J = 0$  in the moduli space.
 For our purposes, there is one main implication of
the vanishing of, say, $z^J$,
 that we should discuss: the metric on the moduli space
is singular at such a point. The easiest way to see this is to make
use of the fact that the moduli space geometry of
Calabi-Yau manifolds is highly constrained. Not only is the
complex structure moduli space (and the K\"ahler moduli
space as well) a K\"ahler manifold
in its own right, it is actually a {\it special} K\"ahler manifold.
We recall from section \ref{sec:bacmod}
that special K\"ahler manifolds
have the property that there is a holomorphic potential $\cal F$ for the
K\"ahler potential, called the prepotential. Letting
$G_J = \partial {\cal F}/ \partial z^J$, 
the K\"ahler potential on the
on the complex structure moduli space can be written
\beq
\label{eKahler}
   K = -\ln(i\, {\overline z}^I\, G_I - i\, z^I\,{\overline G}_I)~.
\eeq
If we knew
the explicit form for $G_J(z)$, we would thus be able to calculate the local
form of the metric near the conifold point. Considerations of monodromy are
sufficient to do this: as we will discuss in greater generality below, if
we follow a path in the moduli space that encircles $z^J = 0$, the period $G_J$
is not single-valued but rather
undergoes a non-trivial monodromy transformation
\beq
\label{emonodromy}
      G_J \rightarrow G_J + z^J~.
\eeq
Near $z^J = 0$, we can therefore write
\beq
\label{elocalform}
  G_J(z^J) = {1 \over 2 \pi i}\, z^J \ln z^J + \hbox{\rm single-valued}~.
\eeq
Using this form one can directly compute that the metric $g_{J \overline J}$
has a curvature singularity
at $z^J = 0$. 

The reason that the singularity of
 the metric on the moduli space is an important
fact is due to its appearance in the Lagrangian
 for the four-dimensional effective
description of the moduli for a string model built on such a Calabi-Yau.
Namely, the (space-time)
 non-linear sigma model Lagrangian for the complex structure moduli
$\phi^K$ is of the form 
$$
  \int\, d^4x\, g_{I \overline J}\,
  \nabla \phi^I\, \nabla \phi^{\overline J}~.
$$
Hence, when the metric on the moduli space degenerates, so apparently does our
physical description.

This circumstance --- a moduli space of theories containing points at
which physical singularities appear to develop
 --- is one that has been discussed
extensively in recent work of Seiberg and Witten \cite{rSW}.
 The natural explanation
advanced for the physical origin of the singularities encountered is that
states which are massive at generic points in the moduli space become massless
at the singular points.
 As the Lagrangian description is that of an effective field
theory in which massive degrees of freedom have been integrated out,
 if a previously
massive degree of freedom becomes massless, then we will be incorrectly
integrating out a massless mode and hence expect a singularity to develop.
In the case studied in  \cite{rSW}, the states that became massless were
BPS saturated magnetic
monopoles or dyons. Strominger proposed that in compactified 
type IIB string theory there
are analogous electrically or magnetically charged black hole states that
become massless at conifold points. The easiest way to understand these
states is to recall that, in ten-dimensional type IIB string theory, there
are $3 + 1$ dimensional extremally charged
extended soliton solutions with a horizon: so-called
black three-branes
\cite{rHS}.
These solitons carry Ramond-Ramond charge that can be detected
by integrating the five-form field strength over
 a surrounding Gaussian five-cycle
$\Sigma_5$: 
$$
 Q_{\Sigma_5} = \int_{\Sigma_5}F^{(5)}~.
$$
 Now, our real interest is
in how this soliton appears after compactification to four dimensions via
a Calabi-Yau three-fold.
 Upon such compactification, the three spatial dimensions
of the black soliton can wrap around non-trivial three-cycles on the Calabi-Yau
and hence appear to a four-dimensional observer as black hole states.
More precisely, they yield an $N = 2$ hypermultiplet of states. The effective
electric and magnetic charges of the black hole state are then obtained
by integrating  $F^{(5)}$ over $A_I \times S^2$ and $B^J \times S^2$.
Explicitly, making the natural assumption of charge quantization,
we can write
\beq
\label{echarges}
   \int_{A_I \times S^2} F^{(5)} = g_5\, n_I~, ~~~~~
   \int_{B^J \times S^2} F^{(5)} = g_5\, m^J~,
\eeq
where $g_5$ is the five-form coupling and $n_I$, $m_J$ are some integers.
 Of prime
importance is the fact that these are BPS saturated states and hence are
subject to the mass relation \cite{rCDFV}
\beq
\label{eBPS}
   M = g_5\, e^{K/2}\, |m^I\, G_I - n_I\, z^I|~.
\eeq
Let us consider the case in which $n_I$ = $\delta_{IJ}$ and $m^I = 0$,
for all $I$ with $J$ fixed. In the conifold limit for which  
$z^J$ goes to zero,
we see that the mass of the corresponding electrically
charged black hole vanishes.
Hence, it is no longer consistent to exclude such states from
 direct representation
in the Wilsonian effective field theory action describing the low energy
string dynamics. 

The claim is that the singularity encountered above is
due precisely to such exclusion. Curing the singularity should therefore
be achieved by a simple procedure: include the black hole hypermultiplet
in the low energy effective action.
There is a simple check to test the validity
of this claim. Namely, 
if we incorrectly integrate out the black hole hypermultiplet from
the Wilsonian action, we should recover the singularity discussed above.
This is not hard to do. 
The structure of $N = 2$ space-time supersymmetry  implies
that the effective
Lagrangian is governed by a geometrical framework which is identical to that
governing Calabi-Yau moduli space. Namely, we can introduce holomorphic
projective coordinates $z^J$ on the moduli space of the physical model;
the model is determined by knowledge of the  holomorphic functions
$G^{\rm phys}_I(z)$ which are in
turn determined by a holomorphic prepotential
${\cal F}^{\rm phys}$. The superscript ``phys" is meant to distinguish these
objects from their geometric counterparts, although we show momentarily
see that they are identical.
The Lagrangian for an $N = 2$ space-time
theory can be written in a manifestly supersymmetric form as
\beq
{\cal L} = {1\over 8\pi}\,  
 \, {\rm Im}\left\lbrack
  \int d^2\theta\, \tau_{IJ}\, W^{I\alpha} W^J_\alpha
   + \dots \right\rbrack~,
  %
  %
  %
\eeq
where
$$
  \tau_{IJ}= \partial_I G^{\rm phys}_J~.
$$
and $W^I$ are certain superfields.
Therefore,  the coupling constant for the $J$-th $U(1)$ is given
by 
$$
  \tau_{JJ}= \partial_J G^{\rm phys}_J~.
$$
 We can turn the latter statement around by noting
that knowledge of the coupling constant effectively allows us to determine
$G_J$. We can determine the  behaviour of the coupling by a simple one-loop 
Feynman diagram, which again by $N = 2$ is all we need consider. Integrating
out a black hole hypermultiplet in a
 neighborhood of the $z^J = 0$ conifold point
yields the standard logarithmic contribution to the running coupling $\tau_{JJ}$
and hence we can write
\beq
\label{ecoupling}
 \tau_{JJ} = {{1 \over 2 \pi i}} \ln z^J + \hbox{\rm single-valued}~.
\eeq
{}From this we determine, by integration, that
\beq
\label{eperiods}
   G^{\rm phys}_J = {{ 1 \over 2 \pi i}} z^J \ln z^J + 
   \hbox{\rm single-valued}~.
\eeq
We note that this is precisely the same form as we found for $G_J$  earlier
via mathematical  monodromy considerations. 
 Now, by special geometry, everything about the mathematics
and physics of the system follows from knowledge of the $G_J$.
The K\"ahler potential and hence metric on the moduli space
are determined by \calle{prepotential}.
 For our purposes,
therefore, the singularity encountered previously (by determining the metric
on the moduli space from the $G_J$),
 has been precisely reproduced by incorrectly
integrating out the massless soliton state.
 This justifies the claim, therefore,
that we have identified the physical origin of the singularity and also that
by including the black hole field in the Wilsonian action (and therefore not
making the mistake of integrating them out when they are light),
 we cure the
singularity.

\subsection{Conifold Transitions and Topology Change}

In the previous section, we have seen how the singularity that arises when
an $S^3$ collapses to a point is associated with the appearance of new massless
states in the physical spectrum. By including these new massless states in
the physical model, the singularity is cured. In this section, we
consider a simple generalization of this discussion which leads to  dramatic
new physical consequences \cite{rGMS}.  
Concretely, we consider a less generic degeneration in
which:

\begin{enumerate}
\item
   More than one, say $P$, three-cycles degenerate.
\item
    These $P$
    three-cycles are not homologically independent but rather 
    satisfy $R$ homology relations.
\end{enumerate}

\noindent As we will now discuss, this generalization implies that:

\begin{enumerate}
\item
   The bosonic potential for the scalar fields in the hypermultiplets 
  that become massless at the degeneration has $R$ flat directions.
\item
  Moving along such flat directions takes us to another branch of
  type II string moduli space, corresponding to string propagation on a
  topologically distinct Calabi-Yau manifold. If the original Calabi-Yau has
  Hodge numbers $h^{1,1}$ and $h^{2,1}$ then the 
  new Calabi-Yau has Hodge numbers
  $h^{1,1} + R$ and $h^{2,1} - P + R$.
\end{enumerate}

\begin{figure}[htbp]
\epsfysize=3cm
\centerline{\epsfbox{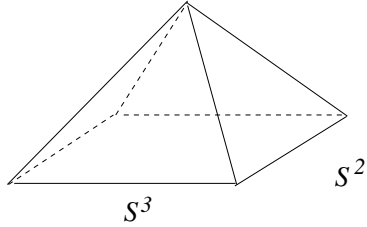}}
\caption{The neighbourhood around a node.}
\label{fig:node}
\end{figure}

In order to understand this result, there are a couple of useful pieces of
background information we should review. First, let us discuss a bit more
precisely the mathematical singularities we are considering
\cite{rLefshetzCandelas}.
 As we have discussed,
the discriminant locus denotes those points in the complex structure moduli
space of a Calabi-Yau where the space fails to be a complex manifold.  
We focus on cases in which the degenerations occur at some number of isolated
points on the Calabi-Yau. In particular, we consider singularities that are
known as ``ordinary double points''.
 These are singular points which can locally
be expressed in the form
\beq
\label{enode}
   \sum_{i=1}^4 w_i^2 = 0
\eeq
in ${\Bbb C}^4$. This local representation is a cone with singular point at the
apex, namely the origin. To identify the base of the cone we intersect it
with a seven-sphere in $\IR^8$, 
$$
  \sum_{i=1}^4 |w_i|^2 = r^2 ~.
$$
Introducing
the complex vector
$\vec{w} = \vec{x} + i \vec{y} = (w_1,w_2,w_3,w_4)$
the equation of the intersection can be expressed as
$$
  \vec{x} \cdot \vec{x} = {r^2\over2}~,
 ~~~~~\vec{y} \cdot \vec{y} = {r^2\over2}~,
 ~~~~~\vec{x} \cdot \vec{y} = 0~.
$$
The first of these is an $S^3$, the latter two equations give an $S^2$ fibered
over the $S^3$.
 As there are no non-trivial such fibrations, the base of the cone
is $S^2 \times S^3$. Calabi-Yau spaces which have such isolated ordinary double point
singularities are known as conifolds and the corresponding point in the moduli
space of the Calabi-Yau is known as a conifold point. The ordinary double point
singularity is also referred to as a node. In figure \ref{fig:node}
we illustrate a neighbourhood of a node.

\begin{figure}[htbp]
\epsfysize=3cm
\centerline{\epsfbox{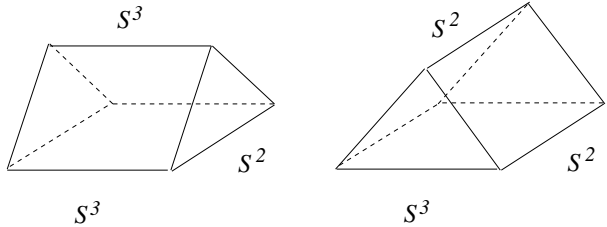}}
\caption{The two possible resolutions of a node.}
\label{fig:nodedef}
\end{figure}

Having described the singularity in this way we immediately discern two distinct
ways of resolving it: either we can replace the apex of the cone
with an $S^3$, known as a deformation of the singularity, or we can replace
the apex with an $S^2$, known as a small resolution of the singularity. 
These two possibilities are illustrated in figure \ref{fig:nodedef}.
The deformation simply undoes the degeneration by re-inflating the
shrunken $S^3$ to positive size.
  The small resolution, on the other hand, has a more
pronounced effect:
 it repairs the singularity in a manner that changes the topology
of the original Calabi-Yau. After all, replacing an
$S^3$ with an $S^2$ is a fairly radical transformation.
We shall find
the {\it physical} interpretation of these two ways of resolving conifold
singularities.

A second piece of background information is a mathematical fact due
to Lefshetz concerning
monodromy. Namely, if $\gamma^a$ for $a = 1,...,k$ are $k$ vanishing
three-cycles at a conifold point in the moduli space, then another 
three-cycle $\delta$ undergoes monodromy
\beq
\label{emonodromy2}
   \delta \rightarrow \delta + \sum_{a=1}^k\, (\delta \cap \gamma^a)\, 
    \gamma^a
\eeq
upon transport around this point in the moduli space.

With this background, we can now proceed to discuss the result quoted at the
beginning of this section. We will do so in the context of a particularly 
instructive example, although it will be clear that the results are general.
We begin with the quintic hypersurface 
in ${\Bbb C}P^4$, which is we have seen to
have  Hodge numbers $h^{2,1} = 101$ and $h^{1,1} = 1$. We then move to a
conifold point by deforming the complex structure to the equation
\beq
\label{esingular}
   x_1\, g(x) + x_2\, h(x) = 0~,
\eeq
where $x$ denotes the five homogeneous ${\Bbb C}P^4$ coordinates
$(x_1,...,x_5)$ and $g$ and $h$ are both generic quartics. We note that
\calle{esingular} and its derivative vanish at the sixteen points
\beq
\label{esixteen}
 x_1 = x_2 = g(x) = h(x) = 0~.
\eeq
 It is straightforward to check,
by examining the second derivative matrix, that for generic
$g$ and $h$ these are sixteen ordinary
double points. And, of  primary importance to our present discussion,
the sixteen singular points lie on the ${\Bbb C}P^2$ contained
in  ${\Bbb C}P^4$ given by $x_1 = x_2 = 0$.
This implies that 
the sixteen vanishing cycles $\gamma^a, a = 1,...,16$ that degenerate to the
double points satisfy the non-trivial homology relation \cite{rCL}
\beq
\label{erelation}
   \sum_{a=1}^{16}\, \gamma^a = 0~.
\eeq
We are thus in the desired situation. To proceed
with the physical analysis, we follow two steps. First,
 we check that inclusion of the appropriate
massless hypermultiplets  cures the singularity, as it did in the simpler
case studied in \cite{rStrom}.
 Second, we  analyze the physical implication
of the existence of a non-trivial homology relation.

\vspace{.25in}
\noindent
i) {\it Singularity resolution:}

\vspace{.15in}
We introduce a symplectic homology basis $\{A_I,B^J\}$, with $I,J = 1,...,204$.
By suitable change of basis we can take our sixteen vanishing cycles
$\gamma^a$ $(a=1,...,16)$ to be $B^1,...,B^{15}$ and $ -\sum_{a=1}^{15} B^a$.
As usual, we define 
$$
  z^I = \int_{B^I} \Omega~,~~~~~G_J = \int_{A_J} \Omega~.
$$
Now, for any cycle $\delta$ we have, as discussed before, the monodromy
$$
 \delta \rightarrow \delta + \sum_{a=1}^{16}\,
 (\delta \cap \gamma^a)\, \gamma^a
  ~.
$$
From this we learn that the local form of the period over $\delta$ is given
by
\beq
\label{eperiod}
\int_{\delta} \Omega = {1 \over 2 \pi i}\, \sum_{a=1}^{16}(\delta
\cap \gamma^a)  \,  ( \int_{\gamma^a} \Omega )
  \,(\ln \int_{\gamma^a} \Omega) + \hbox{\rm single-valued} ~.
\eeq
Specializing this general expression to $\delta = A_J$, we therefore see
\beq
\label{eGperiod}
 G_J = {1 \over 2 \pi i}\,z^J \ln z^J + 
  {1 \over 2 \pi i}\,(\sum_{I=1}^{15} z^I)\, (\ln
  \sum_{I=1}^{15} z^I) + \hbox{\rm single-valued}~.
\eeq
By special geometry, this latter expression determines the properties of
the singularity associated with the conifold degeneration being studied. Thus,
the question we now seek to answer is: if we incorrectly integrate out the black
hole states which become massless at this conifold point, do we reproduce
the form \calle{eGperiod}?

To address this issue we must identify the precise number and charges of
the states that are becoming massless at the degeneration point. As discussed
in \cite{rStrom},
 the counting of black hole states is a delicate issue for which
there is as yet no rigorous algorithm. In \cite{rStrom}, one homology class in
$H^3$ degenerated at the conifold singularity and it was hypothesized that
this implies one fundamental black hole state --- the one of minimal charge ---
needs to be included in the Wilsonian action. In the present example, though,
we have sixteen three cycles in
fifteen homology classes in $H^3$ degenerating. In \cite{rGMS}
 it was argued that
this should imply sixteen fundamental black hole fields need to be included in
the Wilsonian action. Physically speaking, the black three-brane can wrap around
any of the sixteen degenerating three-cycles, which at
 large overall radius of the
Calabi-Yau would be widely separated.
It thus seems sensible that even though there
are only fifteen homology classes degenerating, we actually get sixteen massless
black hole states. Subsequent analyses by \cite{BBS,OOY}
 have shown
 that one expects a massless hypermultiplet
for each {\it supersymmetric} three-cycle in a given degenerating homology
class. It can be a subtle undertaking to find all such supersymmetric
cycles, but the physical reasoning given above for the example at hand
is quite convincing, so we will pursue the implications of having sixteen
new massless hypermultiplets.

The charges of these states are easy to derive. If we let
$H^a$ be the black hole hypermultiplet associated with the vanishing cycle
$\gamma^a$ then the charge of $H^a$ under the $I$-th $U(1)$ is given by
$$
   Q^a_I = A_I \cap \gamma^a~,
$$
 where  we write $F^{(5)}$ as the self-dual part
of $\sum_I \alpha^I F_I^{(2)}$, with $\alpha^I$ dual to $A_I$. We immediately
learn from this that the black holes states have charges
\beq
\label{echarges2}
 Q^a_I = \delta^a_I~,~~~ 1 \le a \le 15 \quad \hbox{ \rm and } \quad
 Q^{16}_I = -1,~~~ 1 \le I \le 15~,
\eeq
with all other charges zero. This is enough data to determine the running of
the gauge couplings:
\beq
\label{etauij}
  \tau_{IJ} = {{1 \over 2 \pi i}} \sum_{a = 1}^{16}\, Q^a_I\, Q^a_J\,
 \ln m^a ~,
\eeq
where the mass $m^a$ of $H^a$ is proportional to $\sum_I Q^a_I z^I.$
Using the above charges we therefore have
\beq
\label{erunning}
  \tau_{IJ} = {1 \over 2 \pi i} \delta_{IJ}\, \ln z^J +
   {{1 \over 2 \pi i}}\, \ln\sum_{k=1}^{15} z^k + 
 \hbox{\rm single-valued}~.
\eeq
Integrating we find therefore
\beq
\label{eGJ}
  G_J^{\rm phys} = {{1 \over 2 \pi i}} z^J\, \ln z^J +
 {{1 \over 2 \pi i}}(\sum_{k=1}^{15} z^k)\,  (\ln
  \sum_{k=1}^{15} z^k) 
  + \hbox{\rm single-valued}~.
\eeq
We note that this matches \calle{eGperiod}
 and hence we have shown that inclusion of
the sixteen black hole soliton
 states which become massless cures the singularity.

Having shown that a slight variant on Strominger's original proposal is
able to cure the singularity found in this more complicated situation,
we now come to the main point of the discussion:

\vspace{.25in}
\noindent
ii) {\it What is the physical significance of  non-trivial homology
relations between vanishing cycles?}

\vspace{.15in}
To address this question we consider the scalar potential governing the black
hole hypermultiplets. It can be written as
\cite{Westbook}:
\beq
\label{epotential}
  V = \sum\, E^I_{\alpha \beta}\,  E^{\alpha \beta}_I~,
\eeq
where
\beq
\label{eE}
  E^I_{\alpha \beta} = \sum_{a=1}^{16}\, Q^I_a\, \epsilon_{\alpha \gamma}
  h^{*(a)}_{\gamma} \, h^{(a)}_{\beta} - (\alpha \leftrightarrow \beta)~,
\eeq
in which the indices satisfy $I = 1,...,15,~ \alpha, \beta, \gamma = 1,2$.
The fields $h^{(a)}_{1}$ and $h^{(a)}_{2}$ are the two complex scalar fields
in the hypermultiplet $H^a$.

We consider the possible flat directions which this potential admits.
The most obvious flat directions are those for which 
$\langle h^{(a)}_{\beta} \rangle = 0$ with non-zero values for the
scalar fields in the vector multiplets. Physically, moving along such
flat directions takes us back to the Coulomb phase in which the black hole
states are massive. Mathematically, moving along such flat directions,
\beq
 {1 \over 2 \pi i}\, (\sum_{k=1}^{15} z^k)\,  \ln\sum_{k=1}^{15} z^k
\eeq
gives positive volume back to the degenerated $S^3$'s and hence resolves
the singularity by deformation. 

The non-trivial homology relation implies that there is  another flat direction.
Since $Q^I_a = A_I \cap \gamma^a$, we see that the homology relation
$\sum_{a = 1}^{16} \gamma^a = 0 $ implies $\sum_a Q^I_a = 0$, for all $I$.
This then implies that we have another flat direction of the form
$\langle h^{(a)}_{\beta} \rangle = v^{\beta}$ for all $a$ with $v$ constant.
In fact, simply counting degrees of freedom shows that this solution is unique
up to gauge equivalence. What happens if we move along this flat direction?
It is straightforward to see that this takes us to a Higgs branch in which
fifteen vector multiplets pair up with fifteen hypermultiplets to become
massive. This leaves over one massless hypermultiplet from the original sixteen
that become massless at the conifold point. We see therefore that the spectrum
of the theory goes from $101$ vector multiplets and $1$ hypermultiplet
(ignoring the dilaton and graviphoton) to $ 101 - 15 = 86$ vector multiplets
and $1 + 1 = 2$ hypermultiplets. 

What is the significance of these numbers? Well, it turns out \cite{rCGH1,rCGH2}
that precisely these Hodge numbers arise
from performing the other resolution of the conifold singularity (besides
the deformation) --- the small resolution described earlier! Hence, we
appear to have found the physical mechanism for effecting a small resolution
and in this manner changing the topology of the Calabi-Yau background.

Although we have focused on a specific example, it is straightforward to work
out what happens in the more general setting of $P$ isolated vanishing cycles
satisfying $R$ homology relations. Following our discussion above, we get
$P$ black hole hypermultiplets becoming massless with $R$ flat directions
in their scalar potential. Performing a generic deformation along these
flat directions causes $P - R$ vectors to pair up with the same number
of hypermultiplets. Hence the Hodge numbers change according to
\beq
\label{eHodge}
 (h^{2,1}, h^{1,1}) \rightarrow (h^{2,1} - (P-R), h^{1,1} + R)~.
\eeq
The Euler characteristic of the variety thus    jumps by $2P$.

So, in answer to the question posed above: homology relations amongst the
vanishing cycles give rise to new flat directions in the scalar black hole
potential. Moving along such flat directions takes us smoothly to  new branches
of the type II string theory moduli space. These other branches correspond to
string propagation on 
topologically distinct Calabi-Yau manifolds.
 We have therefore apparently physically
realized the Calabi-Yau conifold transitions
 discussed some years ago --- without
a physical mechanism --- in insightful papers of Candelas,
 Green and H\"ubsch \cite{rCGH1,rCGH2}.
 In the type II string moduli space we thus see that
we can smoothly go from one Calabi-Yau manifold to another by varying the
expectation values of appropriate scalar fields.

\subsection{Black Holes, Elementary Particles and a New Length Scale}

There are two other aspects of these topology changing transitions which are
worthy of emphasis. First, in the Coulomb phase, the black hole soliton states
are massive. At the conifold point they become massless. As we move into
the Higgs phase, some number of them are eaten by the Higgs mechanism with
the remainder staying massless. Now, with respect to the topology of the
new Calabi-Yau in the Higgs phase,
 these massless degrees of freedom are associated
with new elements of $H^{1,1}$.
 Such states, as is well known, are perturbative
string excitations --- commonly referred to as elementary ``particles". Thus,
a massive black hole sheds its mass, becomes massless and then re-emerges
as an elementary particle-like excitation.
 There is thus no invariant distinction
between black hole states
 and elementary perturbative string states: they smoothly
transform into one another through the conifold transitions. This
realizes an old suspicion linking black holes and elementary particles
in a quite explicit manner.

The second point we wish to
 emphasize is one made in 
\cite{She} and discussed in
 Shenker's lectures of this school.
Let us consider a bit more carefully the origin of the logarithmic running
in \calle{ecoupling}.
 We have a charged black hole hypermultiplet of mass (putting
all of the couplings in place): 
$$
   m_{\rm bh} \sim |z| m_s/g_s~,
$$
 where
$m_s \sim 1/\sqrt{\alpha'}$. The one-loop charge screening
interaction therefore contributes 
\beq
\label{scale}
  \int^{\Lambda} {d^4k \over (2 \pi)^4} \,
  {1 \over k^2 + m_bh^2}
   \sim -\ln {\Lambda\over m_{\rm bh} }~,
\eeq
where $\Lambda$ is the ultraviolet scale above which an effective description
in terms of a point-like black hole state is no longer accurate.
At first sight, one's natural guess is that $\Lambda$ should be on the order
of $m_s$ since above this energy scale we expect the extended nature
of the string to have a dominant effect. An examination of
\calle{scale} reveals that this expectation does not appear to be correct.
Rather, in order for \calle{scale} to be $g_s$ independent (as it must
by the decoupling of vector multiplets and neutral hypermultiplets
like the dilaton), we see that $\Lambda \sim m_s/g_s$, so that
\calle{scale} gives the desired $\ln z$ behavior. In other
words, the point-like description of these solitons persists
all the way down to length scales of order $g_s \sqrt{\alpha'}$. For small
string coupling, this is much less than the previous conception of
the string scale setting a ``lower'' length limit to physical processes.
Study of such non-perturbative string degrees of freedom thus reveals
the possibility of a whole new sub-string scale geometrical world --- a realm
that is presently under instense study.

\section{The Basics of Toric Geometry}
\label{toricintro}

Toric geometry has played an important role in a number of developments
in quantum geometry. Projective and weighted projective spaces are examples of
toric varieties and hence a huge number of Calabi-Yau manifolds can
be realized as subspaces of toric varieties. As we shall see, the toric
structure provides a systematic framework for understanding detailed
properties of Calabi-Yau manifolds
constructed in this manner. Our goal in this chapter
is to give an elementary discussion of toric geometry
emphasizing those points most relevant to the present lectures. For more
details and proofs the reader should consult \cite{rOda,rFulton}.
We will then apply this formalism to better understand some of the
details of flop transitions discussed earlier as well as to
extend and generalize the conifold transitions of the last chapter.

\subsection{Intuitive Ideas}
\label{sec:3.1}

Toric geometry describes the structure of a certain class of geometrical
spaces in terms of simple combinatorial data. When a space admits
a description in terms of toric geometry, many basic and essential
characteristics of the space --- such as its divisor classes, its
intersection form and other aspects of its cohomology --- are neatly coded
and easily deciphered from analysis of corresponding lattices. We will
describe this more formally in the following subsections. Here
we outline the basic ideas.

The complex projective space ${\Bbb C}P^n$ can be expressed as
\beq
{\Bbb C}P^n = {\IC^{n+1} - \{0,0,...,0\} \over \IC^* }
 ~,
\eeq
where the $\IC^*$ action by which we quotient is
$(z_1,...,z_{n+1}) \rightarrow  (\lambda z_1,...,\lambda z_{n+1})$.
A toric variety is a generalization of this in which rather than
just removing the origin, we remove a point set $F_{\Delta}$ depending
on certain data $\Delta$ and we quotient by a number of $\IC^*$ actions.
That is, a toric variety $V$  can be expressed in the form 
\beq
    V = {\IC^m - F_{\Delta} \over (\IC^*)^p}~.
\eeq

As we shall see in the sequel, the action of $(\IC^*)^p$ and the
form of $F_{\Delta}$ will be determined in terms of certain simple
combinatorial data. In building up to this description, it is worthwile
to recall another feature of ${\Bbb C}P^n$: it contains $(\IC^*)^n$
as a dense open subset and can therefore be thought of as a compactification
of $(\IC^*)^n$. To keep a concrete example in mind, ${\Bbb C}P^1$
is nothing but the sphere $S^2$. The latter is a compactification of
$\IC^*$ in which we add the points at zero and infinity. These points,
in fact, can be thought of as limiting points of the natural action
of $\IC^*$ on itself.
In general, this gives 
another way of thinking about toric varieties: partial or complete
compactifications of  products of $\IC^*$ where the boundary points
included are derived from limits of an action of the $\IC^*$ factors
on themselves. We will begin
with this perspective.

Following the above remarks, a
 toric variety  $V$ over $\IC$ (one can work over other fields but that
shall not concern us here) is a complex geometrical space which
contains the {\it algebraic torus} 
$$
    T =  \underbrace{\IC^* \times\ldots\times \IC^*}_n
       \cong (\IC^*)^n
$$
 as a dense open subset. Furthermore, there is an action
of $T$ on $V$; that is, a map
 $T \times V \rightarrow V$ which extends the natural
action of $T$ on itself.
The points in $V-T$ can be regarded as limit points for the action of
$T$ on itself; these
serve to give a partial compactification
of $T$. Thus, $V$ can be thought of as a $(\IC^*)^n$ together with
additional limit points which serve to partially (or completely)
compactify the space\footnote{As we
shall see in the next subsection, this discussion is a bit
na\"\i ve ---
these spaces need not be smooth, for instance.
Hence it is not enough just to say what points are added ---
we must also specify the local structure near each new point.
}.
Different toric varieties, therefore, are
distinguished by their different compactifying sets. The latter, in turn,
are distinguished
by restricting the limits of the allowed action of $T$ and these
restrictions can be encoded
in a convenient combinatorial structure as we now describe.

In the framework of an action $T \times V \rightarrow V$, we can focus
our attention on one-parameter subgroups of the full $T$
action\footnote{We
use subgroups depending on one {\it complex} parameter.}.
Basically, we follow all possible holomorphic curves in $T$ as they act
on $V$ and ask whether
or not the action has a limit point in $V$.
As the algebraic torus $T$ is a commutative algebraic group,
all of its one-parameter
subgroups  are labeled by points in a lattice $N\cong \BZ^n$ in the following
way. Given $(n_1,\ldots,n_n) \in \BZ^n$ and $\lambda \in \IC^*$, we consider
the one-parameter group $\IC^*$ acting on $V$ by
\beq
\label{eone}
   \lambda \cdot
   (z_1,\ldots,z_n) \rightarrow (\lambda^{n_1}z_1,\ldots,\lambda^{n_n}z_n)~,
\eeq
where $(z_1,\ldots,z_n)$ are local holomorphic coordinates on $V$
(which may be thought of as residing in the open dense $(\IC^*)^n$ subset
of $V$). Now, to describe all of $V$ (that is, in addition to $T$)
we consider the action of \calle{eone} in the limit that
$\lambda$ approaches zero (and thus moves from $\IC^*$ into $\IC$).
It is these limit points which supply the partial compactifications of
$T$ thereby yielding the toric variety $V$.
 The limit points
obtained from the action \calle{eone} depend upon the explicit vector of
exponents
$(n_1,\ldots,n_n) \in \BZ^n$, but many different exponent vectors
can give rise to the same limit point.
We obtain different toric varieties by
imposing different restrictions on the allowed choices of $(n_1,\ldots,n_n)$
and by grouping them together (according to common limit points) in
different ways.

We can describe these
restrictions and groupings in terms of
 a ``fan'' $\Delta$ in $N$ which is a collection
of {\it strongly convex rational polyhedral cones\/} $\sigma_i$ in the real
vector space $N_{\IR}  =N \otimes_{\BZ} \IR$. (In simpler language, each
$\sigma_i$ is a convex
cone with apex at the origin spanned by a finite number
of vectors which live in the lattice $N$ and such that any angle
subtended by these vectors at the apex is less than $180^\circ$.)
The fan $\Delta$ is a collection
of such cones
which satisfy the requirement that the face of any cone in $\Delta$ is also
in $\Delta$.
 Now, in constructing $V$, we associate a coordinate patch of
$V$ to each {\it large\/} cone\footnote{There
are also coordinate patches for the smaller cones, but
we ignore these for the present.}
 $\sigma_i$ in $\Delta$,
where large refers to a cone spanning an $n$-dimensional
subspace of $N$.
This  patch consists
of $(\IC^*)^n$ together with all the limit points of the action \calle{eone}
for $(n_1,\ldots,n_n) $ restricted to lie in $\sigma_i \cap N$.
There is a single point which serves as the common limit point for
all one-parameter group actions with vector of exponents
$(n_1,\ldots,n_n) $ lying in the {\it interior\/} of $\sigma_i$;
for exponent vectors on the boundary of $\sigma_i$, additional
families of limit points must be adjoined.

We glue these
patches together in a manner dictated by the precise way in which the
cones $\sigma_i$ adjoin each other in the collection $\Delta$.
Basically, the patches are glued together in a manner such that the
one-parameter group actions in the two patches agree along the common
faces of the two cones in $\Delta$.
We will
be more precise on this point shortly. The toric variety $V$ is therefore
completely determined by the  combinatorial data of $\Delta$.

\begin{figure}[htbp]
\centerline{\epsfxsize=7cm\epsfbox{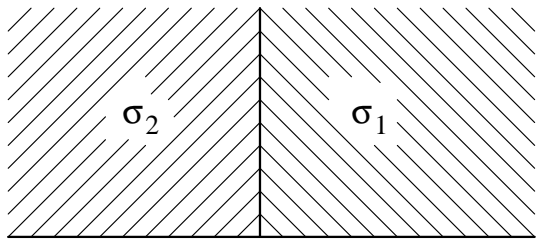}}
\caption{A simple fan.}
\label{figure:4}
\end{figure}

For a simple illustration of these ideas, consider the two-dimensional
example of $\Delta$ shown in \ref{figure:4}. As $\Delta$ consists of two
large cones, $V$ contains two coordinate patches. The first patch ---
corresponding to the cone $\sigma_1$ --- represents $(\IC^*)^2$ together
with all limit points of the action
\beq
\label{etwo}
   \lambda \cdot (z_1,z_2) \rightarrow
   (\lambda^{n_1}z_1,\lambda^{n_2}z_2)~,
\eeq
with $n_1$ and $n_2$ non-negative
as $\lambda$ goes to zero.
We add the single point $(0,0)$ as the limit point when $n_1>0$ and
$n_2>0$; we add points of the form $(z_1,0)$ as limit points when
$n_1=0$ and $n_2>0$; we finally add the points of the form $(0,z_2)$
as limit points when $n_1>0$ and $n_2=0$.\footnote{Note
that it is possible to extend this reasoning to the case
$n_1=n_2=0$:  the corresponding trivial group action has as ``limit points''
all points $(z_1,z_2)$, so even the points interior to $T$ itself
can be considered as appropriate limit points for actions by subgroups.}

Clearly, this patch corresponds to $\IC^2$.
By symmetry, we also associate a $\IC^2$ with the cone $\sigma_2$.
Now, these two copies of $\IC^2$ are glued together in a manner
dictated by the way $\sigma_1$ and $\sigma_2$ adjoin each other.
Explaining this requires that we introduce some more formal machinery
to which we now turn.

\subsection{The $M$ and $N$ Lattices}
\label{sec:3.2}

In the previous subsection, we have seen how a fan $\Delta$ in $N_{\IR}$
serves to define a toric variety $V$. The goal of this subsection is
to make the connection between lattice data and $V$ more explicit by showing
how to derive the transition functions between the patches of $V$. The first
part of our presentation will be in the form of an algorithm that
answers the question: given
the fan $\Delta$ how do we explicitly define coordinate patches and
transition functions for the toric variety $V$? We will then briefly
describe   the mathematics underlying the algorithm.

Towards this end, it proves worthwhile to introduce a second lattice
defined as the dual lattice to $N$, $M=\Hom(N,\BZ)$,
where $\Hom$ denotes homomorphism. We denote the
dual pairing of $M$ and $N$ by $\langle~,~\rangle$. Corresponding to the fan
$\Delta$ in $N_{\IR}$, we define a collection of dual cones
$\check\sigma_i$
in $M_{\IR}$ via
\beq
\label{edualc}
    \check\sigma_i = \{\bfm \in M_{\IR}: \langle
     \bfm,\bfn\rangle \ge 0 ~, \forall \bfn
    \in \sigma_i \}~. 
\eeq
Now, for each dual cone $\check\sigma_i$ we choose a
finite set of elements $\{\bfm_{i,j} \in M, ~j=1,\dots,r_i\}$ such that
\beq
\label{espan}
    \check\sigma_i \cap M = \BZ_{\ge 0 }\, \bfm_{i,1} +\ldots +
    \BZ_{\ge 0}\, \bfm_{i,r_i}~.
\eeq
We then find a finite set of relations
\beq
\label{erelations}
    \sum_{j = 1}^{r_i} \, p_{s,j}\, \bfm_{i,j} = 0~,
\eeq
with $s=1,\dots,R,$ such that {\it any} relation
\beq
\label{eanyrelation}
     \sum_{j = 1}^{r_i}\, p_j\,  \bfm_{i,j} = 0
\eeq
can be written as a linear combination of the given set with integer
coefficients.  That is, 
$$
    p_j=\sum_{s=1}^R\, \mu_s \, p_{s,j}~,
$$ 
for some integers
$\mu_s$.
We associate a coordinate patch $U_{\sigma_i}$ to the cone
$\sigma_i \in \Delta$ by
\beq
\label{econstraints}
    U_{\sigma_i} = \{(u_{i,1},\ldots,u_{i,r_i} )\in \IC^{r_i}
    \ |\
    u_{i,1}{}^{p_{s,1}}u_{i,2}{}^{p_{s,2}}\ldots u_{i,r_i}{}^{p_{s,r_i}} = 1
    \hbox{ for all }s\}~,
\eeq
the equations representing constraints on the variables $u_{i,1}$, \dots,
$u_{i,r_i}$.
We then glue these coordinate patches $U_{\sigma_i}$ and $U_{\sigma_j}$
together by finding a complete set of relations of the form
\beq
\label{eglue}
   \sum_{l = 1}^{r_i}\, q_l \, \bfm_{i,l}  
 + \sum_{l = 1}^{r_j}\, q_l' \, \bfm_{j,l}
   =0~,
\eeq
where the $q_l$ and $q_l'$ are integers.
For each of these relations we impose
the coordinate transition relation
\beq
\label{etransition}
     u_{i,1}{}^{q_1}\, 
     u_{i,2}{}^{q_2}\ldots 
     u_{i,r_i}{}^{q_r}\,
     u_{j,1}{}^{q_1'}\,
     u_{j,2}{}^{q_2'}\ldots
      u_{j,r_j}{}^{q_r'} = 1~.
\eeq
This algorithm explicitly shows how the lattice data encodes the defining
data for the toric variety $V$.

Before giving a brief description of the mathematical meaning
behind this algorithm, we pause to illustrate it in two examples.

\vspace{.2in}
\noindent
{\it Example 1}:

\vspace{.13in}
Let us return to the fan $\Delta$ given in figure \ref{figure:4}.
It is straightforward to see that the dual cones, in this case, take
precisely the same form as in \ref{figure:4}. We have $\bfm_{1,1} = (1,0),
  ~\bfm_{1,2} = (0,1),~ \bfm_{2,1} = (-1,0),~ \bfm_{2,2} = (0,1)$. As the basis
vectors within a given patch are linearly independent, each patch consists
of a $\IC^2$. To glue these two patches together we follow
\calle{eglue} and write the set of relations
\beqn
\label{eexample}
      \bfm_{1,1} + \bfm_{2,1}  &=&  0~, \\
\label{eexamplet}
      \bfm_{1,2} - \bfm_{2,2}  &=&  0~.
\eeqn
These yield the transition functions
\beq
\label{etransitionex}
    u_{1,1}  =  u_{2,1}^{-1}~,
   ~~~~~
    u_{1,2}  =  u_{2,2} ~.
\eeq
These  transition functions imply that the corresponding
toric variety $V$ is the space ${\Bbb C}P^1 \times \IC$.

\begin{figure}[htbp]
\centerline{\epsfxsize=5cm\epsfbox{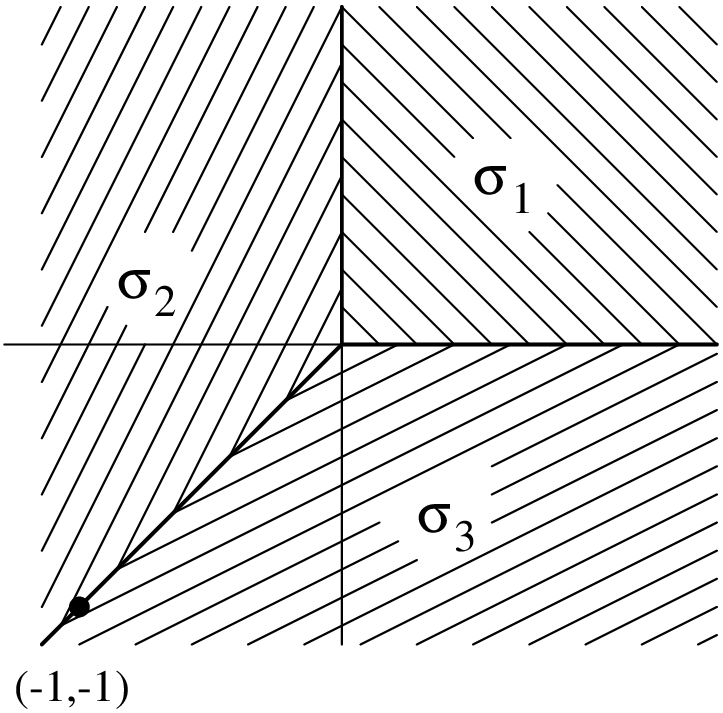}}
\caption{The fan for $\IC P^2$.}
\label{figure:5}
\end{figure}

\begin{figure}[htbp]
\centerline{\epsfxsize=5cm\epsfbox{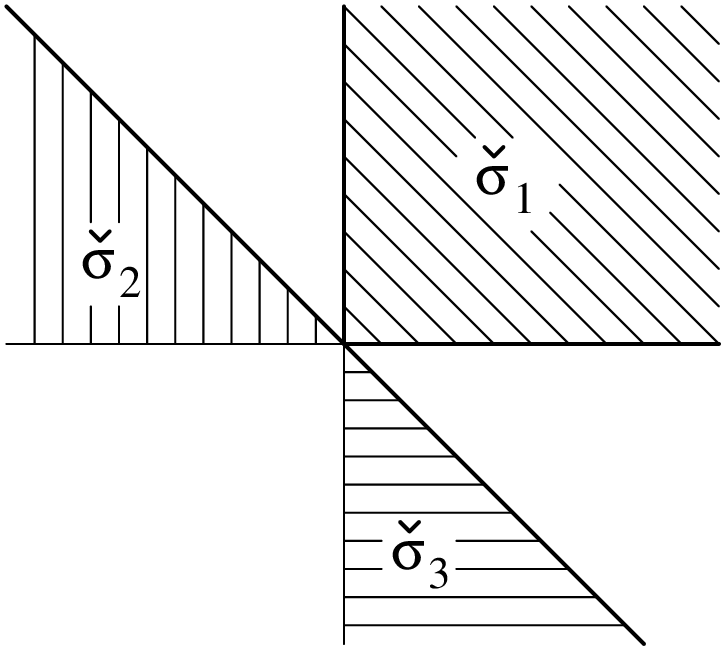}}
\caption{The dual cones for $\IC P^2$.}
\label{figure:6}
\end{figure}

\vspace{.2in}
\noindent
{\it Example 2}:

\vspace{.13in}
Consider the fan  $\Delta$ given in figure
\ref{figure:5}. It is
straightforward to determine that the dual cones in $M_{\IR}$ take the form
shown in \ref{figure:6}. Following the above procedure we find that the
corresponding toric variety $V$ consists of three patches with
coordinates related by
\beq
\label{eptwo}
    u_{1,1} = u_{2,1}^{-1}~, ~~~~~ u_{1,2} = u_{2,2}\, u_{2,1}^{-1}
     ~,
\eeq
and
\beq
\label{eptwot}
     u_{2,2} = u_{3,2}^{-1}~, ~~~~~ u_{2,1} = u_{3,1}\, u_{3,2}^{-1}~.
\eeq
These transition functions imply that the toric variety $V$ associated
to the fan $\Delta$ in \ref{figure:5} is ${\Bbb C}P^2$.

The mathematical machinery behind this association of lattices and
complex analytic spaces relies on a shift in perspective regarding
what one means by a geometrical space. Algebraic geometers
identify geometrical spaces by means of the {\it rings of
functions} that are well defined on those spaces%
\footnote{This can be done in a number of different contexts. Different
kinds of rings of functions, (such as continuous functions, smooth
functions or algebraic functions), lead to different kinds of geometry
(topology, differential geometry or algebraic geometry in the three
cases mentioned).  We will concentrate on rings of algebraic functions,
and algebraic geometry.}.
 To make this concrete
we give two illustrative examples. Consider the space $\IC^2$.
It is clear that the ring of functions on $\IC^2$ is isomorphic to
the polynomial ring $\IC[x,y]$, where $x$ and $y$ are formal symbols,
but may be thought of as {\it coordinate functions\/} on $\IC^2$.
By contrast, consider the space $(\IC^*)^2$. Relative to  $\IC^2$,
we want to eliminate geometrical points  either of whose coordinates
vanishe. We can do this by {\it augmenting\/} the ring $\IC[x,y]$ so
as to include functions that are not well defined on such geometrical
points. Namely, $\IC[x,y, x^{-1}, y^{-1}]$ contains functions only
well defined on $(\IC^*)^2$. The ring $\IC[x,y, x^{-1}, y^{-1}]$ can
be written more formally as  $\IC[x,y,z,w]/\{zx - 1, wy - 1\}$
where the denominator denotes modding out by the  ideal generated by
the listed functions.
In a bit more highbrow mathematical language, one says
\beq
\label{especone}
    \IC^2 \cong\Spec\IC[x,y] ~,
\eeq
and
\beq
\label{espectwo}
    (\IC^*)^2 \cong\Spec{ \IC[x,y,z,w]\over \{zx - 1, wy - 1\} }
    ~,
\eeq
where the term ``$\Spec$'' may intuitively be thought of as
``the minimal
space of points where the following function ring is well defined''.

With this terminology, the coordinate patch $U_{\sigma_i}$ corresponding
to the cone $\sigma_i$ in a fan $\Delta$ is given by
\beq
\label{especpatch}
    U_{\sigma_i}\cong\Spec\IC[\check\sigma_i \cap M] ~,
\eeq
where by $\check\sigma_i \cap M$ we refer to the monomials in
local coordinates that are naturally assigned to lattice points in
$M$ by virtue of its being the dual space to $N$. Explicitly,
a lattice point $(m_1,\ldots,m_n)$ in $M$ corresponds to the monomial
$z_1^{m_1}z_2^{m_2}\ldots z_n^{m_n}$. The latter is sometimes referred
to as {\it group character} of the algebraic group action given by
$T$.

Within a given patch, $\Spec\IC[\check\sigma_i \cap M]$
is a polynomial ring generated by the monomials associated to
the lattice points in $\check\sigma_i$. By the map given in the previous
paragraph between lattice points and monomials, we see that linear
relations between lattice points translate into multiplicative
relations between monomials. These relations are precisely those
given in \calle{erelations}.

Between patches,
if $\sigma_i$ and $\sigma_j$ share a face, say $\tau$,
then $\IC[\check\sigma_i \cap M]$ and $\IC[\check\sigma_j \cap M]$
are both subalgebras of $\IC[\check\tau \cap M]$. This provides
a means of identifying elements of $\IC[\check\sigma_i \cap M]$ and
elements of $\IC[\check\sigma_j \cap M]$ which translates into a map
between $\Spec\IC[\check\sigma_i \cap M]$ and
$\Spec\IC[\check\sigma_j \cap M]$. This map is precisely that given in
\calle{etransition}.

\subsection{Singularities and their Resolution}
\label{sec:3.3}

In general, a toric variety $V$ need not be a smooth space. One advantage of
the toric description is that a simple analysis of the lattice data
associated with $V$ allows us to identify singular points. Furthermore,
simple modifications of the lattice data allow us to construct from $V$
a toric variety $\tilde V$ in which all of the singular points are
repaired. We now briefly describe these ideas.

The essential result we need is as follows:

\begin{theorem}\ \\
Let $V$ be a toric variety associated to a fan $\Delta$ in $N$.
$V$ is smooth if for each cone $\sigma$ in the fan we can find 
an integral basis $\{\bfn_1,\ldots,\bfn_n\}$ of $N$ and an integer $r \le n$ such
that $\sigma = \IR_{\ge 0}\,  \bfn_1 + \ldots  + \IR_{\ge 0}\,\bfn_r$.
\end{theorem}

For a proof of this statement the reader should consult, for
example, \cite{rOda} or \cite{rFulton}.
The basic idea behind the result is as follows.
If $V$ satisfies the criterion in the proposition, then the
dual cone $\check\sigma$ to $\sigma$ can be expressed as
\beq
\label{edualcone}
    \check\sigma = \sum_{i = 1}^r \IR_{\ge 0}\, \bfm_i +
    \sum_{i = r+1}^n \IR \, \bfm_i~,
\eeq
where $\{\bfm_1,\ldots,\bfm_n\}$ is the dual basis to 
$\{\bfn_1,\ldots,\bfn_n\}$.
We can therefore write
\beq
\label{elattice}
   \check\sigma \cap M =  \sum_{i = 1}^r \BZ_{\ge 0}\, \bfm_i +
  \sum_{i = r+1}^n \BZ_{\ge 0}\, \bfm_i 
 + \sum_{i = r+1}^n \BZ_{\ge 0}\, (-\bfm_i)~.
\eeq
From our prescription of subsection \ref{sec:3.2}, this patch is
therefore isomorphic to
\beq
\label{espece}
        \Spec{{\IC[x_1,\ldots,x_r,y_{r+1},\ldots,y_n]}\over
	\{x_iy_i - 1,~i=1,\dots,n\} }~.
\eeq
In plain language,
 this is simply $\IC^r \times (\IC^*)^{n-r}$, which
is certainly non-singular. The key to this patch being smooth is
that $\sigma$ is $r$-dimensional and it is spanned by
$r$ linearly independent lattice vectors in $N$. This implies,
via the above reasoning, that there are no ``extra'' constraints
on the monomials associated with basis vectors in the patch
(see \calle{econstraints}),
hence leaving a smooth space.

Of particular interest are toric varieties $V$
whose fan $\Delta$ is {\it simplicial}.  This means that each
cone $\sigma$ in the fan can be written in the form
$\sigma = \IR_{\ge 0}\,  \bfn_1 + \ldots  + \IR_{\ge 0}\, \bfn_r$
for some linearly independent vectors $\bfn_1,\dots,\bfn_r\in N$.
(Such a cone is itself called {\it simplicial}.)
When $r=n$, we define a ``volume'' for simplicial cones as follows:
choose each $n_j$ to be the first non-zero lattice point on the
ray $\IR_{\ge0}\, n_j$ and define $\vol(\sigma)$ to be the volume of
the polyhedron with vertices $O,\bfn_1,\dots,\bfn_n$.  (We normalize our
volumes so that the unit simplex in $\IR^n$ (with respect to the
lattice $N$) has volume 1.
Then the volume of $\sigma$ coincides with the index $[N:N_\sigma]$,
where $N_\sigma$ is the lattice generated by $\bfn_1,\dots,\bfn_n$.)
Note that the coordinate chart $U_\sigma$ associated to a simplicial
cone $\sigma$ of dimension $n$ is smooth at the origin precisely when
$\vol(\sigma)=1$.

\begin{figure}[htbp]
\centerline{\epsfxsize=7cm\epsfbox{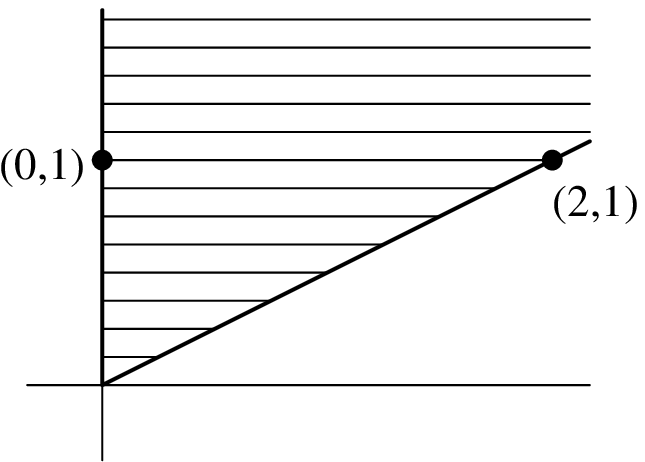}}
\caption{The fan for $\IC/\BZ_2$.}
\label{figure:7}
\end{figure}

To illustrate this idea, we consider the fan $\Delta$ of
figure \ref{figure:7}
which gives rise
to a singular variety. This fan has one big (simplicial)
cone of volume 2 generated by
$v_1 = (0,1) = \bfn_1$ and $v_2 = (2,1) = \bfn_1 + 2\, \bfn_2$.
The dual cone $\check\sigma$
is generated by $w_1 = (2,-1) = 2\, \bfm_1 - \bfm_2$ and $w_2 = (0,1) = 
\bfm_2$.
In these expressions, $\bfn_i$ and $\bfm_j$ are the standard basis vectors.
It is clear that this toric variety is not smooth since it does not
meet the conditions of the proposition. More explicitly, following
\calle{econstraints},
we see that $\check\sigma \cap M = \BZ_{\ge 0}\, (2\, \bfm_1 - \bfm_2)
+ \BZ_{\ge 0}\, \bfm_2 +  \BZ_{\ge 0}\, \bfm_1 $ and hence
$$
   V =  \Spec {\IC[x,y,z] \over \{ z^2 - xy = 0 \}  } ~.
$$
In plain language,
$V$ is the vanishing locus of $z^2 - xy $ in $\IC^3$. This is singular
at the origin, as is easily seen by the transversality test.
 Alternatively, a simple change of variables; $z = u_1u_2$,
$x = u_1^2, y = u_2^2$, reveals that $V$ is in fact $\IC^2/\BZ_2$
(with $\BZ_2$ generated by the action $(u_1,u_2) \rightarrow (-u_1, -u_2)$)
which is singular at the origin as this is a fixed point.
Notice that the key point leading to this singularity is the fact
that we require three lattice vectors to span the two-dimensional sublattice
$\check\sigma \cap M$.

The proposition and this discussion suggest a procedure to follow
in order
to modify any such $V$ so as to repair  singularities which it might
have. Namely, we construct a new fan $\tilde \Delta$ from the original
fan $\Delta$ by {\it subdividing}: first subdividing all cones into
simplicial ones and then subdividing the cones $\sigma_i$ of volume
greater than 1 until
the stipulations of the  non-singularity
proposition are met.
 The new fan $\tilde \Delta$
will then be the toric data for a non-singular {\it resolution\/} of
the original toric variety $V$. This procedure is called {\it blowing-up}.
We illustrate it with our previous example of $V = \IC^2/\BZ_2$.
Consider constructing $\tilde \Delta$ by subdividing the cone in $\Delta$
into two pieces by a ray passing through the point $(1,1)$. It is then
straightforward to see that each cone in $\tilde \Delta$ meets the
smoothness criterion. By following the procedure of subsection 
\ref{sec:3.2}
one can derive the transition functions on $\tilde V$ and find that it is
the total space of the line bundle ${\cal O}(-2)$ over $\IC P^1$
(which is smooth). This is the well known blow-up of the quotient
singularity $\IC^2/\BZ_2$.

If the volume of a cone as defined above
behaved the way one might hope, i.e. whenever dividing
a cone of volume $v$ into other cones, one produced new cones whose
volumes summed to $v$, then subdivision would clearly be a finite
process. Unfortunately this is not the case\footnote{If the closest lattice
point to the origin on the subdividing ray does not lie on a face of
the polyhedron $\langle O,\bfn_1,\dots,\bfn_n\rangle$ then the new polyhedra
will be unrelated to the old and the volumes will not add.}
 and in general one can
continue dividing any cone for as long as one has the patience. This
corresponds to the fact that one can blow up any point on a manifold to
obtain another manifold. In our case, however, we will utilize the fact that
string theory demands that the
canonical bundle of a target space is trivial, that is,
we must have vanishing first Chern class. The \CY\ manifold will
not be the toric variety itself (as we will see),
 but we
do require that any resolution of singularities adds nothing new to
the canonical class of $V$.
This will restrict the allowed blow-ups rather severely.

In order to have a resolution which adds nothing new to the canonical
class, the singularities must be what are called {\it canonical
Gorenstein singularities} \cite{rReid}.  A characterization of which toric
singularities have this property was given by Danilov and Reid.  To
state it, consider a cone $\sigma$ from our fan $\Delta$ and examine
the one-dimensional edges of $\sigma$.  As we move away
from $O$ along any of these edges we eventually reach a point in $N$. In this
way, we
associate a collection of points  $\SS\subset N$ with $\sigma$. (These points
will lie in the boundary of a polyhedron $P^\circ$ which we will
discuss in more detail in subsection \ref{sec:3.5}.)
The fact we
require is given in the following definition.

\begin{definition}\ \\
The singularities of the affine toric variety $U_\sigma$
are canonical Gorenstein singularities if
all the points in $\SS$ lie in an affine hyperplane $H$ in $N_{\IR}$ of the
form
\beq
\label{hyper}
   H=\{\bfx\in N_{\IR}\ |\ \langle \bfm,\bfx\rangle=1\} ~,
\eeq
for some $\bfm\in M$ and if there are no lattice points 
$\bfx\in{\sigma\cap N}$
with $0<\langle \bfm,\bfx\rangle<1.$
\end{definition}

Furthermore, in order to avoid adding anything new to the canonical class,
we must choose all one-dimensional cones used in subdividing $\sigma$
from among rays of the form $\IR_{\ge0}\, \bfx$ where $\bfx\in\sigma\cap N$ lies
on the hyperplane $H$ (i.e., $\langle \bfm,\bfx\rangle=1$).

If we have a big simplicial cone, then
the $n$ points in $N_{\IR}$ associated to
the one-dimensional subcones of this cone always define an affine hyperplane in
$N_{\IR}$. If we assume the singularity is canonical and Gorenstein,
then this hyperplane is one integral unit away from the origin and volumes
can be conveniently calculated on it.  In particular,
if the volume of the big cone is greater than
$1$, then this hyperplane
will intersect more points in $\sigma\cap N$. These additional points define
the
one-dimensional cones that can be used for further subdivisions of the
cone that do not affect the canonical class. Since volumes are calculated
in the hyperplane $H$, the volume property
behaves well under such resolutions, i.e., the sum of the volumes of
the new cones is equal to the volume of the original cone that was subdivided.

In some cases, there will not be enough of these additional points to complete
subdivide into cones of volume $1$.  However, in the cases
of primary interest to the present lectures
 (in which $V$ is a four-dimensional toric
variety
which contains three-dimensional Calabi-Yau varieties as hypersurfaces),
we {\it can\/} achieve a partial resolution of singularities
which leaves only isolated singularities on $V$.  Happily, the Calabi-Yau
hypersurfaces will avoid those isolated singularities, so their singularities
are completely resolved by this process.

For simplicity of exposition, we shall henceforth assume that our toric
varieties $V$ have the following property: {\it if we partially resolve by means
of a subdivision which makes all cones simplicial and  divides
simplicial cones into cones of volume $1$, adding nothing new to the
canonical class, then we obtain a smooth variety}.  This property holds
for the example  considered in section \ref{sec67},
 which we shall return to shortly.
We will point out
from time to time the modifications which must be made when this property
is {\it not} satisfied; a systematic exposition of the general case is
given in \cite{rAGMII}.

\begin{figure}[htbp]
\centerline{\epsfxsize=9cm\epsfbox{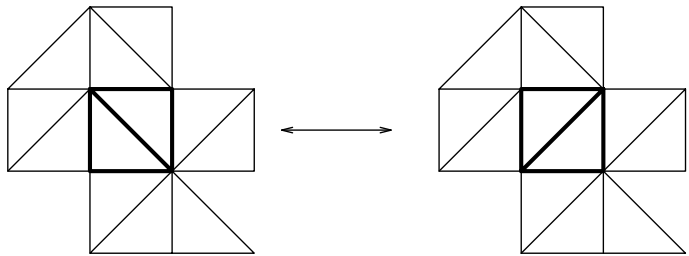}}
\caption{A flop in toric geometry.}
\label{figure:8}
\end{figure}

An important point for our study is the fact that, in general, there is
no unique way to  construct
$\tilde \Delta$ from the original
fan $\Delta$.  On the contrary, there are often numerous ways of subdividing
the cones in $\Delta$ so as to conform to the volume $1$ and canonical
class conditions.
Thus, there are numerous smooth varieties that can arise from different ways
of resolving the singularities on the original singular space.
These varieties are birationally equivalent but will, in general, be
{\it topologically distinct}.
For three-dimensional Calabi-Yau varieties such topologically distinct
resolutions can always be related by a sequence of flops \cite{rflops}.
For the simplest kind of flops\footnote{In fact, these are the
 only kinds of flops
that we need \cite{rflopsbis}.},
a small neighbourhood of the $\IC P^1$ being flopped
is isomorphic to an open subset of a three-dimensional
toric variety and that flop can be given a toric description as follows.
To a three-dimensional toric variety we associate a fan in $\IR^3$. If
this variety is smooth we can intersect the fan with an $S^2$
enclosing the origin to obtain a triangulation of $S^2$ or part of
$S^2$.
(Different
smooth models will correspond to different triangulations of $S^2$.) We
show a portion of two such triangulations in figure \ref{figure:8}.
In this figure, one sees
that if two neighbouring triangles form a convex quadrilateral, then
this quadrilateral can be triangulated the other way around to give a
different triangulation. Any two triangulations can be related by a
sequence of such transformations. When translated into toric geometry,
the reconfiguration of the fan shown in figure \ref{figure:8}
is precisely a flop, as we discuss in the next section.

\subsection{Compactness and Intersections}
\label{sec:3.4}

Another feature of the toric variety $V$ which can be directly
determined from the data in $\Delta$ is whether or not it is compact.
Quite simply, $V$ is compact if $\Delta$ covers all of $\IR^n$.
For a more precise statement and proof the reader is referred to
\cite{rOda}. This condition on $\Delta$ is intuitively clear. Recall that
we have associated points in $N$ with one-parameter group actions on
$V$. Those points in $N$ which also lie in $\Delta$ are special in that
the limit points of the corresponding group actions are part of $V$.
Now, if every point in $N$ lies in $\Delta$, then the limit points
of all one-parameter group actions are part of $V$. In other words,
$V$ contains all of its limit points --- it is compact.
The examples we have given illustrate this point. Only the fan of
figure \ref{figure:6} covers all of $N$ and hence only its corresponding
toric variety ${\Bbb C}P^2$ is compact. Note that a
compact toric variety cannot be a \CY\ manifold. This does not stop
toric geometry being useful in the construction of \CY\ manifolds however,
as we shall see.

This picture of complete fans corresponding to compact varieties can
be extended to analyze parts of the fan and gives one a good idea of
how to interpret a fan just by looking at it. If we consider an
$r$-dimensional cone $\sigma$ in the interior of a fan, then there is a
$(n-r)$-dimensional complete fan surrounding this cone (in the normal
direction). Thus we can
identify an $(n-r)$-dimensional compact toric subvariety
$V^\sigma\subset V$ associated to
$\sigma$. For example, each one-dimensional cone in $\Delta$ is
associated to a codimension one holomorphically embedded subspace of
$V$, i.e., a {\it divisor}.

We can take this picture further. Suppose an $r$-dimensional cone
$\sigma_r$ is part of an $s$-dimensional cone $\sigma_s$, where $s>r$.
When we interpret these cones as determining subvarieties of $V$, we see that
$V^{\sigma_s}\subset V^{\sigma_r}\subset V$. Now suppose we take two
cones $\sigma_1$ and $\sigma_2$ and find a maximal cone
$\sigma_{1,2}$ such that $\sigma_1$ and $\sigma_2$ are both contained
in $\sigma_{1,2}$. The toric interpretation tells us that
\beq
\label{eXti}
   V^{\sigma_{1,2}}\cong V^{\sigma_1}\cap V^{\sigma_2}~.
\eeq
If no such $\sigma_{1,2}$ exists, then $V^{\sigma_1}$ and
$V^{\sigma_2}$ do not intersect. If we take $n$ one-dimensional cones
$\sigma_i$ that form the one-dimensional edges of a big cone, then
the divisors $V_{\sigma_i}$ intersect at a point.

Thus we see that the fan $\Delta$ contains information about the
intersection form of the divisors within $V$. Actually, the fan
$\Delta$ contains also the information to determine self-intersections 
and thus all the intersection numbers are determined by $\Delta$.

Referring back to figure 
\ref{figure:8} we can describe a flop in the language of
toric geometry. To perform the transformation in figure \ref{figure:8}, we first
remove the diagonal bold line the in middle of the diagram. This
line is the base of a two-dimensional cone (which thus has
codimension one).
The only one-dimensional compact toric variety is ${\Bbb C}P^1$, so this line
we have removed represented a rational curve. After removing this line,
we are left with a square-based cone in the fan which is not simplicial,
so the resulting toric
variety is singular. We then add the diagonal line in the other
direction to resolve this singularity thus adding in a new rational
curve. This is precisely a flop --- we blew down one rational curve
to obtain a singular space and then blew it up with another rational
curve to resolve the singularity.

\subsection{Hypersurfaces in Toric Varieties}
\label{sec:3.5}

Our interest is not with toric varieties, {\it per se}, but rather
with Calabi-Yau spaces. The preceding discussion is useful in
this domain because a large class of Calabi-Yau spaces can
be realized as hypersurfaces in toric varieties. 
 The toric varieties of greatest relevance here are
weighted projective spaces. We have seen how ordinary projective
spaces (in particular ${\Bbb C}P^2$) are toric varieties and the same is
true for weighted projective spaces.

To illustrate this point, let us construct the weighted projective space
$W{\Bbb C} P^{2}(3,2,1)$. As in the case of ${\Bbb C}P^2$,
 there are three patches
for this space. The explicit transition functions between these patches
are:
\beq
\label{epwtwo}
    u_{1,1} = u_{2,1}^{-1}~,
    \quad
    u_{1,2}^3 = u_{2,2} u_{2,1}^{-2}
\eeq
and
\beq
\label{epwtwot}
   u_{2,2} = u_{3,2}^{-1}~, \quad u_{2,1}^2 =
   u_{3,1}^3u_{3,2}^{-1}~.
\eeq

\begin{figure}[htbp]
\centerline{\epsfxsize=7cm\epsfbox{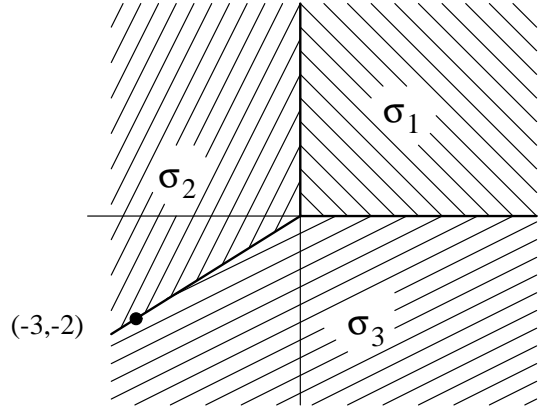}}
\caption{The fan for $W{\Bbb C}P^{2}(3,2,1)$.}
\label{figure:9}
\end{figure}

Consider the fan  $\Delta$ of figure \ref{figure:9}. By following the procedure
of subsection \ref{sec:3.2}, one can directly determine that this fan yields
the same set of transition functions. Notice that
$W{\Bbb C}P^{2}(3,2,1)$ is not smooth, by the considerations of 
subsection \ref{sec:3.2}.
This is as expected since the equivalence relation of
\calle{wtedspace}
 has non-trivial fixed points. All higher dimensional weighted
projective spaces can be constructed in the same basic way.

Now, how do we represent a hypersurface in such a toric variety?
In our discussion we shall follow \cite{rBatyrev}.
A hypersurface is given by a homogeneous polynomial of degree
$d$ in the homogeneous weighted projective space coordinates.
Recall that points in the $M$ lattice correspond to monomials
in the {\it local} coordinates associated to the particular patch
in which the point resides. Consider first the
subspace of $M$ in which all lattice coordinates are positive.
We specify the {\it family} of
degree $d$ hypersurfaces by drawing a polyhedron $P$
defined as the minimal convex polyhedron that
surrounds all lattice points corresponding to (the local representation of)
monomials of degree $d$. By sliding this polyhedron along the coordinate
axes of $M$ such that one vertex of $M$ is placed at the origin, we
get the representation of these monomials in the other weighted
projective space patches --- a different patch for each vertex.
To specify a particular hypersurface
(i.e.\ a particular degree $d$ equation), one would need to give more
data than is encoded in this lattice formalism ---
the values of the {\it coefficients} of each degree $d$ monomial in
the defining equation of the hypersurface would have to be specified. However,
the toric framework is particularly
well suited to studying the whole family of such hypersurfaces.

\begin{figure}[htbp]
\centerline{\epsfxsize=5cm\epsfbox{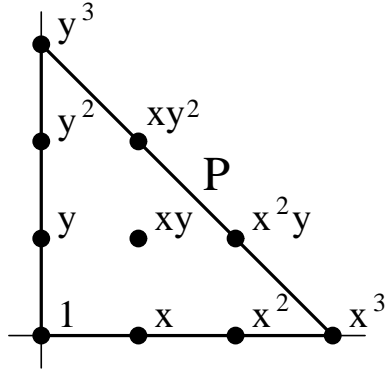}}
\caption{The polyhedron of monomials.}
\label{figure:10}
\end{figure}

As a simple example of this, consider the cubic hypersurface in
$\IC P^2$ which has homogeneous coordinates $[z_1,z_2,z_3]$. In local
coordinates, say $x = z_1/z_3$ and $y = z_2/z_3$ (the patch in which
$z_3 \ne 0$) the homogeneous cubic monomials are
$1,x,y,x^2,y^2,xy,x^3,y^3,x^2y,xy^2$. (One multiplies each of these
by suitable powers of $z_3$ to make them homogeneous of degree three.)
These monomials all reside in the polyhedral region of $M$ as shown
in figure \ref{figure:10}.

As shown in \cite{rBatyrev}, there is a simple condition on
$P$ to ensure that the resulting hypersurface is Calabi-Yau.
This condition consists of two parts. First, $P$ must contain
precisely one interior point. Second, if we call this interior point
$\bfm_0$ then it must be the case that $P$ is {\it reflexive\/} with
respect to $M$ and $\bfm_0$. This means the following.

\begin{definition}\ \\
Given $P\subset M_{\IR}$ we can construct
the polar polyhedron $P^\circ\subset N_{\IR}$ as
\beq
\label{eXpp}
  P^\circ = \{(x_1,\ldots,x_n)\in N_{\IR};\quad\sum_{i=1}^nx_iy_i
  \geq -1 \hbox{ for all }(y_1,\ldots,y_n)\in P\}~,
\eeq
where we have shifted the position of $P$  in $M$ so that
$\bfm_0$ has coordinates $(0,\ldots,0)$.
If the vertices of $P^\circ$ lie in $N$ then $P$ is called reflexive.
\end{definition}

\begin{figure}[htbp]
\centerline{\epsfxsize=5cm\epsfbox{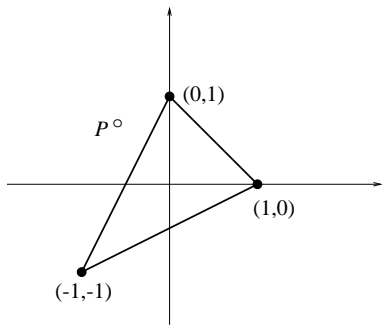}}
\caption{The dual of the polyhedron $P$.}
\label{figure:polarp2}
\end{figure}

\noindent The
origin $O$ of $N$ will then be the unique element of $N$ in
the interior of $P^\circ$.
For the example at hand, the polar polyhedron is easily computed
to have vertices $(1,0),$ $(0,1),$ $(-1,-1)$ and we draw this dual polyhedron
in figure 
\ref{figure:polarp2}.

Also note that given a reflexive $P^\circ \subset N_{\IR}$, we can construct
$\Delta$ by building the fan comprising the cones over the faces,
edges and vertices of $P^\circ$ based on $O$. In general, such
a fan may contain cones of volume greater than 1. However, all the other points
in $P^\circ\cap N$, except $O$, are on faces and edges of $P^\circ$
and can thus be used to resolve the singularities of $V$ without
affecting the canonical class. The exceptional divisors introduced
into $V$ by this resolution of singularities can intersect the
hypersurface to produce exceptional divisors in the \CY\ manifold.
(However, this does not necessarily happen in all cases.
If we consider a point in the
interior of a codimension one face of $P^\circ$ then the exceptional
divisor induced in $V$ would not intersect the hypersurface and
therefore would give
no contribution to the \CY\ manifold.)

To finish specifying $\Delta$ we must say which sets of the one-dimensional
cones are to be used as the set of edges of a larger cone.  Phrased in
terms of the relevant lattice points in $P^\circ\cap N$, what we need
to specify is
 a triangulation of $P^\circ$, with vertices in $P^\circ\cap N$,
each simplex of which includes $O$.  Replacing each simplex with the
corresponding cone whose vertex lies at $O$, we produce the fan $\Delta$.
Conversely, if we are given $\Delta$ then intersecting the cones of $\Delta$
with the polyhedron $P^\circ$ produces a triangulation of $P^\circ$.

\subsection{K\"ahler and Complex Structure Moduli}
\label{sec:3.6}

Having seen how Calabi-Yau hypersurfaces in  a weighted projective
space are described in the language of toric geometry we now
indicate how the complex structure and K\"ahler structure moduli
on these spaces are represented.

In general, not all such moduli have a representation in toric
geometry. Let's begin with complex structure moduli. As discussed in
subsection \ref{sec:moduliI}, 
such moduli are associated to elements in
$H^{d-1,1}(X)$, where $X$ is the Calabi-Yau space, and under favorable
circumstances \cite{rGH}
 some of these can be represented
by monomial perturbations of the same degree of homogeneity as $X$.
By our discussion of the previous subsection, these are the lattice
points contained within $P$. Thus, those complex structure
deformations with a monomial representation have a direct realization
in the toric description of $X$.

The other set of moduli are associated with the K\"ahler structure of
$X$. Note that an arbitrary element of $H^{1,1}(X)$ can, by Poincar\'e
duality, be represented as a $(2d-2)$-cycle in $H_{2d-2}(X)$.
As explained in subsection \ref{sec:3.4}
 and above, divisors in $X$ are given by
some of the
one-dimensional cones in $\Delta$ which, in turn, correspond to points
in $P^\circ\cap N$. To be more precise, by this method,
every point in $P^\circ\cap N$
except $O$ and points in codimension one faces of $P^\circ$ gives
a (not necessarily distinct) class in $H_{2d-2}(X)$.

Unfortunately, it does not follow that $H_{2d-2}(X)$ is generated by such
points in $P^\circ$. In general an exceptional divisor in $V$ may
intersect $X$ in several isolated regions. This leads to many classes
in $H_{2d-2}(X)$ being identified with the same point in $P^\circ$. If we
define the K\"ahler form on $X$ in terms of the cohomology of $V$, we
thus restrict to only part of the moduli space of K\"ahler forms on
$X$.  We will do this in the next subsection and study K\"ahler forms
directly on $V$; these always induce K\"ahler forms on\footnote{This is
not completely obvious when $V$ is singular, but it is verified in 
\cite{rAGMII}.} $X$,
and will produce only part of the K\"ahler moduli space of $X$.
As we will see however, restricting to this part of the moduli space
 will not cause any problems for our
analysis of the mirror property.

\subsection{Holomorphic  Quotients}
\label{sec:3.7}

There are two other related ways of building a toric variety $V$
from a fan $\Delta$, in addition to the method we have discussed to
this point.
For a more detailed discussion of
the approach of this subsection, the reader is referred to
\cite{rCox}.

As we discussed in section \ref{sec:3.1},
an $n$-dimensional toric variety
$V$ can be realized as
\beq
\label{eintuitholo}
   { {\IC}^{n+h^{1,1}(V)} - F_{\Delta}\over (\IC^*)^{h^{1,1}(V)} }~,
\eeq
where $F_{\Delta}$ is a subspace of ${\IC}^{n + h^{1,1}(V)}$ determined by
$\Delta$. One might wonder why the particular form in
\calle{eintuitholo} arises. We shall explain this shortly, however, we note that
first, being a toric variety, $V$ contains a $(\IC^*)^n$ as a dense open
set (as in \calle{eintuitholo}) and second,
 without removing
$F_{\Delta}$ the quotient is badly behaved (for example it may
not be Hausdorff). For a clear
discussion  of the latter issue we  refer the reader to pages
190--193 of \cite{WP}.
This is called a holomorphic quotient.
Alternatively, the quotient in \calle{eintuitholo} can be carried
out in two stages: thinking of each $\IC^*$ as $\IR_+ \times U(1)$,
we can first quotient by $\IR_+$ and then by $U(1)$.
The first step is accomplished by introducing a
 `moment map' $\mu: {\IC}^{n+h^{1,1}(V)} \to {\IR}^{h^{1,1}(V)}$
and restricting to one of its level sets. The second step is then
directly accomplished by taking the quotient by the remaining
$(S^1)^{h^{1,1}(V)}$.  (There is a way
to determine which fan $\Delta$ corresponds to each specified value
of the moment map --- see for example \cite{rAudin}.)
This latter construction is referred to as taking the
symplectic quotient. The reader should pause and return to the
discussion of the linear sigma model in section \ref{sec43}.
The imposition of the D-term constraints is nothing but restricting
to a level set of a particular moment map, after which we further
quotient by $U(1)$. In other words, the linear sigma model is a physical
realization of a particular class of symplectic quotients.

The groups $(\IC^*)^k$ by which we take quotients are often constructed
out of a lattice of rank $k$.  If $L$ is such a lattice, we let $L_{\IC}$
be the complex vector space constructed from $L$ by allowing complex
coefficients.  The quotient space $L_{\IC}/L$ is then an algebraic group
isomorphic to $(\IC^*)^k$.  A convenient way to implement the quotient by
$L$ is to exponentiate vectors componentwise (after multiplying by $2\pi i$).
For this reason, we adopt the notation $\exp(2\pi i\,L_{\IC})$ to indicate
this group $L_{\IC}/L$.

Let us consider the holomorphic quotient in greater detail.
To do so we need to introduce a number of definitions.
Let $\AA$ be the set of points in $P^\circ\cap N$.
{\it We assume henceforth that $\AA$ contains no point which lies in the
interior of a codimension one face of $P^\circ$.}
(The more general case is treated in \cite{rAGMII}.)
Denote by $r$ the
number of points in $\AA$. Let $\AAO$ be the set $\AA$ with $O$
removed which is
isomorphic to the set of one-dimensional cones in the fully resolved
fan $\Delta$.
To every point
$\rho \in \AAO$ associate a formal variable $x_{\rho}$. Let
$\IC^{\AAO} \equiv \Spec\IC[x_{\rho}, \rho \in {\AAO} ]$.
$\IC^{\AAO}$ is simply $\IC^{r-1}$.
Let us define  the  polynomial ideal $B_0$ to be
generated by
 $\{ x^{\sigma}, \sigma\hbox{ a cone in }P^\circ\}$,
 with  $x^{\sigma}$ defined as
$\prod_{\rho \notin \sigma} x_{\rho}$.
Let us introduce the lattice $A_{n-1}(V)$ of divisors modulo linear
equivalence on $V$. (On a smooth toric variety, linear
equivalence is the same thing as homological equivalence. See, for
example, \cite{rFulton}, p.64, for a fuller explanation.)
This group may also be considered as $H_{2d-2}(V,\BZ)$,
if $V$ is compact and smooth, which we will assume for the rest of this
section.
Finally, define
\beq
\label{Gee}
   G \equiv \Hom(A_{n-1}(V),\IC^*)\cong\exp(2\pi i\,A_{n-1}(V)^\vee)
   \cong (\IC^*)^{h^{1,1}(V)}~,
\eeq
where $A_{n-1}(V)^\vee$ denotes the dual lattice of $A_{n-1}(V)$.
Then, it can be shown that $V$ can be realized as the
holomorphic quotient
\beq
\label{eholo}
   V \cong {\IC^{\AAO}  - F_\Delta \over G} ~,
\eeq
where $F_{\Delta}$ is the vanishing locus of the elements in the ideal
$B_0$.

To give an idea of where this representation of $V$ comes from,
consider the exact sequence
\beq
\label{eexactoric}
    0 \tto M \tto \BZ^{\AAO} \tto
    A_{n-1}(V) \tto 0 ~,
\eeq
where $\BZ^{\AAO}$ is the free group over $\BZ$ generated by
the points (i.e.\ toric divisors) in $\AAO$. To see why this is
exact, we explicitly consider the maps involved. Elements in
$\BZ^{\AAO}$  may be associated with integer valued
functions defined on the points
in $\AAO$. Any such function $f$ is given by its value on the
$r-1$ points in $\AAO$. The map from
$\BZ^{\AAO} \rightarrow
A_{n-1}(V)$ consists of
\beq
\label{emap}
    f \mapsto \sum_{\rho \in \AAO} f(\rho) D_{\rho} 
    ~,
\eeq
where $D_{\rho}$ is the divisor class in $V$ associated to
the point $\rho$ in $\Delta$. Clearly, every toric divisor can be so
written.  It is known \cite{rFulton} that the toric divisors generate all of
$A_{n-1}(V)$ and hence this map is surjective.

Any element $\bfm \in M$
is taken into  $\BZ^{\AAO}$ by the mapping
\beq
\label{emapa}
   \bfm \mapsto \langle \cdot , \bfm\rangle ~.
\eeq
This map
is injective as any two linear functions which agree on
$\AAO$ agree on $N$ (by our assumption that the points of
$\AAO$ span $N$). $M$ is the kernel
of the second map because the points $\bfm \in M$ correspond to global
meromorphic functions (by their group characters, i.e. monomials as
discussed in subsection \ref{sec:3.2}
and hence give rise to divisors linearly equivalent to 0.

Taking the exponential of the
dual of \calle{eexactoric} we have
\beq
\label{eexactorica}
   1 \tto (\IC^*)^{h^{1,1}(V)}
   \tto  (\IC^*)^{\AAO} \tto \exp(2\pi i\,N_{\IC}) \tto
    1.
\eeq
Notice that  $\exp(2\pi i\,N_{\IC})=N_{\IC}/N$
is the algebraic torus $T$
from which $V$ is obtained by partial compactification. We have now
seen that $T$ arises as a holomorphic quotient of
\beq
\label{eholob}
    (\IC^*)^{\AAO}
\eeq 
by
\beq
\label{eholoc}
    (\IC^*)^{h^{1,1}(V)} ~.
\eeq
To represent $V$ in a
similar manner, we therefore need to partially compactify this quotient.
The data for so doing, of course, is contained in the fan $\Delta$
(just as it was in  our earlier approach to building $V$). As shown
in \cite{rCox}, the precise way in which this partial compactification
is realized in the present setting is to use $\Delta$ to replace
\calle{eholob} by
the numerator on the right hand side of
\calle{eholo}.

\subsection{Toric Geometry of the Partially and
 Fully Enlarged K\"ahler Moduli Space}
\label{sec:3.8}

The orientation of our discussion of toric geometry to this point has been to
describe the structure of certain Calabi-Yau hypersurfaces.
It turns out that the {\it moduli\/} spaces of these Calabi-Yau
hypersurfaces are themselves realizable as toric varieties.
Hence, we can make use of the machinery we have outlined to not
only describe the target spaces of our non-linear $\sigma$-model
conformal theories but also their associated moduli spaces.

In this subsection, we shall outline how the partially and fully
enlarged K\"ahler moduli space
is realized as a toric variety and in the next subsection
we will do the same for the complex structure moduli space.

First we require a definition for the partially enlarged moduli space.
The region of moduli space we are first interested in is the region where one
approaches a large radius limit. Let us therefore partially compactify our
moduli space of complexified K\"ahler forms by adding points
corresponding to large radius limits.

In the discussion above, we showed how a toric variety could be
associated to a fan $\Delta$. If the moduli space of the toric variety
is also a toric variety itself, then we can describe it in terms of
another fan (in a
different space). This is called the {\it secondary fan\/} and
will be denoted $\Sigma$. $\Sigma$ is a complete fan and thus
describes a compact moduli space.
At first we will not
study the full fan $\Sigma$ but rather the fan $\Psf\subset\Sigma$,
the {\it partial secondary fan}. We describe this fan in detail below. This
fan will specify our partially enlarged moduli space which we will,
from now on, denote by $\MM_{\Psf}$.

Recall the exact sequence we had for the group $A_{n-1}(V)$ of
divisors on $V$ modulo linear equivalence:
\beq
\label{eexactmod}
   0 \tto M \tto \BZ^{\AAO } \tto
   A_{n -1}(V) \tto 0~.
\eeq
By taking the dual and exponentiating we obtained \calle{eexactorica} and
thus realized $V$ as a holomorphic quotient. Suppose we repeat this
process with \calle{eexactmod} except this time we {\it do not\/} take the
dual. This will lead to an expression of our toric variety, $\MM_{\Psf}$, as
a compactification of
$(\IC^*)^\AAO/(\IC^*)^n\cong(\IC^*)^{h^{1,1}(V)}$. This is just the
right form for a moduli space of ``complexified K\"ahler forms'' as
discussed earlier in subsection \ref{sec:complexJ}.

To actually specify the compactification of the above dense open subset
of $\MM_{\Psf}$, we recall our discussion of subsection \ref{sec:3.2}.
There we indicated that compactifications in toric geometry are
specified by following particular families of one-parameter paths
out towards infinity. The limit points of such paths become part of
the compactifying set. The families of paths to be followed are specified
by cones in the associated fan, as we have discussed. This formalism
presents us with a tailor-made structure for compactifying the (partially)
enlarged K\"ahler moduli space: take the cones in the associated
fan $\Psf$ to be the K\"ahler cone of $V$ adjoined with the
K\"ahler cones of its neighbours related by flops. The interior of each such
cone, now interpreted as a component of $\Psf$, gives rise to one  point
in the compactification of the partially enlarged K\"ahler moduli space.
This point is clearly the infinite radius limit of the Calabi-Yau space
corresponding to the chosen K\"ahler cone. These are the marked points
in figure \ref{fig:manyboxes}. 

The final point of discussion, therefore, is the construction of the
K\"ahler cone of $V$ and its flopped neighbours. Now, in most common
situations, these various birational models
will all arise as different desingularizations of a single underlying
singular variety $V_s$. As discussed in subsection \ref{sec:3.3},
these desingularizations
can be associated with different fans,
$\Delta$, and are all related by flops of rational curves.
Furthermore, from subsection \ref{sec:3.5},
 the construction of $\Delta$ amounts to
a triangulation of $P^\circ$ with vertices lying in the set
$\AAO$. Thus, we expect that the
cones in $\Psf$ will be in some kind of correspondence with the
triangulations based in $\AAO$. We will now describe the precise construction
of
$\Psf$.

To understand the construction of the partial secondary fan, we will need one
technical result
which we now state without proof after
some preliminary definitions.
(The proofs can be found in \cite{rOda,rFulton} for the smooth case,
and in \cite{rAGMII}
for the singular case.)
We can consider the intersection of the fan $\Delta$ with $P^\circ$ to
determine a triangulation of $P^\circ$ with vertices taken from
the set of points $P^\circ\cap N$.
We recall that this is a special kind of triangulation (this point
will be important a little later). That is, there is a point
$O$ in the interior of $P$ which is a vertex of every simplex in the
triangulation.
For each $n$-dimensional simplex $\beta$ in $\Delta$, we define
a real linear function by specifying its value at each of the $n$
vertices of $\beta$ except for $O$. The linear function vanishes at $O$.
Let us denote by $\psi_\beta$ such a function defined on
$\beta \in \Delta$. We can extend $\psi_\beta$ by linearity to
a smooth function on all of $N_{\IR}$ which we shall denote by
the same symbol.
Now, we can also define a continuous (but generally not smooth) function
$\psi_{\Delta} : N_{\IR} \rightarrow \IR$ simply by assigning a real number
to each point in $\Delta \cap N$ except $O$ and
within each cone over a simplex $\beta$ defining
the value of $\psi_{\Delta}$ to be $\psi_\beta$ extended beyond $P$
by linearity. In general, this construction
will yield ``corners'' in $\psi_{\Delta}$ at the boundaries between cones.

We say that $\psi_{\Delta}$ is {\it convex\/}
if the following inequality holds:
\beq
\label{inequal}
   \psi_\beta(p) \ge \psi_{\Delta}(p)~,
\eeq
for all points $p \in N_{\IR}$.
Similarly, $\psi_\Delta$ is {\it strictly convex\/} if the equality
is true only for points within the cone containing $\beta$.
The theorem alluded to above which we shall need states that
\beq
\label{eKfms}
  \hbox{Space of K\"ahler forms on $V_\Delta$}\cong
  {\hbox{Space of strictly convex $\psi_\Delta$}\over\hbox{Space of smooth
   $\psi_\Delta$}}~.
\eeq
When $V_\Delta$ is singular,
we must interpret the ``K\"ahler forms'' in this theorem
in an orbifold sense \cite{rAGMII}.

This theorem can also be used to determine whether $V_\Delta$ is K\"ahler
or not. If $V_\Delta$ is not K\"ahler then its K\"ahler cone will be
empty. Thus
\beq
\label{eXVK}
   \hbox{$V_\Delta$ is K\"ahler} \Leftrightarrow \hbox{$\Delta$ admits a
   strictly convex $\psi_\Delta$}~.
\eeq
Such a fan is called {\it regular}.

Given $\Delta$, then, we can in principle determine the structure of
the K\"ahler cone $\xi_\Delta$ associated to this particular desingularization.
Now consider two smooth toric varieties $X_{\Delta_1}$ and $X_{\Delta_2}$
which are obtained from two different fans $\Delta_1$ and $\Delta_2$
whose intersections with $P^\circ$ give triangulations based on
the same set $\AAO$. This will give two
cones $\xi_{\Delta_1}$ and $\xi_{\Delta_2}$ within the space
$A_{n-1}(V)_\IR$. A function which is strictly convex over $\Delta_1$
cannot be strictly convex over $\Delta_2$ and so
$\xi_{\Delta_1}$ and $\xi_{\Delta_2}$ can only
intersect at their boundaries.
Thus, the K\"ahler cones of different birational models fill out
different regions of $A_{n-1}(V)_\IR$ \cite{rOP}.
We can define the partial secondary fan $\Psf$ to consist of all such
cones $\xi_\Delta$ together with all of their faces.

\begin{figure}[htbp]
\centerline{\epsfxsize=10cm\epsfbox{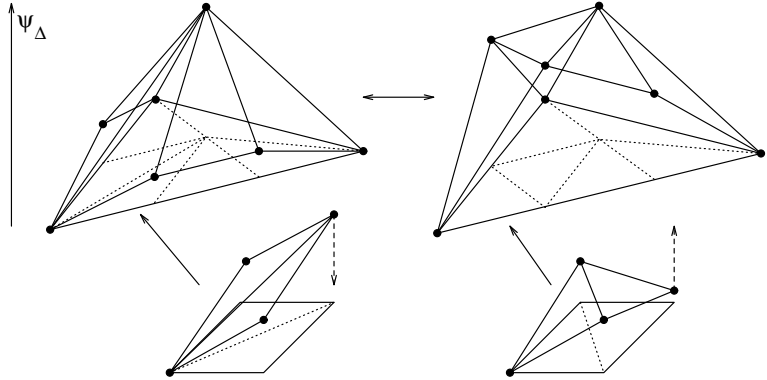}}
\caption{A flop in terms of the function $\psi_\Delta$.}
\label{figure:11}
\end{figure}

If we take $X_{\Delta_1}$ and $X_{\Delta_2}$ to be related by a flop,
then $\xi_{\Delta_1}$ and $\xi_{\Delta_2}$ touch each other
on a codimension 1 wall. One can persuade oneself of this fact by
carefully studying figure \ref{figure:11}.
 The base of the polytope in each case
in this figure
is a section of the fan and the value of $\psi_\Delta$ is mapped out
over this base. The condition that $\psi_\Delta$ is convex is simply
the statement that the resultant surface is convex in the usual sense.
One can move through the space $A_{n-1}(V)_\IR$ by varying the
heights of the solid dots above the base. Note that the flop
transition can be achieved by changing the value of $\psi_\Delta$ at
just one of the points in $\AAO$. (One can mod out by smooth
affine functions by fixing the solid dots at the edge of the base to
be at height zero.)

This shows that the smooth resolutions of $V_s$
correspond to cones in $A_{n-1}(V)_\IR$ touching each other along
codimension 1 walls if they are related by flops.
We now want to explicitly find these cones.
In practice, the authors of \cite{rOP,rBFS} have given a simple algorithm
for carrying out the procedure described below.

Define an $n\times(r-1)$ matrix $A$ whose columns are the coordinates of
the elements of $\AAO$ in $N$. Define an integer matrix $B$ as a matrix whose
columns span the kernel of $A$. We will denote the row vectors of $B$
as $b_i$, $i=1\ldots r-1$. These vectors are vectors in the lattice
$A_{n-1}(V)$.

Each big cone $\sigma\in \Delta$ is specified by its one-dimensional subcones
and thus $n$ elements of $\AAO$, say $\rho_i$, $i\in I$. We can then
specify a big cone $\xi_\sigma$ in $A_{n-1}(V)_\IR$ as the cone
which has one-dimensional edges given by $\{b_j\}$, where $j$ runs over
the {\it complement\/} of the set $I$. We then describe the cone
$\xi_\Delta$ associated to $\Delta$ as
\beq
\label{eXlitc}
  \xi_\Delta = \bigcap\limits_{\sigma\in\Delta}\xi_\sigma~.
\eeq
The cones $\xi_\Delta$ for different resolutions of singularities fit
together to form a fan --- the partial secondary fan. As its name
suggests, this fan is not complete and thus does not yield a compact
moduli space. This is the algorithm necessary to fill in the details
of the calculation of section \ref{sec67}, as we shall shortly see.

All  cones in the secondary fan for the partially enlarged K\"ahler
moduli space are geometric in origin, by construction. Namely, each is a
K\"ahler cone on a particular Calabi-Yau manifold. These distinct cones
can be thought of as different ``phases'' of the model, all differing
by flops of rational curves.
 One might, at first glance,
think that this enlargement is the end of the story as far as distinct points in
the moduli space. However, let us think back to our discussion of Witten's
linear sigma model in section \ref{sec43}.
 There, we found that a simple model ---
the quintic hypersurface --- has a K\"ahler parameter space with two phases.
One is the usual geometric K\"ahler moduli space, which we associated to
the region of positive $r$. The other, which is connected to this region, is
associated with a Landau-Ginzburg model. In this example, then, the moduli
space has two phases but only one has a familiar geometric interpretation.

In fact, there {\it is} a geometric structure that can naturally be
given to this example. The field that we called $P$ in section 
\ref{sec43}
has charge $-5$, while the coordinate fields have charge $1$. This means, by
definition,
that $P$ can be thought of as a section of the line bundle ${\cal O}(-5)$
over the toric base ${\Bbb C}P^4$.
(For a discussion of such bundles the reader
can consult \cite{GH}.)
As is well known, this bundle is a blow-up
of the singular space $\IC^5 / \IZ_5$. The two, therefore, are birational
to one another, since they only differ in a codimension one neighborhood of
the origin.
 Notice that $\IC^5 / \IZ_5$ {\it is}
the configuration space of the Landau-Ginzburg orbifold which constitutes
the $r <0$ phase: we have five fluctuating fields with a $\IZ_5$ orbifold
identification.
In other words, Witten's linear sigma model
construction is telling us that we should not only allow a given model
to undergo birational transformations associated with the compact
toric variety in which it is embedded
 (which in the case of ${\Bbb C}P^4$
would yield just one phase) but, rather, also allow for birational 
transformations
involving a particular bundle over this toric variety.

\begin{figure}[htbp]
\epsfysize=4cm
\centerline{\epsfbox{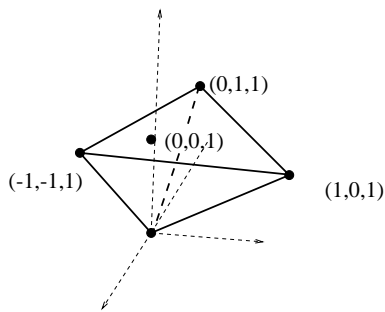}}
\caption{The polyhedron $P^\circ$ embeded in one dimension higher.}
\label{fig:polarp3}
\end{figure}

More precisely, notice
that anomaly cancellation requires that the sum of the charges of the fields
in the linear sigma model vanishes. Geometrically, this is interpretable
as the first Chern class of the total space (base and line bundle) being
equal to zero. As we have discussed in \ref{sec:3.4}, a non-compact toric
variety of dimension $n$
 (such as a bundle over a compact base) is associated with a fan
that does not fill all of $\IR^n$. Furthermore, vanishing first Chern class
requires the toric data for such a variety to lie in a hyperplane. Thus,
the toric varieties which we are led to study have toric data
 points of this sort.
Given the toric data for a compact toric variety, how can we produce
such a data set? It is simple:
 take the $n$-dimensional polar polyhedron $P^\circ$ and
embed it as a polyhedron lying on a hyperplane in one dimension higher.
A simple way to do this is simply to add an $(n+1)$-st coordinate to each point
in $P^\circ$ which has value $1$. For example, consider the polar
polyhedron for $\IC P^2$ in figure 
\ref{figure:polarp2}. We can lift this to three dimensions
in the manner indicated, yielding the pyramid
shaped figure \ref{fig:polarp3}.

\begin{figure}[htbp]
\epsfysize=3cm
\centerline{\epsfbox{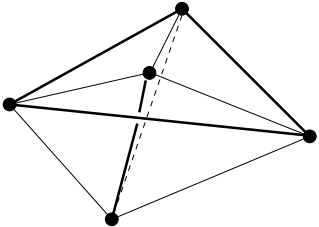}}
\caption{The triangulation with base points $(1,0,1),$
         $(0,1,1),$ $(-1,-1,1),$ $(0,0,1)$. }
\label{fig:pyramid}
\end{figure}

If we now consider this toric variety which lies
in one dimension higher, we note
that there are two classes of triangulations of its points: those for which
{\it every} cone has the ``old'' $n$-dimensional origin (the point
now with coordinates (0,0,...,0,1)) as a vertex, and those triangulations
which do not. (Notice that triangulations in this one dimension higher space
now all involve the ``new'' origin $(0,0,...,0)$. ) The first
class of triangulations are in one-to-one correspondence with those of
the original $n$-dimensional geometry. The others are genuinely new.
Let us continue with our example in $\IC P^2$ to understand what they are.
The triangulation of figure \ref{fig:pyramid}
is the toric data for the line
bundle ${\cal O}(-3)$ over $\IC P^2$. To see this, note that by virtue of
the form of the base of the pyramid, we directly see that the total toric
variety contains $\IC P^2$.

\begin{figure}[htbp]
\epsfysize=3cm
\centerline{\epsfbox{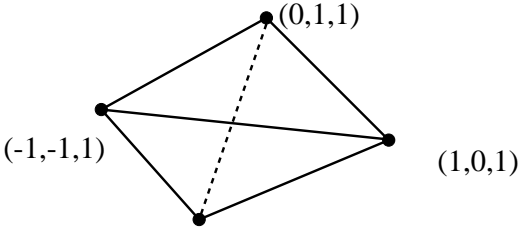}}
\caption{The triangulation which does not use the point $(0,0,1)$.}
\label{fig:dual4}
\end{figure}

 Now, since the toric data does not fill out $\IR^3$
we know that the space is non-compact, and furthermore, since the points all
lie on a hyperplane we know this non-compact space
 has vanishing first Chern class.
From our discussion in section
\ref{sec:Some Classical Geometry} we know that the Chern classes
of $\IC P^2$ come from expanding $(1 + J)^3$, and hence it has first Chern class
$3J$. To cancel this, we need to use the line bundle ${\cal O}(-3)$.
If we consider the other kind
 of triangulation in which we do not use $(0,0,1)$
as in figure \ref{fig:dual4},
 we see that there is only one cone with volume three.
This means the space is singular and, in fact, by following our discussion in
\ref{sec:3.3} it is not hard to see that the space is $\IC^3/\IZ_3$.
Thus, by passing to this one dimension higher toric variety, we are able
to clearly see the birational transition between these two spaces.

It is straightforward to see that if we perform exactly the same analysis in
five dimensions instead of three we will
 interpolate from ${\cal O}(-5)$ over $\IC P^4$
to $\IC^5/\IZ_5$, exactly reproducing the phases analysis of the linear
sigma model. Thus, the story which emerges from this discussion 
is the following:

\begin{enumerate}

\item
To construct the partially enlarged K\"ahler moduli space, we follow
the algorithmic procedure for producing the secondary fan given above,
working with the toric data for the compact toric variety in which our
Calabi-Yau is embedded.

\item
To construct the fully enlarged K\"ahler moduli space, we follow
the algorithmic procedure for producing the secondary fan working with
the toric data for the total space of a line bundle over the above
compact toric variety, which is a non-compact
 manifold with vanishing first Chern class.
A simple way to get this toric data is to embed the compact toric data
in a hyperplane in one higher dimension, as discussed. The new triangulations
associated with passing to one dimension higher correspond to physical models
whose interpretation generally is not in terms of a Calabi-Yau sigma model.
The Landau-Ginzburg example encountered above is one such possibility and
there are others as discussed in \cite{WP,AGM}.

\end{enumerate}

We note that our discussion has focused on Calabi-Yau hypersurfaces
in toric varieties, but can easily be extended to more general circumstances.
The interested reader can consult \cite{CG}
and references
therein for details.

\subsection{Toric Geometry of the Complex Structure Moduli Space}
\label{sec:3.9}

Let us consider the moduli space of complex structures on a
hypersurface within a weighted projective space.
We will use the $n+1$ homogeneous coordinates $[z_0,\ldots,z_n]$.
If we write down the
most general form of the equation defining the hypersurface (i.e.
include all terms compatible with the weight of each coordinate) we
obtain something like
\beq
\label{eGp}
   p=a_0\, z_0z_1\ldots z_n+\ldots+
   a_s\, z_0^{p_0}+a_{s+1}\, z_1^{p_1}+\ldots+a_d\, z_d^{n_d}+\ldots+
   a_i\, z_1^{n^i_1}z_2^{n^i_2}+\ldots=0~.
\eeq
Let $k$ be the number of terms in this polynomial.
As we vary the complex coefficients $a_i$ we may or may not vary the
complex structure of the hypersurface. Some of the variations in $a_i$
give nothing more than reparametrizations of the hypersurface and so
cannot affect the complex structure. (Moreover, sometimes not all of the
possible deformations of complex structure can be achieved by
deformations of the above type. This always happens for K3 surfaces
for example and can happen in complex dimension 3 if the algebraic
variety has not been embedded
in a large enough ambient space
\cite{rGH}.)
A simple reparametrization of the hypersurface is given by the
$(\IC^*)^{n+1}$ action
\beq
\label{eCdd}
   (\IC^*)^{n+1}:(z_0,z_1,\ldots,z_n)\mapsto(\alpha_0z_0,\alpha_1z_1,
   \ldots,\alpha_nz_n),\quad\alpha_i\in\IC^*~.
\eeq
We will consider the case where the only local deformations\footnote{Note
that we have left open
the possibility that there are other, more global, deformations which
do not affect the complex structure.  These would take the form of
discrete symmetries preserving the equation \calle{eGp}.  We shall ignore
such symmetries for the purposes of this paper; their effects on the
analysis of the moduli space are discussed in detail in \cite{rAGMII}.}
 of the
polynomial \calle{eGp} which fail to give a deformation of complex structure
are deformations which amount to a reparametrization of the form
\calle{eCdd}.
One should note that this is quite a strong requirement and
excludes, for example, the quintic hypersurface in 
${\Bbb C}P^4$ which has a group of
reparametrizations isomorphic to $GL(5,\IC)$ rather than $(\IC^*)^5$.
Also, if some of the deformations of complex structure are
obstructed in the sense of \cite{rGH}, (as indeed will happen in our
example), then we only recover a lower dimensional subspace of the
moduli space.

If we first assume that none of the $a_i$'s vanish, then we describe
an open subset $\MM_0$ of our moduli
space of the form $(\IC^*)^k/(\IC^*)^{n+1}\cong (\IC^*)^{k-n-1}$. By allowing
some of the $a_i$'s to vanish we can (partially) compactify this space. It
would
thus appear that our moduli space is a toric variety.

In direct correspondence with the partially enlarged and fully enlarged K\"ahler moduli
spaces of the last section, we note that if we impose the condition
\beq
\label{eaisone}
    a_0=1~,
\eeq
we get the partial complex structure moduli space.
This constraint reduces the $(\IC^*)^{n+1}$
 invariance to $(\IC^*)^n$; we have
used the other $\IC^*$ to rescale the entire equation, in setting
$a_0=1$.  Now as
explained earlier, each monomial in \calle{eCdd} is represented by a point
in $P$ in the lattice $M$. The condition that all reparametrizations are
given by \calle{eCdd} may be stated in the form that there are no points
from $P\cap M$ in the interior of codimension one faces on $P$.
It can be
seen that the $(\IC^*)^n$ action on any monomial is given by the
coordinates of this point in $M$. This gives
rise to the following exact sequence
\beq
\label{ecses}
  1\tto \exp(2\pi i\,N_{\IC})\tto(\IC^*)^{k-1}\tto\MM_0\tto1
\eeq
which gives another description of $\MM_0$.
The $(\IC^*)^{k-1}$ is the space of polynomials with non-zero coefficients
and $\MM_0$ is the resultant open subset of the moduli space in which no
coefficient vanishes. This
open set can then be compactified by adding suitable regions derived
from places where some of the coefficients vanish. Thus, we have again
arrived at something resembling a toric variety.

The full complex structure moduli space is obtained by relaxing the
constraint \calle{eaisone}, thereby obtaining a toric variety of one
higher dimension.

How does one explicitly realize the full complex structure moduli
space as a toric variety? In our discussion of the partially
enlarged K\"ahler moduli space, we began with a natural interpretation
of cones (K\"ahler cones of flopped models) and then augmented this
with other ``cones'' associated to other physical models which are
connected with these geometrical sigma models. How can we introduce
a cone-like structure for the complex structure moduli space in order
to build a secondary fan for it? The answer arises from the work of
\cite{GKZ,GKZ2}
and is described in some detail in 
\cite{AGMsd}. Briefly put,
the discriminant locus of the hypersurface corresponding to $p = 0$
is some complicated expression in terms of its defining coefficients.
As we let the magnitude of these coefficients run to infinity in
a generic manner, typically one term in the descriminant polynomial
will dominate over all others. However, if we let the coefficients run
to infinity in a different direction, another term may dominate. In this
way, we can partition a real section of the complex structure moduli space
into cones which are distinguished by the particular monomial in the
descriminant polynomial which dominates when a ray in that cone is followed out to
infinity. These cones form the secondary fan for the complex structure
moduli space. It can be shown that the algorithm for
explicitly generating these cones is precisely the same as for the
fully enlarged K\"ahler moduli space discussed in the last section.
The only difference is that we make use of the fan   obtained from the
cones over the faces of $P$ instead of $P^\circ$. Different triangulations
of this fan once again translate into different cones in the secondary fan.
Unlike the case of the enlarged K\"ahler moduli space in which (some of) these
cones correspond to birationally equivalent but topological different
manifolds, the different cones in the complex structure moduli space have
a far less dramatic interpretation: as above, they are simply regions in
which different monomials in the discriminant polynomial are dominant.

\newsection{Applications of Toric Geometry}
\label{sec:toricapl}

In this section we shall make use of the toric formalism to better
understand various details of mirror symmetry and spacetime topology
change. An important question, and one which has not been fully settled
as of this writing, concerns how sensitively our conclusions depend upon
our working with Calabi-Yau's embedded in toric varieties. That is,
which properties of quantum geometry that we have discussed in
previous sections and continue to study here are truly intrinsic to
Calabi-Yau string theory, and which reflect special properties of Calabi-Yau's
in toric varieties? We do not have a full answer to this question, and it
is one that should be borne in the back of one's mind when considering
this material.

\subsection{Mirror Manifolds and Toric Geometry}

\label{sec:toricmirror}

In the previous section we have given some background on how one
realizes certain families of Calabi-Yau spaces in the formalism
of toric geometry. As is clear from that discussion, many of the detailed
properties and desired manipulations of these spaces are conveniently
encoded in combinatorial lattice data. We now describe how aspects
of mirror symmetry can also be formulated using toric methods.

It was originally discovered by S.-S. Roan \cite{rflopsbis}
 that the mirror manifold
construction  discussed in section
\ref{MirrorManifolds} has a simple and natural description in
toric geometry. Roan found that when the orbifolding occurring in \cite{GP}
was described in toric terms, it led to an identification between
the $N$ lattice of $X$ and the $M$ lattice of its mirror $Y$.
From this,
he could show mathematically that the Hodge numbers of the pairs constructed
in \cite{GP} satisfy the appropriate equalities. The results of
Roan, therefore,
 indicate that toric methods provide the correct mathematical language
to discuss mirror symmetry.

After Roan's work, Batyrev \cite{rBatyrev}, Batyrev and Borisov 
\cite{rSomemore2}
 and others
(see \cite{BK})
have further pursued the application of toric
methods to mirror symmetry and successfully generalized Roan's results.
The essential idea is based on the fact
that for a Calabi-Yau
hypersurface in a toric variety the polyhedron
$P$ in the $M$ lattice contains
the data associated with the complex structure deformations and the
polar polyhedron $P^\circ$ in the $N$ lattice contains data associated with
the K\"ahler structure. Since mirror symmetry interchanges these data
it is natural to suspect that if $X$ and $Y$ are a mirror pair,
and if each has a realization as  a toric hypersurface, then
the polyhedron $P$ associated to $X$, say $P_X$,
and its polar $P^\circ_X$ should be isomorphic to $P^\circ_Y$
and $P_Y$, respectively.
In fact,  Batyrev   has shown that for any Calabi-Yau hypersurface  $X$ in
a toric variety described by the  polyhedra $P$ and $P^\circ$ in
$M$ and $N$ respectively, if we construct a new hypersurface $Y$
by interchanging the roles of $P$ and $P^\circ$ then the result is
also Calabi-Yau and furthermore has Hodge numbers consistent with
$Y$ being the mirror of $X$.
This result of Batyrev agrees with that of Roan in the special case
of quotients of Fermat-type hypersurfaces in which
$X$ and $Y$ are related by orbifolding, but goes well beyond this
class of examples.  It must be borne in mind, though, that true mirror symmetry
involves much more than these equalities between Hodge numbers.
While it seems quite certain that the new pairs
 constructed by these toric means
are mirrors, establishing this would require showing that both
members of a proposed pair correspond to isomorphic conformal theories.
Progress in this direction has been made in 
\cite{MP},
but a complete argument has yet to be found.
We therefore confine our attention to the use of toric methods for
those examples in which the latter conformal field theory
requirement has been established --- namely those of \cite{GP}.

\subsection{Complex Structure vs. K\"ahler Moduli Space}

In section \ref{sec:3.6}, we saw how the complex
structure and K\"ahler structure moduli spaces can be built as toric
varieties by constructing their respective secondary fans. Now, if
mirror symmetry implies that $P_X$ of $X$ is to be identified with
$P_Y^\circ$ of its mirror $Y$ (and vice versa), we immediately conclude
that the full complex structure moduli space of $X$ is isomorphic to the
fully enlarged K\"ahler moduli space of $Y$ and vice versa. Very simply,
the secondary fans are constructed with these toric data sets (and their
triangulations) as input. The identification between $P_X$ and $P_Y^\circ$
implies that the inputs are identified and hence the secondary fan outputs
are identified as well.

In other words, we now see the resolution to the problem raised in section
\ref{sec:implications}.
 Namely, although a single K\"ahler
cone of Calabi-Yau $Y$ is not isomorphic to the complex structure moduli
space of its mirror $X$, if we pass to the fully enlarged K\"ahler moduli
space of of $Y$ then this {\it is} isomorphic to the (full) complex
structure moduli space of $X$, and
vice versa. (The partially enlarged K\"ahler moduli
space of $Y$ is similarly isomorphic
 to the partial complex structure moduli space
of $X$.)

Beyond this important abstract conclusion, describing the complex and K\"ahler
moduli spaces in terms of toric varieties gives a simple way of establishing
explicit maps between them. Recall that this was necessary, for instance,
to carry out the explicit check on the {\it physical} need to introduce
the other phases in the enlarged K\"ahler moduli space.
In the next section we briefly make use of this toric formalism to
fill in the details of that calculation by explicitly carrying out
the procedure of section \ref{sec:3.8}.

\subsection{An Example}
\label{sec:anexample}
   
\subsubsection{Asymptotic Mirror Symmetry and The Monomial-Divisor Mirror Map}

Given a specific manifold $X$ (which we shall take to be the Calabi-Yau
studied in section \ref{sec67})
 at some large radius limit our aim is to
determine precisely which ``large complex structure limit'' its mirror
partner $Y$ has attained. This can be achieved if we can find the
{\it mirror map} between the
complexified K\"ahler moduli space of $X$ and the moduli
space of complex structures of $Y$.
We have already seen in the preceding sections that
in the cases we are considering, both these spaces are isomorphic to
toric varieties which are (compactifications of)
$(\IC^*)^\AAO/(\IC^*)^n$.

In the case of the moduli space of complexified K\"ahler forms on $X$,
$\AAO$ represented the set of toric divisors on $X$. The $(\IC^*)^n$
action represents linear equivalence and is determined by the
arrangement of the points corresponding to $\AAO$ in the lattice $N$.
In the case of the moduli space of complex structures on $Y$, using
the results of subsections \ref{sec:toricmirror}, 
$\AAO$ now represents the set of
monomials in the defining equation for $Y$ (with the exception of the
$a_0$ term) and the $(\IC^*)^n$ action represents reparametrizations
determined by the arrangement of the points of $\AAO$ in the lattice
$N$.

We have thus arrived at a natural proposal for the mirror map,
namely to simply identify the divisors of $X$ given by $\AAO\subset N$
with the monomials of $Y$ also given by $\AAO\subset N$. The induced map
between the moduli spaces is called
the {\it monomial-divisor map} and it is unique up to symmetries of
the point set $\AAO\subset N$.
It turns out that although
this proposal for a
mirror map has the correct asymptotic behavior near the large radius
limit points, it 
differs from the actual mirror map
away from large radius limits. This is a point which has been studied
in some detail and the reader can consult \cite{AGMsd}  for details.
As our only concern for the calculation carried out in this
section is with large radius limits,  we may take this na\"{\i}ve
identification of the two moduli spaces as an approximation of
 the true mirror map which is adequate for our purposes.
For a more mathematical discussion of these points see
\cite{rAGMII}.

In order to determine the large radius limits of $X$, we now
consider compactifications of $(\IC^*)^\AAO/(\IC^*)^n$. In terms of
the K\"ahler form $J$ on $X$, we are studying a limit in which $e^{2\pi
i(B+iJ)}\to0$ by taking $J\to\infty$ inside the K\"ahler cone $\xi_X$
of $X$\footnote{More precisely, $\xi_X$ represents that part of the K\"ahler
cone of $X$ which comes from the ambient space $V$; it is in reality
the K\"ahler cone of $V$ that we study.}.
 In the language of toric geometry, this point added to the
moduli space is given by the cone $\xi_X\subset A_{n-1}(V)_\IR$. In
this way we determine a compactification of the
space of complexified K\"ahler forms on $X$ which includes all large
radius limits. It is the toric variety given by the K\"ahler cone of
$X$ and its neighbours with respect to the lattice $A_{n-1}(V)$.

A large radius limit of $X$ can now be translated into a large complex
structure limit of $Y$. The fact that $J$ remains within $\xi_X$
dictates the relative growth of the coefficients $a_i$ of the
monomials as they are taken to $\infty$ (or 0).
Any path in the moduli space with the property
 that the coefficients of the corresponding
family of hypersurfaces obey these growth properties will approach the
large complex structure limit point specified by $\xi_X$.
We will now demonstrate how
this can be done explicitly by an example.

\subsubsection{A Calculation}

In section \ref{sec:flop}, we showed the result of such an identification for
the case of a hypersurface 
$X_s$  in $V\cong W\IC P^4(6,6,3,2,1)$ given by
$$
  f=z_0^3+z_1^3+z_2^6+z_3^9+z_4^{18}=0 
$$
and its mirror.
In particular, we showed that correlation functions calculated in
corresponding limits are in fact equal, thus providing strong support
for the picture presented. Although we shall not have time nor space
to discuss how the correlation functions themselves are calculated
(see \cite{AGM} and references therein for details on this aspect),
we would like to show how corresponding limit points are explicitly found.
As we will only consider geometric phases, we shall only work with
the partially enlarged K\"ahler moduli space.

As in section \ref{sec67},
let $X_s$ be a hypersurface in $V\cong W \IC P^4(6,6,3,2,1)$ given by
\beq
\label{eXfm}
   f=z_0^3+z_1^3+z_2^6+z_3^9+z_4^{18}=0~.
\eeq
The reason we choose to work with  $X_s$ and its mirror is that
we require an example sufficiently complicated to exhibit flops.
This is true for $X_s$. In particular,
$X_s$ has two curves of $\BZ_2$ and
$\BZ_3$ singularities respectively (from the
weighted projective space identifications) and these intersect at three points
which locally have the form of $\BZ_6$ singularities.
These singularities are the same as the singularities studied in 
\cite{rAorb}.
Any blow-up of
these singularities to give a smooth $X$
gives an exceptional divisor with 6 irreducible
components, thus $h^{1,1}(X)=7$. When one resolves the singularities in
$W\IC P^4(6,6,3,2,1)$ one only obtains an exceptional set with 4
components. One of these components intersects $X$ in regions around
the 3 former $\BZ_6$ quotient singularities. Thus 3 elements of
$H^2(X)$ are being produced by a single element of $H^2(V)$.

In terms of K\"ahler form moduli space one can picture this as
follows. Each of the three $\BZ_6$ quotient singularities contributes a
component of the exceptional divisor. As far as the K\"ahler cone of $X$
is concerned the volume of these three divisors can be varied
independently. If we wish to describe the K\"ahler form on $X$ in
terms of a K\"ahler form on $V$ however, these three volumes had
better be the same since they all come from one class in $H^2(V)$.
Thus we are restricting to the part of the moduli space of K\"ahler
forms on $X$ where these three volumes are equal. An important point
to notice is that even though we are ignoring some directions in
moduli space, we can still get to a large radius limit where {\it all\/}
components of the exceptional divisor in $X$ are large.

The toric variety $W\IC P^4(6,6,3,2,1)$ is given by complete fan around
$O$ whose one dimensional cones pass through the points
\beqn
\alpha_5&=&(1,0,0,0)~,\nonumber\\
\alpha_6&=&(0,1,0,0)~,\nonumber\\
\alpha_7&=&(0,0,1,0)~,
\label{eXvx}\\ 
\alpha_8&=&(0,0,0,1)~,\nonumber\\
\alpha_9&=&(-6,-6,-3,-2)
  ~.\nonumber
\eeqn
This data uniquely specifies the fan in this case. (The reason for the
curious numbering scheme will become apparent.) This fan is comprised of
five big cones most of which have volume greater than $1$. For example, the cone
subtended by $\{\alpha_5,\alpha_7,\alpha_8,\alpha_9\}$ has volume 6.
The sum of the volumes of these 5 cones is 18 and thus we need to
subdivide these 5 cones into 18 cones to obtain a smooth \CY\
hypersurface. The extra points on the boundary of $P^\circ$
which are available to help us do this are
\beqn
\alpha_1&=&(-3,-3,-1,-1)~,\nonumber\\
\alpha_2&=&(-2,-2,-1,0)~,\nonumber\\
\alpha_3&=&(-4,-4,-2,-1)~,\nonumber\\
\alpha_4&=&(-1,-1,0,0)~.\nonumber
\eeqn
Note that, as required, none
of these points lies in the interior of a codimension
one face of $P^\circ$.
Any complete fan $\Delta$ of simplicial cones having {\it all\/}
of the lines through
$\{\alpha_1,\ldots,\alpha_9\}$ as its set of one-dimensional cones will
consist of 18 big cones and specify a smooth \CY\ hypersurface, but the
data $\{\alpha_1,\ldots,\alpha_9\}$ does not uniquely specify this
fan.

\begin{figure}[htbp]
\centerline{\epsfxsize=10cm\epsfbox{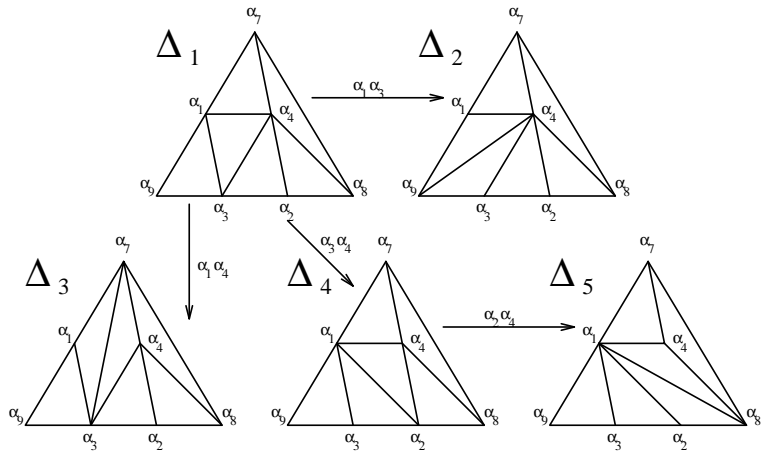}}
\caption{The five smooth models.}
\label{figure:12}
\end{figure}

\begin{figure}[htbp]
\centerline{\epsfxsize=10cm\epsfbox{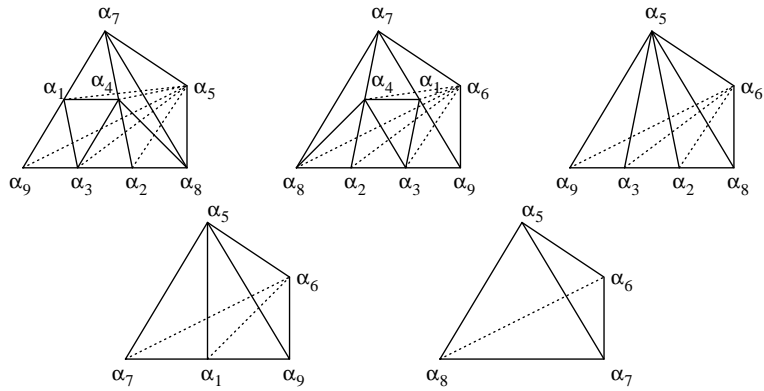}}
\caption{The tetrahedra in resolution $\Delta_1$.}
\label{figure:13}
\end{figure}

A little work shows that there are 5 possible fans consistent with
this data, all of which are regular. That is, all 5 possible toric
resolutions of $W\IC P^4(6,6,3,2,1)$ are K\"ahler. We can uniquely
specify the fan $\Delta$ just by specifying the resulting
triangulation of the face $\{\alpha_7,\alpha_8,\alpha_9\}$. The
possibilities are shown in figure \ref{figure:12} and in figure 
\ref{figure:13} the
three-dimensional simplices are shown for the resolution $\Delta_1$.

To obtain $Y$ as a mirror of $X$, we divide $X$ by the largest
phase symmetry consistent with the trivial canonical bundle
condition, as discussed in section \ref{sec:conmir}.
 This is given by the following generators:
\beqn
 \lbrack z_0,z_1,z_2,z_3,z_4\rbrack&\to&
  \lbrack\omega z_0,z_1,z_2,z_3,\omega^2z_4\rbrack~,
\nonumber \\
 \lbrack z_0,z_1,z_2,z_3,z_4\rbrack&\to&
 \lbrack z_0,\omega z_1,z_2,z_3,\omega^2z_4\rbrack~,
\label{eXpss}\\
 \lbrack z_0,z_1,z_2,z_3,z_4\rbrack&\to&
  \lbrack z_0,z_1,\omega z_2,z_3,\omega^2z_4\rbrack~,
\nonumber
\eeqn
where $\omega=\exp(2\pi i/3)$.
This produces a whole host of quotient singularities but since we are
only concerned with the complex structure of $Y$ we can ignore this
fact.

\begin{figure}[htbp]
\centerline{\epsfxsize=10cm\epsfbox{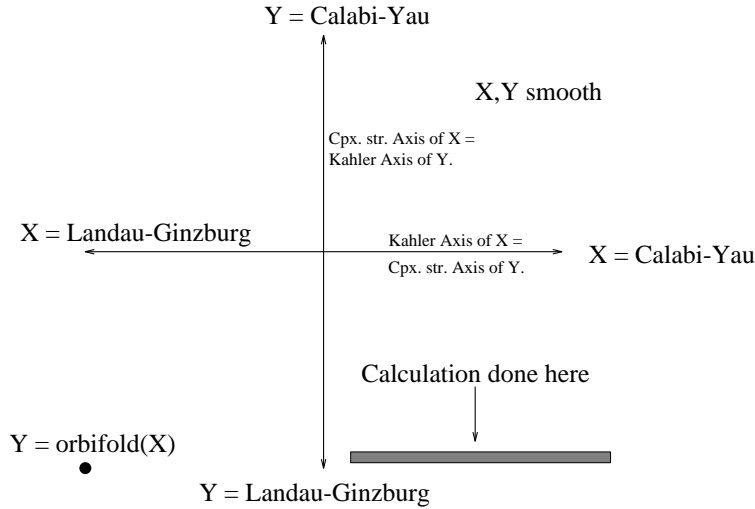}}
\caption{Area of K\"ahler sector of $X$ and $Y$ where we
perform calculation.}
\label{figure:14}
\end{figure}

In light of the results of \cite{WP} and the discussion in section 
\ref{sec43},
we should
actually be more careful in our use of language here. To be more
precise, given the {\it Landau-Ginzburg\/} model $X_{\rm LG}$ whose
superpotential is specified in \calle{eXfm}, we can construct another
Landau-Ginzburg theory $Y_{\rm LG}$ as the orbifold of
$X_{\rm LG}$ by the group generated by \calle{eXpss} and having  the same
superpotential \calle{eXfm}. $Y_{\rm LG}$ is the mirror of
$X_{\rm LG}$. Using our discussion of section 
\ref{MirrorManifolds}, we know that 
the smooth \CY\ manifolds occupy a different
region of the same moduli space as the Landau-Ginzburg theory. Thus if
we deform {\it both\/} of our mirror pair $X_{\rm LG}$ and
$Y_{\rm LG}$, then we can obtain two smooth mirror manifolds $X$
and $Y$. If we wanted to compare all correlation functions of the
conformal field theories of $X$ and $Y$ then we would have to do this.
All we are going to do in this section however is to compare
information concerning the K\"ahler sector of $X$ with the complex
structure sector of $Y$. Information concerned with the complex
structure of $Y$ as a smooth manifold is isomorphic to that of
$Y_{\rm LG}$. Thus, there is no real need to deform $Y_{\rm LG}$
into a smooth \CY\ manifold. In figure \ref{figure:14},
we show very roughly the
slice in which we do the calculation in this section. Note that this
figure is very oversimplified since the moduli space typically splits
into many more regions and indeed the whole point of this calculation
is to show that the area concerned spans more than one region.

The most general deformation of \calle{eXfm} consistent with this
$(\BZ_3)^3$ symmetry group is
\beqn
W &=&a_0 z_0z_1z_2z_3z_4 +
a_1 z_2^3z_4^9 + a_2 z_3^6z_4^6 + a_3 z_3^3z_4^{12} +
a_4 z_2^3z_3^3z_4^3\nonumber\\
  && +a_5z_0^3+a_6z_1^3+a_7z_2^6+a_8z_3^9+a_9z_4^{18} = 0~.
\label{eComplex}
\eeqn
One can show that $h^{2,1}(Y)=7$. The group of reparametrizations of
\calle{eComplex} is indeed $(\IC^*)^5$ as required which shows that we
obtain 5 deformations of complex structure induced by deformations of
\calle{eComplex}. Note that for both $X$ and $Y$ we had 7 deformations of
which only 5 will be analyzed via toric geometry. It is no
coincidence that these numbers match --- it follows from the
monomial-divisor mirror map.

\subsection{The Moduli Spaces}

Let us now build the cones in $A_{n-1}(V)_\IR$ to form the partial
secondary fan. The method was outlined in subsection \ref{sec:3.8}.
Again, we shall only carry this out for the partially enlarged moduli space,
although it is not much harder to do the fully enlarged case.
 We first build
the $4\times9$ matrix $A$ with columns $\alpha_1,\ldots,\alpha_9$.
{}From this we build the $9\times5$ matrix $B$ whose columns span
the kernel of $A$.
The rows of $B$ give vectors in $A_{n-1}(V)_\IR$. Note that a change of
basis of the kernel of $A$ thus corresponds to a linear transformation
on $A_{n-1}(V)_\IR$. In order for us to translate the coordinates in
$A_{n-1}(V)_\IR$ into data concerning the coefficients $a_i$ in the
complex structure of $Y$ we need to chose a specific basis in
$A_{n-1}(V)_\IR$.

We have already fixed $a_0=1$. We still have a $(\IC^*)^4$ action on the
other $a_1,\ldots,a_9$ by which can fix 4 of these coefficients equal
to one. Let us choose $a_5=a_6=a_7=a_8=1$ and denote the matrix that
corresponds to this choice as $B_1$. Our 5 degrees of
freedom are given by $\{a_1,a_2,a_3,a_4,a_9\}$. We want that a
point with coordinates $(b_1,b_2,b_3,b_4,b_5)$ in $A_{n-1}(V)_\IR$
corresponds to $\{a_1=e^{2\pi i(c_1+ib_1)},a_2=e^{2\pi i(c_2+ib_2)},
\ldots,a_9=e^{2\pi i(c_5+ib_5)}\}$ for some value of the $B$-field
$(c_1,\ldots,c_5)$. This means our matrix $B_1$ should be of the form
$$B_1=\left(\matrix{1&0&0&0&0\cr0&1&0&0&0\cr0&0&1&0&0\cr0&0&0&1&0\cr
  B_{5,1}&B_{5,2}&B_{5,3}&B_{5,4}&B_{5,5}\cr
  B_{6,1}&B_{6,2}&B_{6,3}&B_{6,4}&B_{6,5}\cr
  B_{7,1}&B_{7,2}&B_{7,3}&B_{7,4}&B_{7,5}\cr
  B_{8,1}&B_{8,2}&B_{8,3}&B_{8,4}&B_{8,5}\cr0&0&0&0&1\cr}\right).
$$
$B_1$ is now completely
determined by the condition that its columns span the kernel of $A$.

For the actual calculation below, we use a slightly different set of
coordinates, choosing $\{a_0,a_1,a_2,a_3,a_4\}$ as the 5 degrees of freedom
and setting
$a_5=a_6=a_7=a_8=a_9=1$.
(We do this to express \calle{eComplex} in the
form: ``Fermat + perturbation'', in order
to more easily apply the calculational techniques of \cite{ALR}.)
The new basis can
be obtained from $B_1$ by using a $\IC^*$ action
$\lambda:z_4\to\lambda z_4$. We obtain the following matrix:
\def\ff#1#2{-{\scriptstyle #1\over #2}}
\def\vsttr{\vphantom{\displaystyle\sum}}
$$B=\left(\matrix{0&1&0&0&0\cr0&0&1&0&0\cr0&0&0&1&0\cr0&0&0&0&1\cr
  \vsttr\ff13&0&0&0&0\cr\vsttr\ff13&0&0&0&0\cr
  \vsttr\ff16&\ff12&0&0&\ff12\cr\vsttr\ff19&0&\ff23&\ff13&\ff13\cr
  \vsttr\ff1{18}&\ff12&\ff13&\ff23&\ff16\cr}\right).
$$

\begin{figure}[htbp]
\epsfxsize=10cm
\centerline{\epsfbox{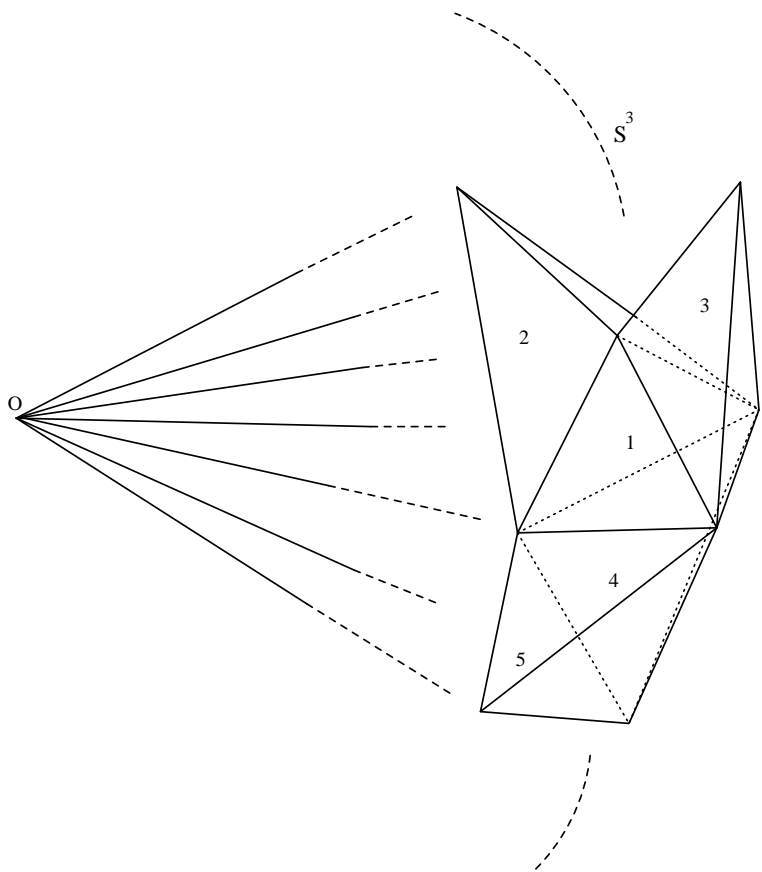}}
\caption{The partial secondary fan.}
\label{figure:15}
\end{figure}

For each of the resolutions $\Delta_1,\ldots,\Delta_5$ we can now
construct the corresponding cone in $\Psf$ following the method in
subsection \ref{sec:3.8} using the $B$ matrix above. 
These five cones are shown
schematically in figure \ref{figure:15}
and the explicit coordinates in table \ref{table:A}.

\begin{table}
\caption{Generators for first cone.}
\label{table:A}
{
\def\ilspace{\omit&height2.5pt&&&&\cr}
$$\vbox{\tabskip=0pt \offinterlineskip
\halign{\strut#&
\vrule#&\hfil\quad$#$\quad\hfil&\vrule#&\hfil\quad$#$\quad\hfil&
\vrule#\cr \tablerule\ilspace
&&v_1&&(-{1\over3},0,0,0,0)&\cr\ilspace
&&v_2&&(-{7\over{18}},-{1\over2},-{1\over3},-{2\over3},
    -{1\over6})&\cr\ilspace
&&v_3&&(-{1\over6},-{1\over2},0,0,-{1\over2})&\cr\ilspace
&&v_4&&(-{2\over9},0,-{1\over3},-{2\over3},-{2\over3})&\cr\ilspace
&&v_5&&(-{1\over9},0,-{2\over3},-{1\over3},-{1\over3})&\cr\ilspace
\tablerule\noalign{\vskip2pt}
}}
$$}
\end{table}

\begin{table}
\caption{Generators for all cones.}
{
\def\ilspace{\omit&height2.5pt&&&&\cr}
$$
\def\ilspace{\omit&height1.5pt&&&&\cr}
\vbox{\tabskip=0pt \offinterlineskip
\halign{\strut#&
\vrule#&\hfil\quad#\quad\hfil&\vrule#&\hfil\quad#\quad\hfil&
\vrule#\cr \tablerule\ilspace
&&Resolution&&Generators&\cr\ilspace
\tablerule\ilspace
&&$\Delta_1$&&$v_1$, $v_2$, $v_3$, $v_4$, $v_5$&\cr\ilspace
&&$\Delta_2$&&$v_1$, $v_1-v_2+v_3+v_4$, $v_3$, $v_4$, $v_5$&\cr\ilspace
&&$\Delta_3$&&$v_1$, $v_2$, $v_2-v_3+v_4$, $v_4$, $v_5$&\cr\ilspace
&&$\Delta_4$&&$v_1$, $v_2$, $v_3$, $v_2+v_3-v_4+v_5$, $v_5$&\cr\ilspace
&&$\Delta_5$&&$v_1$, $v_2$, $v_3$, $v_2+v_3-v_4+v_5$, &\cr
\omit&&&&
\hphantom{$v_1$, $v_2$, }
$v_3+(v_2+v_3-v_4+v_5)-v_5$&\cr\ilspace
\tablerule\noalign{\vskip2pt}
}}$$}
\end{table}

Note that as expected the fan we generate, $\Psf$, is not a complete
fan and does not therefore correspond to a compact K\"ahler moduli space.
This is a reflection of our working only with the partially enlarged
moduli space --- the geometrical models only fill out part of
the full moduli space.

We now wish to translate this moduli space of K\"ahler forms into the
equivalent structure in the moduli space of complex structures on $Y$.
The way that we picked the basis in the $B$ matrix in this section
tells us exactly how to proceed.
For a point
$(u_0,u_1,\ldots, u_4)$ in $A_{n-1}(V)_{\IR}$ we define
\beq
\label{eXek}
  w_k=e^{2\pi i(c_k+iu_k)}~,
\eeq
for any real $c_k$. This is then mapped to \calle{eComplex} by
\beqn
\label{eXmm}
 a_i&=&w_i^{-1},\quad i=0,\ldots,4~,\\
  a_i&=&1,\quad i=5,\ldots,9~.
\eeqn

This map explicitly tells us how to approach infinity in the complex
structure moduli space of $Y$ to arrive at the putative mirror
of a given large radius point in $X$. The five directions so generated
are precisely those used in the calculation of section \ref{sec67}.
As we showed in that section, ratios of correlation functions agree exactly
in corresponding limits.

Although we will not discuss it in these lectures, we note for
completeness that the fully enlarged K\"ahler moduli space of $X$
has 95 other regions in addition to these five geometric ones.
One of these is a Landau-Ginzburg region, 27 are singular
Calabi-Yau regions (strings propogating on Calabi-Yau's with orbifold
singularities) and 67 are regions which are ``hybrids'' of these
others. As yet, no one has studied this hybrid models in any significant detail.

\section{ The Web of Connected Calabi-Yau Manifolds}
\label{sec:web}

In  section \ref{topologychange2},
we have seen that type II string  theory provides
us with a mechanism for physically realizing topology changing transitions
through conifold degenerations. This naturally raises two related questions:

\begin{enumerate}

\item
 Are all Calabi-Yau manifolds
            interconnected through a web of such
            transitions?
\item
    Are there other kinds of singularities,
    besides the ordinary double
    points discussed above, which might have qualitatively different physics and
    which might also have an important role in extending the Calabi-Yau web?

\end{enumerate}

In this section we discuss some work 
presented in \cite{CGGK} which is  relevant to these two questions. Related
work can be found in \cite{Candelas,DKVdelpezzo,MSdelpezzo}
 and references therein.
The background we have developed in toric geometry plays a central role.

\subsection{Extending the Mathematical Web of Calabi-Yau Manifolds}

In an important series of papers \cite{rCGH1,rCGH2}, 
it was argued some time ago
that all Calabi-Yau manifolds realized as complete intersections in products of
(ordinary) projective spaces are mathematically connected through conifold
degenerations. As we mentioned above, although an intriguing prospect, it
previously seemed that string theory did not avail itself of these topology
changing transitions --- as discussed, perturbative string theory is
inconsistent at conifold points. The recent work described above shows
that inclusion of non-perturbative effects cures the physical inconsistencies,
at least in type II string theory, and hence the physical theory does allow such
topology changing transitions to occur.

 Since the time of \cite{rCGH1,rCGH2}, the class
of well studied Calabi-Yau manifolds has grown. Initially inspired by work of
Gepner 
\cite{GEP},
 the class of hypersurfaces in weighted projective
four-dimensional spaces has received a significant amount of attention
\cite{rmany1,rmany2,CLS}.
It was shown in  \cite{rS1,rS2}
 that there are 7555 Calabi-Yau spaces
of this sort. Inspired by mirror symmetry,
 another class of Calabi-Yau spaces (containing
these 7555  hypersurfaces) that have been under detailed
study are complete
intersections in toric varieties
\cite{AGM,WP}. Understanding
the structure of the moduli space of type II vacua requires that we
determine if all of these Calabi-Yau spaces
are interconnected through a web of  topology changing transitions. 

In the following we will  describe a systematic
procedure for finding transitions
between Calabi-Yau manifolds realized as complete intersections in toric
varieties. The method is  elementary although at the present time
there aren't any general results on its range of applicability. Rather,
the usefulness of this method becomes apparent by directly applying it to 
a subclass of the Calabi-Yau spaces realized in this manner.
For instance, using the
approach discussed below,
 all 7555 Calabi-Yau hypersurfaces in weighted projective 
four-dimensional space
are {\it mathematically} connected to the 
web\footnote{A similar conclusion has
been reached by P. Candelas and collaborators using different
methods \cite{Candelas}.}.
We say mathematically because the
transitions this procedure yields are not all of the conifold sort. Rather,
there are Calabi-Yau spaces
connected through more complicated singularities than
the ordinary double points used in \cite{rGMS}, as
described in section \ref{topologychange2}. For example,
some of these singularities
are such that electrically {\it and} magnetically charged black hole states
become simultaneously massless giving us an analog of the phenomenon discussed
in 
\cite{rAD}.
Arguing for physical transitions through these theories requires
more care than those involving conifold points.  Whereas the term conifold
transition refers
to Calabi-Yau spaces linked through conifold degenerations, the
term {\it extremal} transitions
\cite{rDMLG}
 refers to analogous links through any of
the singularities at finite distance 
encountered on the discriminant locus. At present there is
only a fully satisfying physical understanding of the conifold subclass of
extremal transitions\footnote{Since giving and writing-up these lectures,
a good deal of progress has been made on sorting out the physics
of other sorts of singularities and the transitions they
involve. The reader can consult \cite{DKVdelpezzo,MSdelpezzo}.}.

The procedure  described
 below is relevant for Calabi-Yau spaces embedded in toric
varieties and this was another motivation for the material in 
section \ref{toricintro}.

 To keep the discussion here concise, we shall focus
on the case of hypersurfaces in weighted projective four-dimensional spaces,
although we shall briefly mention some generalization
at the end of this section. As discussed in
\cite{rBatyrev} and reviewed in section \ref{toricintro},
 the data describing such Calabi-Yau manifolds is:

\begin{enumerate}

\item
   A lattice $N \simeq {\Bbb Z}^4$ and its real extension 
  $N_{\IR} = N \otimes_{\IZ} {\Bbb R}$.

\item
   A lattice $M = {\rm Hom}(N,{\Bbb Z})$ and its real extension 
  $M_{\Bbb R} = M \otimes_{\Bbb Z} {\Bbb R}$.

\item
   A reflexive polyhedron $P \subset M_{\Bbb R}$.

\item
  The dual (or polar) polyhedron $P^{\circ} \subset N_{\Bbb R}$.

\end{enumerate}

Now, given the above sort of toric data for two different families of
Calabi-Yau spaces in two different weighted projective four-dimensional spaces,
 how might
we perform a transition from one to the other? Well, given the polyhedra
$(P, P^{\circ})$
for one Calabi-Yau and $(Q,Q^{\circ})$ for the other, one has the  natural 
manipulations of set theory  to relate them: namely, the operations
of taking intersections and unions.
 Consider then, for instance, forming new toric
data by taking the intersection 
$$
  R = {\rm convex~hull}\Big( (P \cap M) \cap (Q \cap M) \Big) ~.
$$
Further assume that $R$ (and its 
dual $R^{\circ}$) are reflexive polyhedra
so that the singularities encountered
are at finite distance in the moduli 
space \cite{rHay}.
How are the three Calabi-Yau spaces
$X, Y, Z$
associated to $(P, P^{\circ})$, $(Q,Q^{\circ})$ and $(R,R^{\circ})$
respectively, related? As discussed in section \ref{toricintro},
the toric data contained in the polyhedron in
$M_{\Bbb R}$ is well known to describe the complex structure deformations of
the associated Calabi-Yau realized via monomial deformations of its
defining 
equation\footnote{By mirror symmetry, of course, it can also
be used to describe the K\"ahler structure on the mirror
Calabi-Yau.}.  Concretely, the lattice points in $P \cap M$
are in one-to-one correspondence with monomials in the defining equation of
$X$, and similarly\footnote{
More precisely, some subset of these points correspond
to the toric complex structure deformations, as mentioned earlier. For details
see \cite{AGM,rAGMII}.}
for $Y$ and $Z$.
 Thus, in going from $X$ to $Z$ we
have specialized the complex structure by restricting ourselves to a subset
of the monomial deformations. This is reminiscent of the example 
studied  earlier,
in which we specialized the complex structure of the quintic from its original
$101$-dimensional moduli space to an $86$-dimensional subspace. This is not
the end of the story. Clearly the dual $R^{\circ}$ 
contains $P^{\circ}$. As discussed earlier,
the toric data contained in the polar polyhedron describes
the K\"ahler structure deformations of the associated Calabi-Yau. Concretely,
lattice points in  $P^{\circ} \cap N$ correspond to toric divisors which are
dual to elements in  
in\footnote{Again, to be more precise some subset of
the lattice points correspond to non-trivial
 elements in $H^2(X, {\Bbb Z})$. For
details see \cite{AGM,rAGMII}.} $H^2(X, {\Bbb Z})$.
Thus, in passing from $X$ to $Z$ we have
also added toric divisors, i.e. we have performed a blow-up. This again is
reminiscent of the example studied earlier: after specializing the complex
structure we performed a small resolution. All of the discussion we have just
had relating $X$ to $Z$ can be similarly applied to relate $Y$ to $Z$.
Hence, by using the toric data associated to $X$ and to $Y$ to construct
the toric data of $Z$, we have found that $Z$ provides a new Calabi-Yau that 
both $X$ and $Y$ are linked to in the web.

Of course, the key assumption in the above discussion is that $(R,R^{\circ})$
provides us with toric data for a Calabi-Yau, i.e. they are reflexive
polyhedra. At present, there isn't a general method for
picking $(P, P^{\circ})$ and  $(Q,Q^{\circ})$ such that this is necessarily the case.
In fact, the toric data for a given Calabi-Yau is not unique but,
for instance,  depends
on certain coordinate choices. Thus the reflexivity of $(R,R^{\circ})$ or
lack thereof 
depends sensitively on the coordinate choices used in representing 
$(P, P^{\circ})$ and  $(Q,Q^{\circ})$. Hence, a more appropriate question is
whether there exists suitable representations
 of $(P, P^{\circ})$ and  $(Q,Q^{\circ})$
such that $(R,R^{\circ})$ is reflexive. In \cite{CGGK}, an exhaustive
search was carried out in the following manner.
The toric data $(P, P^{\circ})$ and $(Q,Q^{\circ})$, was arbitrarily chosen
from the set of 7555 hypersurfaces. A variety of 
coordinate representations  for each (related
by $SL(5,\BZ)$ transformations and coordinate permutations) were
considered and directly
checked to see if $(R,R^{\circ})$ obtained from their intersection is reflexive.   
When such an $(R,R^{\circ})$ is reflexive, we learn that
$X$ and $Y$  are (mathematically) connected
through the Calabi-Yau $Z$. We note that, in general, $Z$ is not associated to
a Calabi-Yau hypersurface in a
 weighted projective space --- but rather a Calabi-Yau
embedded in a more general toric variety.

In this manner, by
direct computer search, it was checked that all 7555 hypersurfaces in
weighted projective four-dimensional space are linked (and through the process described
we have actually linked them up to numerous other Calabi-Yau spaces  --- the
 $Z$-type Calabi-Yau spaces above).
The main physical question, then, is what is the nature of the singularities
encountered when we specialize the complex structure in the manner dictated
by the intersection of $P$ and $Q$.
Analysis of the simplest examples shows that one often encounters
singularities which are qualitatively different from the
well understood case of several ordinary double points 
studied in \cite{rGMS}, considered previously. 

To illustrate this point, and the discussion of this section more
 generally, let us 
consider two explicit examples.

\subsection{Two Examples}

\vspace{.2in}
\noindent
{\it Example 1:}

\vspace{.1in}
Let us take $X$ to be  the family of quintic Calabi-Yau hypersurfaces
in ${\Bbb C}P^4$ and $Y$ to be
 the family of Calabi-Yau hypersurfaces of degree 6 in 
 $W{\Bbb C}P^4(1,1,1,1,2)$.
The Hodge numbers of $X$ are
$(h^{2,1}_X,h^{1,1}_X) = (101,1)$ and those of $Y$ are 
$(h^{2,1}_Y,h^{1,1}_Y) = (103,1)$.
  Following the procedure described above and using the discussion
of chapter \ref{toricintro}
$P^{\circ} \cap N$ is given by 
\beq
\label{etoricdatI}
\matrix{ (1 & 0 & 0 & 0  ) ~ , \cr
         (0 & 1 & 0 & 0  ) ~, \cr
         (0 & 0 & 1 & 0  ) ~ , \cr
	 (0 & 0 & 0 & 1  ) ~ , \cr
         (-1&-1&-1&-1)     ~,\cr 
       }
\eeq
and $Q^{\circ} \cap N$ by
\beq
\label{etoricdatII}
\matrix{ (1&0&0&0)~, \cr
         (0&1&0&0)~, \cr
         (0&0&1&0)~, \cr
         (0&0&0&1)~, \cr
         (-1&-1&-1&-2)~, \cr
       }
\eeq
From these polyhedra
we find that the toric  data for family $Z$, $R^{\circ}$, is the convex hull of
\beq
\label{etoricdatIII}
\matrix{ (1&0&0&0)~,\cr
         (0&1&0&0)~,\cr
         (0&0&1&0)~,\cr
         (0&0&0&1)~,\cr
         (-1&-1&-1&-1)~,\cr
         (-1&-1&-1&-2)~. \cr
       }
\eeq
Note that for ease of presentation we are taking unions of data in $N$ space
which is dual to taking intersections in $M$ space\footnote{The duality is
only generally
valid when considering intersections and unions in $\IR^4$ instead of
${\Bbb Z}^4$.},
 discussed above.
Consider first the transition from $Y$ to $Z$.
 One can show that the singular subfamily
obtained by specializing the complex structure of $Y$, in the manner discussed
above, consists of Calabi-Yau spaces which generically have 20 ordinary
 double points all
lying on a single ${\Bbb C}P^2$ and hence obeying
 one non-trivial homology relation.
This, therefore, is another example of the conifold transitions described in
section \ref{topologychange2}.
 Thus, we can pass from $Y$ to $Z$ in
the manner discussed and the Hodge numbers change to 
$(h^{2,1}_Z,h^{1,1}_Z) = (103 - 20 + 1, 1 + 1) = (84,2)$. The relation between
$X$ and $Z$, though, is more subtle. In specializing the complex structure of
$X$ dictated by the toric manipulation,
 we find a singular family of Calabi-Yau spaces,
each generically having one singular point. The local description of this singularity,
however, is {\it not} an ordinary double point, but rather takes the form
\beq
\label{eVBDP}
  x^2 + y^4 + z^4 + w^4 = 0~.
\eeq
This singularity is characterized by Milnor number $27$ which means
that there are 
$27$ homologically independent $S^3$'s,
 simultaneously vanishing at the singular
point. Thus, the singularity encountered is
quite unlike the case of ordinary double points.

 Using standard methods of
singularity theory \cite{rGab},
 one can show that the intersection
matix of these $S^3$'s is {\it non-trivial} and has rank 20.
Mathematically, it is straightforward to show that the transition from
$X$ to $Z$ through such a degeneration causes the Hodge numbers to make the
appropriate change.

Physically, in contrast to the previous cases, not only are $A$-type cycles
shrinking down, but some dual $B$-type cycles are shrinking down as well.
From this we see a phenomenon akin to that studied in 
\cite{rAD}: we appear to have
 electrically and magnetically charged states simultaneously becoming 
massless\footnote{In \cite{rWittenIII}
 it was independently noted that the phenomenon of 
 \cite{rAD} could be embedded in
string theory in such a manner.}. 
It is such degenerations that require more care in establishing
the existence of physical transitions. 
This also raises the interesting question
of whether the web of Calabi-Yau spaces
 requires such transitions for its connectivity,
or if by following suitable paths conifold transitions would suffice.

\vspace{.2in}
\noindent
{\it Example 2:}

\vspace{.15in}
We take $X$ to be the family of
quintic Calabi-Yau hypersurfaces
in ${\Bbb C}P^4$ and we take $Y$ to
 be the family of Calabi-Yau hypersurfaces of degree 8 in 
 $W{\Bbb C}P^4(1,1,1,1,4)$.
 As in the previous example, the transition from $Y$ to $Z$
just involves ordinary double points, so the discussion of 
\cite{rGMS} suffices.
However, in passing from $X$ to $Z$ we encounter another type of singularity,
known as a triple point. Namely, the generic Calabi-Yau in the subfamily of $X$
obtained by specialization of
 the complex structure contains a single singular point
whose local description is
\beq
\label{eTriple}
  x^3 + y^3 + z^3 + w^3 = 0~.
\eeq
The Milnor number for this singularity is equal to 16, and thus in
this example we have 16 vanishing three-cycles (homological to $S^3$'s)
simultaneously shrinking to one point. The intersection  matrix in this case
has rank 10 and we thus again are dealing
 with a physical situation with massless
electrically and magnetically charged particles.

\vspace{.3in}

\subsection{Remarks}

For ease, in our discussion above, we have focused on hypersurfaces in
weighted projective four-dimensional space
 (which naturally led to hypersurfaces in
more general toric varieties). We can carry out the same program on
codimension $d$ Calabi-Yau spaces. For these
it is best to use the full reflexive Gorenstein cone associated with
the Calabi-Yau, but basically the idea is the same. For instance, the
union of the Gorenstein toric fan (in the N lattice) for 
$W{\Bbb C}P^5(3,3,2,2,1,1)$ $(5,7)$ and $W{\Bbb C}P^5(3,3,2,2,2,1)$ $(5,8)$ is 
Gorenstein with index 2. Hence, these codimension two Calabi-Yau
 spaces are linked
through such transitions.

In this manner links have been established between numerous
Calabi-Yau spaces of
codimension two and between Calabi-Yau spaces of codimension three.
Furthermore, as in each of these classes it is not hard  to construct 
Calabi-Yau spaces
with simple toric representations of various codimension
(toric representations, of course, are not unique), we can link together
the webs of different codimension as well.
 For instance, the quintic hypersurface,
which is a member of the 7555 hypersurface web, is also linked to
the web of complete intersections in products
 of ordinary projective spaces. Hence,
all such Calabi-Yau spaces are so linked.

We therefore do not know the full answer to the two questions that
motivated the discussion of this chapter, but some insight has been gained
into each.

\subsection{Summary}

\begin{figure}[htbp]
\epsfysize=10cm
\centerline{\epsfbox{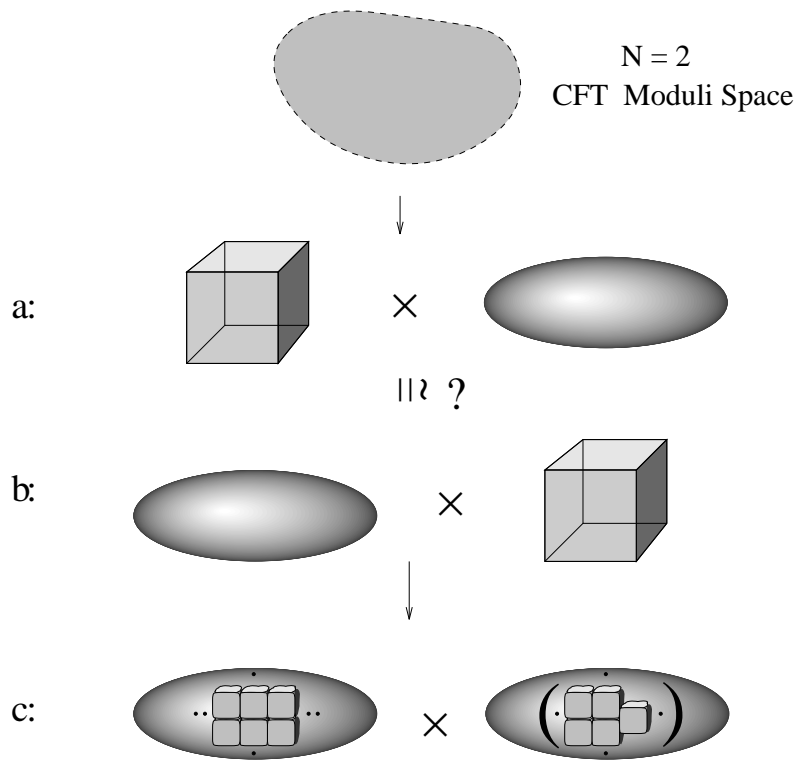}}
\caption{Models of the moduli space}
\label{fig:models}
\end{figure}

We can summarize the major developments described in these lectures
by the following two figures. In part (a) figure \ref{fig:models},
we see the abstract form of a connected component of the moduli space
of an $N=2$, $c=9$ conformal field theory, and its geometrical interpretation
in terms of the complexified K\"ahler cone and the complex structure
moduli space  of the associated Calabi-Yau manifold. This is the  picture
which was accepted for some time. 
When mirror symmetry was  found, a new
piece of this picture became apparent --- given in part (b) of
figure \ref{fig:models} --- in which
the abstract moduli space is also interpretable in terms of the complex
structure and complexified K\"ahler structure of the mirror to the original
Calabi-Yau. This, as we have discussed raised a puzzle since figures
(a) and (b) in figure \ref{fig:models},
which are supposed to be the geometrical incarnation of one and the same
conformal field theory moduli space, are not isomorphic. The resolution, as
we have discussed, is to enlarge the complexified K\"ahler moduli space
in both figures (a) and (b) so that each now takes the form given
in part (c) of figure \ref{fig:models}.
 That is, the left hand side of part (a) is augmented to
the left hand side of part (c),
which is now isomorphic to the left hand side
of part (b). Similarly for the right hand sides.
 We put parantheses  around one  part of the figure to
note that in either geometrical interpretation ( (a) or (b) ) the phase
structure has a natural physical interpretation on one side
of the story --- the K\"ahler sector.
The physical interpretation of these phases, as we discussed at the
end of section \ref{InterrelationsBetweenSCFT}\ is deepened by the work of
\cite{AGMsd} in which evidence is presented that each
phase which appears to lack a direct geometric interpretation actually
does have one so long as we analytically continue from an 
appropriate large radius Calabi-Yau region.

\begin{figure}[htbp]
\epsfysize=10cm
\centerline{\epsfbox{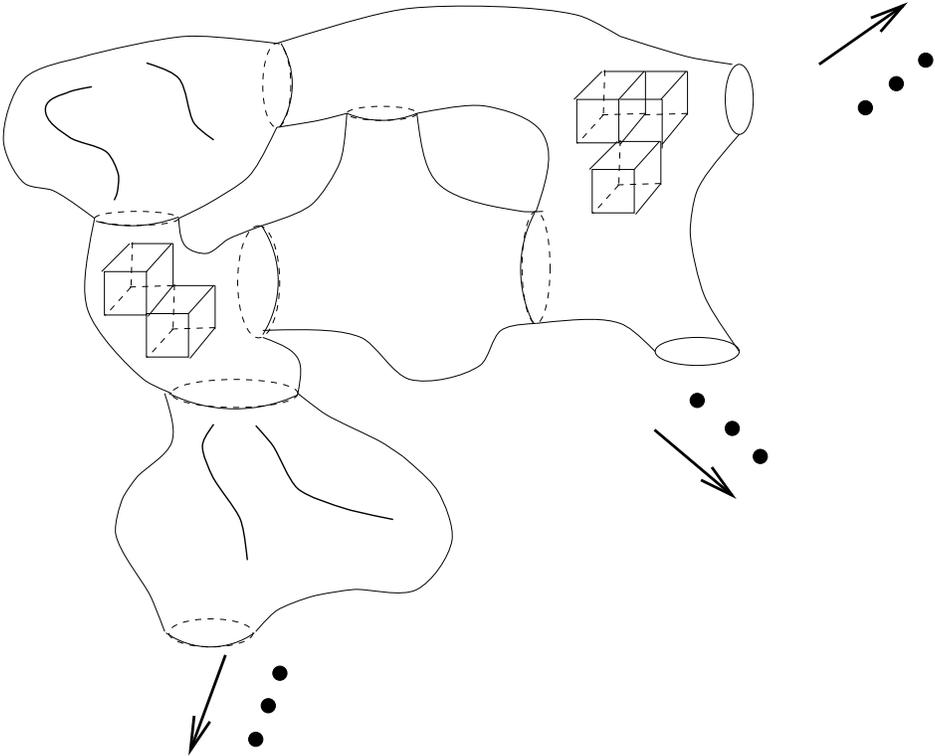}}
\caption{The structure of the moduli space as has emerged from these lectures.}
\label{fig:overview}
\end{figure}

The non-perturbative results of this section build on this picture even further.
In particular, we now see that the moduli space of non-perturbative
string theory has a geometric interpretation that is far more rich
than one would expect from perturbative considerations. It appears
that all of these distinct connected components of conformal field theory
moduli space join together into a single component of type II string
theory moduli space. This is heuristically sketched in  
figure
\ref{fig:overview}, which is the picture that has finally emerged.

\section{Conclusions}

In these lectures, we have sought to give the reader some understanding of the 
emerging field of quantum geometry. The basic philosophy we have followed is to
allow the {\it physics} of $N = 2$ space-time supersymmetric string theory to 
be our guide towards the
correct geometrical framework for describing string theory. We have seen that
this analysis has naturally led us to some unexpected consequences. Foremost
amongst these is the realization that one underlying string model may, in fact,
have two distinct non-linear sigma model realizations --- that is, with two
distinct target spaces. When the explicit isomorphism between these two
realizations involves flipping the sign of one
of the $U(1)$ charges in the $N = 2$
superconformal algebra, we call the two target spaces mirror manifolds. From the
viewpoint of classical geometry, these two manifolds are distinct objects; for
instance, they are topologically distinct. From the viewpoint of quantum
geometry, as we have discussed, they are identical. This is a prime example of
how classical and quantum geometry differ.

This distinction between classical and quantum geometry is
dramatically augmented by the realization that topology change ---
an operation which by fundamental definition in classical geometry is
discontinuous --- can be perfectly continuous and smooth in the
quantum geometry of string theory. We have seen this in the context
of mild topology changing flop transitions which occur even at the
level of classical perturbative string theory, and, strikingly, through
the drastic topology changing conifold transitions. These, as we have
seen, require non-perturbative string effects in an essential way.

We began these lectures by emphasizing that string duality does not
respect the decomposition of perturbative/non-perturbative effects ---
in fact, that is the source of its tremendous power. In the case at hand,
for instance, string duality allows certain
 conifold transitions in type II string
theory to be mapped to heterotic transitions on $K3 \times T^2$
as shown by \cite{KachruVafa} and as reviewed by Aspinwall in this volume.
Remarkably, in the heterotic language, the transitions are perturbative 
in nature.

Clearly, string theory is really forcing us to broaden our
understanding of the way in which geometrical data determines observable
 physics.
One can't help feeling that we are only catching glimpses of
the proverbial iceberg's tip --- understanding the  mathematics
and physics underlying string duality is certain to expose it further and
deeply affect our conceptions of this remarkable unified theory.


\section*{Acknowledgments}

I would 
like to thank P. Aspinwall, T. Chiang, J. Distler, M. Gross, Y.
Kanter,
D. Morrison and R. Plesser for
collaborations which yielded in some of the results described here.

I also thank A. Greenspoon and C. Lazariou who proof-read the lectures
during their preparation and caught many typos. I would also
like to thank Costas Efthimiou for his tireless effort in 
assisting me with the preparation of these notes.

This work has been
supported by a National Young Investigator award,
by the Alfred P. Sloan foundation and by the National Science Foundation.

\end{document}